\documentclass[12pt]{article}

\usepackage{stmaryrd}
\usepackage{cite} 
\usepackage{array} 
\usepackage{amssymb}
\usepackage{graphics,graphpap}
\usepackage{graphicx}
\usepackage{color}
\usepackage{graphicx}
\usepackage{dcolumn}
\usepackage{epsfig}
\usepackage{epstopdf}
\usepackage{bbm}
\usepackage{amsmath}
\usepackage{amsfonts}
\usepackage{textcomp}
\usepackage{setspace}
\usepackage{slashed}

\newcommand{\beq}{\begin{eqnarray}}
\newcommand{\eeq}{\end{eqnarray}}

\newcommand{\bmp}{\noindent\begin{minipage}{16cm}}
\newcommand{\emp}{\end{minipage}\vskip 7mm} 

\def\drawbox#1#2{\hrule height#2pt
        \hbox{\vrule width#2pt height#1pt \kern#1pt
              \vrule width#2pt}
              \hrule height#2pt}

\def\Asym#1#2{\vcenter{\vbox{\drawbox{#1}{#2}
              \kern-#2pt 
              \drawbox{#1}{#2}}}}

\def\simge{\mathrel{%
   \rlap{\raise 0.511ex \hbox{$>$}}{\lower 0.511ex \hbox{$\sim$}}}}

\def\simle{\mathrel{
   \rlap{\raise 0.511ex \hbox{$<$}}{\lower 0.511ex \hbox{$\sim$}}}}

\def\s#1{\setbox0=\hbox{$#1$}%
\rlap{\ifdim\wd0>.7em\kern.22\wd0\else\kern.1\wd0\fi /}#1}

\newcommand{\Slash}[1]{\slashed{#1}}

\begin{document}

\begin{titlepage}
\title{\vspace*{-2.0cm}
\bf\Large
The Power of Neutrino Mass Sum Rules for Neutrinoless Double Beta Decay Experiments\\[5mm]\ }

\author{
Stephen F.~King\thanks{email: \tt S.F.King@soton.ac.uk},~~~Alexander Merle\thanks{email: \tt A.Merle@soton.ac.uk},~~~and~~Alexander J.~Stuart\thanks{email: \tt A.Stuart@soton.ac.uk}
\\ \\
{\normalsize \it Physics and Astronomy, University of Southampton,}\\
{\normalsize \it Southampton, SO17 1BJ, United Kingdom}\\
}
\date{\today}
\maketitle
\thispagestyle{empty}

\begin{abstract}
\noindent
Neutrino mass sum rules relate the three neutrino masses within generic classes of flavour models, leading to restrictions on the effective mass parameter measured in experiments on neutrinoless double beta decay as a function of the lightest neutrino mass. We perform a comprehensive study of the implications of such neutrino mass sum rules, which provide a link between model building, phenomenology, and experiments.  After a careful explanation of how to derive predictions from sum rules, we discuss a large number of examples both numerically, using all three global fits available for the neutrino oscillation data, and analytically wherever possible. In some cases, our results disagree with some of those in the literature for reasons that we explain. Finally we discuss the experimental prospects for many current and near-future experiments, with a particular focus on the uncertainties induced by the unknown nuclear physics involved. We find that, in many cases, the power of the neutrino mass sum rules is so strong as to allow certain classes of models to be tested by the next generation of neutrinoless double beta decay experiments. Our study can serve as both a guideline and a theoretical motivation for future experimental studies.
\end{abstract}

\end{titlepage}

\section{\label{sec:intro}Introduction}

Since the first experimental evidence for neutrino oscillations by the Super-Kamiokande experiment~\cite{Fukuda:1998mi}, we have come a long way to measure all leptonic mixing angles. This enterprise was completed by the discovery of a non-zero reactor mixing angle $\theta_{13}$ in 2013 by the Daya Bay~\cite{An:2012eh} and RENO~\cite{Ahn:2012nd} experiments. In particular, we have learned that the leptonic mixing angles are quite large compared to their analogues in the quark sector~\cite{Beringer:1900zz}, which still appears to be very puzzling from a theoretical point of view.

At the moment, probably the best guess we have to explain such mixing patterns is by so-called \emph{flavour symmetries} (see Refs.~\cite{King:2013eh,Morisi:2012fg,Grimus:2011fk,Altarelli:2010gt} for recent reviews), although other origins could be possible as well (see, e.g., Refs.~\cite{Haba:2000be,Adulpravitchai:2009re}). While some flavour models predict a range of neutrino masses and mixings, others are strong enough to predict correlations between several observables, such as leptonic mixing angles and neutrino masses. These are the key outputs for a model to be testable. A typical form of such correlations are so-called \emph{sum rules}, which can appear for neutrino mixing parameters~\cite{King:2005bj,Masina:2005hf,Antusch:2005kw,Antusch:2007rk} or for neutrino masses, the latter being the case to be investigated in this paper.

One of the first occasions where the term \emph{neutrino mass sum rule} has been used in the meaning we refer to was in Ref.~\cite{Altarelli:2009kr}, and the importance of such relations has been stressed as well in, e.g., Refs.~\cite{Chen:2009um,BarryRodejohann-Classification}. In particular, it has been recognized that sum rules could considerably constrain the so-called effective neutrino mass $|m_{ee}|$ as measured in neutrinoless double beta decay experiments. First studies of the implications of models leading to such correlations have been provided in, e.g., Refs.~\cite{Altarelli:2008bg,Hirsch:2008rp}, and the first systematic study of a few cases has been done in Ref.~\cite{Bazzocchi:2009da}. However, these references have in fact not mentioned the term sum rule. Nevertheless, in particular Ref.~\cite{Bazzocchi:2009da} has provided analyses of some of the cases studied here, too, so that we will refer to it at some places. To our knowledge, the first systematic study of neutrino mass sum rules as such has been provided by Ref.~\cite{Barry:2010zk}, which in particular introduced the illustrative geometrical interpretation of sum rules as triangle equations. A further study followed~\cite{Dorame:2011eb}, which discussed the general types of sum rules which had appeared in the literature by then.

\begin{figure}[t]
\centering
\includegraphics[width=9cm]{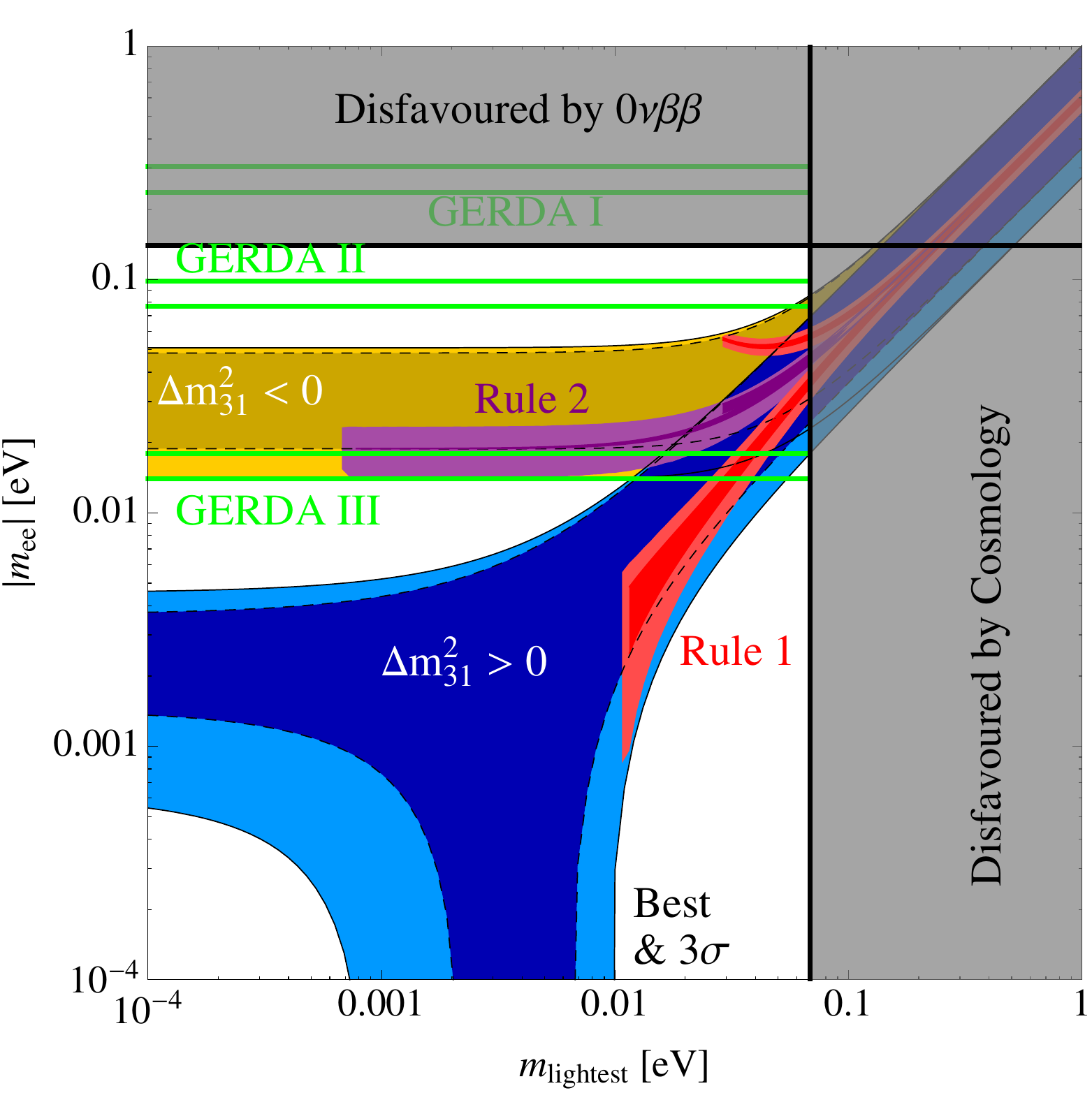}
\caption{\label{fig:power}The power of sum rules. Two example sum rules (rule~1: $\tilde m_1^{-1} + \tilde m_2^{-1} = \tilde m_3^{-1}$ indicated by the forked red region, rule~2: $\tilde m_1 + \tilde m_2 = \tilde m_3$ indicated by the violet region) are displayed, along with the result of GERDA phase~I~\cite{Agostini:2013mzu} and the maximum sensitivities of GERDA~\cite{Abt:2004yk,JanicskoCsathy:2009zz} for phase~II and phase~III, \emph{including} the nuclear physics uncertainties which generate the gaps between the horizontal green lines. Even with these uncertainties, the inverted ordering region of rule~1 is clearly falsifiable, thereby illustrating the power of the sum rules. Technical details will be given later in the text.}
\end{figure}

In this paper, we aim at extending the previous studies in the light of the newest global fit values of the neutrino oscillation parameters~\cite{Tortola:2012te,Fogli:2012ua,GonzalezGarcia:2012sz}, with a particular focus on the prospects of near future experiments on neutrinoless double beta decay. We take into account the uncertainties imposed by the nuclear physics involved. Our study may help to advance the state of the field for several reasons. First of all, in particular the handling of phases in sum rules can be a bit subtle, which has led to several incorrect results in the literature, e.g., wrong predictions of the allowed regions of $|m_{ee}|$. This could be disastrous, since it could potentially lead to wrong conclusions if a non-zero rate of neutrinoless double beta decay was observed. Thus it is worth to carefully discuss this point and to correct some of the results obtained previously. Secondly, it is known that a relatively large value of $\theta_{13}$, such as the one measured, can considerably influence the allowed regions for $|m_{ee}|$~\cite{Lindner:2005kr}, which is particularly true when additional constraints such as mass sum rules are imposed, and which makes an updated study worthwhile. Thirdly, the studies performed up to now have focused on the phenomenology of $|m_{ee}|$, without a complete discussion of the experimental prospects, in particular in what regards the nuclear physics uncertainties. We close all these gaps by not only providing a detailed study of all neutrino mass sum rules we were able to find in the literature, but we also discuss the prospects of many current and future experiments on neutrinoless double beta decay, thereby taking into account nine different methods to calculate the so-called nuclear matrix elements. We also provide a complete classification of all flavour models known to us which lead to neutrino mass sum rules, so that a fairly complete picture of all combinations of symmetries and neutrino mass generation mechanisms is obtained.

There is no study available which is comparatively complete as the one presented here, and in fact our results could be used by all three communities, model builders, phenomenologists, and experimentalists. The text will be useful if the reader would like to, e.g., know the detailed predictions of the sum rule obtained in their model or study which models could be distinguished by their experiment. To give all potential readers a flavour of the power of neutrino mass sum rules, we have depicted an example result in Fig.~\ref{fig:power}: as can be seen, the two example sum rules predict regions for the effective mass $|m_{ee}|$ which are so distinct that they can potentially be distinguished by on-going experiments such as GERDA~\cite{Abt:2004yk,JanicskoCsathy:2009zz}. The remarkable point is that this statement remains true even if the nuclear physics uncertainties (indicated by the splittings between the green horizontal double lines) are taken into account. More details, and the corresponding numbers for this and other experiments, will be given later in the text.

The paper is structured as follows. In Sec.~\ref{sec:sumrules}, we give an illustrative discussion of the emergence of sum rules in neutrino flavour models. We then discuss the most general sum rules possible in Sec.~\ref{sec:terms}, before illustrating how to carefully derive the constraints imposed by them on the effective neutrino mass in Sec.~\ref{sec:effective}. Next, we discuss the two cases of trivial and non-trivial sum rules in Secs.~\ref{sec:trivial} and~\ref{sec:non-trivial}, respectively, before investigating all sum rules we have found in the literature in Sec.~\ref{sec:concrete}, where we also provide a systematic classification of all relevant flavour models known to us. We discuss in detail the experimental prospects for neutrinoless double beta decay as well as the impact of the nuclear physics uncertainties in Sec.~\ref{sec:experiments}, and we finally summarise the numerical predictions obtained for all sum rules in Sec.~\ref{sec:predictions}. Our conclusions are presented in Sec.~\ref{sec:conc}.

\section{\label{sec:sumrules}Neutrino mass sum rules in family symmetry models}

We first consider the emergence of neutrino mass sum rules from family symmetry models in which neutrino masses arise from the Weinberg operator~\cite{Weinberg:1979sa}. Without assuming right-handed neutrinos, small neutrino masses can be generated by the dimension five operator which breaks both the total and the individual lepton numbers,
\begin{equation}
\mathcal{L}^{\rm Weinberg }_{LL} = y_{ij} \frac{\overline{\ell_{i}^{c}} \ell_{j} HH}{\Lambda_{L}} \; ,
\end{equation}
where
$\ell_{i}$ represents the three lepton doublets  ($i=1,2,3$), $H$ represents the Standard Model (SM) Higgs doublet, $y_{ij}$ are dimensionless couplings, and $\Lambda_{L}$ is the cut-off scale for the lepton number violation operator. After the Higgs fields develop their vacuum expectation values (VEVs), these operators lead to the physical neutrino mass matrix $M_{\nu}$. In general, $M_{\nu}$ is a complex symmetric matrix which depends on 6 independent complex parameters $(a,b,c,d,e,f)$,
\begin{equation}\label{eq:csd-tbm3}
M_{\nu} =
\left(\begin{array}{ccc}
 a & \ \ \  b \ \ \  & c \\
. & d & e \\
. & . & f
\end{array}\right).
\end{equation}
When $M_{\nu}$ is diagonalised it has three eigenvalues, the complex neutrino masses $\tilde{m}_i$, which in general are unrelated.

However, in certain models, $M_{\nu}$ can be written in terms of two complex parameters, leading to a model dependent relation between the $\tilde{m}_i$. For example, in the notation of Ref.~\cite{Chen:2009um}, if the three lepton doublets transform as a triplet of a family symmetry $A_{4}$, and the three right-handed charged leptons are assigned to be singlets under $A_{4}$,
\begin{equation}
L =
\left(\begin{array}{c} \ell_{1} \\ \ell_{2} \\ \ell_{3}
\end{array}\right) \sim \mathbf{3}  \;, \ \ e_R \sim \mathbf{1}  \;, \ \ \mu_R \sim \mathbf{1''}  \;, \ \ \tau_R \sim \mathbf{1'}  \;.
\end{equation}
The Lagrangian that gives rise to neutrino masses in that model is
\begin{equation} \mathcal{L}_{LL} = \frac{\overline{L^{c}}LHH}{\Lambda_{L}}
\left(\frac{\left< \phi_{S} \right>}{\Lambda} + \frac{\left< u
\right>}{\Lambda}\right) \; ,
\end{equation}
where $\Lambda$ ($\neq \Lambda_{L}$ in general) is the cutoff scale of the $A_{4}$ symmetry. The triplet flavon field, $\phi_{S} \sim \mathbf{3}$, and the singlet flavon field, $u \sim \mathbf{1}$, acquire the following complex VEVs,
\begin{equation}
\frac{\left< \phi_{S} \right>}{\Lambda} =
\left(\begin{array}{c} 1 \\ 1 \\ 1 \end{array}\right) \alpha_{s}  \; ,
\quad \frac{\left< u \right>}{\Lambda} = \alpha_{0}  \; .
\end{equation}
The VEV $\left<\phi_{S}\right>$ breaks the $A_{4}$ symmetry down to $G_{S}$, which is the subgroup of $A_{4}$ generated by the group element $S$. Upon electroweak symmetry breaking, the following effective neutrino mass matrix is generated,
\begin{equation}\label{mr}
M_{\nu}= \left( \begin{array}{ccc} 2\alpha_{s} + \alpha_{0} & -\alpha_{s} & -\alpha_{s} \\ -\alpha_{s} & 2\alpha_{s} &
-\alpha_{s} + \alpha_{0} \\ -\alpha_{s} & -\alpha_{s} + \alpha_{0} & 2\alpha_{s} \end{array}\right)
\frac{v^{2}}{\Lambda_{L}}\; ,
\end{equation}
where $v$ is the SM Higgs VEV. This mass matrix is form-diagonalizable
\cite{Chen:2009um}, i.e., it is always diagonalized, independent of the values for the parameters $\alpha_{s}$ and $\alpha_{0}$, by the tri-bimaximal mixing matrix $U_{\mbox{\scriptsize TB}}$~\cite{Harrison:2002er},
\begin{equation}
U_{\mbox{\scriptsize TB}}^{T} M_{\nu} U_{\mbox{\scriptsize TB}}
= \mbox{diag}(3\alpha_{s}+\alpha_{0}, \; \alpha_{0}, \; 3\alpha_{s}-\alpha_{0}) \cdot \frac{v^{2}}{\Lambda_{L}}
= {\rm diag} (\tilde m_1, \tilde m_2, \tilde m_3) \; .
\end{equation}
Because the three mass eigenvalues $\tilde{m}_{1,2,3}$ are determined by two parameters, up to an overall scale, there is a sum rule among the three light masses,
\begin{equation}
\tilde{m}_{1} - \tilde{m}_{3} = 2 \tilde{m}_{2} \; .
\end{equation}
More generally, in other models of this kind, the complex neutrino masses $\tilde{m}_i$ may be related by a sum rule of the form,
\beq
\alpha \tilde{m}_1+\beta  \tilde{m}_2 = \tilde{m}_3, \label{masssum}
\eeq
where $\alpha, \beta $ are model dependent complex constants. 

We now consider the emergence of sum rules from family symmetry models based on the type~I seesaw mechanism~\cite{Minkowski:1977sc,Ramond:1979py,Yanagida:1979as,GellMann:1980vs,Glashow:1979nm,Mohapatra:1979ia}, for an example $A_4$ model of tri-bimaximal mixing. In such a model the three right-handed neutrinos may transform as a triplet of $A_{4}$,
\begin{equation}
N = \left(\begin{array}{c}
N_{1} \\ N_{2} \\ N_{3}
\end{array}\right) \sim \mathbf{3} \; ,
\end{equation}
and the right-handed neutrino Majorana mass matrix is generated by 
$\overline{N^{c}} N ( \left< \phi_{S} \right> + \left< u \right>)$,
\begin{equation}\label{eq:mr-1}
M_R
= \left( \begin{array}{ccc}
2\alpha_{s} + \alpha_{0} & -\alpha_{s} & -\alpha_{s} \\
-\alpha_{s} & 2\alpha_{s} & -\alpha_{s} + \alpha_{0} \\
-\alpha_{s} & -\alpha_{s} + \alpha_{0} & 2\alpha_{s}
\end{array}\right) \Lambda \; .
\end{equation}
The Dirac neutrino mass matrix is generated by the interaction $y H \overline{L} N$,
\begin{equation}\label{eq:md-1}
M_{D} = \left(\begin{array}{ccc}
1 & 0 & 0
\\
0 & 0 & 1
\\
0 & 1 & 0
\end{array}\right) yv \; .
\end{equation}
After the seesaw mechanism takes place, the resulting effective neutrino mass matrix is
\begin{equation}
M_{\nu}^{\rm eff} =  - M_{D} M_R^{-1} M_{D}^{T}.
\end{equation}
This effective neutrino mass matrix is diagonalized by $U_{\rm TB}$ with the complex mass eigenvalues being
\begin{equation}
\label{m}
\mbox{diag}(\tilde{m}_{1}, \tilde{m}_{2}, \tilde{m}_{3}) = \left(
\frac{1}{3\alpha_{s}+\alpha_{0}}, \; \frac{1}{\alpha_{0}}, \; \frac{1}{3\alpha_{s}-\alpha_{0}}\right) \frac{y^{2} v^{2}}{\Lambda} \; .
\end{equation}
Because the three mass eigenvalues $\tilde{m}_{1,2,3}$ are determined by two parameters, up to an overall scale, there is a sum rule among the three light masses. However, in this seesaw realization, the sum rule clearly involves the inverse of the three light neutrino masses and is given from Eq.~\eqref{m} by,
\begin{equation}
\frac{1}{\tilde{m}_{1}} - \frac{1}{\tilde{m}_{3}} = \frac{2}{\tilde{m}_{2}} \; ,
\label{eq:inverse_example}
\end{equation}
which can lead to both normal and inverted mass orderings. More generally, in seesaw models of this kind, the right-handed neutrino masses may be similarly related as in Eq.~\eqref{masssum}, leading to inverse relationships between light physical neutrino masses of the form,
\beq
\frac{\gamma}{\tilde{m}_1}+\frac{\delta}{\tilde{m}_2} = \frac{1}{\tilde{m}_3},  \label{inversemasssum}
\eeq
where $\gamma, \delta $ are model dependent constants.\footnote{Note that $\delta$ here has got nothing to do with the CP violating phase denoted by the same Greek letter.}

The examples of neutrino mass sum rules discussed above are for $A_4$ family symmetry models which give rise to tri-bimaximal neutrino mixing. However, this prediction is phenomenologically problematic, as a vanishing reactor mixing angle $\theta_{13}$ is by now known to be excluded by data~\cite{Tortola:2012te,Fogli:2012ua,GonzalezGarcia:2012sz}. It is possible to maintain the neutrino mass sum rules while allowing for a non-zero reactor angle by invoking charged lepton mixing angle corrections. Technically this corresponds to a violation of the symmetry corresponding to the $T$ generator which is preserved in the charged lepton sector. This enforces the diagonality of the charged lepton mass matrix in the basis discussed above. This strategy (introducing reactor mixing via charged lepton corrections) may be applied to any family symmetry model of tri-bimaximal mixing, and will always maintain the neutrino mass sum rule, which justifies the treatment of such models in this paper. In addition it will lead to mixing angle sum rules, which however are not our principal concern here. 

As an example of such a model where the sum rule is maintained in the presence of charged lepton corrections, we briefly discuss a $\Delta (96)$ model. In the ``Grand'' $\Delta (96)$ Model~\cite{King:2012in}, a $\Delta(96)$ flavour symmetry is first applied to generate ``bi-trimaximal'' (BT) mixing in the neutrino sector, i.e.,
\begin{eqnarray}
\begin{array}{ccc}
\theta_{12}^{\nu}\approx36.2^{\circ}, &\theta_{13}^{\nu}\approx 12.2^{\circ}, &\theta_{23}^{\nu}\approx 36.2^{\circ}.
\end{array}
\label{eq:BT}
\end{eqnarray}
Notice that these predictions fall outside the $1\sigma$ ranges for the lepton mixing angles of the PMNS mixing matrix, assuming a diagonal charged lepton mixing matrix~\cite{Tortola:2012te,Fogli:2012ua,GonzalezGarcia:2012sz}. Thus, this motivates going beyond a simple $\Delta(96)$ model of leptons to a Grand Unified Theory of flavour where the charged lepton mass matrix $M_e$ is only approximately diagonal, leading to a slightly non-diagonal charged lepton mixing matrix $U_e$. Then, the non-diagonal $U_e$ leads to small corrections in the BT predictions of Eq.~\eqref{eq:BT}, yielding
\begin{eqnarray}
\begin{array}{ccc}
\theta_{12}\approx 32.7^{\circ}, &\theta_{13}\approx 9.6^{\circ}, &\theta_{23}\approx 36.9^{\circ}.
\end{array}
\end{eqnarray}
Even though the above angles have received a correction from the charged lepton sector, this correction is independent of the mixing originating from the neutrino sector, i.e., the neutrino mass matrix remains unchanged. Hence the sum rule will be unchanged by the charged lepton corrections, and it remains as
\begin{equation}
\frac{1}{\tilde m_3}\pm \frac{2i}{\tilde m_2}=\frac{1}{\tilde m_1},
\end{equation}
a subset of the sum rules given in Eq.~\eqref{inversemasssum}. Note that this inverse neutrino mass sum rule arises because of the application of the type~I seesaw mechanism to a trivial Dirac mass matrix and a non-trivial heavy neutrino mass matrix, i.e., flavons couple only to $\overline{N^{c}} N$.
 
Now that we have seen one example where corrections do not affect the sum rule, it is insightful to consider a model in which the higher order corrections to the mass matrices themselves also leave the sum rule unchanged. As an example of this we discuss a Golden Ratio $A_5$ model in the presence of a minimal next-to-leading-order (NLO) correction~\cite{Cooper:2012bd}. In the ``Golden Model'' at leading order (LO), Golden Ratio (GR) mixing is predicted, i.e.,
\begin{eqnarray}
\begin{array}{ccc}
\label{eq:GRmixing}
\theta_{12}= \tan^{-1} \left( \frac{1}{\phi_g} \right)=31.7^{\circ}, &\theta_{13}= 0^{\circ}, &\theta_{23}= 45^{\circ},
\end{array}
\end{eqnarray}
where $\phi_g=(1+\sqrt{5})/2$ is the Golden Ratio. The model also predicts leading order complex light neutrino masses given by
\begin{eqnarray}
\label{redmass}
\begin{array}{ccc}
\tilde m_1^{\rm LO}=\frac{ \beta }{6\phi_g-2+4 e^{i \delta } \xi},& \ \tilde m_2^{\rm LO}=\frac{ \beta }{4 e^{i \delta } \xi-(\frac{6}{\phi_g} +2) }, & \ \tilde m_3^{\rm LO}=\frac{ \beta }{2 \left(1+4 e^{i \delta } \xi \right)},
\end{array}
\end{eqnarray}
where $\delta,\beta,\xi$ are parameters involving the Higgs VEV, complex flavon VEVs, and coupling constants, cf.\ Ref.~\cite{Cooper:2012bd} for exact definitions. Observe that these complex masses obey the inverse neutrino mass sum rule
\begin{equation}
\label{sumruleLO}
\frac{1}{\tilde m_1^{\rm LO}}+\frac{1}{\tilde m_2^{\rm LO}}=\frac{1}{\tilde m_3^{\rm LO}}.
\end{equation}
Unfortunately, from Eq.~\eqref{eq:GRmixing} it can be seen that GR mixing is also excluded by the measurement of a large reactor mixing angle~\cite{An:2012eh,Ahn:2012nd}. Therefore, it is worthwhile to consider the effect of NLO corrections to the Golden Model.

In this model, the NLO corrections manifest themselves as a correction to the heavy neutrino mass matrix, $M_R=M_R^{\rm LO}+\Delta M_R$. The importance of this minimal correction $\Delta M_R$ is that it preserves the sum rule for the heavy neutrinos, i.e., $M_1^{\rm NLO}+M_2^{\rm NLO}=M_3^{\rm NLO}$. This preserved sum rule will then translate to the light neutrino masses (after application of the type~I seesaw mechanism). This can be seen by inspection of the Golden Model's light neutrino masses to NLO. They are given by
\begin{eqnarray}
 \tilde m_1^{\rm NLO} &\approx& \tilde m_1^{\rm LO}-\frac{9 \beta  }{2\phi_g\sqrt{30}   \left(1-3 \phi_g -2 \xi e^{i \delta }  \right)^2}\epsilon, \nonumber\\
 \tilde m_2^{\rm NLO} &\approx& \tilde m_2^{\rm LO}-\frac{9\beta \phi_g }{2\sqrt{30} \left(2-3 \phi_g +2 \xi e^{i \delta }  \right)^2}\epsilon,\nonumber\\
 \tilde m_3^{\rm NLO} &\approx& \tilde m_3^{\rm LO}-\frac{9\beta}{2\sqrt{6}(1+4\xi e^{i\delta})^2}\epsilon.
 \label{eq:analNLOmass}
\end{eqnarray}
From Eq.~\eqref{eq:analNLOmass} it is clearly seen that 
\begin{equation}
\label{sumruleNLO}
\frac{1}{\tilde m_1^{\rm NLO}}+\frac{1}{\tilde m_2^{\rm NLO}} \approx \frac{1}{\tilde m_3^{\rm NLO}}
\end{equation}
to first order in the small parameter $\epsilon$. Hence, the minimal correction to the heavy neutrino mass matrix allows for preservation of the light neutrino mass sum rule to NLO.

As seen in the examples on how sum rules arise, the following empirical observation can be made: the most general sum rule one can envisage involves powers of neutrino masses $p$, and has the schematic form (ignoring complex coefficients which multiply each term in the sum rule and which will be included later),
\begin{equation}
 \tilde m_1^p  + \tilde m_2^p  + \tilde m_3^p  = 0,
 \label{eq:gen_rule_0}
\end{equation}
The origin of the ``power of the sum rule'' $p$ is then as follows. Whenever the light neutrino mass matrix $M_\nu$ is proportional to the power $n$ of a certain (inverse) mass matrix $M$, where $M$ contains the two decisive flavon couplings and all other matrices in the product are trivial (i.e., either the unit matrix or proportional to a matrix without any free-parameters, up to an overall scale), then the power $p$ in the sum rule will be given by $1/n$,
\begin{equation}
 M_\nu \propto M^n \Rightarrow p = \frac{1}{n}.
 \label{eq:rule_power}
\end{equation}
In the type~I seesaw example discussed above, the decisive matrix is $M=M_R$, cf.\ Eq.~\eqref{eq:mr-1}, while the Dirac mass matrix $M_D$ from Eq.~\eqref{eq:md-1} is trivial. This suggests that $n = p = -1$, which is indeed realized in Eq.~\eqref{eq:inverse_example}. This intuitive observation has proven to be correct in all the examples we have found in the literature, which involve powers $p = \pm 1/2$ and $p = \pm 1$. However, it also justifies the study of other powers, such as $p = \pm 1/3$ or $p = \pm 1/4$. At least by these easy observations, higher powers $p > 1$ do not seem to be realistic. Nevertheless we will discuss one such example to see the effect of the higher power. Such exotic sum rules could still be justified in a certain framework, e.g., by more complicated models or simply by phenomenology.

Equipped with an intuitive picture of how sum rules can arise, we will now study the more technical aspects of sum rules in detail.

\section{\label{sec:terms}Generalised neutrino mass sum rules}

Let us now enter the technical details. Using the complex neutrino mass eigenvalues $\tilde m_i$, the most general sum rule possible, reinstating the complex coefficients ignored in Eq.~\eqref{eq:gen_rule_0}, is given by
\begin{equation}
 A_1 \tilde m_1^p e^{i\chi_1} + A_2 \tilde m_2^p e^{i \chi_2} + A_3 \tilde m_3^p e^{i \chi_3} = 0.
 \label{eq:gen_rule_1}
\end{equation}
In this equation, we have $p \neq 0$, $\chi_i \in [0, 2\pi)$, and hence $A_i > 0$, since any phase of $A_i$ could be absorbed into the phase $\chi_i$.\footnote{One could mathematically also have $A_i = 0$, but this would only lead to a \emph{partial sum rule} which does not involve all three masses.} The power $p$ is the central ingredient characterizing the sum rule, and it is always known in a given model. Note that the phase $\chi_i$ is \emph{not} a Majorana phase, but rather a phase coming from the sum rule itself (e.g., a minus sign). In other words, the phases $\chi_i$ are \emph{fixed}, and we always \emph{know} their value for a given sum rule (at least if the model from which the sum rule originates is powerful enough to predict them). For example, in the model from Ref.~\cite{Cooper:2012bd}, discussed above, the sum rule was given to next-to-leading order by, cf.\ Eq.~\eqref{sumruleNLO},
\begin{equation}
 \frac{1}{\tilde m_1} + \frac{1}{\tilde m_2} = \frac{1}{\tilde m_3}.
 \label{eq:Golden_Rule}
\end{equation}
Comparing this concrete sum rule to the general form given in Eq.~\eqref{eq:gen_rule_1} yields:
\begin{equation}
 p=-1,\ \ A_1 = A_2 = A_3 = 1,\ \ \chi_1 = \chi_2 = 0,\ \ {\rm and}\ \ \chi_3 = \pi.
 \label{eq:Golden_Rule_comp}
\end{equation}
Note that we can always do this comparison if we attempt to translate a sum rule from a concrete model into the general language presented here.

We can proceed by writing the complex masses in polar form, $\tilde m_i = m_i e^{i \phi_i}$, where $m_i \geq 0$ are the physical mass eigenvalues and $\phi_i \in [0, 2 \pi )$ are phases which we will later on prove to be identical to the Majorana phases. Dividing Eq.~\eqref{eq:gen_rule_1} by $A_1 > 0$, and abbreviating $B_i \equiv A_i/A_1$ as well as $\tilde \phi_i \equiv \chi_i + p \phi_i$, one can rewrite:
\begin{equation}
 m_1^p e^{i\tilde \phi_1} + B_2 m_2^p e^{i\tilde \phi_2} + B_3 m_3^p e^{i\tilde \phi_3} = 0,
 \label{eq:gen_rule_2}
\end{equation}
where $B_{2,3} > 0$. We will discuss Eq.~\eqref{eq:gen_rule_2} shortly, after defining our terminology.

In order to know what we are talking about, we suggest the following terminology for sum rules, which will be further motivated later on:
\begin{itemize}

\item \emph{trivial sum rules}: All coefficients have a modulus of 1, $B_2 = B_3 = 1$.\\
$\hookrightarrow$ Examples: $\tilde m_1 + \tilde m_2 - \tilde m_3 = 0$ or $\tilde m_1^{-1} + \tilde m_2^{-1} + i \tilde m_3^{-1} = 0$.

\item \emph{non-trivial sum rules}: There are coefficients with a modulus different from 1.\\
$\hookrightarrow$ Examples: $2 \tilde m_1 + 2 \tilde m_2 - \tilde m_3 = 0$ or $2 \tilde m_1^{-1} + 2 \tilde m_2^{-1} + i \tilde m_3^{-1} = 0$.

\end{itemize}

Equipped with these definitions, we first discuss the most subtle part of the game, namely the physical parametrisation of the effective mass.

\section{\label{sec:effective}Effective mass vs.\ sum rules: parametrisation issues to be understood}

Neutrinoless double beta decay ($0\nu\beta\beta$) is a lepton number violating process where a nucleus $(A,Z)$ decays into another one by the emission of two electrons:
\begin{equation}
 (A,Z) \to (A,Z+2) + 2 e^- .
 \label{eq:0nbb}
\end{equation}
The violation of lepton number is immediate, since the final state contains two leptons while the initial state contains none.\footnote{Note that, however, the relation to the Majorana nature of the neutrino might be more subtle~\cite{Schechter:1981bd,Duerr:2011zd}.} In the simplest case of light neutrino exchange, the amplitude for the process is proportional to a quantity called the \emph{effective mass} $m_{ee}$.

The ``problem'' with this quantity is that it can be parametrised in several ways, which may at times look confusing. To unambiguously clarify these points, we will here in some detail review how the effective mass is obtained, thereby pointing out some important subtleties. Although these issues are in principle known, we chose to give a detailed explanation in order to prevent any confusion.

We start with the Feynman diagram for the process, which looks like:
\begin{center}
\includegraphics[width=9cm]{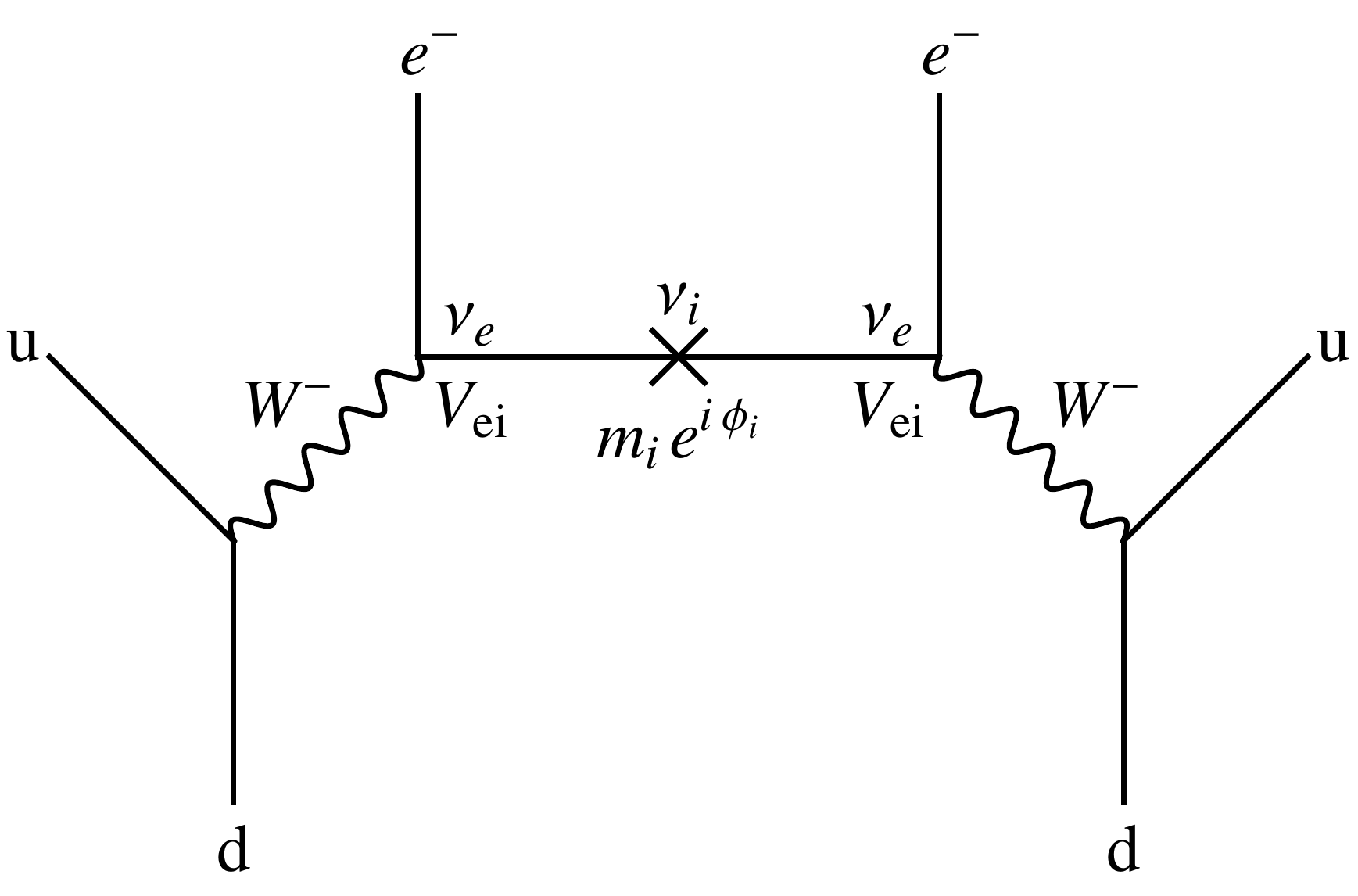}
\end{center}
Note that we have assumed the most simple version of the process, i.e., there are only left-handed SM-like $W$-bosons, and the exchange particle is a light active Majorana neutrino. Then the propagator of the fermion line is a Majorana propagator which contains a charge conjugation matrix $C$~\cite{Pal:2010ih}, which by the Majorana condition $\nu_i^c = C (\overline{\nu_i})^T = e^{i \phi_i} \nu_i$ translates into a Majorana phase $\phi_i$ for the mass eigenstate $\nu_i$. Note that we have already used the same notation $\phi_i$ for the Majorana phase as done in the sum rules, cf.\ Sec.~\ref{sec:terms}. However, we still have to show that this is actually correct.

If there are three active neutrino mass eigenstates $\nu_{1,2,3}$, one obtains the following proportionality in the amplitude:
\begin{equation}
 A_{ee} \propto \sum_{i=1}^3 P_L V_{ei} e^{i \phi_i} \frac{\Slash{p} + m_i}{p^2 - m_i^2} V_{ei} P_L,
 \label{eq:amp_1}
\end{equation}
where $V$ denotes the CKM-equivalent part of the PMNS-matrix ($V$ is the same as $U$ with all Majorana phases set to zero). Note that, due to the Majorana nature of the exchanged neutrino, the two vertices are \emph{indistinguishable}, i.e., the amplitude must be proportional to $V_{ei}^2$ instead of $|V_{ei}|^2$, the use of the latter sometimes being part of the confusion.

In order to arrive at the effective mass, two more steps are necessary. First, due to the two projection operators $P_L$ which originate from the SM-like $W$-bosons, one can rewrite $P_L (\Slash{p} + m_i) P_L = m_i P_L$. Second, since the average nuclear momentum transfer is much larger than the neutrino mass, $\sqrt{\langle p^2 \rangle} = \mathcal{O}(100~{\rm MeV}) \gg m_i$, one can neglect the term $m_i^2$ in the denominator. Hence, the proportionality in Eq.~\eqref{eq:amp_1} reduces to
\begin{equation}
 A_{ee} \propto \sum_{i=1}^3 V_{ei}^2 e^{i \phi_i} m_i \equiv m_{ee},
 \label{eq:amp_2}
\end{equation}
which serves as a definition of the effective mass $m_{ee}$. The final step is to realize that a detection of $0\nu\beta\beta$ could only constrain the absolute value $|m_{ee}|$, which means that the decay rate it can only depend on two phases. Multiplying Eq.~\eqref{eq:amp_2} by $e^{- i \phi_1}$ and defining $\alpha_{i1} \equiv \phi_i - \phi_1$ ($i=2,3$) then leads the final form of the effective mass
\begin{equation}
 |m_{ee}| = |m_1 V_{e1}^2 + m_2 V_{e2}^2 e^{i \alpha_{21}} + m_3 V_{e3}^2 e^{i \alpha_{31}}|.
 \label{eq:mee_1}
\end{equation}
Note that this expression is nearly independent of the parametrisation, except for the choice to remove the phase from the first term instead of choosing any of the other two.

We can now insert the PDG parametrisation, cf.\ Eq.~(13.79) of Ref.~\cite{Beringer:1900zz},
\begin{equation}
 U_{\rm PMNS}^{\rm PDG} =
 \underbrace{
 \begin{pmatrix}
 c_{12} c_{13} & & & & s_{12} c_{13} & & & & s_{13} e^{-i\delta} \\
 - s_{12} c_{23} - c_{12} s_{23} s_{13} e^{i\delta} & & & & c_{12} c_{23} - s_{12} s_{23} s_{13} e^{i\delta} & & & & s_{23} c_{13}\\
 s_{12} s_{23} - c_{12} c_{23} s_{13} e^{i\delta} & & & & -c_{12} s_{23} - s_{12} c_{23} s_{13} e^{i\delta} & & & & c_{23} c_{13}
 \end{pmatrix}
 }_{\equiv V_{\rm PMNS}^{\rm PDG}}
 \begin{pmatrix}
 1 & & & & 0 & & & & 0 \\
 0 & & & & e^{i\alpha_{21}/2} & & & & 0\\
 0 & & & & 0 & & & & e^{i\alpha_{31}/2}
 \end{pmatrix},
 \label{eq:PDG_PMNS}
\end{equation}
into Eq.~\eqref{eq:mee_1} and thereby exactly reproduce the PDG parametrisation of the effective mass, cf.\ Eq.~(13.84) in Ref.~\cite{Beringer:1900zz},
\begin{equation}
 |m_{ee}|_{\rm PDG} = |m_1 c_{12}^2 c_{13}^2 + m_2 s_{12}^2 c_{13}^2 e^{i \alpha_{21}} + m_3 s_{13}^2 e^{i (\alpha_{31} - 2\delta)}|.
 \label{eq:mee_2}
\end{equation}
Now it suddenly appears as if also the Dirac CP phase $\delta$ showed up in the effective mass. This dependence came in through the PMNS matrix element $V_{e3} = s_{13} e^{- i \delta}$. Of course there can still only be two physical phases inside $|m_{ee}|$, which are $\alpha_{21}$ and $(\alpha_{31} - 2\delta)$, which is why some authors choose to \emph{redefine} the mass $\tilde m_3$ in such a way that the Dirac CP phase $\delta$ does not appear in the formula for $|m_{ee}|$ (see, e.g., Refs.~\cite{Lindner:2005kr,Rodejohann:2011mu}). This step is convenient -- and always perfectly justified -- since we can choose any combination of phases to be \emph{physical} as long as there are in total three independent combinations (in the case of a $3\times 3$ Majorana mass matrix~\cite{Beringer:1900zz}). However, there is one point we have to be careful with if we want to investigate sum rules: the redefinition of phases is, in fact, nothing else than \emph{a redefinition of the Majorana phase $\phi_3$}, and by this it will \emph{modify the neutrino mass sum rule} under consideration. This is easy to see, since redefining $\alpha_{31} - 2\delta \to \alpha_{31}$ in Eq.~\eqref{eq:mee_2} is equivalent to redefining $\phi_3 \to \phi_3 + 2\delta$, which would then show up in the steps following Eq.~\eqref{eq:gen_rule_1}. While in general, without any sum rule at work, this redefinition does not show up anywhere else except for $|m_{ee}|$, it does appear when a sum rule is studied in addition. Hence, we have to be careful when applying any redefinition to a Majorana phase, since such a redefinition will, in general, also redefine the sum rule involved.\footnote{The only exception is factoring out an overall phase, as we will see for the example of $\phi_1$ in a second.} Thus we have to be careful when aiming to determine \emph{which} phases are actually constrained by the sum rule. In order to do that in a consistent way, we will in our calculations always stick to the PDG parametrisation [i.e., to Eq.~\eqref{eq:mee_2}], \emph{without} redefining any phases.

Note that one can also think of the effective mass $m_{ee}$ geometrically, as a sum of three vectors, by simply interpreting the complex numbers as vectors in the complex plane~\cite{Lindner:2005kr}:
\begin{center}
\includegraphics[width=9cm]{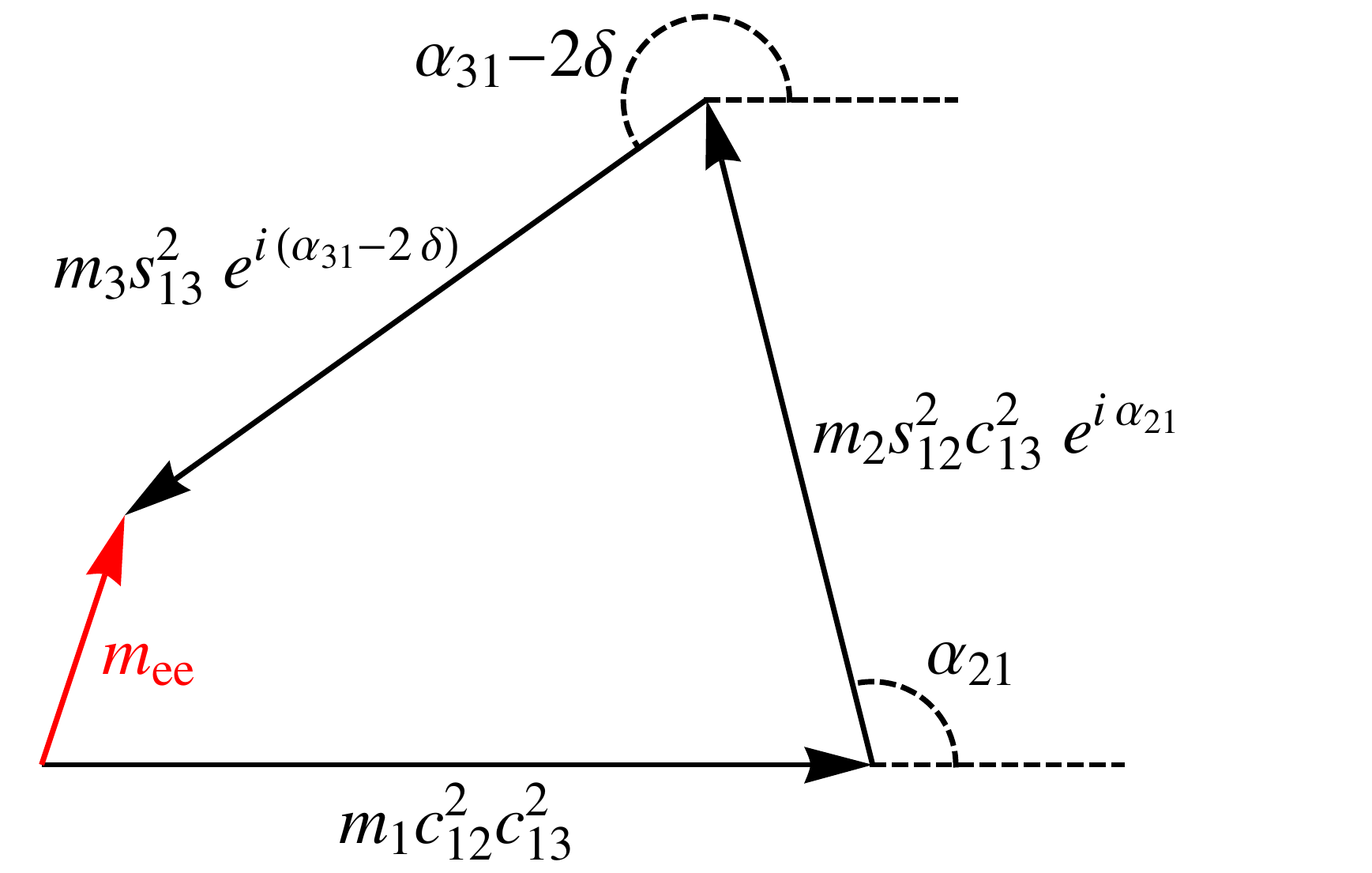}
\end{center}
This picture makes it obvious how $m_{ee}$ can vanish: if the three vectors can form a triangle by adjusting the phases $\alpha_{21}$ and $(\alpha_{31}-2\delta)$, then the resulting ``vector'' $m_{ee}$ will have zero length. If this is not possible, either due to the three pieces having inappropriate lengths or due to some external constraints on the phases, just as imposed by the existence of a certain sum rule, then the resulting vector (and by this $|m_{ee}|$) will be finite.

Before closing this section, we will first comment on an alternative parametrisation of the PMNS matrix, and we will furthermore show why we can identify the Majorana phases $\phi_i$ in the sum rule with those in the effective mass.

First, to make the dependence of $|m_{ee}|$ on only two phases more immediate, one can make use of the so-called \emph{symmetric parametrisation}~\cite{Schechter:1981bd,King:2002nf,Rodejohann:2011vc}, in which each of the three rotation matrices making up the PMNS matrix is taken to be complex,
\begin{equation}
 U^{\rm sym}_{\rm PMNS} = \omega_{23} (\theta_{23}, \phi_{23}) \omega_{13} (\theta_{13}, \phi_{13}) \omega_{12} (\theta_{12}, \phi_{12}),
 \label{eq:par_1}
\end{equation}
where the complex rotation matrices are given by
\begin{eqnarray}
 && \omega_{12} =
 \begin{pmatrix}
 c_{12} & & & & \hfill s_{12} e^{-i\phi_{12}} & & & & 0 \\
 \hfill -s_{12} e^{i\phi_{12}} & & & & c_{12} & & & & 0 \\
 0 & & & & 0 & & & & 1
 \end{pmatrix},\ \ \omega_{13} =
 \begin{pmatrix}
 c_{13} & & & & 0 & & & & \hfill s_{13} e^{-i\phi_{13}} \\
 0 & & & & 1 & & & & 0 \\
 \hfill -s_{13} e^{i\phi_{13}} & & & & 0 & & & & c_{13}
 \end{pmatrix},\nonumber \\
 && \omega_{23} =
 \begin{pmatrix}
 1 & & & & 0 & & & & 0 \\
 0 & & & & c_{23} & & & & \hfill s_{23} e^{-i\phi_{23}} \\
 0 & & & & \hfill -s_{23} e^{i\phi_{23}} & & & & c_{23}
 \end{pmatrix}.
 \label{eq:par_2}
\end{eqnarray}
In this parametrisation we have
\begin{equation}
 |m_{ee}|_{\rm sym} = |m_1 c_{12}^2 c_{13}^2 + m_2 s_{12}^2 c_{13}^2 e^{2 i \phi_{21}} + m_3 s_{13}^2 e^{2 i \phi_{31}}|,
 \label{eq:par_3}
\end{equation}
which makes it immediately clear that only two phases can go inside the effective mass. In turn, the Jarlskog invariant in this parametrisation is given by
\begin{equation}
 J^{\rm sym}_{\rm CP} = {\rm Im} (U_{e1}^* U_{\mu 3}^* U_{e3} U_{\mu 1}) = \frac{1}{8} \sin (2\theta_{12}) \sin (2\theta_{13}) \sin (2\theta_{23}) \cos \theta_{13} \sin (\phi_{13} - \phi_{12} - \phi_{23}),
 \label{eq:par_4}
\end{equation}
which makes it obvious that all three generations are involved in the Dirac CP violation, as they should.\footnote{Note that $J^{\rm PDG}_{\rm CP} = J^{\rm sym}_{\rm CP}|_{\phi_{13} - \phi_{12} - \phi_{23} \to \delta}$.} Furthermore, it is clear from Eqs.~\eqref{eq:par_3} and~\eqref{eq:par_4} that the \emph{information content} inside the Majorana and the Dirac phases is different, which implies that $|m_{ee}|$ \emph{cannot} depend on all the information contained in $J_{\rm CP}$. This holds unless some additional information constrains a Majorana phase in terms of the Dirac phase, which is just what happens for sum rules. This is exactly the reason for the PDG parametrisation being a little more transparent when trying to combine the information cast in a sum rule with the information content of the effective mass.

Of course, both parametrisations -- if applied correctly -- lead in the end to the same result. In fact, one can unambiguously translate the two sets of phases into each other:
\begin{equation}
 \begin{pmatrix}
 \delta\\
 \alpha_{21}\\
 \alpha_{31}
 \end{pmatrix} = Q
 \begin{pmatrix}
 \phi_{12}\\
 \phi_{13}\\
 \phi_{23}
 \end{pmatrix},\ \ \
 \begin{pmatrix}
 \phi_{12}\\
 \phi_{13}\\
 \phi_{23}
 \end{pmatrix} = Q^{-1}
 \begin{pmatrix}
 \delta\\
 \alpha_{21}\\
 \alpha_{31}
 \end{pmatrix},
 \label{eq:par_5}
\end{equation}
where the transformation matrix and its inverse are explicitly given by
\begin{equation}
 Q =
 \begin{pmatrix}
 \hfill -1 & & & \hfill 1 & & & \hfill -1\\
 \hfill 2 & & & \hfill 0 & & & \hfill 0\\
 \hfill -2 & & & \hfill 4 & & & \hfill -2
 \end{pmatrix},\ \ \ Q^{-1}=
 \begin{pmatrix}
 \hfill 0 & & & \hfill 1/2 & & & \hfill 0\\
 \hfill -1 & & & \hfill 0 & & & \hfill 1/2\\
 \hfill -2 & & & \hfill -1/2 & & & \hfill 1/2
 \end{pmatrix}.
 \label{eq:par_6}
\end{equation}
As to be expected, both parametrisations are equivalent if the translation is done correctly.

Now let us comment on the definition of the Majorana phases $\phi_i$. First, we can in any setting go to a basis where the charged lepton mass matrix is diagonal, $M_e = {\rm diag}(m_e, m_\mu, m_\tau)$. Then, the translation between the neutrino flavour basis $\nu_f = (\nu_e, \nu_\mu, \nu_\tau)$ and the mass basis $\tilde \nu=(\tilde  \nu_1, \tilde \nu_2, \tilde \nu_3)^T$ where the neutrino mass eigenvalues are still complex is given by exactly the CKM-part of the PMNS matrix (analogous to quarks, with the only difference that the resulting eigenvalues are in general still complex):
\begin{equation}
 \begin{pmatrix}
 \nu_e\\
 \nu_\mu\\
 \nu_\tau
 \end{pmatrix} = V
 \begin{pmatrix}
 \tilde \nu_1\\
 \tilde \nu_2\\
 \tilde \nu_3
 \end{pmatrix}.
 \label{eq:phase_1}
\end{equation}
For the light neutrino Majorana mass term this implies
\begin{equation}
 \mathcal{L} = - \overline{\nu_f^c} M_\nu \nu_f + h.c. = - \nu_f^T M_\nu \nu_f + h.c. = - {\tilde \nu}^T \tilde D_\nu \tilde \nu + h.c.,
 \label{eq:phase_2}
\end{equation}
where $\tilde D_\nu = {\rm diag} (\tilde m_1, \tilde m_2, \tilde m_3) = {\rm diag} (e^{i \phi_1} m_1, e^{i \phi_2} m_2, e^{i \phi_3} m_3)$. Since there is already one phase inside $V$, one of the three phases $\phi_{1,2,3}$ can still be absorbed as global phase inside of $\tilde \nu$. If we choose $\phi_1$ to be absorbed, we can redefine $\nu = e^{i \phi_1/2} \tilde \nu$, which leads to
\begin{eqnarray}
 \mathcal{L} &=& - {\tilde \nu}^T
 \begin{pmatrix}
 \tilde m_1 & 0 & 0\\
 0 & \tilde m_2 & 0\\
 0 & 0 & \tilde m_3
 \end{pmatrix}
 \tilde \nu + h.c. = - {\tilde \nu}^T
 \begin{pmatrix}
 m_1 & 0 & 0\\
 0 & m_2 & 0\\
 0 & 0 & m_3
 \end{pmatrix}
 \begin{pmatrix}
 e^{i \phi_1} & 0 & 0\\
 0 & e^{i \phi_2} & 0\\
 0 & 0 & e^{i \phi_3}
 \end{pmatrix}
 \tilde \nu + h.c.\nonumber\\
 &=& - \nu^T
 \begin{pmatrix}
 m_1 & 0 & 0\\
 0 & m_2 & 0\\
 0 & 0 & m_3
 \end{pmatrix}
 \begin{pmatrix}
 1 & 0 & 0\\
 0 & e^{i \alpha_{21}} & 0\\
 0 & 0 & e^{i \alpha_{31}}
 \end{pmatrix}
 \nu + h.c.
 \label{eq:phase_3}
\end{eqnarray}
Relating this to the full PMNS matrix, it can be recast as follows:
\begin{eqnarray}
 \mathcal{L} &=& - \nu^T
 \begin{pmatrix}
 1 & 0 & 0\\
 0 & e^{i \alpha_{21}/2} & 0\\
 0 & 0 & e^{i \alpha_{31}/2}
 \end{pmatrix}
 \begin{pmatrix}
 m_1 & 0 & 0\\
 0 & m_2 & 0\\
 0 & 0 & m_3
 \end{pmatrix}
 \begin{pmatrix}
 1 & 0 & 0\\
 0 & e^{i \alpha_{21}/2} & 0\\
 0 & 0 & e^{i \alpha_{31}/2}
 \end{pmatrix}
 \nu + h.c.\nonumber\\
 &\equiv& - \nu_f^T U^T
 \begin{pmatrix}
 m_1 & 0 & 0\\
 0 & m_2 & 0\\
 0 & 0 & m_3
 \end{pmatrix}
 U \nu_f + h.c.
 \label{eq:phase_4}
\end{eqnarray}
This exactly reproduces the PDG parametrisation,
\begin{equation}
 U \equiv V
 \begin{pmatrix}
 1 & 0 & 0\\
 0 & e^{i \alpha_{21}/2} & 0\\
 0 & 0 & e^{i \alpha_{31}/2}
 \end{pmatrix},
 \label{eq:phase_5}
\end{equation}
which coincides with Eq.~(13.79) from Ref.~\cite{Beringer:1900zz}.

In order to evaluate the effective mass, cf.\ Eq.~\eqref{eq:mee_2}, it is easiest to rewrite the mass eigenvalues for normal (NO) and inverted (IO) mass ordering. Denoting the smallest neutrino mass eigenvalue as $m_{\rm lightest}$, we have:
\begin{eqnarray}
 {\rm NO:\ \ \ }&& m_1 = m_{\rm lightest} < m_2 = \sqrt{m_{\rm lightest}^2 + \Delta m_\odot^2} < m_3 = \sqrt{m_{\rm lightest}^2 + \Delta m_A^2},\label{eq:orderings}\\
 {\rm IO:\ \ \ }&& m_3 = m_{\rm lightest} < m_1 = \sqrt{m_{\rm lightest}^2 + \Delta m_A^2} < m_2 = \sqrt{m_{\rm lightest}^2 + \Delta m_\odot^2 + \Delta m_A^2},\nonumber
\end{eqnarray}
where $\Delta m_\odot^2 \equiv \Delta m_{21}^2$ and $\Delta m_A^2 \equiv |\Delta m_{31}^2|$.

When we want to plot the effective mass $|m_{ee}|$ from Eq.~\eqref{eq:mee_2} versus the smallest mass eigenvalue $m_{\rm lightest}$, we must use the current knowledge on the neutrino oscillation parameters. Looking at the literature, there are currently three global fits on the most recent data available. We will list them in the order in which they appeared:
\begin{itemize}

\item The FTV-fit~\cite{Tortola:2012te} by Forero, Tortola, and Valle.

\item The FLMMPR-fit~\cite{Fogli:2012ua} by Fogli, Lisi, Marrone, Montanino, Palazzo, and Rotunno.

\item The GMSS-fit~\cite{GonzalezGarcia:2012sz} by Gonzalez-Garcia, Maltoni, Salvado, and Schwetz.
 
\end{itemize}

\begin{figure}[t]
\hspace{-1.0cm}
\begin{tabular}{lcr}
\includegraphics[width=5.5cm]{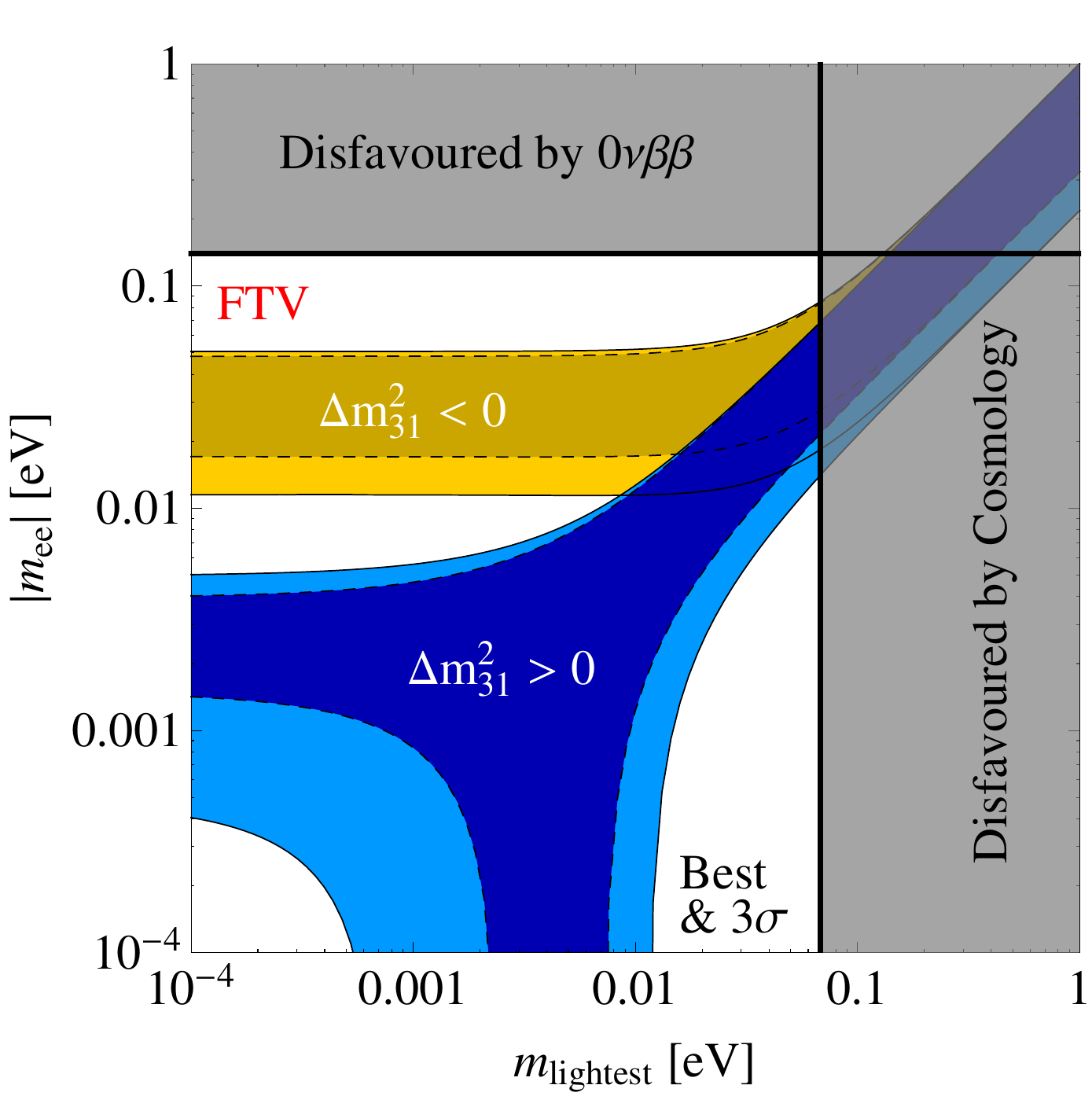} & \includegraphics[width=5.5cm]{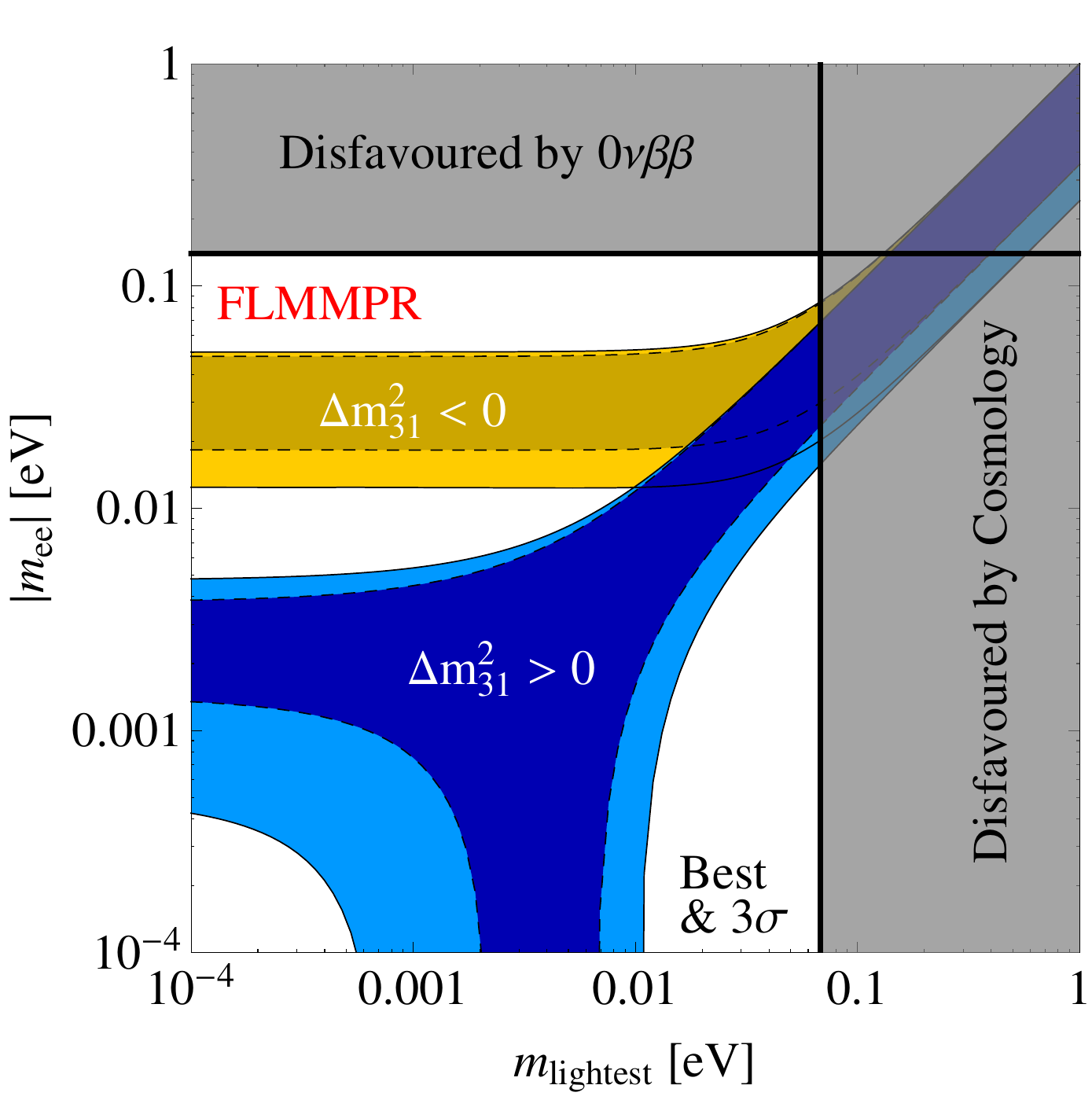} & \includegraphics[width=5.5cm]{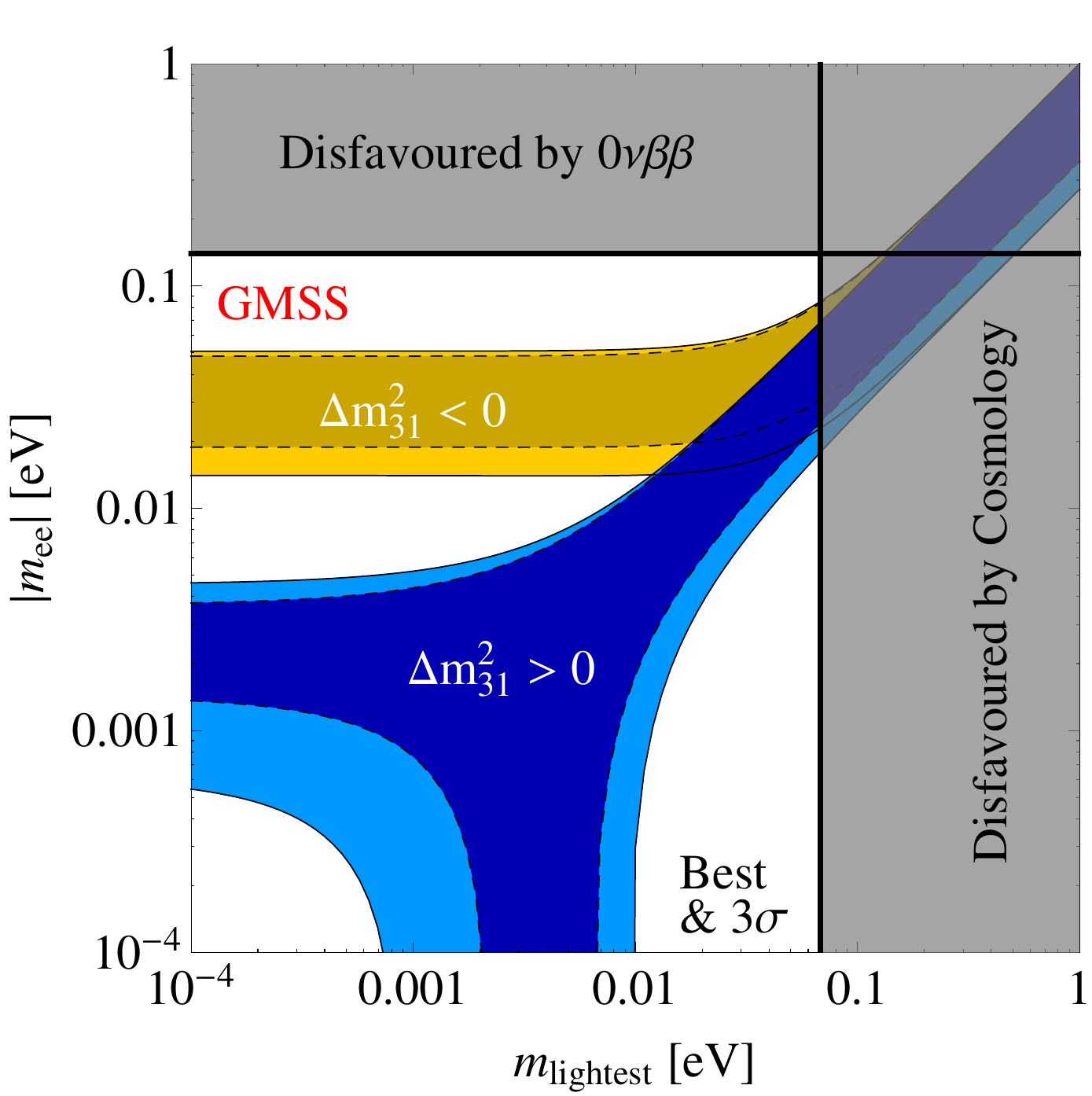}
\end{tabular}
\caption{\label{fig:mee} The effective mass as a function of the smallest mass eigenvalue for the three global fits. As can be seen, not too much is changing from one fit to another.}
\end{figure}

We can plot $|m_{ee}|$ vs.\ $m_{\rm lightest}$ for all three fits, which is displayed in Fig.~\ref{fig:mee}.\footnote{Note that we will in the concrete cases to be discussed, which can be experimentally probed, always use all three global fits. However, for the hypothetical sum rules used to illustrate the discussions, we have refrained from doing so to save space. For these cases, we have decided to plot only the data obtained by using the most recent GMSS-fit, for the simple reason that this fit is kept up to date online, see {\tt http://www.nu-fit.org/}. The corresponding plots using the other fits can be obtained from the authors upon request.} Apparently, there is not too much of a difference between the different fits, although some small deviations are visible by eye. Note that we have also indicated the regions which are disfavoured by searches for neutrinoless double beta decay (the current most optimistic limit of $|m_{ee}|<0.140$~eV coming from the EXO-200 experiment~\cite{Auger:2012ar}) and by cosmological limits on the sum $\Sigma = m_1 + m_2 + m_3$ (the most stringent limit of $\Sigma < 0.230$~eV having been obtained by the Planck data combined with external CMB and BAO measurements~\cite{Ade:2013lta}).\footnote{Note that we have averaged the derived bounds on the lightest active neutrino mass for NO and IO.} It should be mentioned that the limit obtained from $0\nu\beta\beta$ involves a translation of a bound on the decay rate into a bound on the effective mass, which involves unknown nuclear physics, see e.g.\ Refs.~\cite{Faessler:2009zz,Vergados:2012xy,Suhonen:2012ii,Rodin:2012gp} and references therein, whose influence will be discussed in detail in Sec.~\ref{sec:experiments}. Also the bound derived from cosmological observations involves certain assumptions, and maybe even unknown systematic errors which could potentially lead to a wrong conclusion about the limit on the neutrino mass scale~\cite{Maneschg:2008sf}. For these reasons, we prefer to mark the corresponding regions as ``disfavoured'' rather than ``excluded''.

After having carefully set the stage, the next step will be to discuss a simplified class of sum rules in greater detail.

\section{\label{sec:trivial}Trivial sum rules}

We first focus on trivial sum rules, where $A_1 = A_2 = A_3 = 1$. While to some extent the distinction between trivial and non-trivial sum rules is artificial, it is nevertheless justified since indeed the analysis is a bit simpler for the trivial case and also because both mass orderings are always possible for trivial sum rules, which may not necessarily be the case for the non-trivial ones.

The \emph{most general trivial sum rule} is given by
\begin{equation}
 m_1^p e^{i\tilde \phi_1} + m_2^p e^{i\tilde \phi_2} + m_3^p e^{i\tilde \phi_3} = 0.
 \label{eq:rule_1}
\end{equation}

From Eq.~\eqref{eq:rule_1}, we can already see that all mass eigenvalues must necessarily be non-zero for the sum rule to hold: if, e.g., $m_1 \equiv 0$, then we have for $p>0$,
\begin{equation}
 m_2^p e^{i\tilde \phi_2} + m_3^p e^{i\tilde \phi_3} = 0 \Rightarrow m_2^p = - m_3^p e^{i (\tilde \phi_3-\tilde \phi_2)} \Rightarrow |m_2| = |m_3| \Rightarrow m_2 = m_3,
 \label{eq:rule_2}
\end{equation}
which could never be brought into accordance with the neutrino oscillation data. If on the other hand $p<0$ holds, then one side of the equation is infinite, which also destroys the validity of the equal-sign. The same thing happens if any of the other two masses is set to zero. Hence we can always assume $m_j \neq 0$ in what follows.

Multiplying Eq.~\eqref{eq:rule_1} by $e^{- i\tilde \phi_1} = e^{- i(p \phi_1 + \chi_1)}$, defining
\begin{equation}
 \Delta \chi_{i1} \equiv \chi_i - \chi_1 \ \ \ (i=2,3)\ \ \ ,
 \label{eq:rule_2p5}
\end{equation}
and again using $\alpha_{i1}$, we obtain
\begin{equation}
 m_1^p + \left( m_2 e^{i \alpha_{21}} \right)^p e^{i \Delta \chi_{21}} + \left( m_3 e^{i \alpha_{31}} \right)^p e^{i \Delta \chi_{31}} = 0.
 \label{eq:rule_3}
\end{equation}
This equation can be easily interpreted geometrically, as done, e.g., in Refs.~\cite{Barry:2010zk,Dorame:2011eb}:
\begin{center}
\includegraphics[width=9cm]{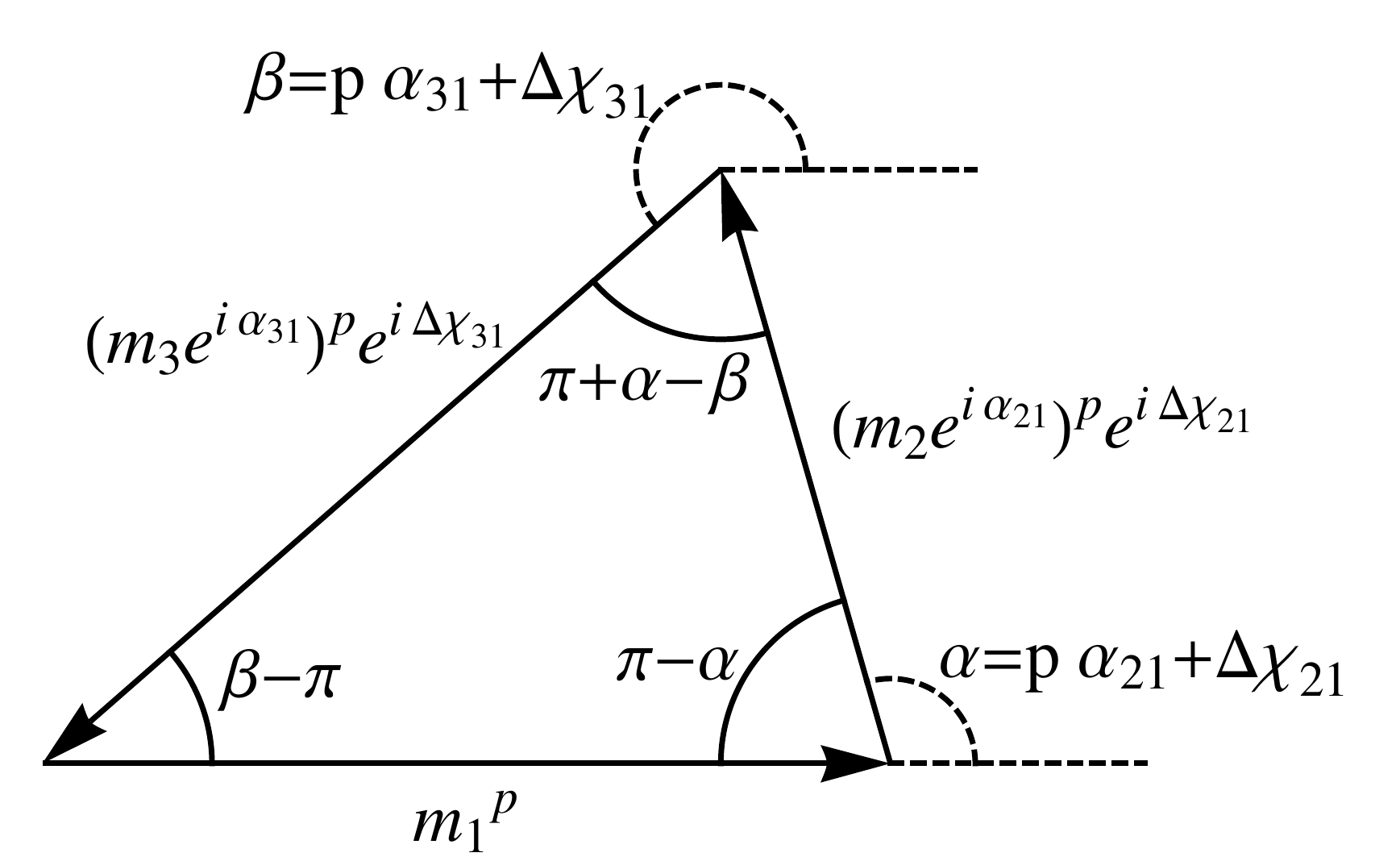}
\end{center}
Apparently, Eq.~\eqref{eq:rule_3} describes nothing else than a sum of three vectors in the complex plane. The first vector is parallel to the real axis, and the fact that the sum is zero imposes the geometrical shape of a triangle. However, of course the orientation of the triangle is irrelevant, which is why only differences of phases appear in Eq.~\eqref{eq:rule_3}.

Since Eq.~\eqref{eq:rule_3} is a complex equation, it must give us two pieces of information, which can be interpreted as the two angles $\alpha \equiv p \alpha_{21} + \Delta \chi_{21}$ and $\beta \equiv p \alpha_{31} + \Delta \chi_{31}$. Applying the law of cosines to the triangle immediately yields:
\begin{equation}
 \cos \alpha = \frac{m_3^{2 p} - m_2^{2 p} - m_1^{2 p}}{2 (m_1 m_2)^p}\ \ \ , \ \ \ \cos \beta = \frac{m_2^{2 p} - m_3^{2 p} - m_1^{2 p}}{2 (m_1 m_3)^p}.
 \label{eq:rule_4}
\end{equation}
The important point about Eqs.~\eqref{eq:rule_4} is that they actually decide about the \emph{validity} of the sum rules: the right-hand sides of the equations can be computed for any values of $(m_1, m_2, m_3, p)$, but real values for $(\alpha, \beta)$ can only be obtained if the right-hand sides are contained in the interval $[-1, +1]$. This procedure works because if a triangle can be formed out of the three sides, it is uniquely determined. In our numerics, however, we have not only applied Eq.~\eqref{eq:rule_4} but we have also for each point probed the validity of the triangle inequality, in order to make sure that our numerical calculation carefully decides about the validity of the sum rule.

\subsection{\label{sec:orderings}Deciding about the mass ordering}

For trivial sum rules, it is straightforward to see that the values of $\alpha$ and $\beta$ reflect the neutrino mass ordering that is present for a given $p$.

\begin{table}[t]
\centering
\begin{tabular}{|c||c|c|}\hline
 $\Delta$  & NO & IO\\ \hline
 $p=+|p|>0$ & $\cos \alpha > -1/2$ \& $|\cos \beta| > 1/2$ & $\cos \alpha < -1/2$ \& $|\cos \beta| < 1/2$ \\ \hline
 $p=-|p|<0$ & $\cos \alpha < -1/2$ \& $|\cos \beta| < 1/2$ & $\cos \alpha > -1/2$ \& $|\cos \beta| > 1/2$ \\ \hline
 \end{tabular}
\caption{\label{tab:Triv_order} NO and IO for trivial sum rules.}
\end{table}

To see this, we introduce the following two useful abbreviations:
\begin{equation}
 x \equiv \frac{m_{\rm lightest}^2}{\Delta m_A^2}\ \ \ {\rm and}\ \ \ \epsilon \equiv \frac{\Delta m_\odot^2}{\Delta m_A^2}.
 \label{eq:abbreviations}
\end{equation}
Then we can rewrite Eqs.~\eqref{eq:rule_4} in a form that is convenient for analytical calculations. Inserting the above abbreviations, one obtains
\begin{equation}
 \left\{
 \begin{matrix}
 {\rm NO}: \hfill & & & \cos \alpha = \frac{(x + 1)^p- (x + \epsilon)^p - x^p}{2 (\sqrt{x (x + \epsilon)})^p} , \hfill \hfill \hfill & & & \cos \beta = \frac{(x + \epsilon)^p- (x + 1)^p - x^p}{2 (\sqrt{x (x + 1)})^p} , \hfill \hfill \hfill\\
 {\rm IO}: \hfill & & & \cos \alpha = \frac{x^p- (x + 1 + \epsilon)^p - (x+1)^p}{2 (\sqrt{x + 1} \sqrt{x + 1 + \epsilon})^p} , \hfill \hfill \hfill & & & \cos \beta = \frac{(x + 1 + \epsilon)^p - x^p - (x + 1)^p}{2 (\sqrt{x} \sqrt{x + 1})^p} . \hfill \hfill \hfill
 \end{matrix}
 \right.
 \label{eq:rule_5}
\end{equation}
To arrive at analytical results, one can neglect $\epsilon \ll 1, x$, which is not a bad approximation given that $\epsilon \simeq 0.05$. In this limit, one obtains:
\begin{equation}
 \left\{
 \begin{matrix}
 {\rm NO}: \hfill & & & \cos \alpha \simeq \frac{1}{2} \left( 1 + \frac{1}{x} \right)^p - 1 , \hfill \hfill \hfill & & & \cos \beta \simeq - \frac{1}{2} \left( \sqrt{ 1 + \frac{1}{x} } \right)^p , \hfill \hfill \hfill\\
 {\rm IO}: \hfill & & & \cos \alpha \simeq \frac{1}{2} \left( \frac{1}{1 + \frac{1}{x}} \right)^p - 1 , \hfill \hfill \hfill & & & \cos \beta \simeq - \frac{1}{2} \left( \frac{1}{\sqrt{1+ \frac{1}{x}}} \right)^p  . \hfill \hfill \hfill
 \end{matrix}
 \right.
 \label{eq:rule_6}
\end{equation}
Depending on whether $p = + |p| > 0$ or $p = - |p| < 0$, one can solve Eqs.~\eqref{eq:rule_6} for $x$ and impose the condition $x > 0$. For example, in the case $p > 0$ one obtains for NO:
\begin{equation}
 x \simeq \frac{1}{[2 (1 + \cos \alpha)]^{1/|p|} - 1}\ \ \ , \ \ \ x \simeq \frac{1}{(-2 \cos \beta)^{2/|p|} - 1} \ \ \ .
 \label{eq:rule_7}
\end{equation}
Imposing the necessary condition $x > 0$ then leads approximately to
\begin{equation}
 \cos \alpha > - \frac{1}{2}\ \ \ {\rm and}\ \ \ |\cos \beta| > \frac{1}{2}.
 \label{eq:rule_8}
\end{equation}
Hence, for positive $p$ we would expect $\cos \alpha$ to be larger than $-1/2$ and $\cos \beta$ to be either larger than $1/2$ or to be smaller than $-1/2$, in case that NO is given.

Similar analysis can be easily done for all other cases, resulting into the domains specified in Tab.~\ref{tab:Triv_order}, which we have verified numerically.

\subsection{\label{sec:trivial_signatures}Signatures of different sum rules}

\begin{figure}[!t]
\begin{tabular}{lcr}
\includegraphics[width=5cm]{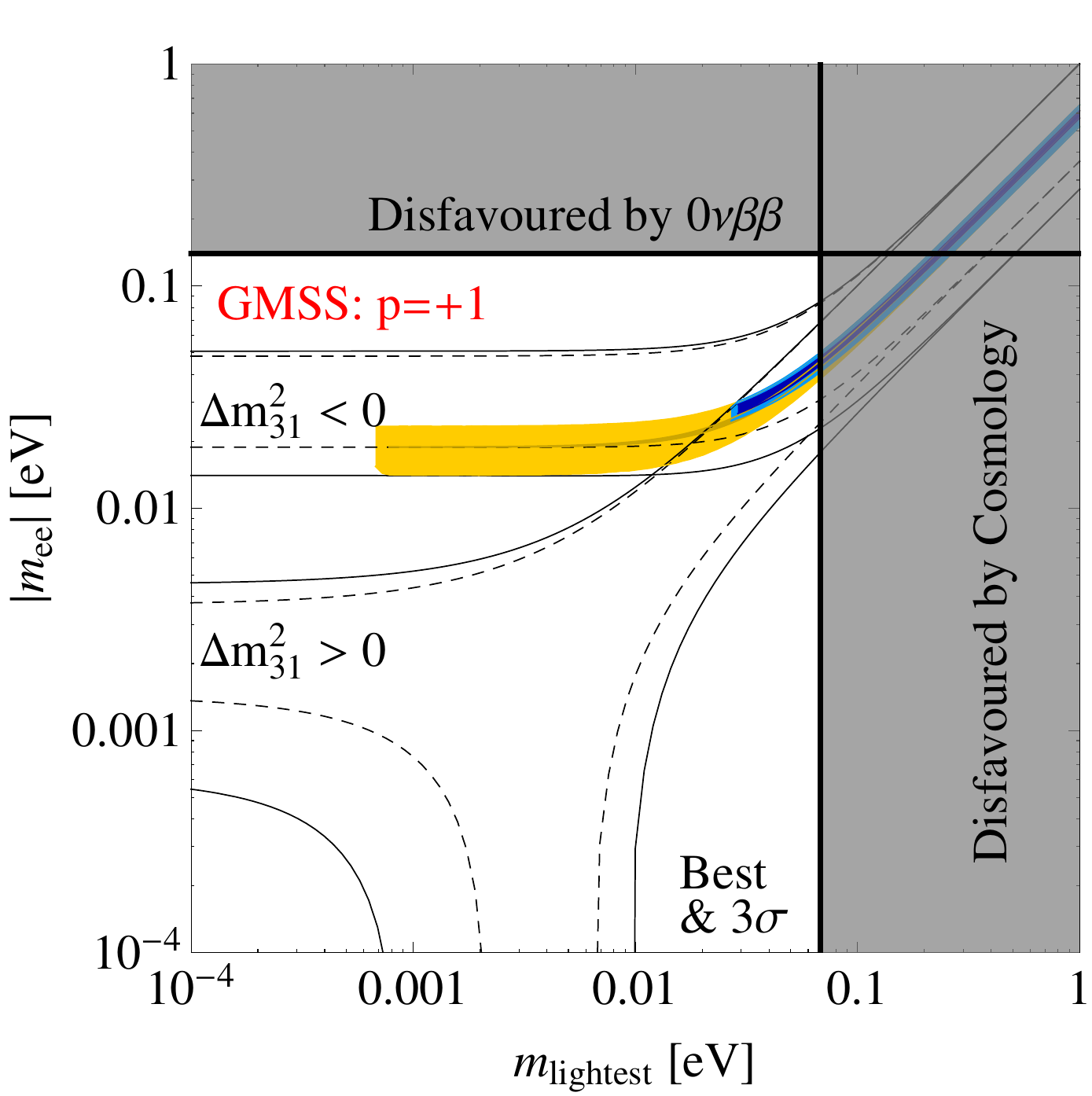}   & \includegraphics[width=5cm]{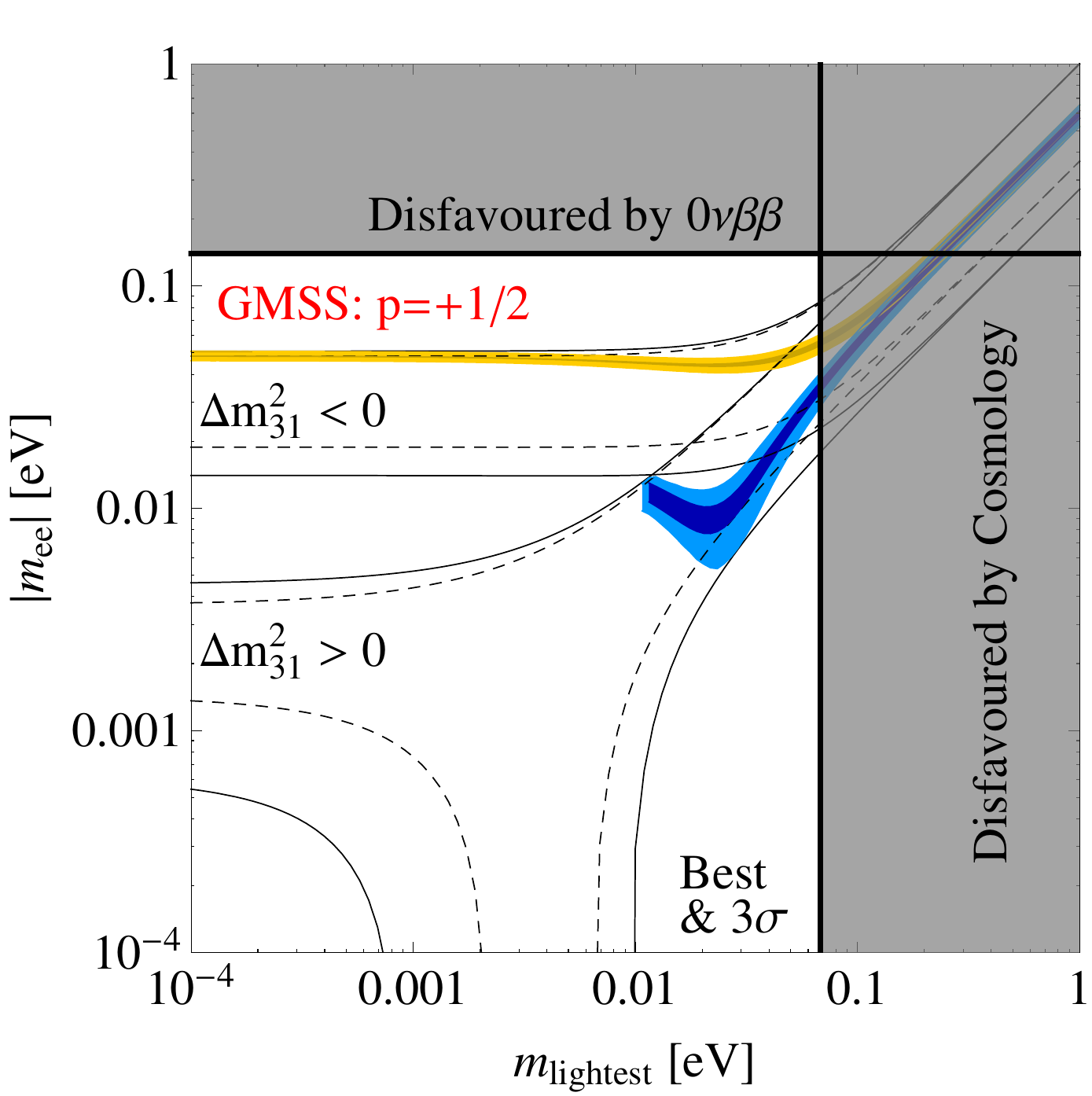} & \includegraphics[width=5cm]{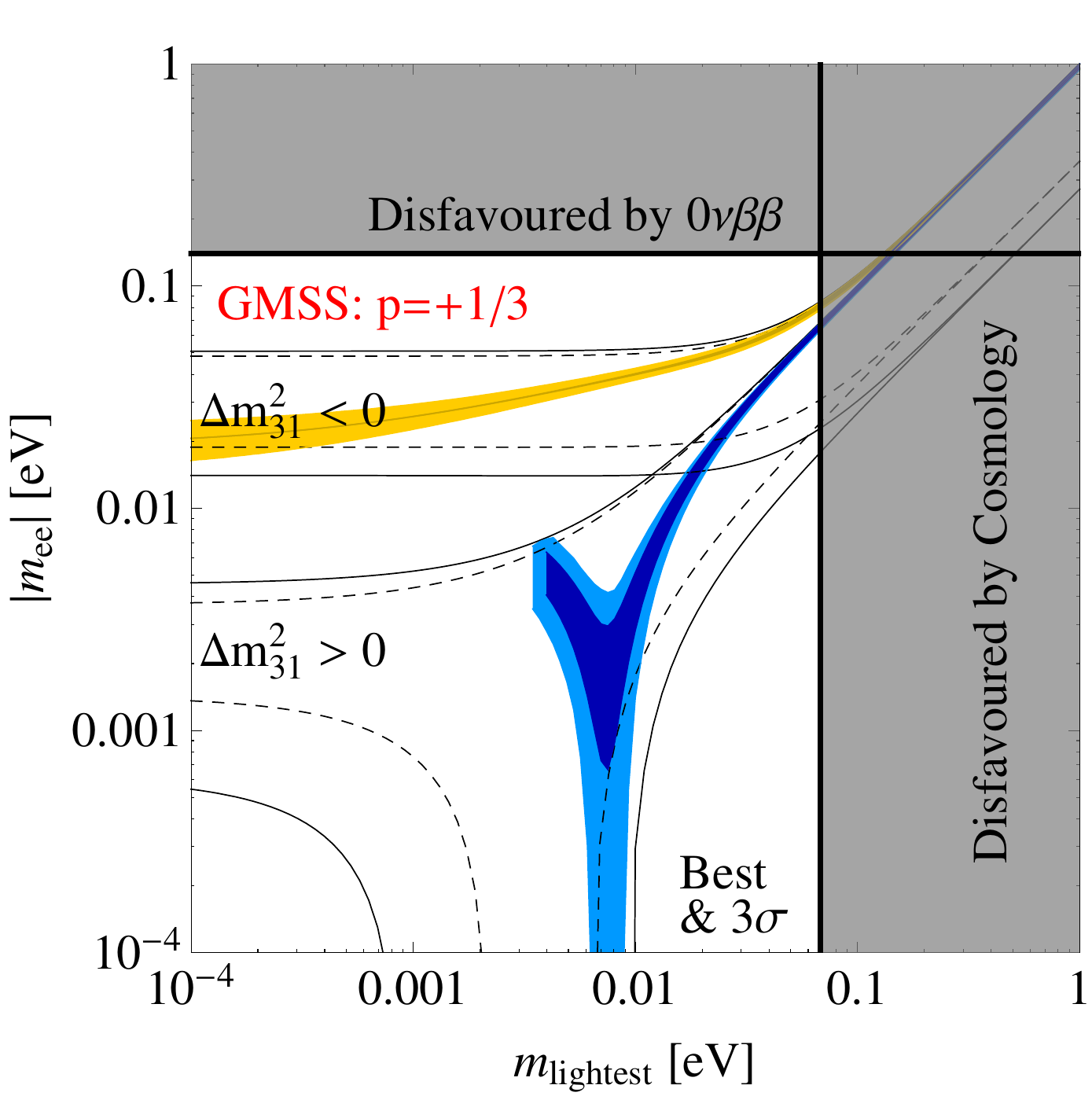}\\
\includegraphics[width=5cm]{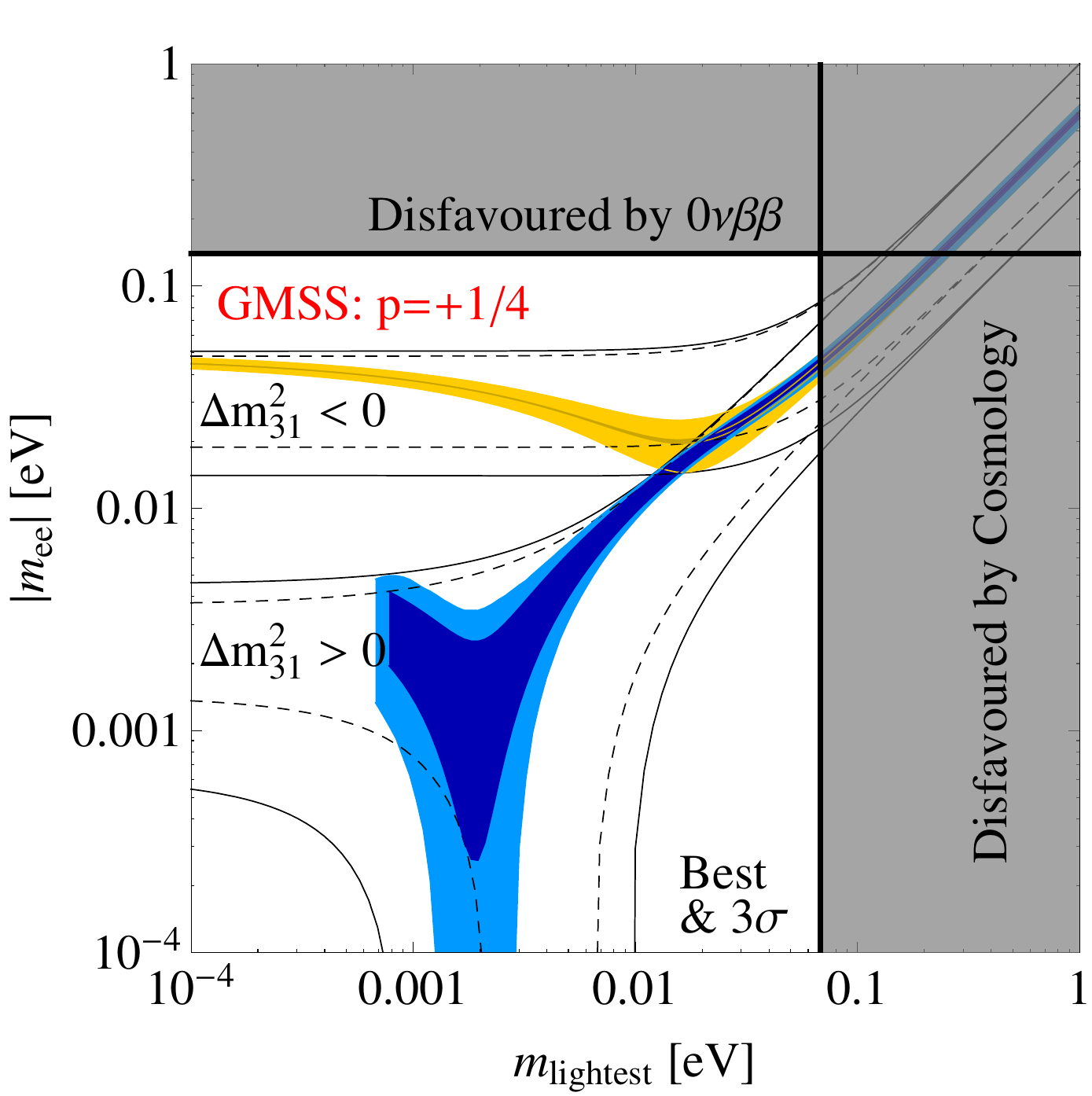} & \includegraphics[width=5cm]{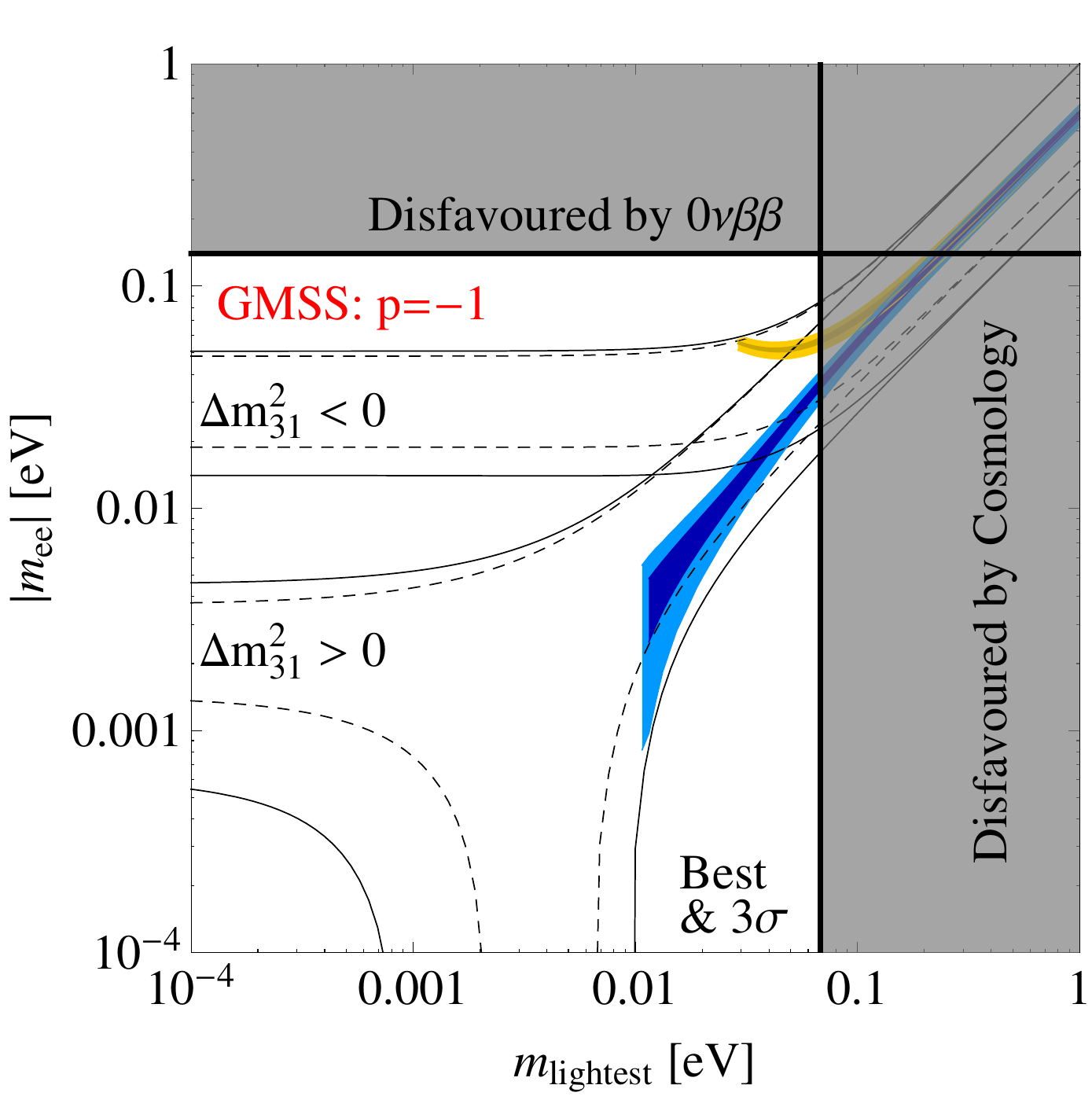} & \includegraphics[width=5cm]{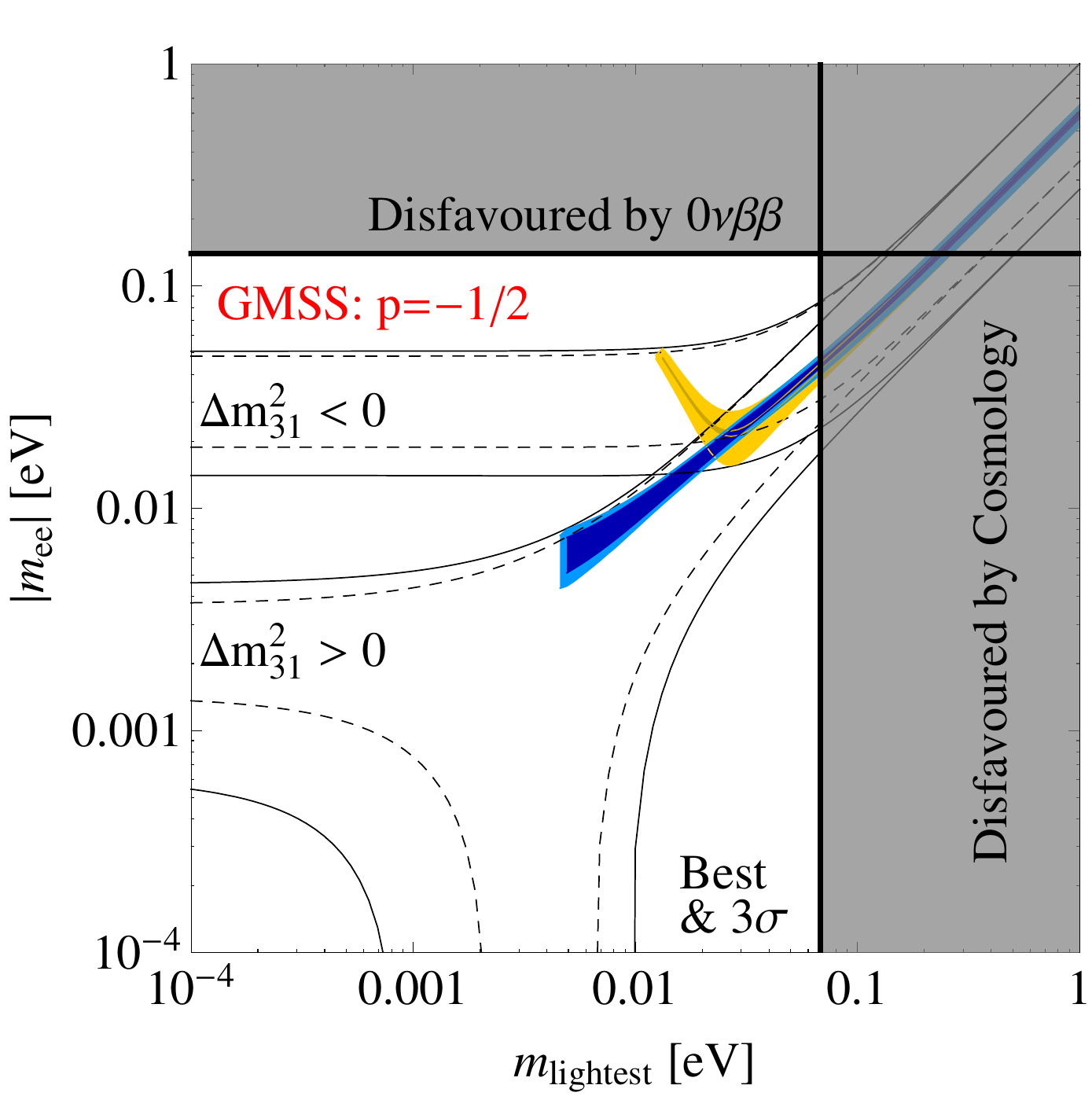}\\
\includegraphics[width=5cm]{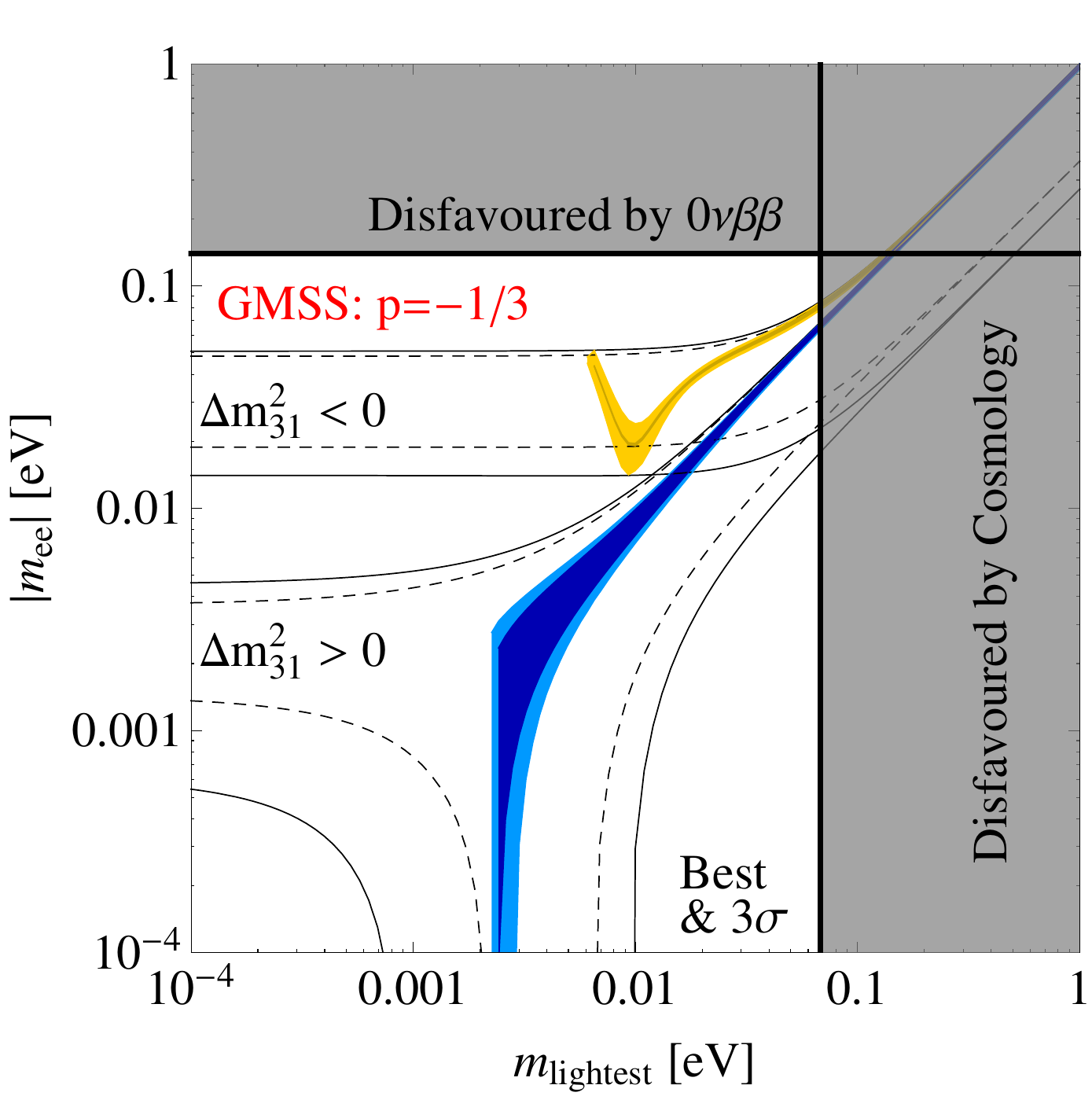} & \includegraphics[width=5cm]{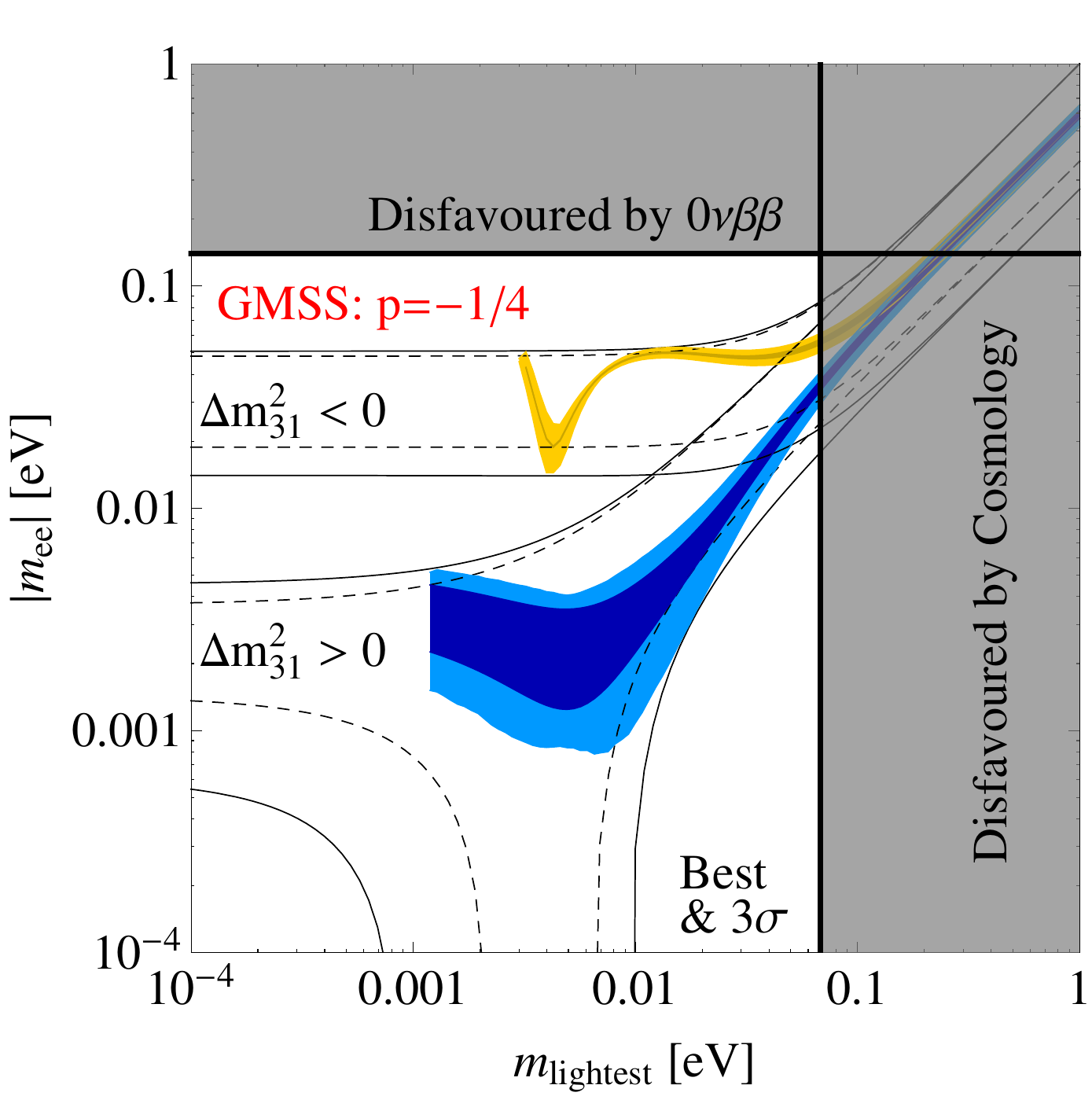} & \includegraphics[width=5cm]{Fig/mee_S.pdf}
\end{tabular}
\caption{\label{fig:Triv_mee_1} Trivial sum rules $ m_1^p + \left( m_2 e^{i \alpha_{21}} \right)^p e^{i \Delta \chi_{21}} + \left( m_3 e^{i \alpha_{31}} \right)^p e^{i \Delta \chi_{31}} = 0$ for $|p| = 1, 1/2, 1/3, 1/4$ with $\Delta \chi_{21} = \Delta \chi_{31} = 0$, compared to the general region.}
\end{figure}

An interesting point is to derive experimental signatures of the different sum rules. In order to do this, we have scanned the values of the smallest neutrino mass eigenvalue\footnote{Note from Eq.~\eqref{eq:orderings} that $m_{\rm lightest} = m_1$ for NO while $m_{\rm lightest} = m_3$ for IO.} $m_{\rm lightest}$ for a range of values between $10^{-4}$~eV and $1$~eV. For any such value $m_{\rm lightest}$, we have checked the validity of the sum rule with the help of Eqs.~\eqref{eq:rule_4} and by making sure that the triangle inequality is fulfilled, thereby either setting all relevant neutrino oscillation parameters to their best-fit values or varying them within their $3\sigma$ ranges. We have fixed the phase differences $\Delta \chi_{21, 31}$, since they are given by a concrete sum rule, while we have varied the Dirac CP phase $\delta$ within $0$ and $2\pi$.\footnote{Note that we have neglected the extremely weak evidence for a range of $\delta$ that could be slightly narrower than $[0, 2\pi)$, since this tiny difference would be hardly visible in the plots and could in any case not be resolved within the current and near-future sensitivity of experiments on $0\nu\beta\beta$.} Note that the Majorana phase differences $\alpha_{21, 31}$ are not varied, since their values are given by the sum rule, cf.\ Eqs.~\eqref{eq:rule_4}. This procedure allows to compute the minimum and maximum allowed values for $|m_{ee}|$.

We have plotted the resulting signatures for the case $\Delta \chi_{21} = \Delta \chi_{31} = 0$ in Fig.~\ref{fig:Triv_mee_1}, for different values of $p$ ($|p| = 1/4$, $1/3$, $1/2$, $1$) and for the GMSS-fit. As we have discussed in Sec.~\ref{sec:sumrules}, due to neutrino masses always being generated by Feynman diagrams, the power $p$ can in ordinary neutrino mass and flavour models only have the absolute value $1$ or $1/n$ with an integer number $n$. However, phenomenologically, other values might be possible. As can be seen from Fig.~\ref{fig:Triv_mee_1}, we have confirmed numerically that both mass orderings are possible for trivial sum rules. Indeed, depending on the exact value of the power $p$, the sum rules can in general lead to very distinctive predictions, see e.g.\ the plots for $-1/2$ or $p=-1/4$. Note that the plots for $p=+1/2, +1/3, +1/4$ seem to contradict a statement given at the beginning of Sec.~\ref{sec:trivial}, since the smallest mass eigenvalue $m_3$ can be very small for IO. However, indeed one finds a non-zero lower bound on $m_3$ also numerically, but it is simply off the plot.
\begin{figure}[!t]
\begin{tabular}{lcr}
\includegraphics[width=5cm]{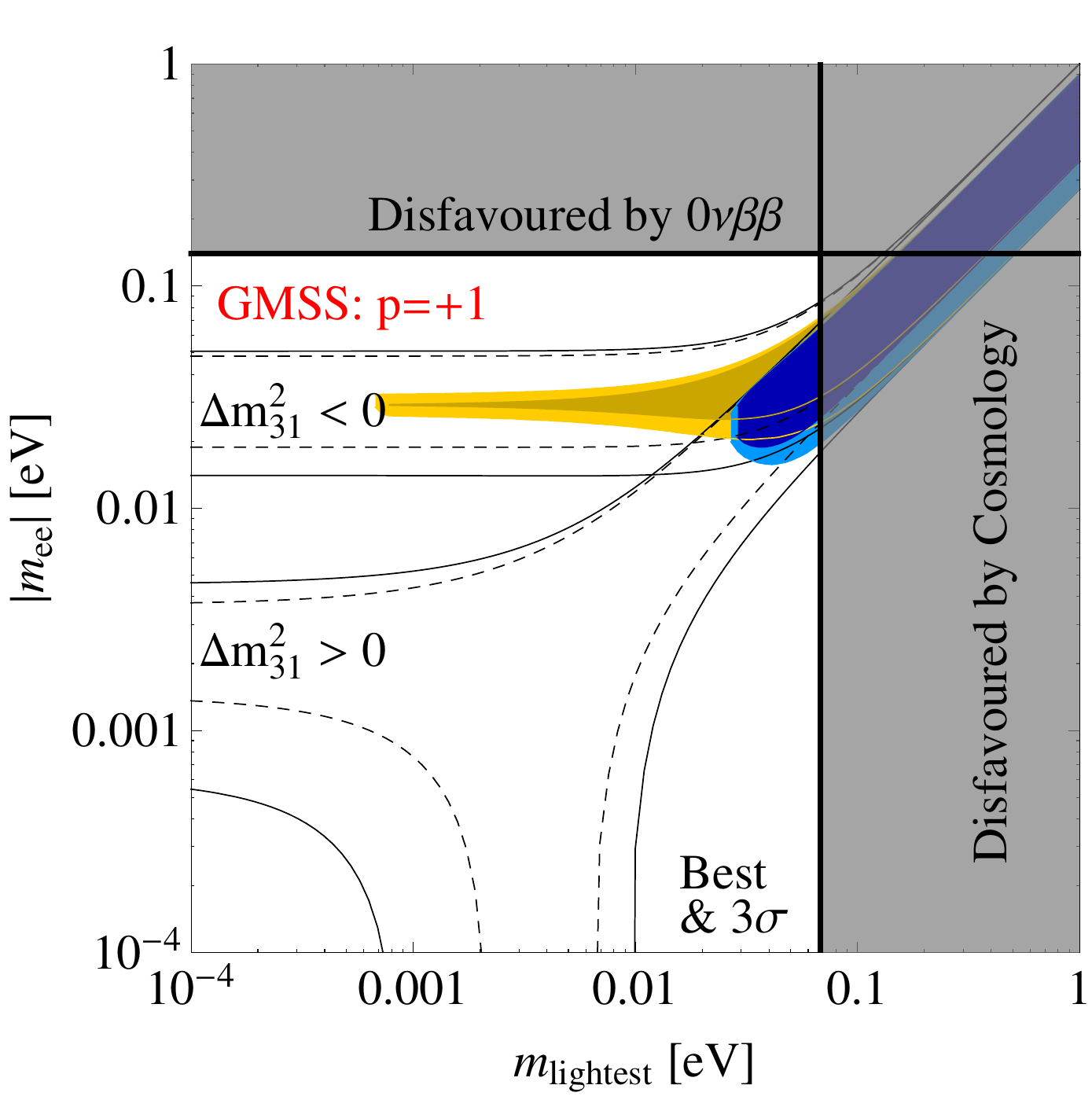}   & \includegraphics[width=5cm]{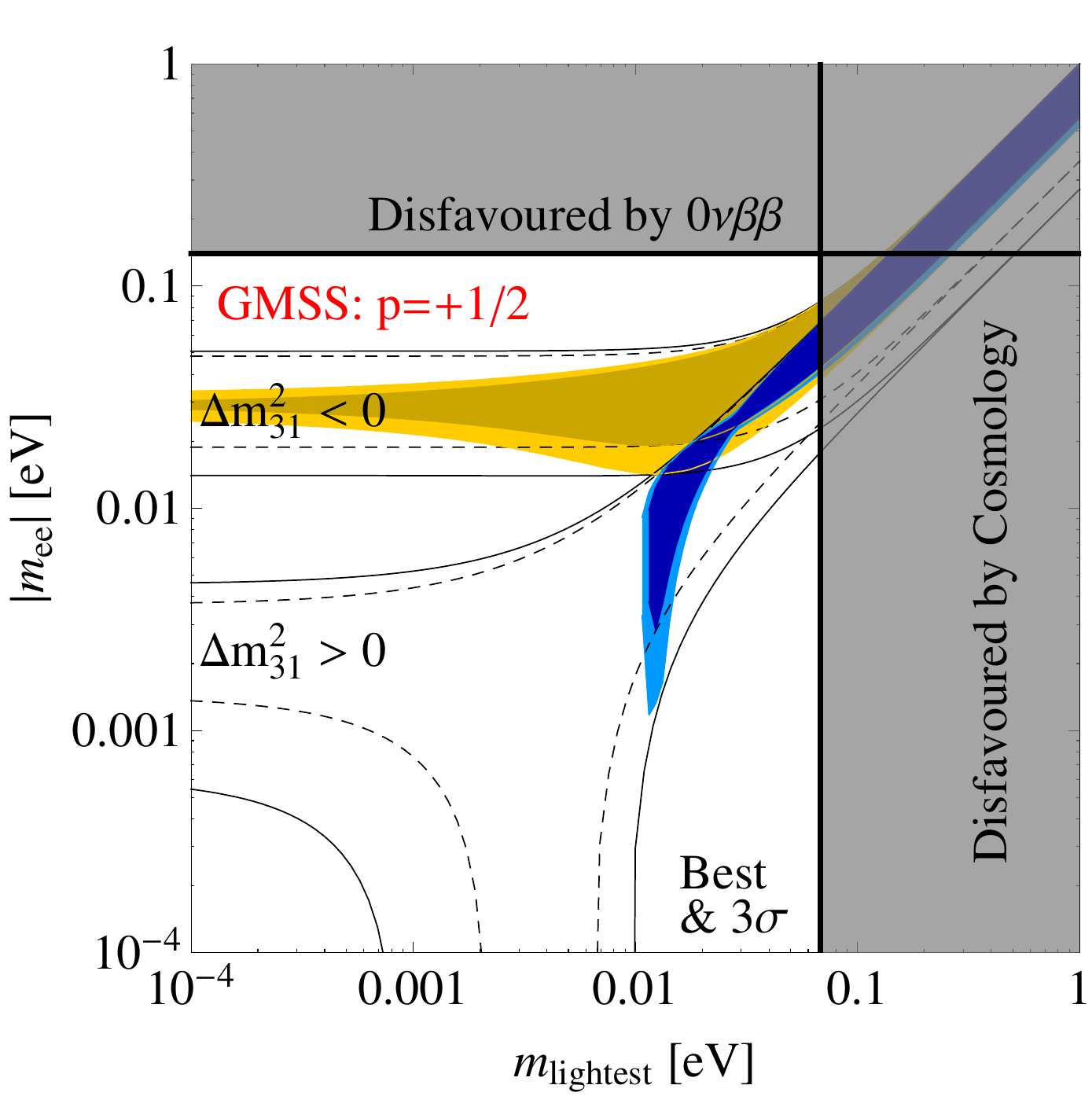} & \includegraphics[width=5cm]{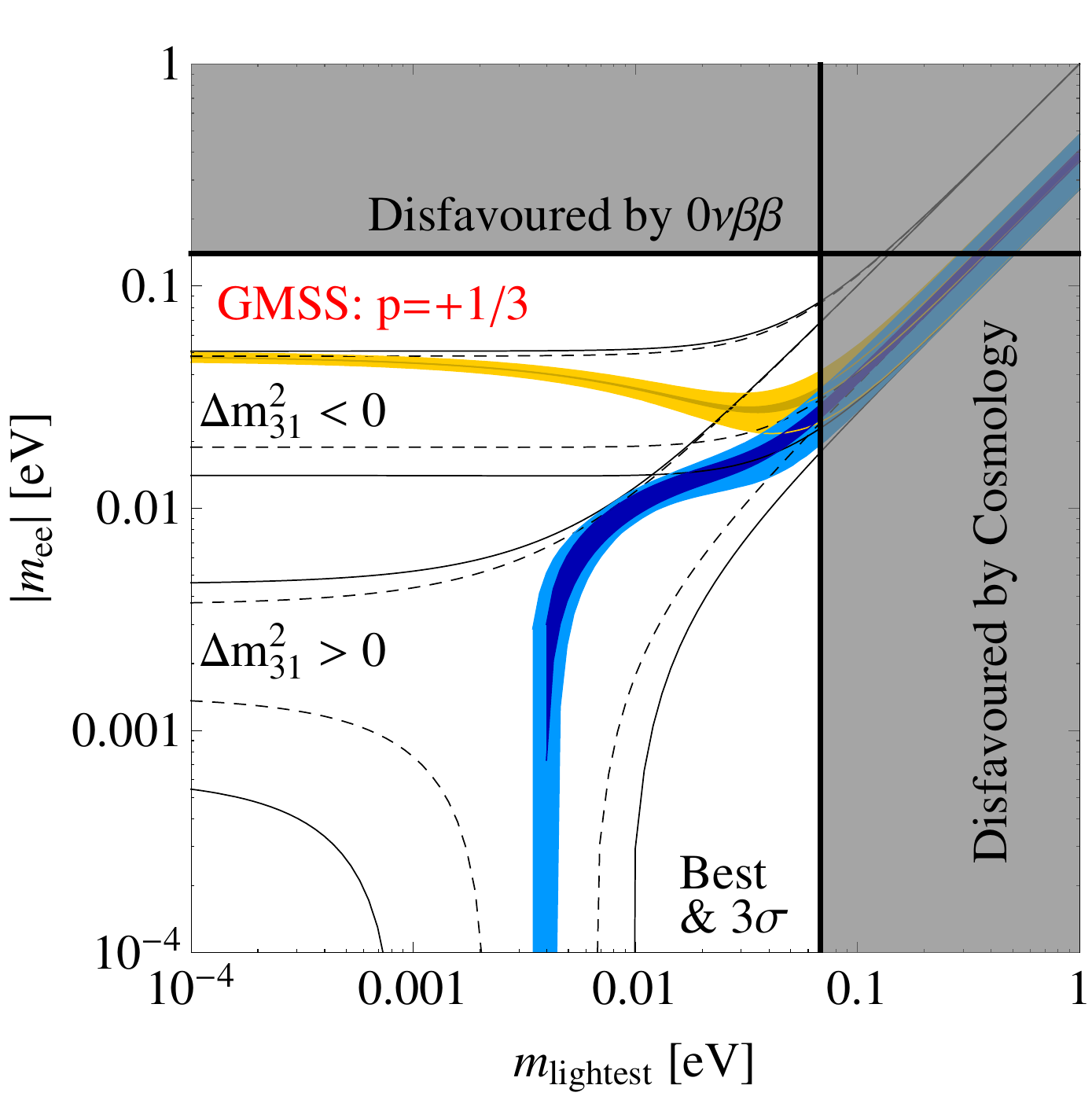}\\
\includegraphics[width=5cm]{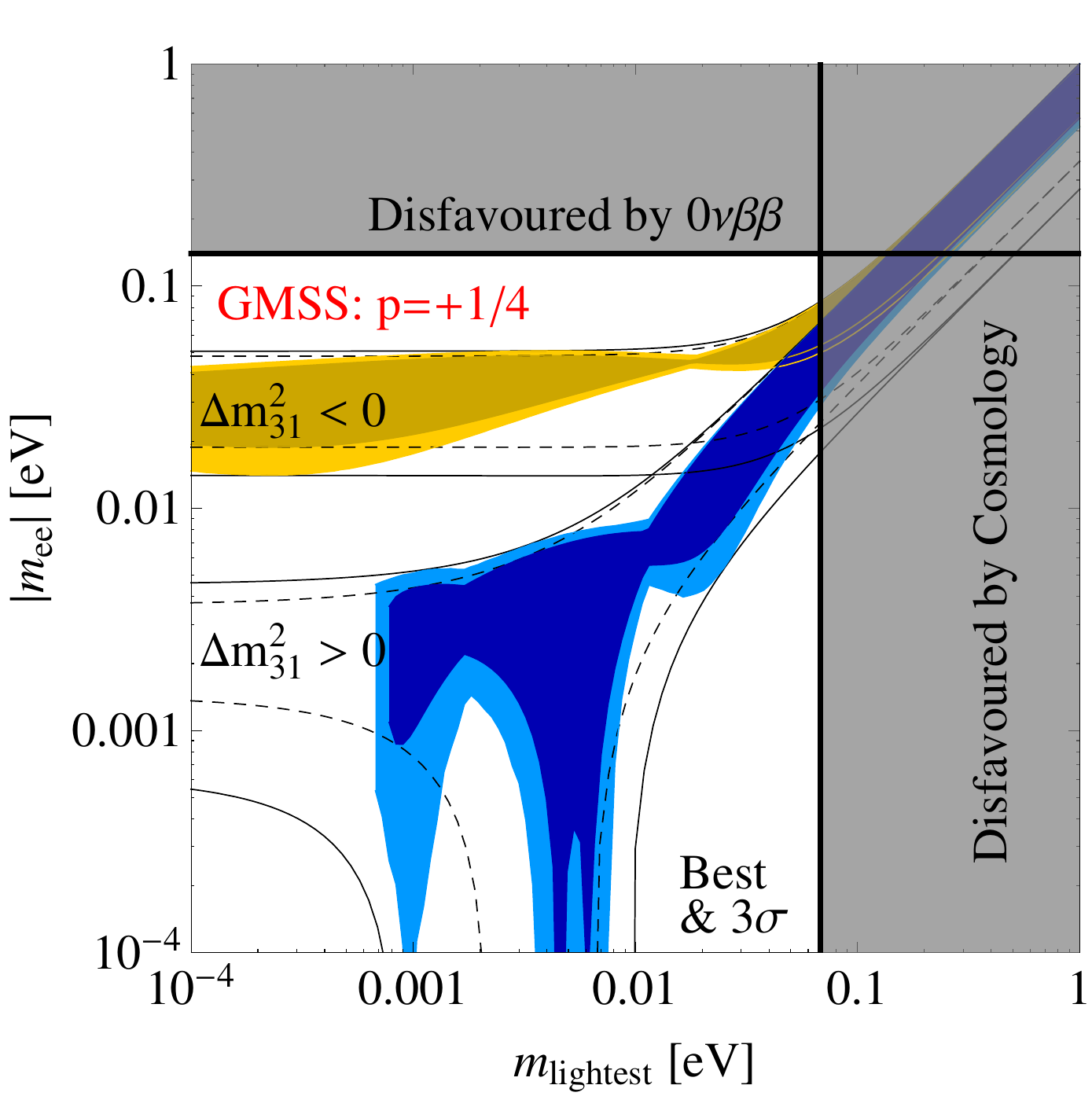} & \includegraphics[width=5cm]{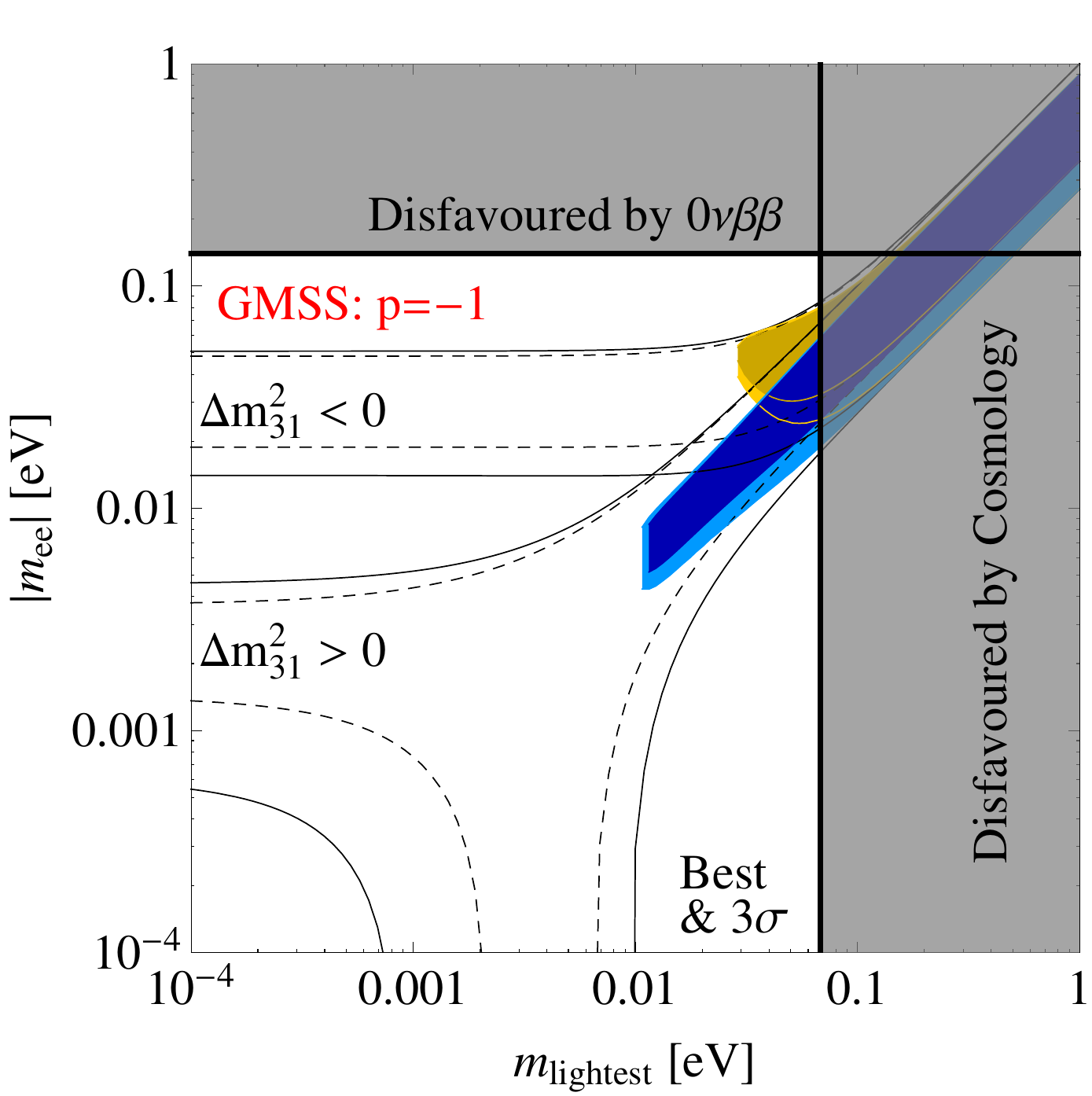} & \includegraphics[width=5cm]{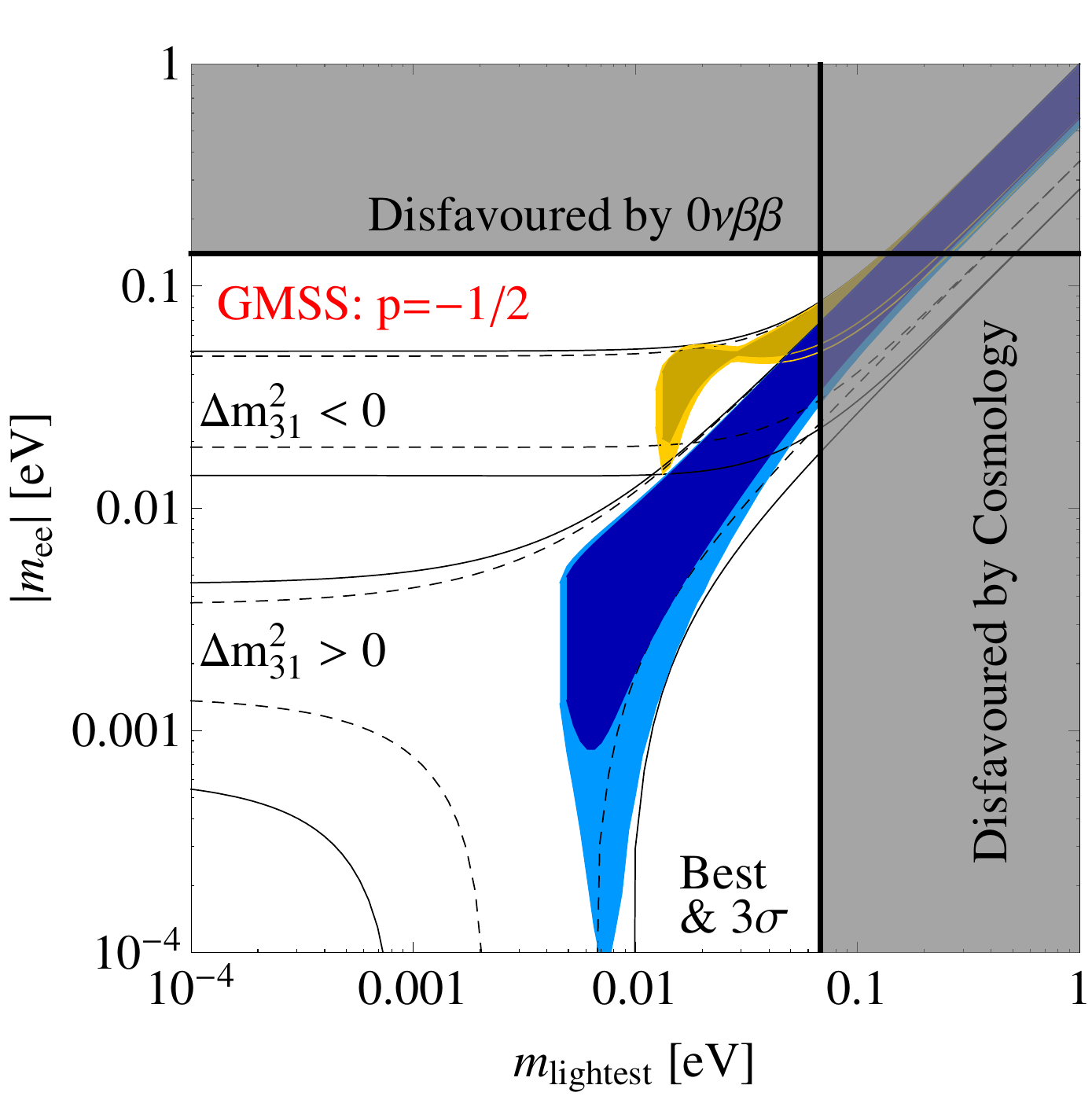}\\
\includegraphics[width=5cm]{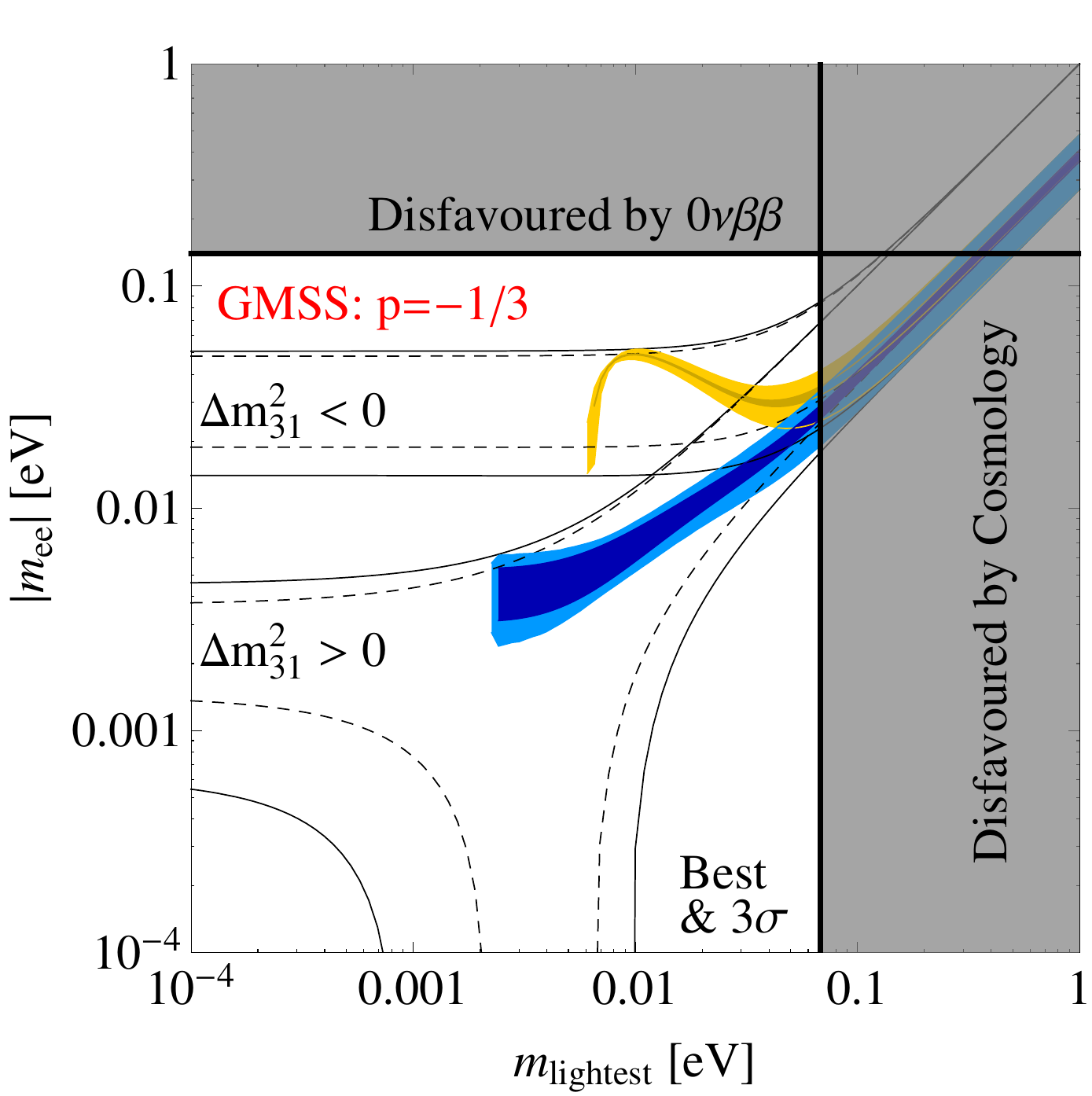} & \includegraphics[width=5cm]{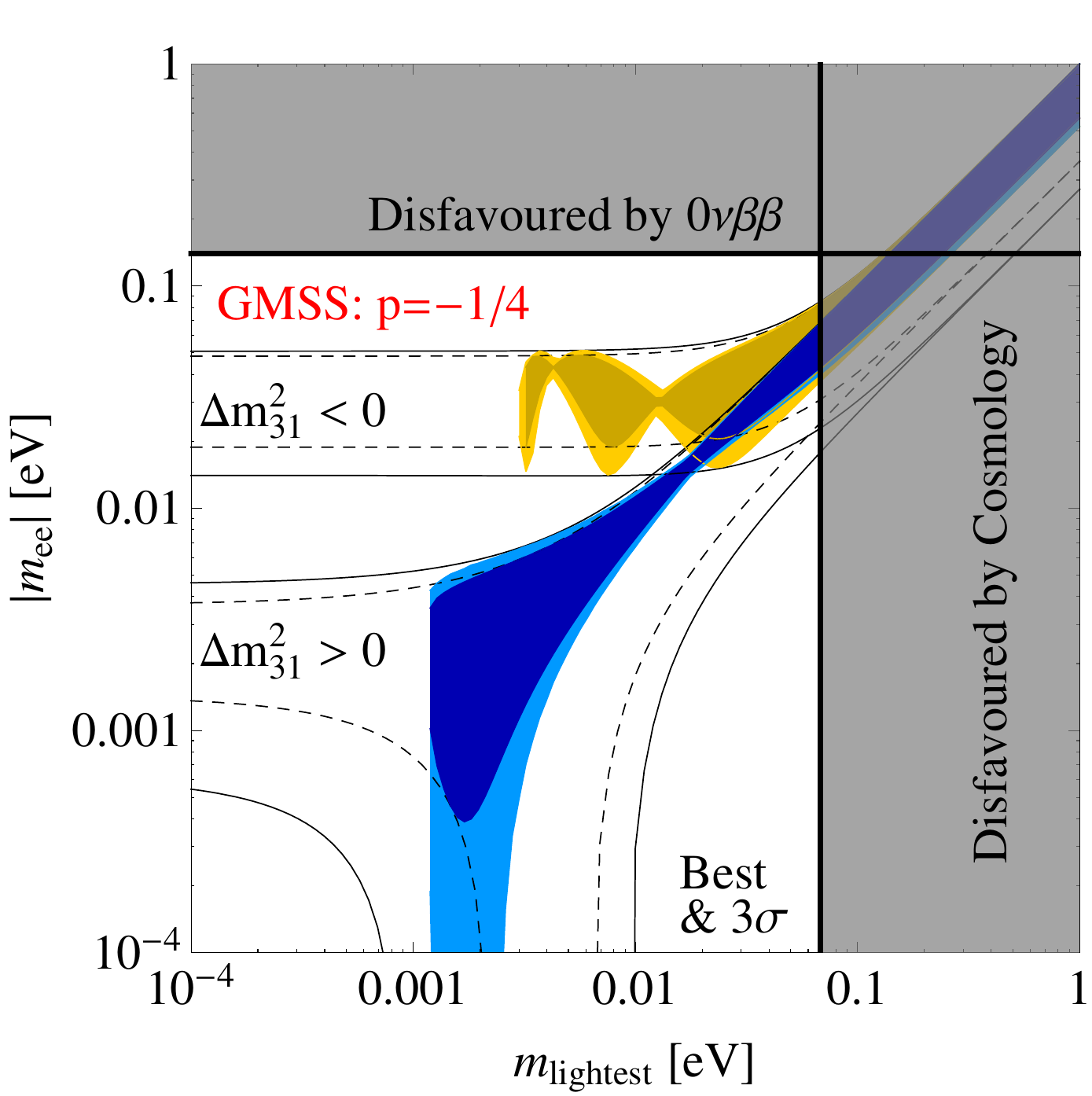} & \includegraphics[width=5cm]{Fig/mee_S.pdf}
\end{tabular}
\caption{\label{fig:Triv_mee_2} Trivial sum rules $ m_1^p + \left( m_2 e^{i \alpha_{21}} \right)^p e^{i \Delta \chi_{21}} + \left( m_3 e^{i \alpha_{31}} \right)^p e^{i \Delta \chi_{31}} = 0$ for $|p| = 1, 1/2, 1/3, 1/4$ with $\Delta \chi_{21} = \pi/3$ and $\Delta \chi_{31} = \pi/5$, compared to the general region.}
\end{figure}
As an alternative example, we have also plotted the signatures for the case $\Delta \chi_{21} = \pi/3$ and $\Delta \chi_{31} = \pi/5$ in Fig.~\ref{fig:Triv_mee_2}, again for $|p| = 1/4$, $1/3$, $1/2$, $1$ and for the GMSS-fit. The resulting signatures are equally characteristic, which shows explicitly that the exact values of the parameters $\Delta \chi_{21}$ and $\Delta \chi_{31}$ do play an important role. In particular, they can considerably shift and/or broaden the allowed ranges of $|m_{ee}|$ for a given value of the smallest neutrino mass eigenvalue $m_{\rm lightest}$, cf.\ the plots for $p=+1/4$ or $p=-1/2$. However, what is not affected by the values of $\Delta \chi_{21}$ and $\Delta \chi_{31}$ is the smallest allowed value of the mass $m_{\rm lightest}$. This is intuitively clear, since this is determined by the lengths of the sides of the triangle, which do not change by varying $\Delta \chi_{21}$ and $\Delta \chi_{31}$.
\begin{figure}[!t]
\centering
\begin{tabular}{lr}
\includegraphics[width=5cm]{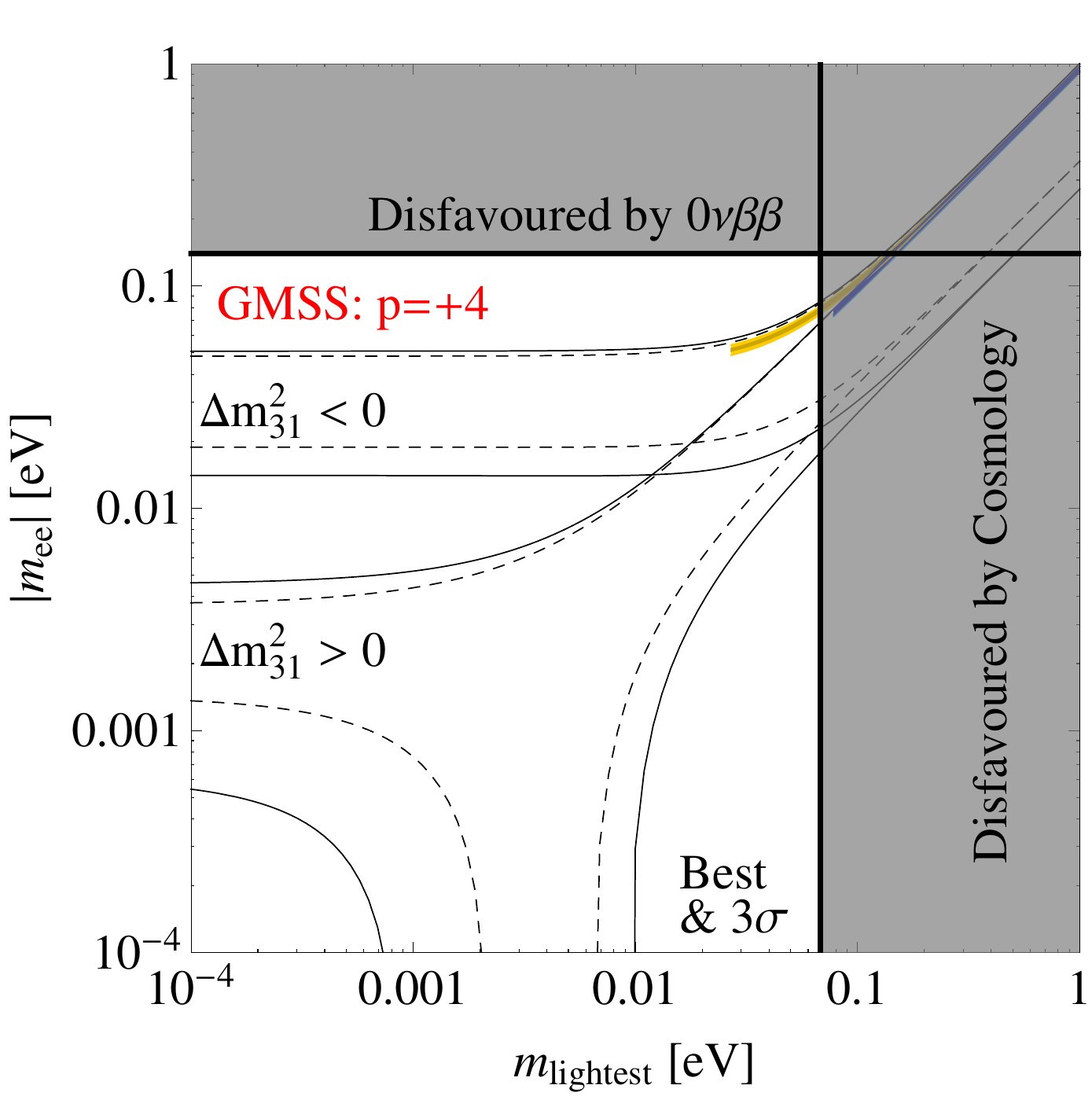} & \includegraphics[width=5cm]{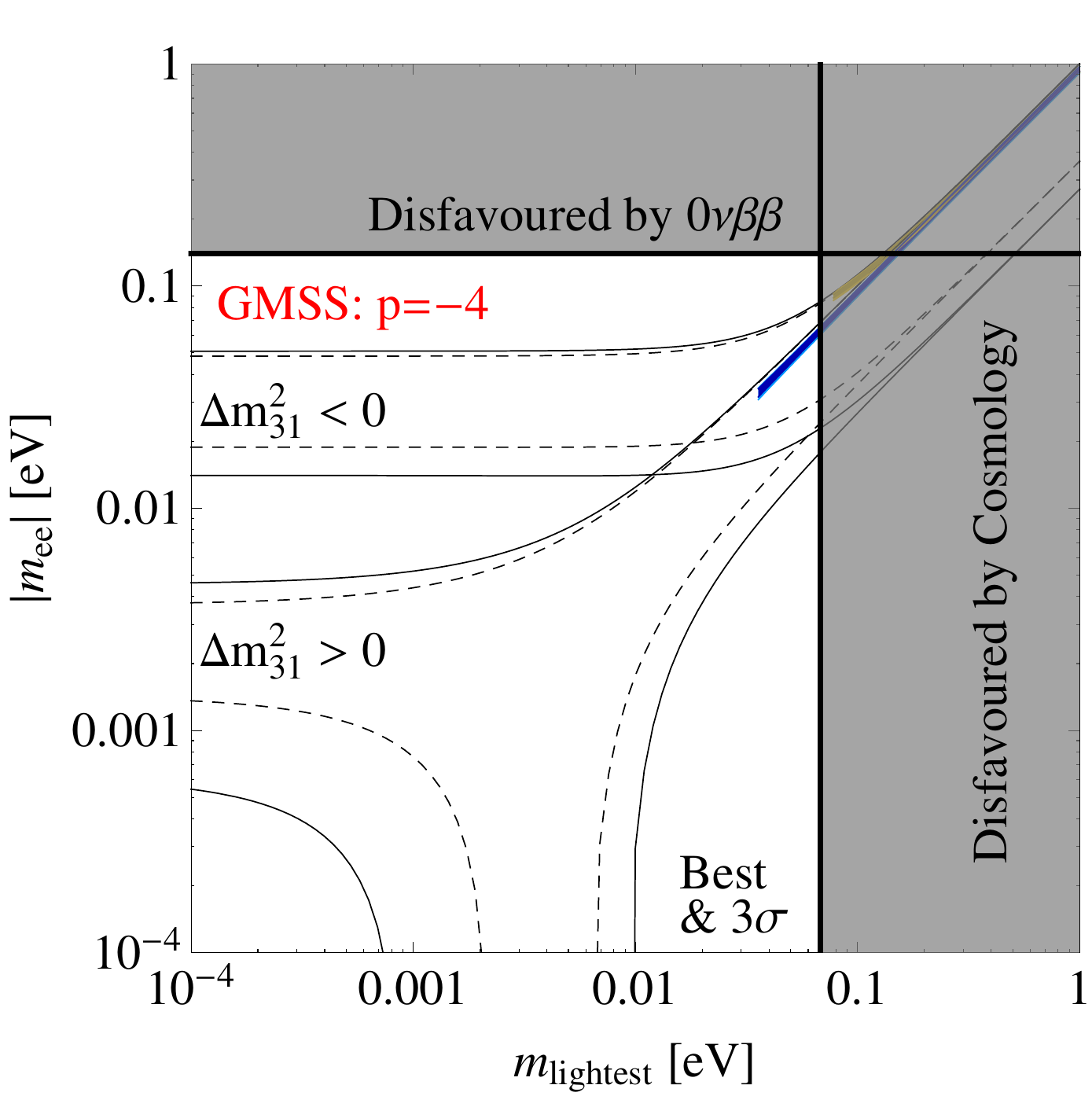}\\
\includegraphics[width=5cm]{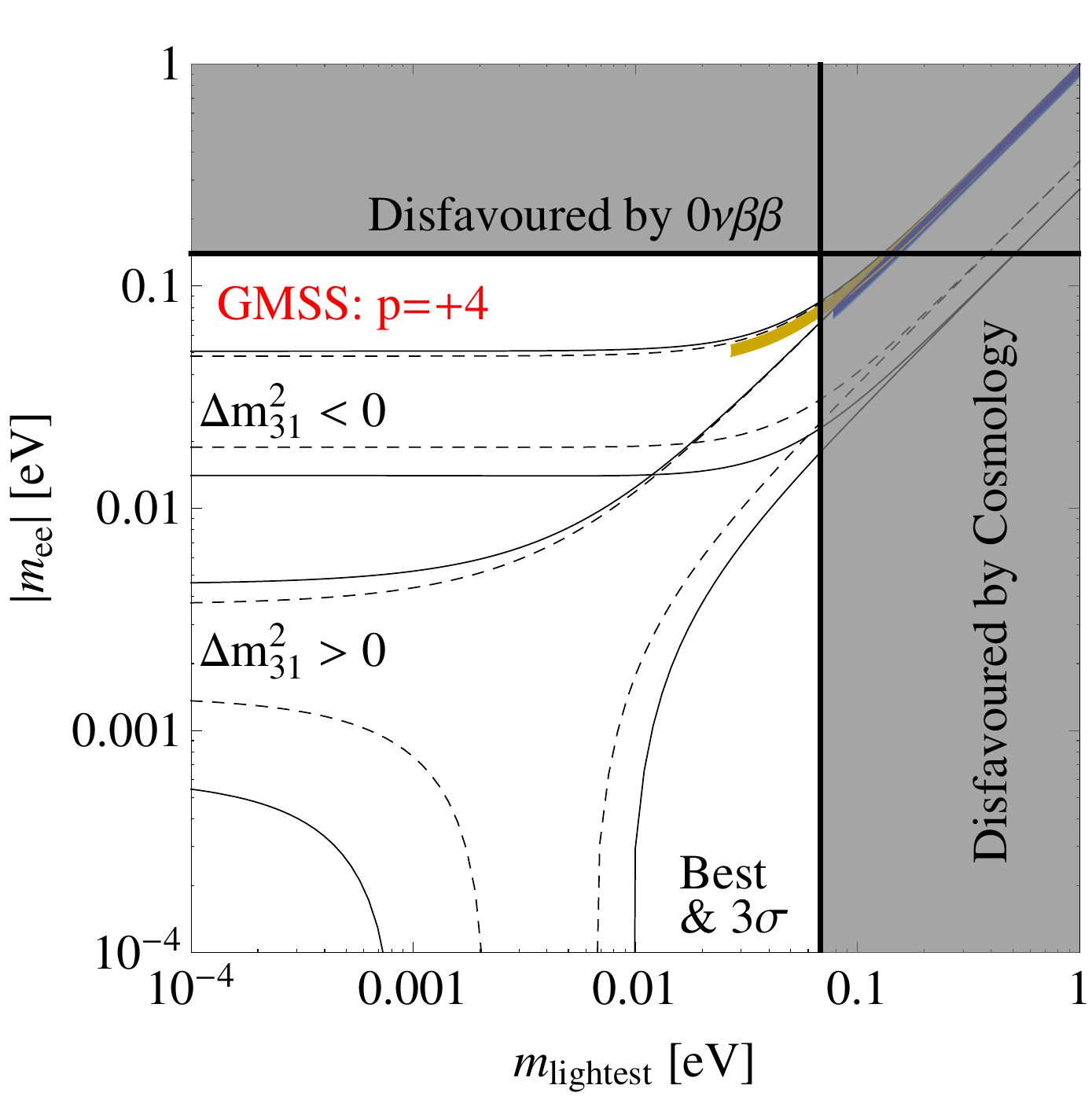} & \includegraphics[width=5cm]{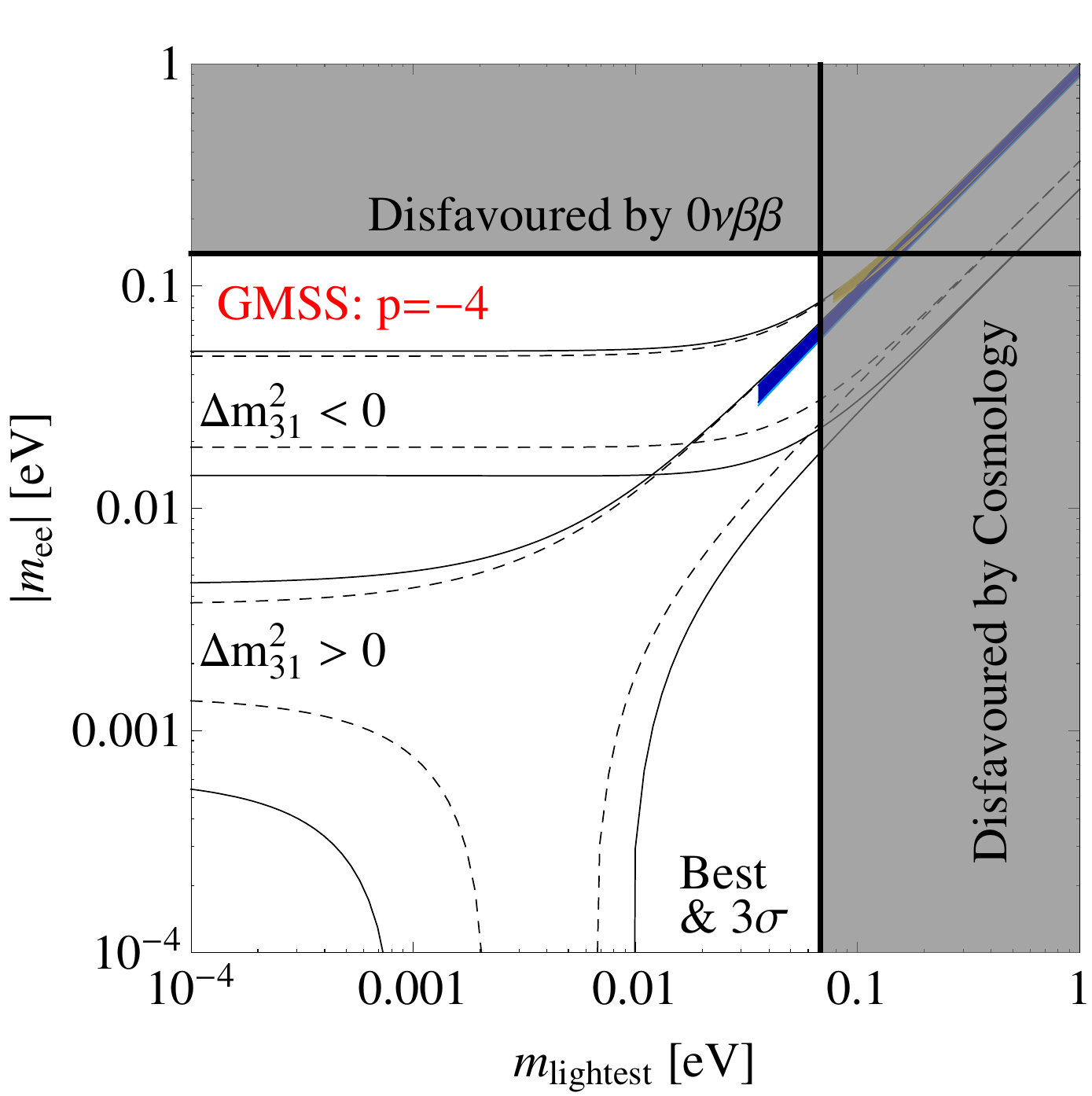}
\end{tabular}
\caption{\label{fig:Triv_mee_4thorder} Trivial sum rules $ m_1^p + \left( m_2 e^{i \alpha_{21}} \right)^p e^{i \Delta \chi_{21}} + \left( m_3 e^{i \alpha_{31}} \right)^p e^{i \Delta \chi_{31}} = 0$ for $|p|=4$. The sum rules start to get into tension with the bound from Planck.}
\end{figure}

Finally, we note that there is an interesting upper ``bound'' on the power $p$. If we leave $p$ free, and simply try to investigate different values from a phenomenological point of view, we start getting into conflict with the cosmological bound on the neutrino mass scale for $|p| \approx 4$. An illustration of this statement is given in Fig.~\ref{fig:Triv_mee_4thorder}, where we can see that, for the trivial sum rules, $p=+4$ ($p=-4$) is strongly disfavoured by the cosmological limit for the case of NO (IO).

\section{\label{sec:non-trivial}Non-trivial sum rules}

We now turn to non-trivial sum rules, where $A_{1,2,3} \neq 1$ in general. Then, the \emph{most general non-trivial sum rule} can be written as
\begin{equation}
 A_1 m_1^p e^{i\tilde \phi_1} + A_2 m_2^p e^{i\tilde \phi_2} + A_3 m_3^p e^{i\tilde \phi_3} = 0.
 \label{eq:NT_rule_1}
\end{equation}
Very similar steps as the ones applied in Sec.~\ref{sec:trivial} lead to
\begin{equation}
 m_1^p + B_2 \left( m_2 e^{i \alpha_{21}} \right)^p e^{i \Delta \chi_{21}} + B_3 \left( m_3 e^{i \alpha_{31}} \right)^p e^{i \Delta \chi_{31}} = 0,
 \label{eq:NT_rule_2}
\end{equation}
where $B_i \equiv A_i / A_1$ for $i=2,3$. Again one can interpret this equation geometrically~\cite{Barry:2010zk,Dorame:2011eb}:
\begin{center}
\includegraphics[width=9cm]{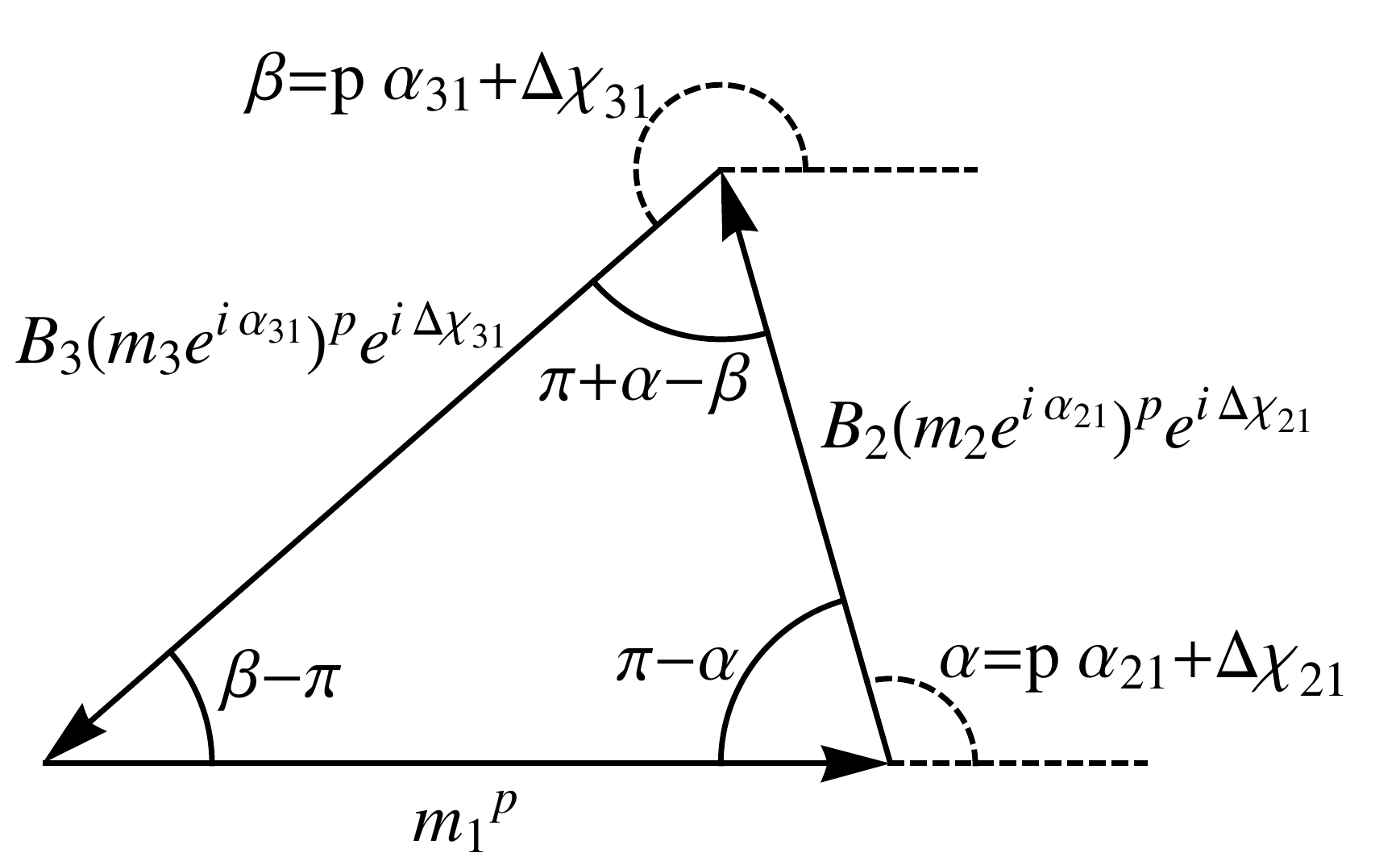}
\end{center}
Then, just as in Eq.~\eqref{eq:rule_4}, one can compute the two angles $\alpha$ and $\beta$ on terms of the parameters of the sum rule,
\begin{equation}
 \cos \alpha = \frac{B_3^2 m_3^{2 p} - B_2^2 m_2^{2 p} - m_1^{2 p}}{2 B_2 (m_1 m_2)^p}\ \ \ , \ \ \ \cos \beta = \frac{B_2^2 m_2^{2 p} - B_3^2 m_3^{2 p} - m_1^{2 p}}{2 B_3 (m_1 m_3)^p}.
 \label{eq:NT_rule_3}
\end{equation}
While these equations look very similar to the ones obtained for trivial sum rules, they actually do imply a very important physical difference: depending on the values of $B_2$ and $B_3$, \emph{non-trivial sum rules may only work for a certain mass ordering}, at least if taken to hold exactly. The decisive point is that NO fundamentally implies that $m_1 < m_2 < m_3$, while IO implies $m_3 < m_1 < m_2$. These relations hold true even in the quasi-degenerate (QD) limit, in which the smallest neutrino mass is much larger than the scales implied by the mass-square differences, $m_{\rm lightest} \gg \sqrt{\Delta m_{A, \odot}}$.

A particularly easy example, to be discussed once more later on in Sec.~\ref{sec:concrete_inverse}, is the sum rule derived in Ref.~\cite{Dorame:2012zv}:
\begin{equation}
 \frac{1}{\sqrt{\tilde m_1}} + \frac{1}{\sqrt{\tilde m_2}} - \frac{2}{\sqrt{\tilde m_3}} = 0,
 \label{eq:Inverse_1}
\end{equation}
which yields $p=-1/2$, $B_2 = 1$, $B_3 = 2$, $\Delta \chi_{21} = 0$, and $\Delta \chi_{31} = \pi$ if compared to Eqs.~\eqref{eq:NT_rule_1} and~\eqref{eq:NT_rule_2}. However, one can easily rewrite Eq.~\eqref{eq:Inverse_1} as
\begin{equation}
 \frac{1}{\sqrt{\tilde m_1}} + \frac{1}{\sqrt{\tilde m_2}} = \frac{2}{\sqrt{\tilde m_3}}.
 \label{eq:Inverse_2}
\end{equation}
Then, the absolute value of the left-hand side of this equation can, in the case of IO ($m_3 < m_1 < m_2$), be estimated as:
\begin{eqnarray}
 |{\rm LHS}| &=& | \frac{1}{\sqrt{\tilde m_1}} + \frac{1}{\sqrt{\tilde m_2}} | = | \frac{1}{\sqrt{m_1}} + \frac{e^{i (\phi_1 - \phi_2)/2}}{\sqrt{m_2}} | \leq \frac{1}{\sqrt{m_1}} + \frac{1}{\sqrt{m_2}}  \nonumber\\
 &<& \frac{1}{\sqrt{m_3}} + \frac{1}{\sqrt{m_3}} = \frac{2}{\sqrt{m_3}} = |{\rm RHS}|.
 \label{eq:Inverse_3}
\end{eqnarray}
Here, the first inequality follows from the possible variation in the phase $(\phi_1 - \phi_2)/2$, while the second inequality simply follows from $m_3 < m_{1,2}$, which implies $\sqrt{m_3} < \sqrt{m_{1,2}}$ and hence $1/\sqrt{m_{1,2}} < 1/\sqrt{m_3}$. Thus, this sum rule can never be fulfilled for IO, even in the QD limit $m_{\rm lightest} \gg \sqrt{\Delta m_{A, \odot}}$, where IO and NO are nearly indistinguishable.

However, one has to note that, in many models, sum rules do only hold to a certain order and it is a priori not clear if they are true beyond that. Some models do yield sum rules also at higher orders (see, e.g., Ref.~\cite{Cooper:2012bd} and the discussion in Sec.~\ref{sec:sumrules}), but it could also be that a sum rule is violated at a certain level (e.g.\ 1\%).\footnote{Note, however, that there is actually no unique way to define the violation of a sum rule by a certain percentage.} In such a case, in particular the QD limit is likely to look very similar for both orderings. Such violations of sum rules at a certain level have been studied in Ref.~\cite{Barry:2010zk}. While in this paper we take on the viewpoint that the sum rule is ``god-given'' and never violated, one has to keep in mind that this is not necessarily true in any context. Hence, our plots show the \emph{maximally possible} phenomenological constraints imposed by sum rules, i.e., a setting where sum rules are violated to some extent would yield larger allowed regions than our plots. This means that, in turn, for a given sum rule no stronger prediction than the one presented here can be obtained, as long as no additional external information (such as, e.g., an experimental determination of the Dirac phase $\delta$) is taken into account. Also if a given model leads to tighter relations between at least some of the parameters involved (as, e.g., relations between different mixing angles which enter $|m_{ee}|$), the resulting prediction could be even stronger.

Note that in the literature one can find examples for both, trivial and non-trivial sum rules. Indeed, it turns out for all the examples we have found, NO and IO are both possible for realistic trivial sum rules. For the non-trivial ones, in turn, some ordering may or may not be allowed, depending on the values of $B_2$ and $B_3$.

\section{\label{sec:concrete}Sum rules in concrete models}

In this section, we discuss several sum rules in various detail. We thereby do not discuss every detail for every sum rule, but rather try to give several illustrative examples for the different features which can appear. This does not imply any ranking of the sum rules, and in particular not of the underlying models. At the end, in Sec.~\ref{sec:concrete_summary}, we give a classification and a summary of all the sum rules we have found in the literature.

\subsection{\label{sec:concrete_golden}The sum rule $\frac{1}{\tilde m_1} + \frac{1}{\tilde m_2} = \frac{1}{\tilde m_3}$}

This sum rule has been found in models based on several symmetries such as $A_4$~\cite{Barry:2010zk}, $S_4$~\cite{Bazzocchi:2009da,Ding:2010pc}, or $A_5$~\cite{Ding:2011cm,Cooper:2012bd}. Even though, as for most sum rules, several possibilities are known to derive them, for this sum rule in particular Ref.~\cite{Cooper:2012bd} is worth mentioning, since in that model the sum rules holds to next-to-leading order (cf.\ discussion in Sec.~\ref{sec:sumrules}). The rule has already been mentioned in Eq.~\eqref{eq:Golden_Rule}, and it is given by:
\begin{equation}
 \frac{1}{\tilde m_1} + \frac{1}{\tilde m_2} = \frac{1}{\tilde m_3}.
 \label{eq:Golden_1}
\end{equation}
Comparing Eq.~\eqref{eq:Golden_1} to Eq.~\eqref{eq:gen_rule_2}, one obtains
\begin{equation}
 p=-1,\ \ B_2 = B_3 = 1,\ \ \Delta \chi_{21} = 0,\ \ {\rm and}\ \ \Delta \chi_{31} = \pi.
 \label{eq:Golden_2}
\end{equation}

\begin{center}
\begin{figure}[h!]
\begin{tabular}{lcr}
\includegraphics[width=5cm]{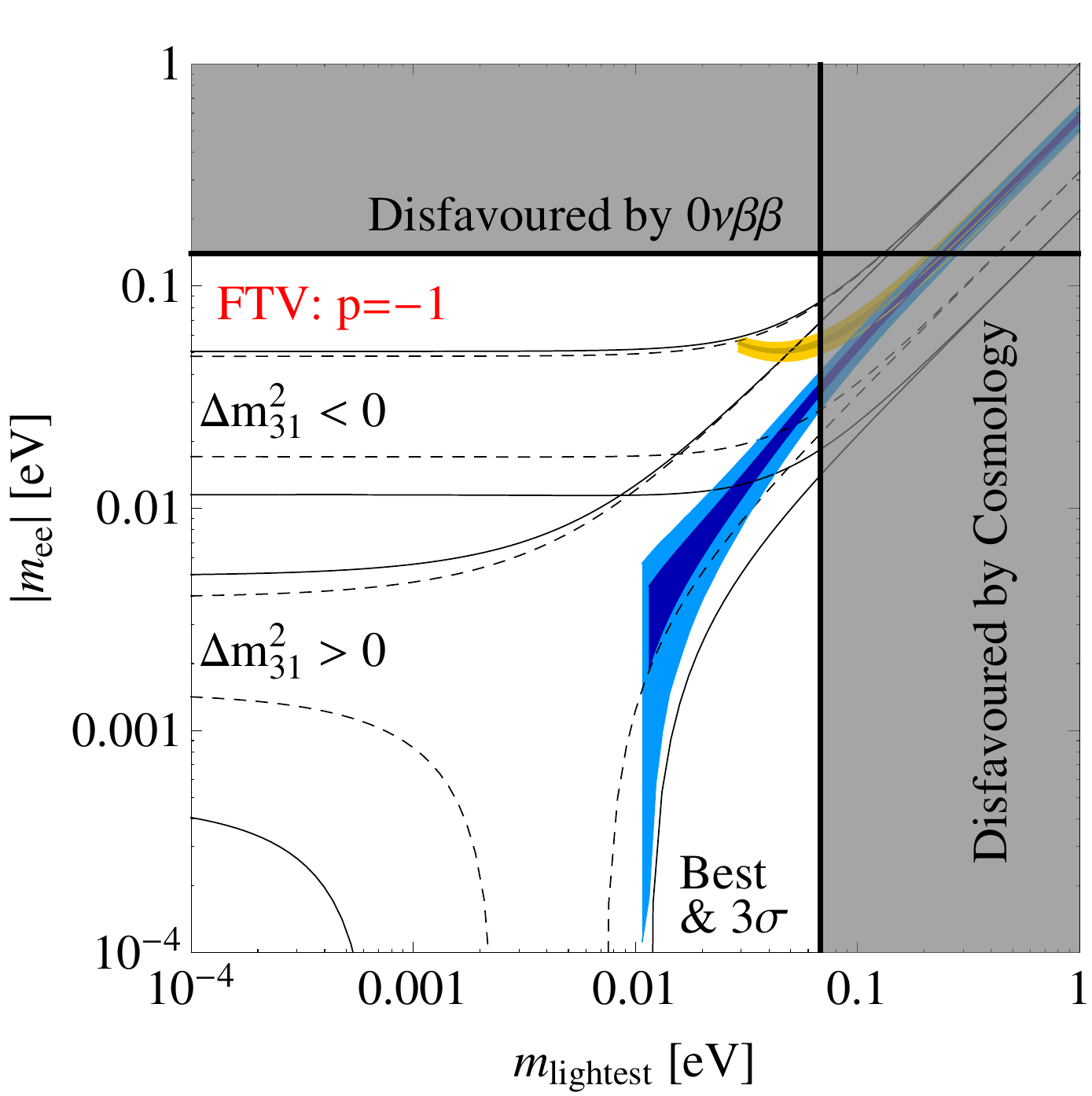} & \includegraphics[width=5cm]{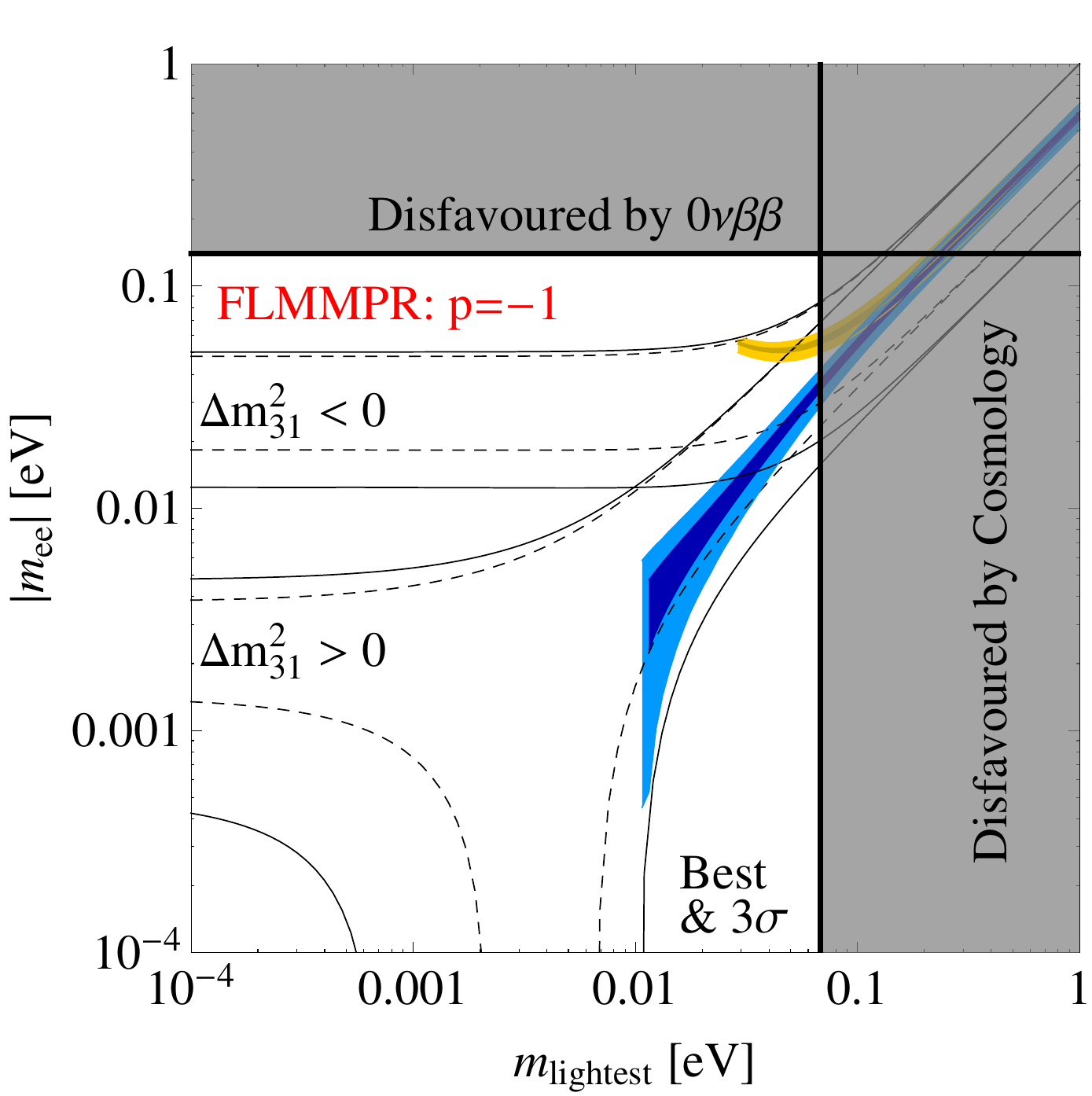} &
\includegraphics[width=5cm]{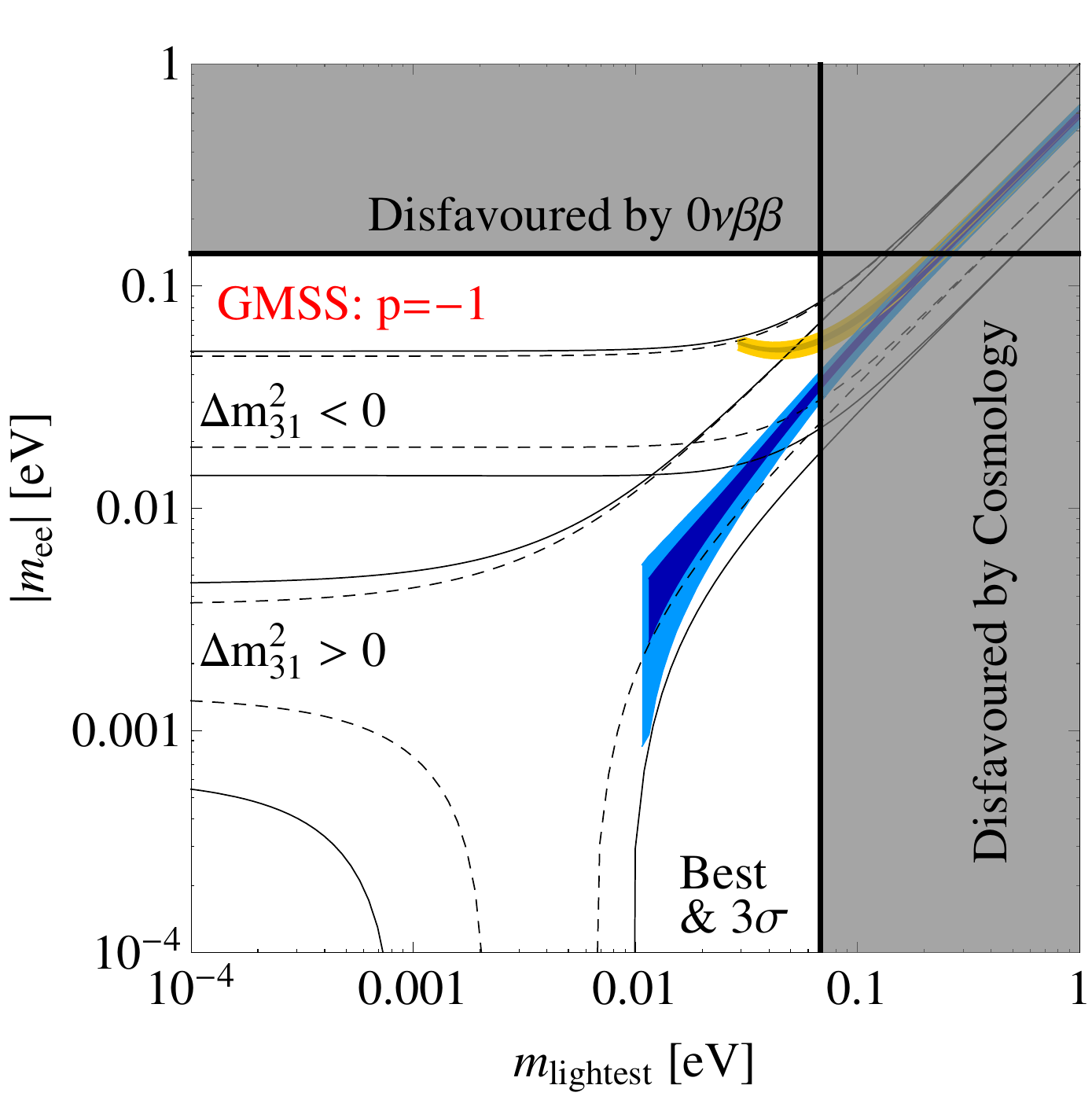}
\end{tabular}
\caption{\label{fig:mee12121066}Allowed regions for the sum rule $\frac{1}{\tilde m_1} + \frac{1}{\tilde m_2} = \frac{1}{\tilde m_3}$.}
\end{figure}
\end{center}

The corresponding allowed regions for the effective mass are displayed in Fig.~\ref{fig:mee12121066}.
Indeed, this plot is in perfect agreement with Fig.~1(a) from Ref.~\cite{Bazzocchi:2009da}, where however the sum rule had not been written down explicitly. Since $B_2 = B_3 = 1$, this is a trivial sum rule and accordingly both orderings should be possible, which is confirmed by our numerical analysis.

\subsection{\label{sec:concrete_Del96}The sum rule $\frac{1}{\tilde m_3} + \frac{2 i (-1)^\eta}{\tilde m_2} = \frac{1}{\tilde m_1}$}

\begin{center}
\begin{figure}[h!]
\begin{tabular}{lcr}
\includegraphics[width=5cm]{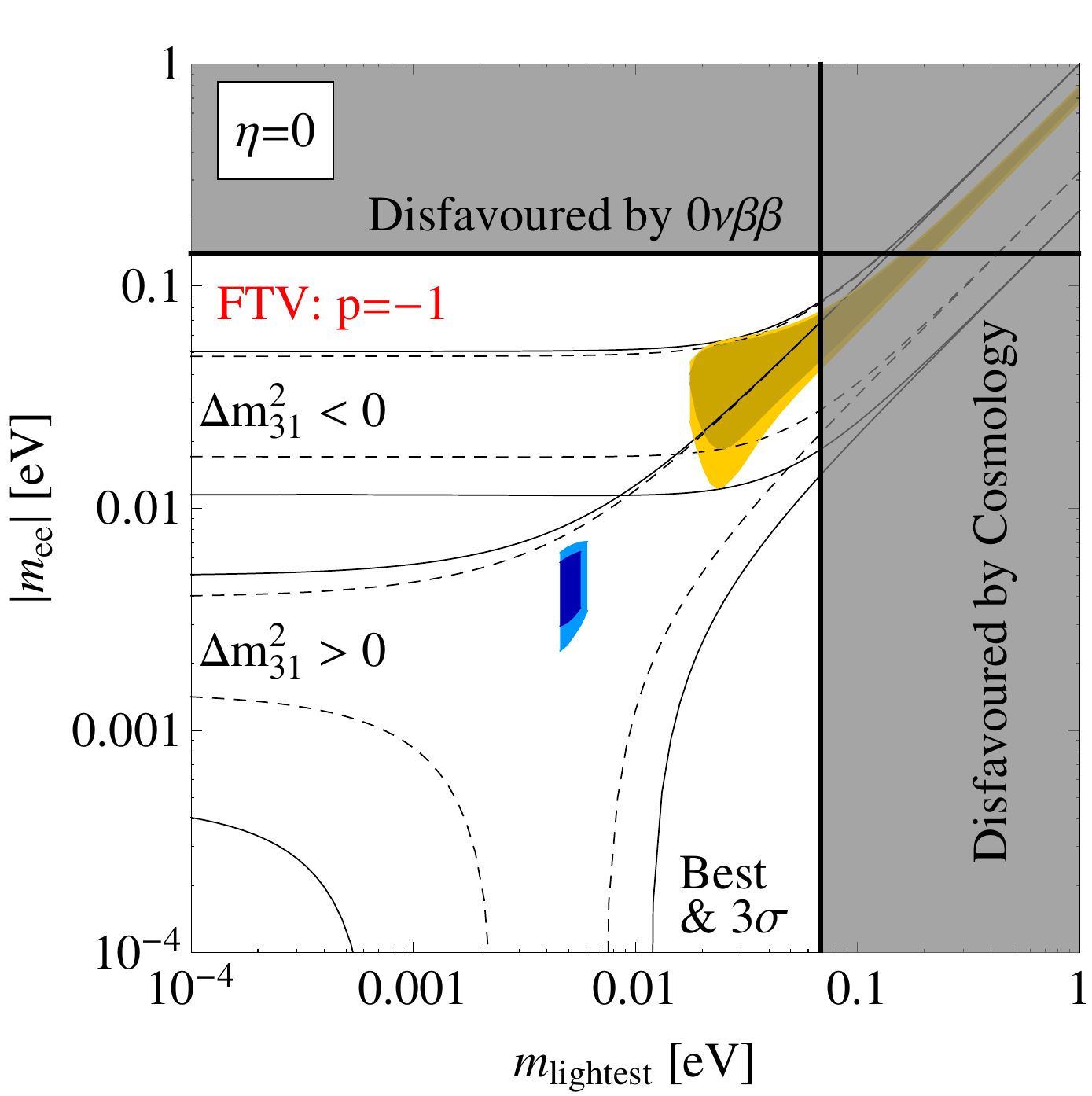} & \includegraphics[width=5cm]{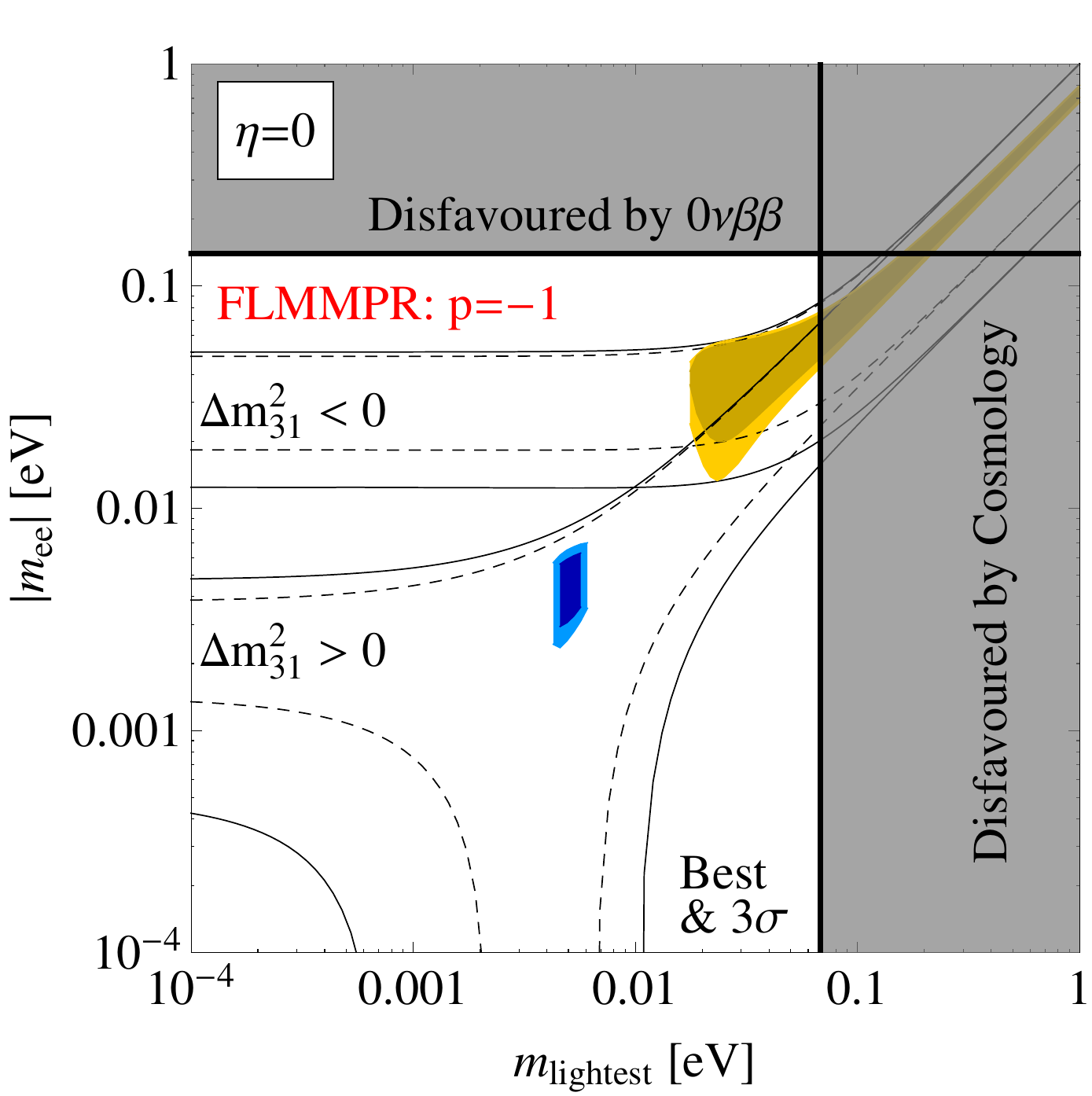} &
\includegraphics[width=5cm]{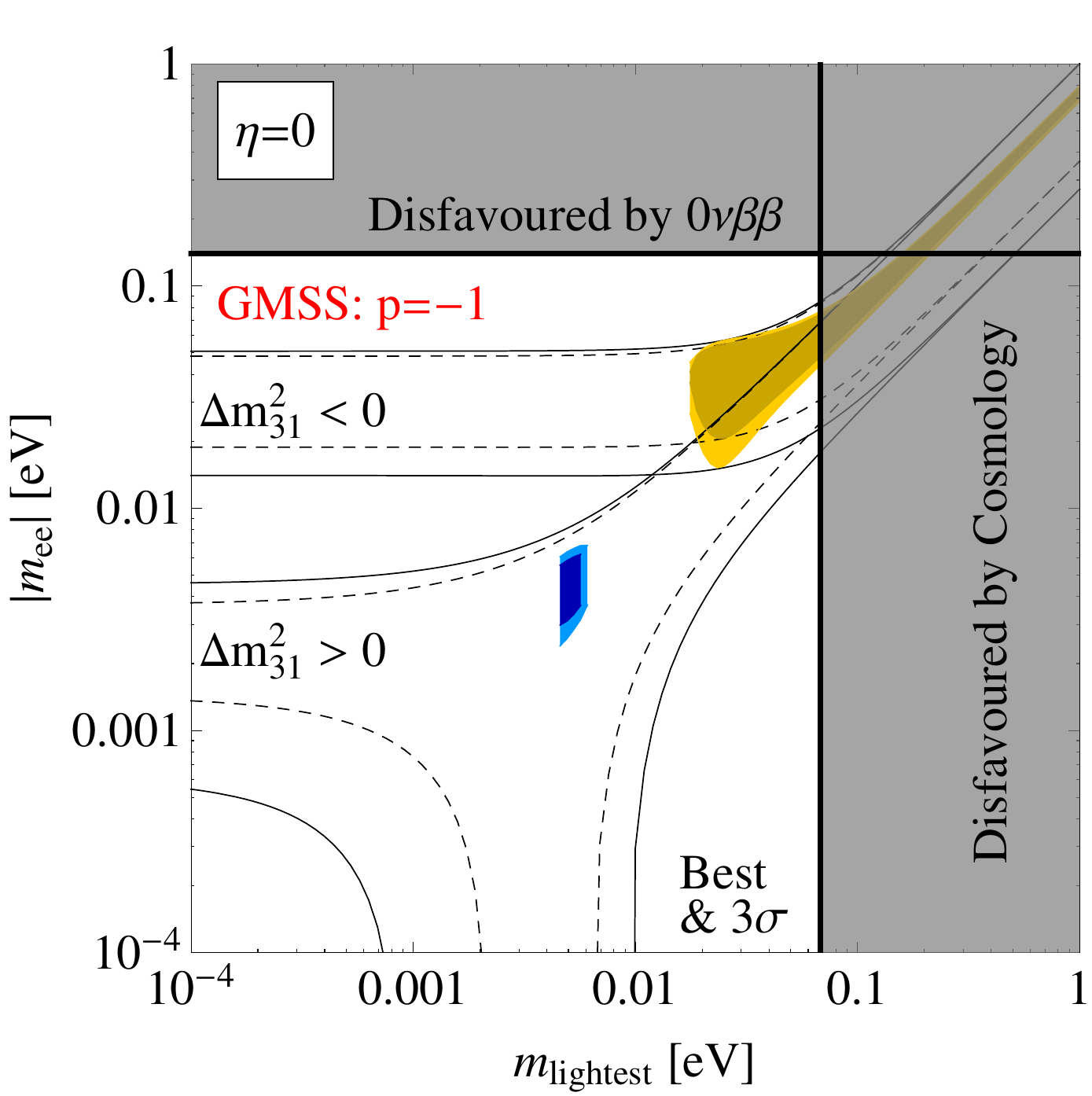}\\
\includegraphics[width=5cm]{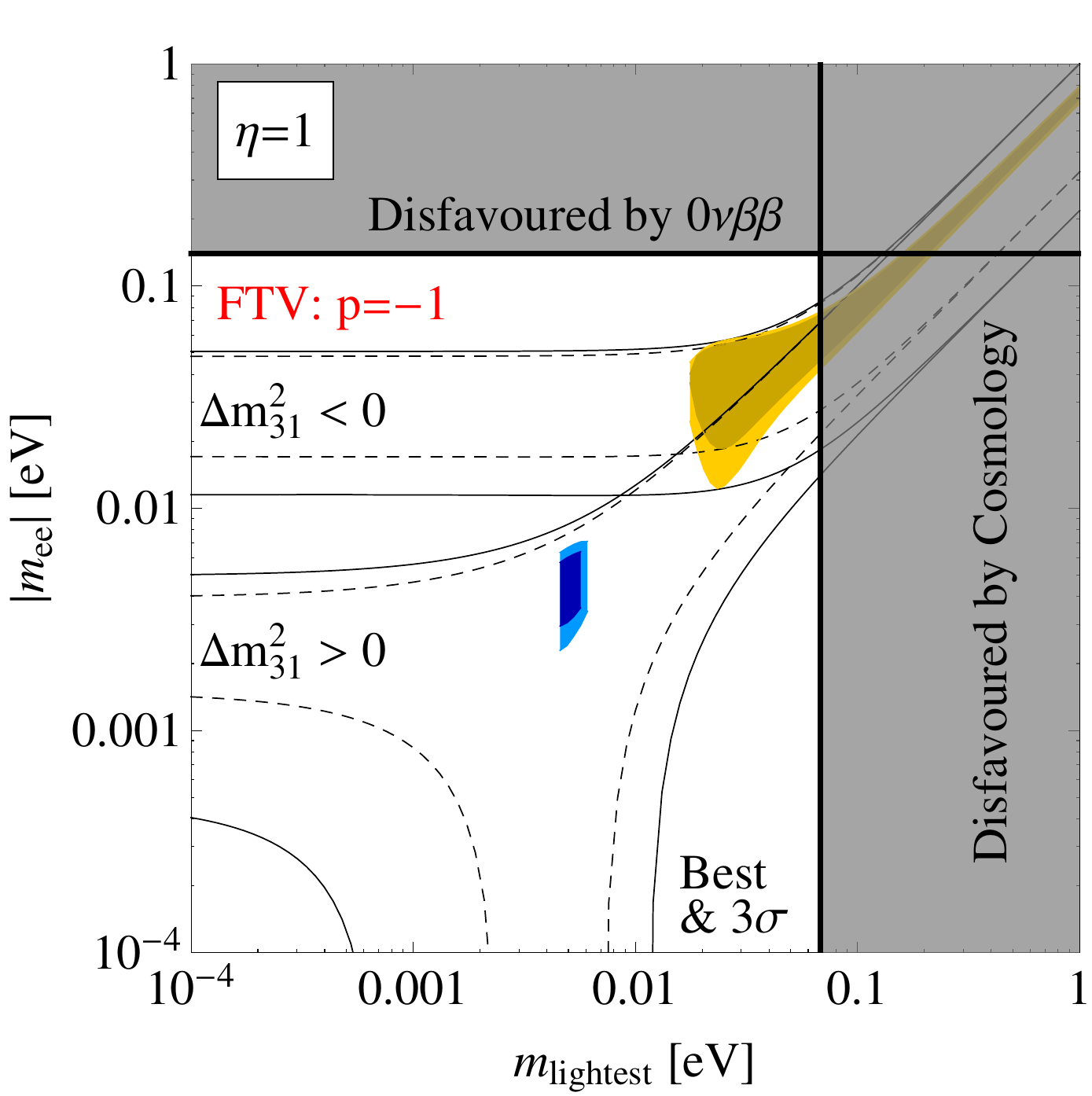} & \includegraphics[width=5cm]{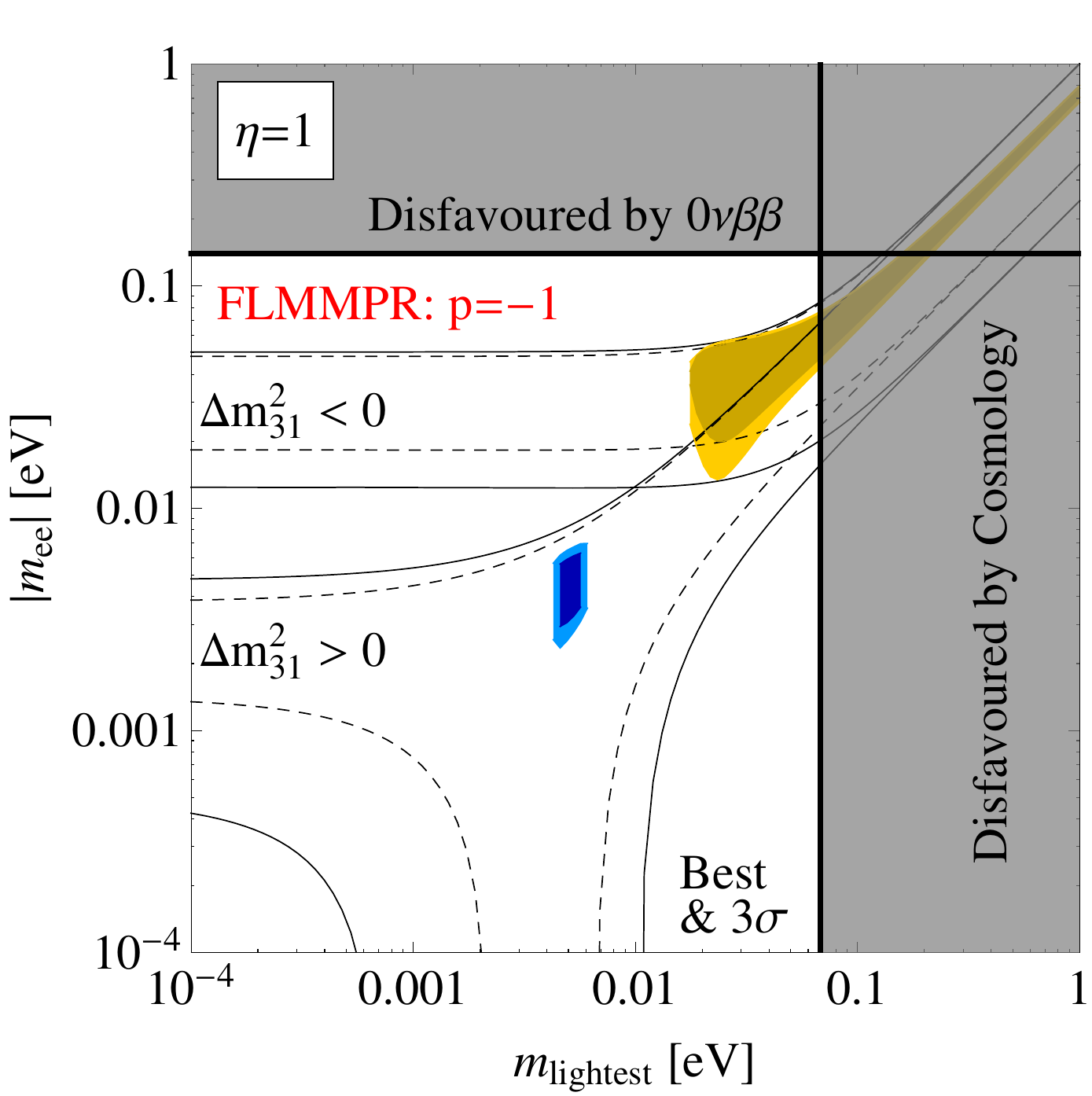} &
\includegraphics[width=5cm]{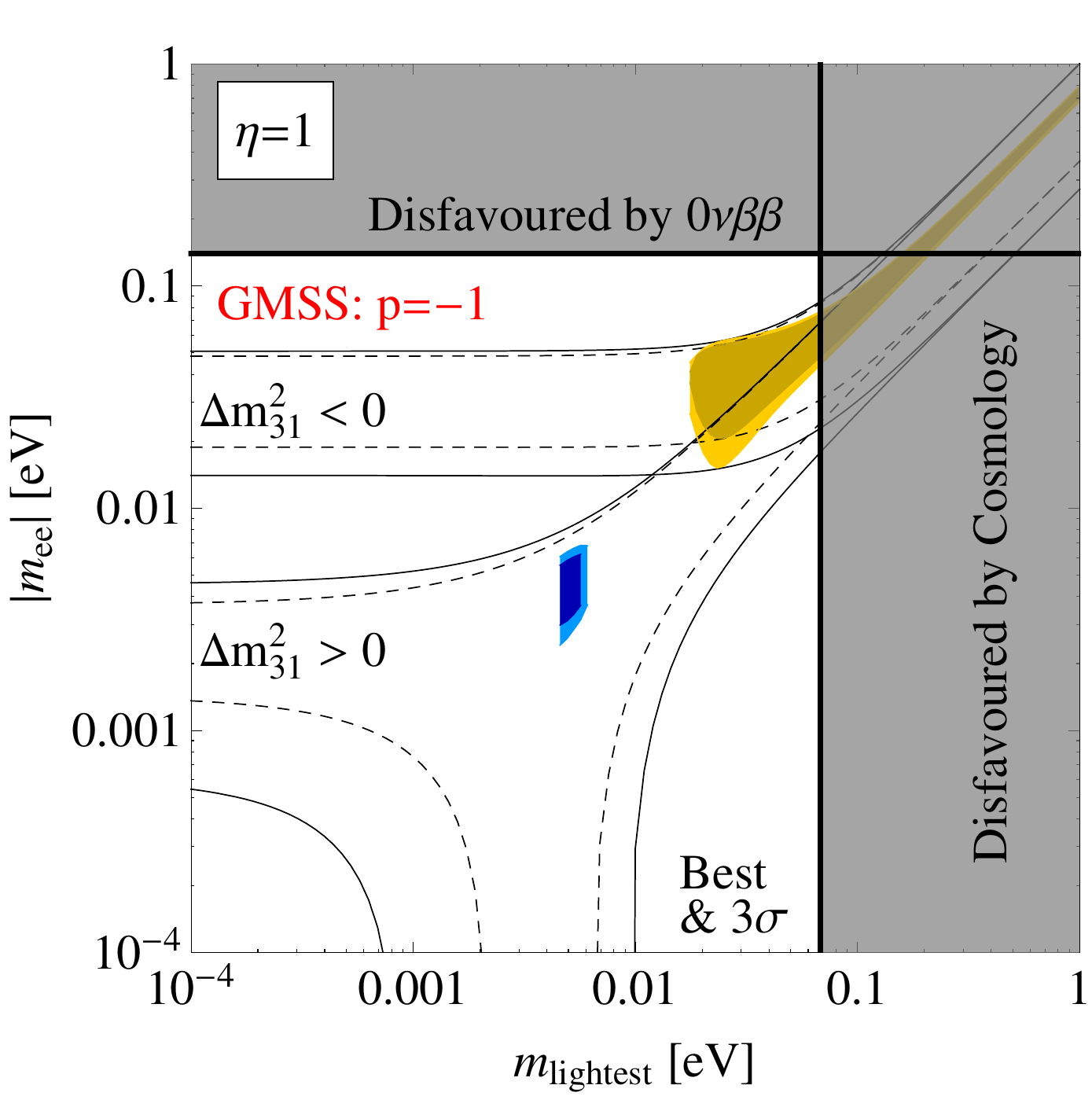}
\end{tabular}
\caption{\label{fig:mee12075741}Allowed regions for the sum rule $\frac{1}{\tilde m_3} + \frac{2 i (-1)^\eta}{\tilde m_2} = \frac{1}{\tilde m_1}$ with $\eta = 0$ (upper row) and $\eta = 1$ (lower row).}
\end{figure}
\end{center}

The model~\cite{King:2012in} leading to the next sum rule is based on an $\Delta (96)$ group. It actually predicts two different sum rules, cf.\ Eq.~(32) of Ref.~\cite{King:2012in}:
\begin{equation}
 \frac{1}{\tilde m_3} + \frac{2 i (-1)^\eta}{\tilde m_2} - \frac{1}{\tilde m_1} = 0,\ \ \ {\rm where}\ \ \ \eta=0,1.
 \label{eq:Del96_1}
\end{equation}
Comparing Eq.~\eqref{eq:Del96_1} to Eq.~\eqref{eq:gen_rule_2}, one obtains
\begin{equation}
 p=-1,\ \ B_2 = 2,\ \  B_3 = 1,\ \ \Delta \chi_{21} = \frac{\pi}{2}, \frac{3 \pi}{2}, \ \ {\rm and}\ \ \Delta \chi_{31} = \pi.
 \label{eq:Del96_2}
\end{equation}
The corresponding allowed regions for the effective mass for the two possible sum rules look like:
The corresponding allowed regions for the effective mass for the two possible sum rules are displayed in Fig.~\ref{fig:mee12075741}.
As can be seen from the plots, the two sum rules are indistinguishable from a phenomenological point of view. This is easy to understand, since the redefinition of $\Delta \chi_{21}$ can alternatively be interpreted as a redefinition of $\alpha_{21}$, which is simply not visible in the resulting value of the effective mass.\footnote{We will later on show how to prove such statements for further example sum rules. For now, however, we first focus on a less subtle property of the sum rules.}

An interesting point to note is that an allowed region for NO only exists for intermediate values of the lightest neutrino mass $m_{\rm lightest}$. This can be understood analytically: abbreviating $\rho_{ij} \equiv m_i / m_j$, one can rewrite the real and imaginary parts of Eq.~\eqref{eq:Del96_1} as
\begin{equation}
 \left\{
 \begin{matrix}
 \hfill \rho_{31} \pm 2 \rho_{32} \sin \alpha_{21} = \cos \alpha_{31}\ ,\\
 \hfill \hfill \hfill \mp 2 \rho_{32} \cos \alpha_{21} = \sin \alpha_{31}\ .
 \end{matrix}
 \right.
 \label{eq:Del96_3}
\end{equation}
Inserting the second equation into the first one, one can easily conclude that
\begin{equation}
 \rho_{31}^2 + 4 \rho_{32}^2 \pm 4 \rho_{31} \rho_{32} \sin \alpha_{21} = 1.
 \label{eq:Del96_4}
\end{equation}
The square root of the left-hand side of this equation can be estimated using the fact that $\sin \alpha_{21} \in [-1, +1]$, while the square root of the right-hand side is always one,
\begin{equation}
 \sqrt{\rm LHS} \in [ |\rho_{31} - 2 \rho_{32}| , \rho_{31} + 2 \rho_{32} ]\ \ \ , \ \ \ \sqrt{\rm RHS} = 1.
 \label{eq:Del96_5}
\end{equation}
Now, for NO we have $\rho_{31} > \rho_{32} > 1$, so that in the hierarchical limit ($m_{\rm lightest} \to 0$) we have $\sqrt{\rm LHS} \to \infty \gg \sqrt{\rm RHS} = 1$ and the equation cannot be fulfilled. In the QD limit, in turn, we know that $\sqrt{\rm LHS} = 2 \rho_{32} - \rho_{31}$, since all $\rho$'s are close to 1, and again using the abbreviations from Eq.~\eqref{eq:abbreviations} we obtain
\begin{equation}
 \rho_{32} = \sqrt{\frac{1+x}{x+\epsilon}}\ \ \ , \ \ \ \rho_{31} = \sqrt{\frac{1+x}{x}}.
 \label{eq:Del96_6}
\end{equation}
The quantity $2 \rho_{32} - \rho_{31}$ can be Taylor expanded in $\epsilon$ to yield
\begin{equation}
 2 \rho_{32} - \rho_{31} \simeq \sqrt{1 + \frac{1}{x}} \left[ 1 - \frac{\epsilon}{x} \right],
 \label{eq:Del96_7}
\end{equation}
whose minimal value tends to $\sqrt{\rm LHS} \simeq \sqrt{1+1}(1-\epsilon) \simeq \sqrt{2} > 1 = \sqrt{\rm RHS}$. Indeed, for NO the sum rule cannot be fulfilled in the QD limit either, and only in between the two limits there is a small region where things work out, as visible in the figure.

For IO, on the other hand, we have $\rho_{32} < \rho_{31} < 1$. In the IH limit we have $m_{\rm lightest} \to 0$ and hence $\rho_{32, 31} \to 0$, which implies $\sqrt{\rm LHS} \to 0 \ll \sqrt{\rm RHS} = 1$. In the QD limit, it is easy to show that $\sqrt{\rm RHS} \in [ \sqrt{\rm LHS}|_{\rm min} , \sqrt{\rm LHS}|_{\rm max} ]$, since
\begin{eqnarray}
 && \sqrt{\rm LHS}|_{\rm min} = 2 \rho_{32} - \rho_{31} = \rho_{32} + \underbrace{(\rho_{32} - \rho_{31})}_{<0} < \rho_{32} - 0 < 1 = \sqrt{\rm RHS}, \ \ \ {\rm and}\nonumber \\
 && \sqrt{\rm LHS}|_{\rm max} = 2 \rho_{32} + \rho_{31} > 2 \rho_{32} + \rho_{32} \simeq 3 > 1= \sqrt{\rm RHS}.
 \label{eq:Del96_8}
\end{eqnarray}
Indeed, the sum rule can be fulfilled in the QD limit in the case where IO is present.

\subsection{\label{sec:concrete_inverse}The sum rule $\frac{1}{\sqrt{\tilde m_1}} = \frac{2}{\sqrt{\tilde m_3}} - \frac{1}{\sqrt{\tilde m_2}}$}

\begin{center}
\begin{figure}[h!]
\begin{tabular}{lcr}
\includegraphics[width=5cm]{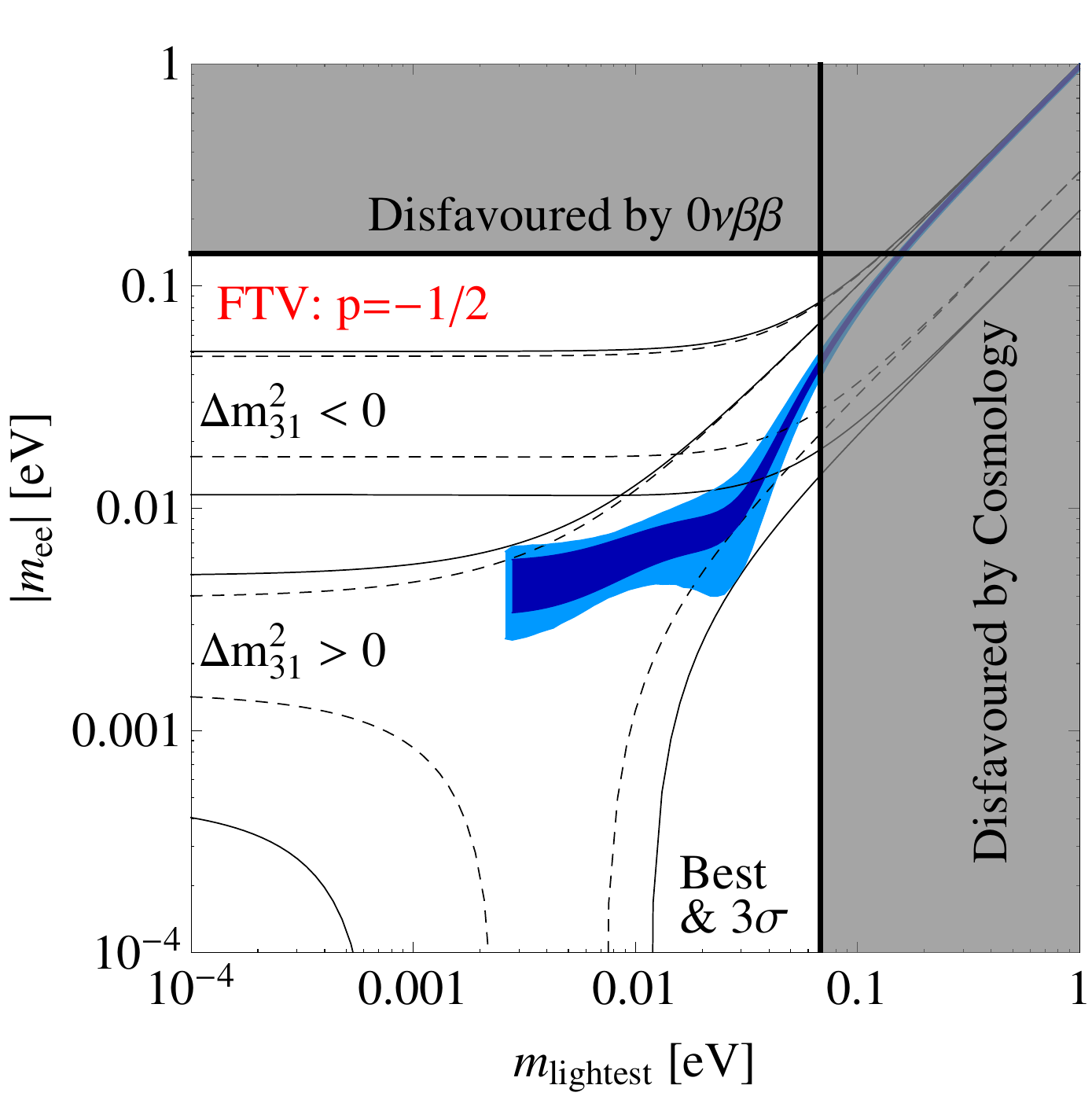} & \includegraphics[width=5cm]{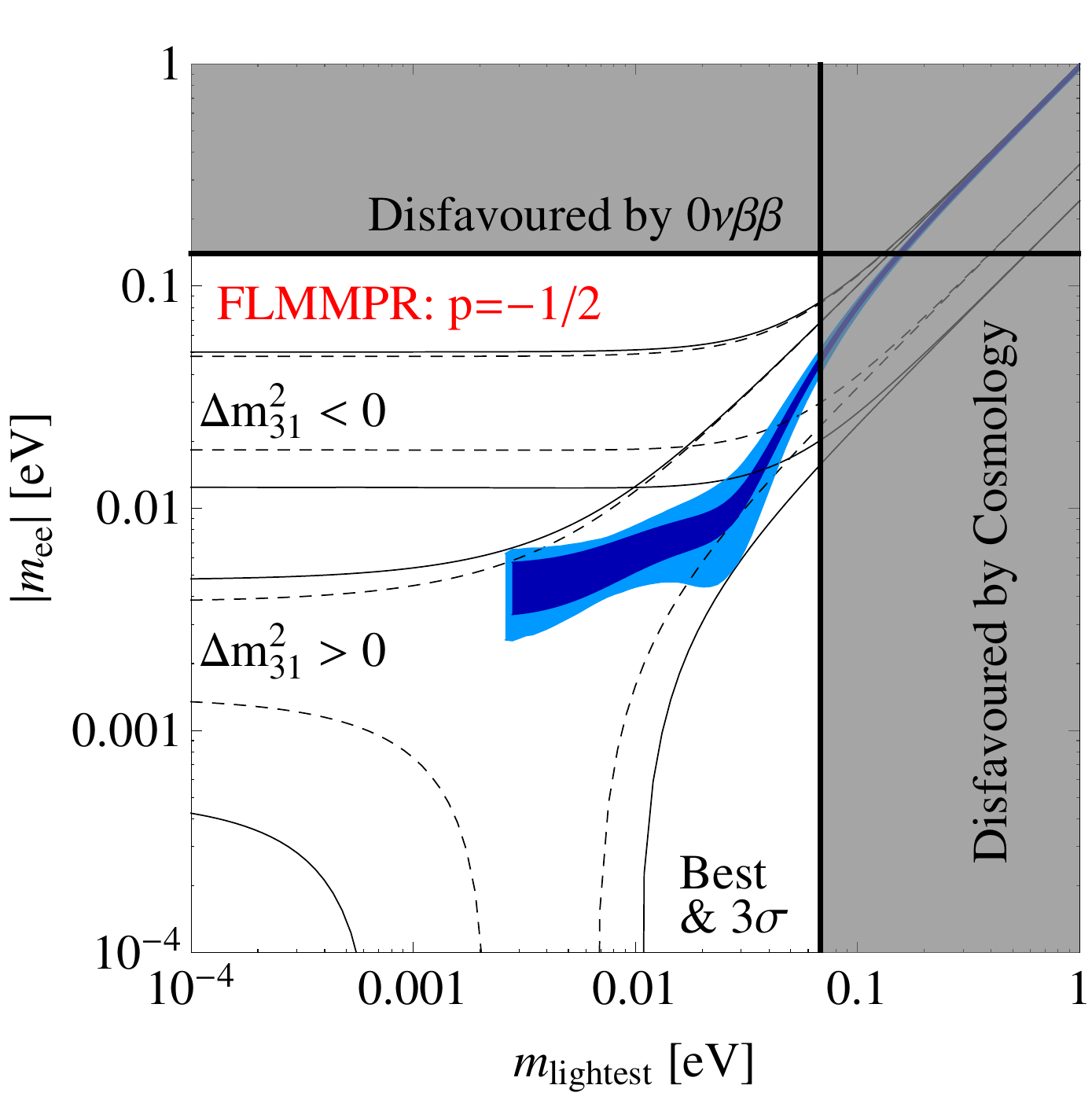} &
\includegraphics[width=5cm]{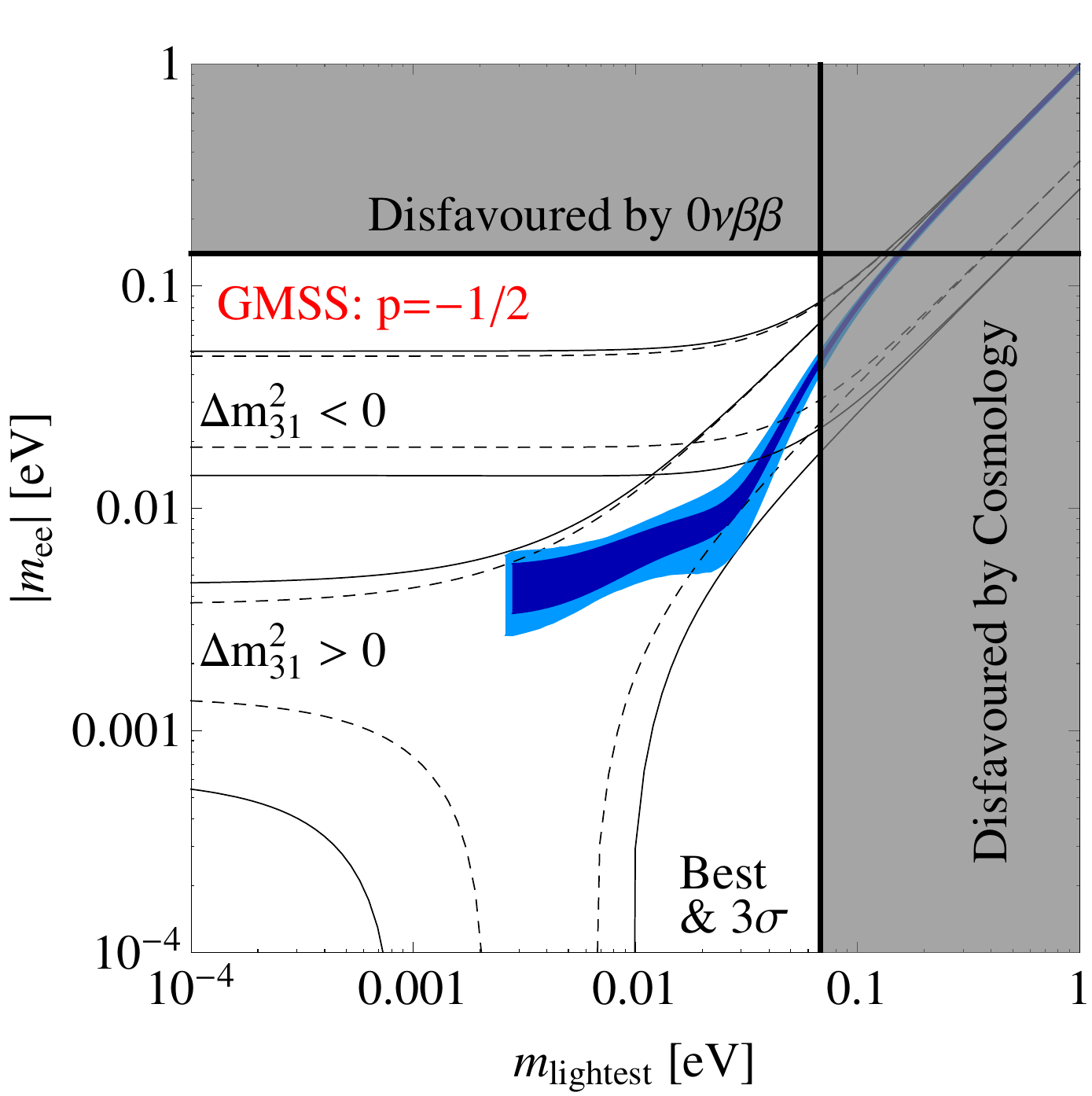}
\end{tabular}
\caption{\label{fig:mee12030155}Allowed regions for the sum rule $\frac{1}{\sqrt{\tilde m_1}} = \frac{2}{\sqrt{\tilde m_3}} - \frac{1}{\sqrt{\tilde m_2}}$.}
\end{figure}
\end{center}

The only model~\cite{Dorame:2012zv} which we are aware of leading to this sum rule is based on the group $S_4$ group and the so-called \emph{inverse seesaw mechanism}~\cite{Mohapatra:1986bd,GonzalezGarcia:1988rw}. The sum rule, cf.\ Eq.~(18) of Ref.~\cite{Dorame:2012zv}, was already mentioned in Eq.~\eqref{eq:Inverse_1}, and it is given by:
\begin{equation}
 \frac{1}{\sqrt{\tilde m_1}} = \frac{2}{\sqrt{\tilde m_3}} - \frac{1}{\sqrt{\tilde m_2}}.
 \label{eq:Inverse_1n}
\end{equation}
A comparison with Eq.~\eqref{eq:gen_rule_2} yields
\begin{equation}
 p=-1/2,\ \ B_2 = 1,\ \ B_3 = 2,\ \ \Delta \chi_{21} = 0,\ \ {\rm and}\ \ \Delta \chi_{31} = \pi.
 \label{eq:Inverse_2n}
\end{equation}
The corresponding allowed regions for the effective mass are displayed in Fig.~\ref{fig:mee12030155}.
As already proven after Eq.~\eqref{eq:Inverse_3}, this sum rule cannot be fulfilled for IO, which is also confirmed by our numerical results but which seems to contradict the region drawn in Fig.~3 of Ref.~\cite{Dorame:2012zv}.

\subsection{\label{sec:concrete_A4Z2}The sum rule $2\sqrt{\tilde m_2} + \sqrt{\tilde m_3} = \sqrt{\tilde m_1}$}

\begin{center}
\begin{figure}[h!]
\begin{tabular}{lcr}
\includegraphics[width=5cm]{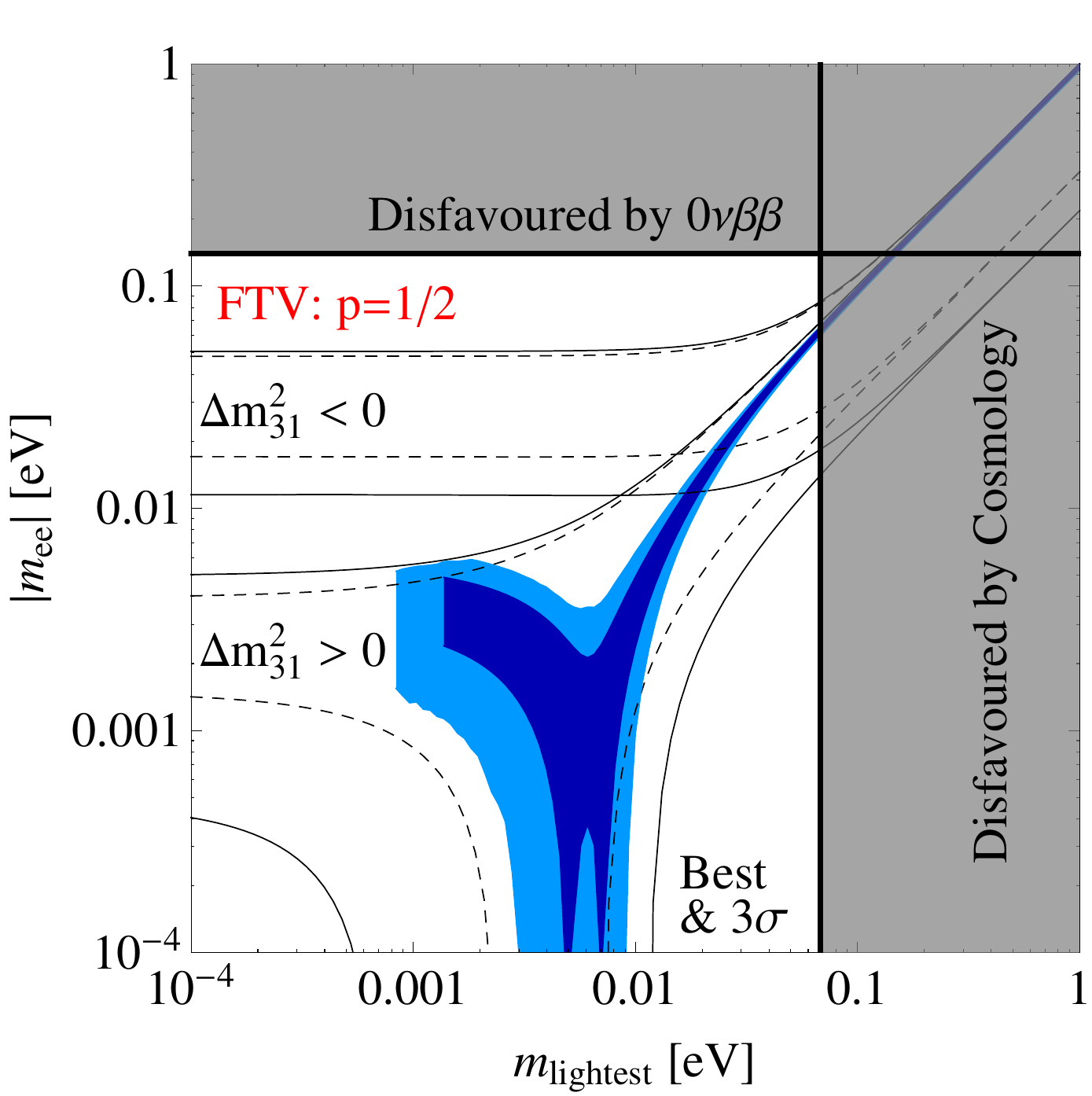} & \includegraphics[width=5cm]{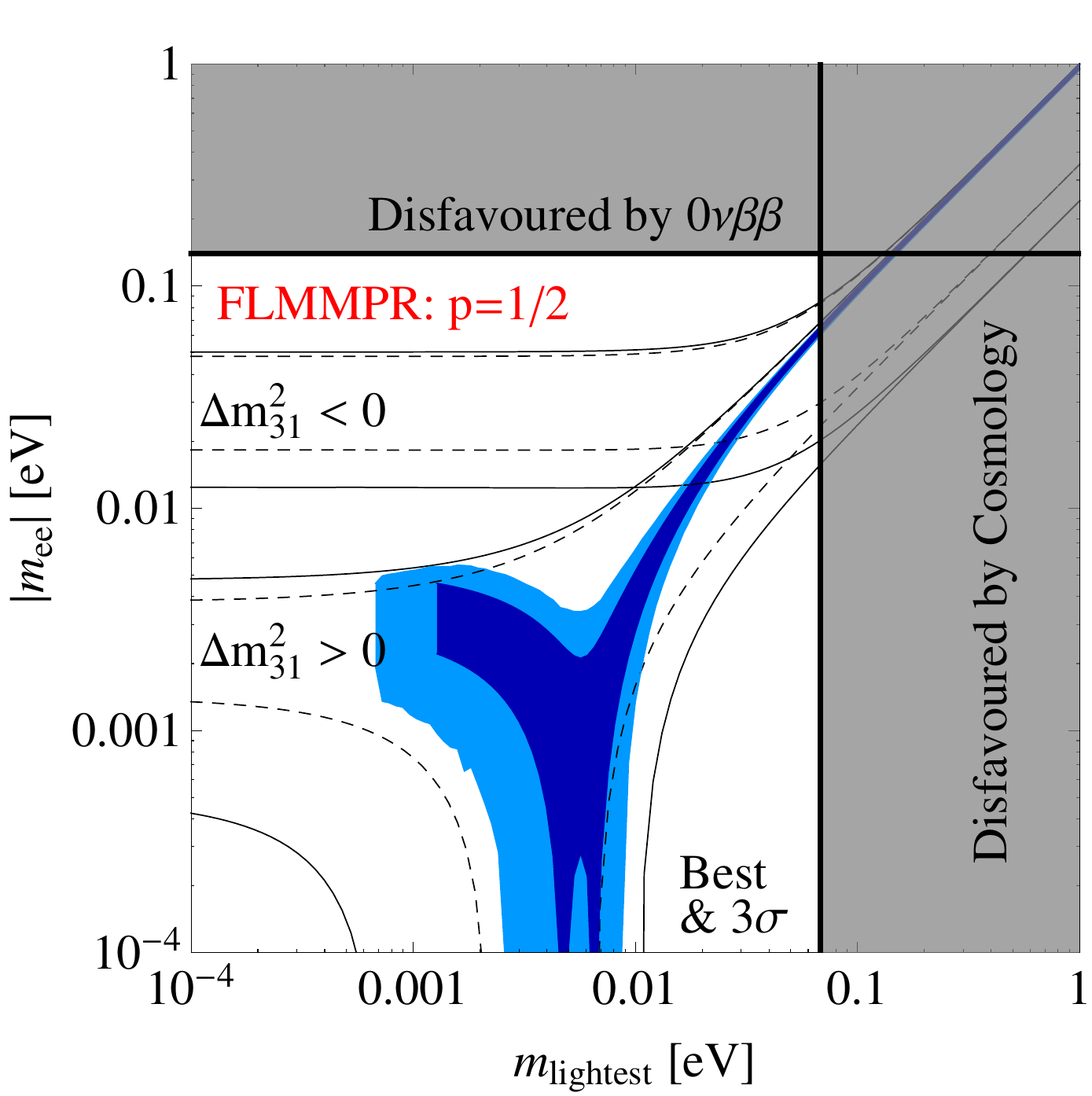} &
\includegraphics[width=5cm]{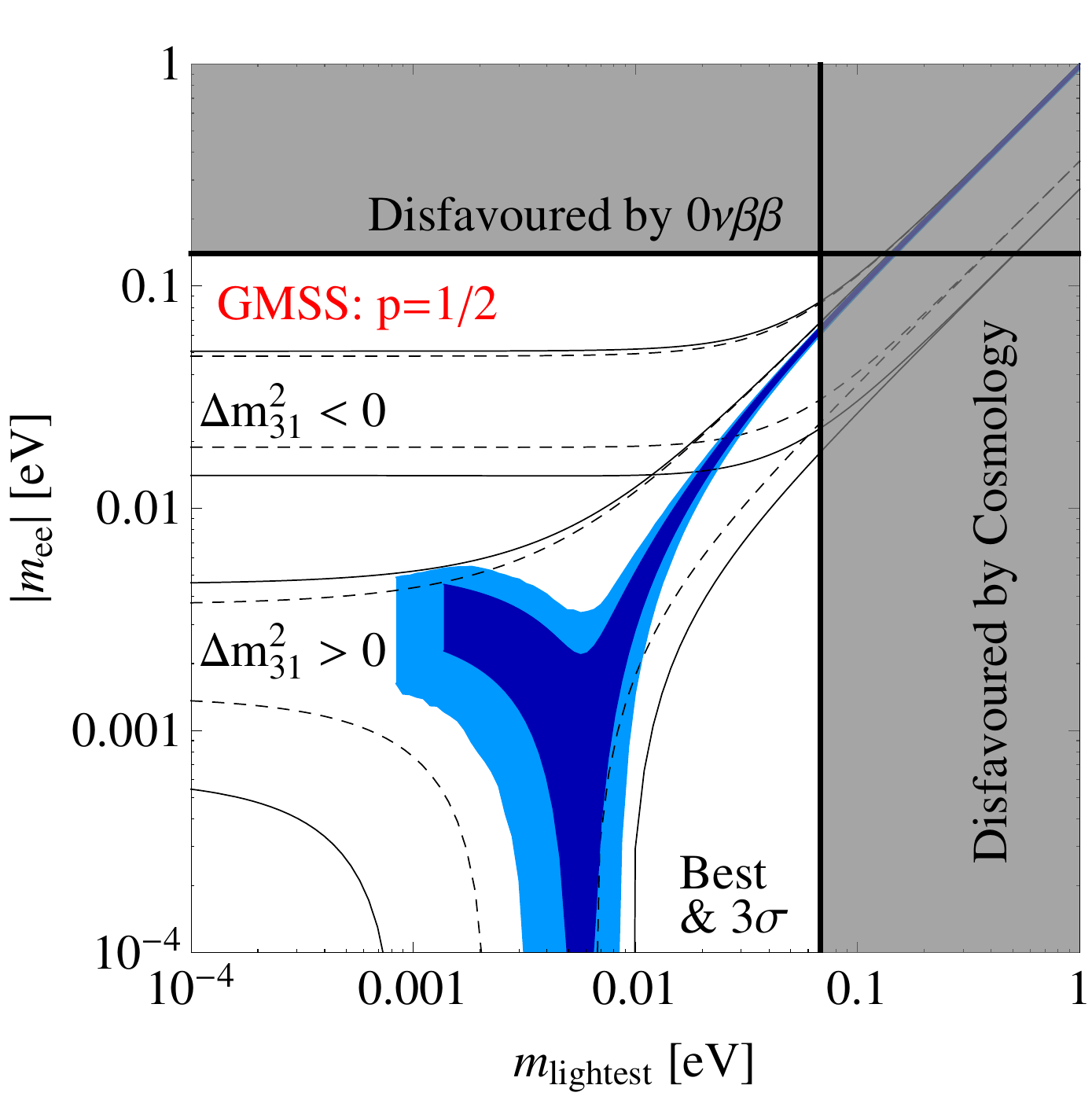}
\end{tabular}
\caption{\label{fig:mee08041521}Allowed regions for the sum rule $2\sqrt{\tilde m_2} + \sqrt{\tilde m_3} = \sqrt{\tilde m_1}$.}
\end{figure}
\end{center}

The model under consideration is based on $A_4\times Z_2$ was already proposed in Ref.~\cite{Hirsch:2008rp}, but the corresponding sum rule was only written down later~\cite{Dorame:2011eb}. The sum rule can be found in the caption of Fig.~3 in Ref.~\cite{Dorame:2011eb}:
\begin{equation}
 2\sqrt{\tilde m_2} + \sqrt{\tilde m_3} = \sqrt{\tilde m_1},
 \label{eq:A4Z2_1}
\end{equation}
which yields
\begin{equation}
 p=1/2,\ \ B_2 = 2,\ \ B_3 = 1,\ \ {\rm and}\ \ \Delta \chi_{21} =\Delta \chi_{31} = \pi.
 \label{eq:A4Z2_2}
\end{equation}
The corresponding allowed regions for the effective mass is displayed in Fig.~\ref{fig:mee08041521}.
We can confirm the corresponding plot in Ref.~\cite{Dorame:2011eb} in the sense that we also only obtain an allowed region for NO (this sum rule forbids IO, which is easy to see since $|2 \sqrt{\tilde m_2}| > |\sqrt{\tilde m_1} - \sqrt{\tilde m_3}|$ if $m_3 < m_1 < m_2$, so that the sum rule can never be fulfilled in that case), but the shape we obtain looks different. We suspect this difference to arise from the different treatment of the Majorana phases applied in Ref.~\cite{Dorame:2011eb}. On the other hand, the implications of this model for $0\nu\beta\beta$ have also been discussed in Ref.~\cite{Bazzocchi:2009da}, and our result is consistent with Fig.~6(a) therein (with the only difference that the latter has only been drawn down to $m_{\rm lightest} = 10^{-3}$~eV).

\subsection{\label{sec:concrete_hypothetical}The hypothetical sum rule $3\sqrt{\tilde m_2} + 3\sqrt{\tilde m_3} = \sqrt{\tilde m_1}$}

\begin{center}
\begin{figure}[h!]
\begin{tabular}{lcr}
\includegraphics[width=5cm]{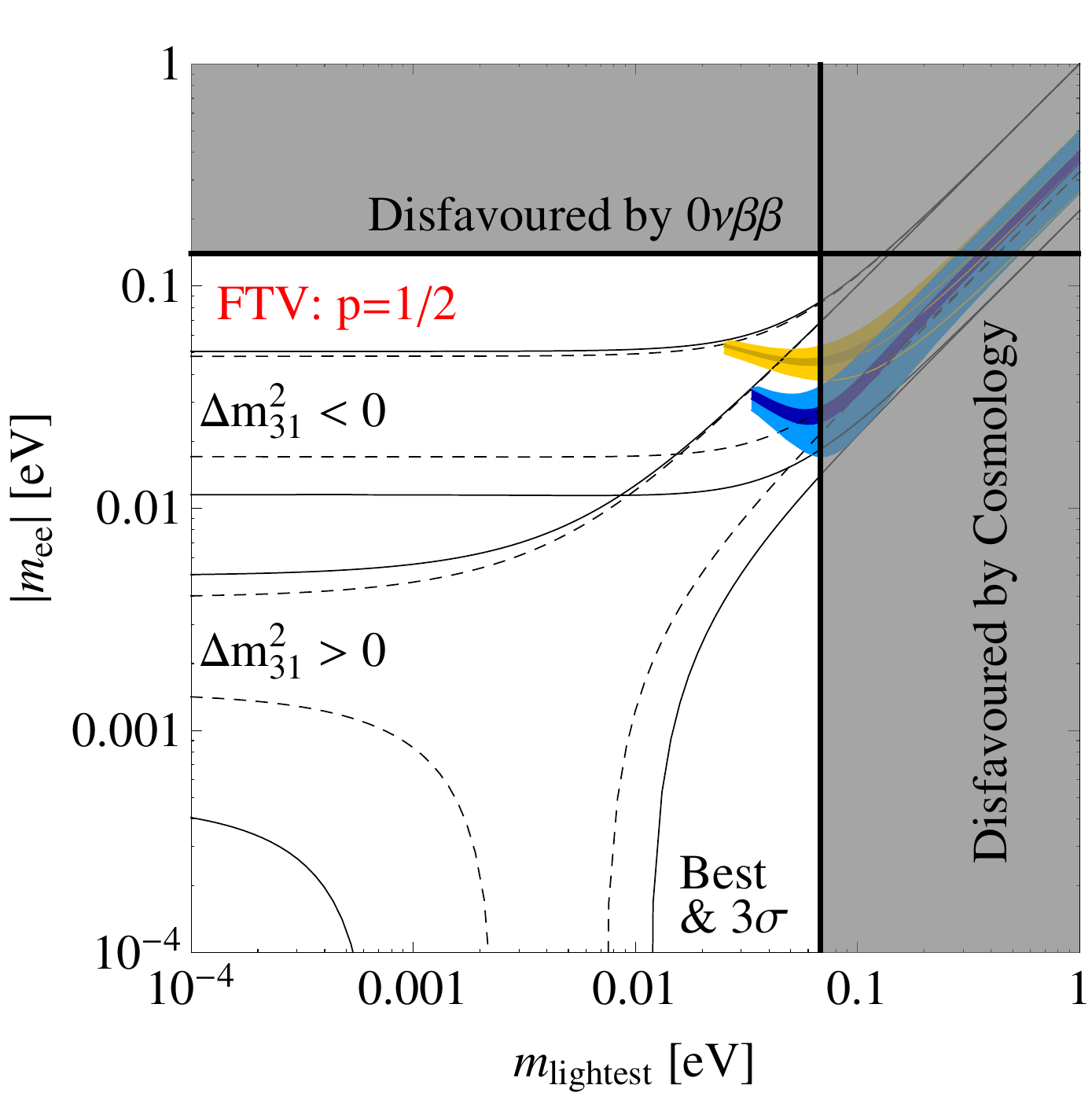} & \includegraphics[width=5cm]{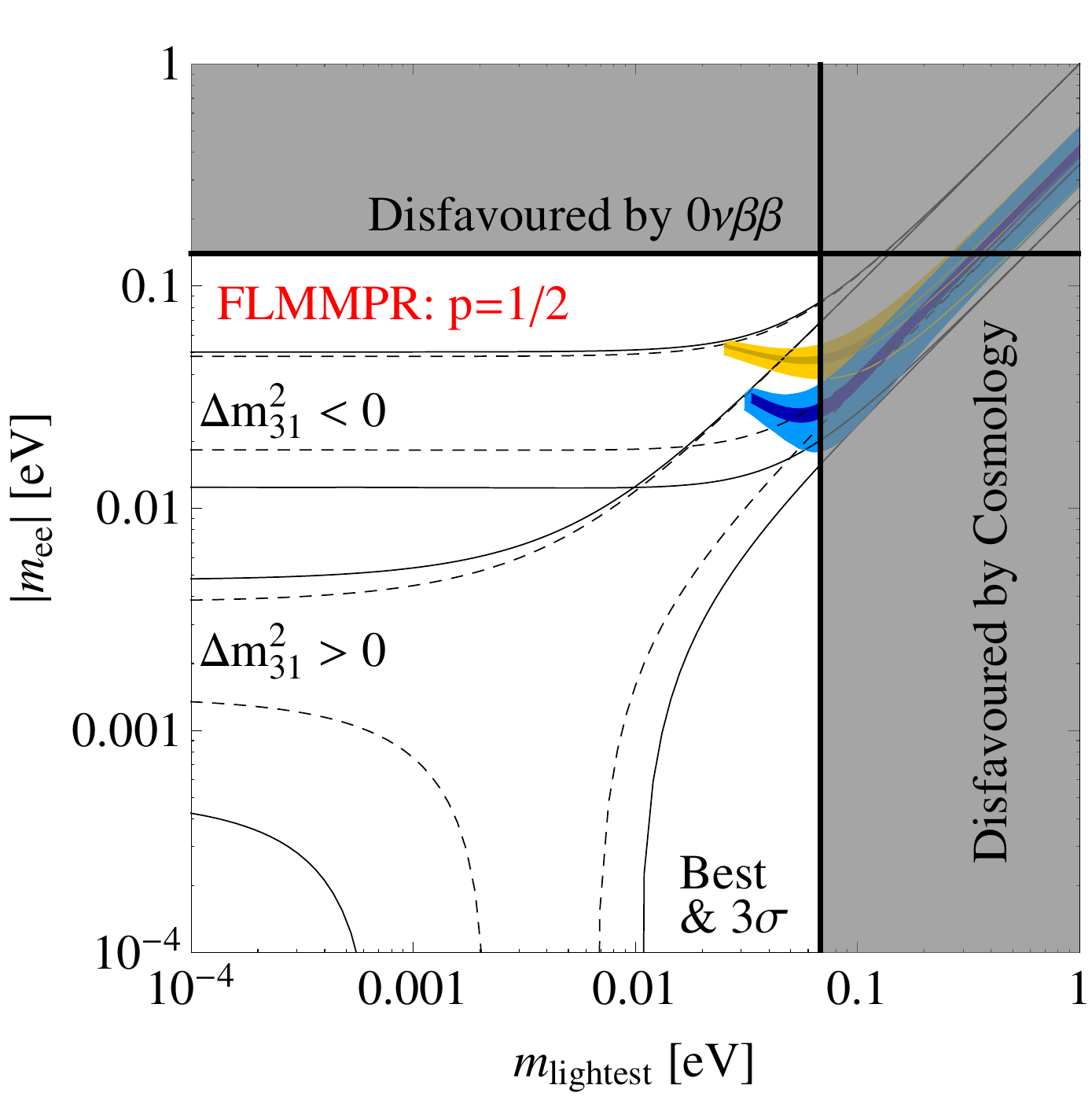} &
\includegraphics[width=5cm]{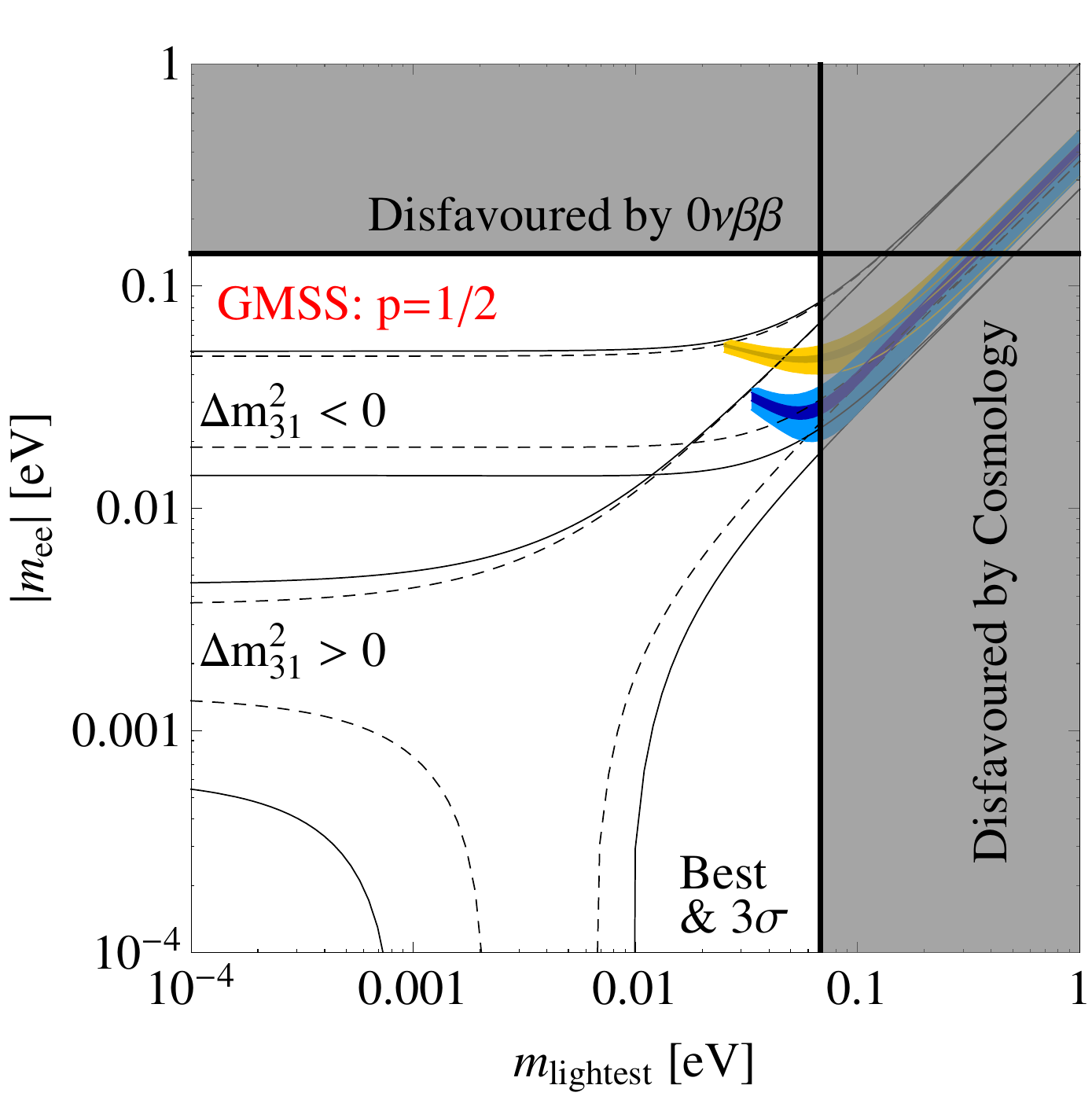}
\end{tabular}
\caption{\label{fig:mee11115614}Allowed regions for the sum rule $3\sqrt{\tilde m_2} + 3\sqrt{\tilde m_3} = \sqrt{\tilde m_1}$.}
\end{figure}
\end{center}

There is another sum rule mentioned in Ref.~\cite{Dorame:2011eb}, which seems not to be based on a concrete model, but it is similar to the one discussed in Sec.~\ref{sec:concrete_A4Z2} and hence it could well stem from a realistic model. This sum rule is stated in the caption of Fig.~3 in Ref.~\cite{Dorame:2011eb}:
\begin{equation}
 3\sqrt{\tilde m_2} + 3\sqrt{\tilde m_3} = \sqrt{\tilde m_1}.
 \label{eq:hypo_1}
\end{equation}
The comparison with Eq.~\eqref{eq:gen_rule_2} implies
\begin{equation}
 p=1/2,\ \ B_2 = B_3 = 3,\ \ {\rm and}\ \ \Delta \chi_{21} =\Delta \chi_{31} = \pi.
 \label{eq:hypo_2}
\end{equation}
The corresponding allowed regions for the effective mass are displayed in Fig.~\ref{fig:mee11115614}.
This result actually agrees with the one obtained in Ref.~\cite{Dorame:2011eb}.\footnote{One has to be careful though, since the order of the corresponding plots is incorrect in the preprint version of Ref.~\cite{Dorame:2011eb}, which however was amended in the published version. The latter is the one we are referring to.}

Let us try to understand the plots analytically. As had been pointed out in Ref.~\cite{Barry:2010zk}, a good tool to estimate the range of validity of a given sum rule is to apply the triangle inequality to the largest side of the triangle. For NO, Eq.~\eqref{eq:orderings} tells us that
\begin{equation}
 3 \sqrt{m_3} > 3 \sqrt{m_2} > \sqrt{m_1},
 \label{eq:hypo_3}
\end{equation}
and the triangle inequality implies that
\begin{equation}
 3 (m_{\rm lightest}^2 + \Delta m_A^2)^{1/4} < 3 (m_{\rm lightest}^2 + \Delta m_\odot^2)^{1/4} + \sqrt{m_{\rm lightest}}.
 \label{eq:hypo_4}
\end{equation}
Indeed, for $m_{\rm lightest}^2 \gg \Delta m_{A,\odot}^2$, this tends to $3 < 4$ which is correct, while for $m_{\rm lightest}^2 \ll \Delta m_{A,\odot}^2$ we would obtain $\Delta m_A^2 < \Delta m_\odot^2$, which is not true. This tells us that the sum rule cannot be fulfilled for a very small $m_{\rm lightest}$. However, we have not yet determined the border of validity of the sum rule. This can be done easily by equating both sides of Eq.~\eqref{eq:hypo_4} and approximating $\Delta m_\odot^2 \approx 0$, which leads to a lower cutoff of
\begin{equation}
 m_{\rm lightest} \simeq \frac{9}{5 \sqrt{7}} \sqrt{\Delta m_A^2} \simeq 0.035~{\rm eV}
 \label{eq:hypo_5}
\end{equation}
for all three fits. This seems to be in excellent agreement with our plots. For IO, in turn, Eq.~\eqref{eq:orderings} implies
\begin{equation}
 3 \sqrt{m_2} > 3 \sqrt{m_2}, \sqrt{m_1},
 \label{eq:hypo_6}
\end{equation}
and the triangle inequality leads to
\begin{equation}
 3 (m_{\rm lightest}^2 + \Delta m_A^2 + \Delta m_\odot^2)^{1/4} < 3 \sqrt{m_{\rm lightest}} + (m_{\rm lightest}^2 + \Delta m_A^2)^{1/4}.
 \label{eq:hypo_7}
\end{equation}
Playing the same game as before, we obtain $3 < 4$ for large $m_{\rm lightest}$ but $3 < 1$ for small $m_{\rm lightest}$, and equating both sides leads to
\begin{equation}
 m_{\rm lightest} \simeq \frac{4}{\sqrt{65}} \sqrt{\Delta m_A^2} \simeq 0.024~{\rm eV},
 \label{eq:hypo_8}
\end{equation}
which is again nearly the same for all three fits. This implies that the validity of the sum rule should go down to slighly lower values of the lightest mass for IO, which is in excellent agreement with our plots.

\subsection{\label{sec:concrete_TpZ2}The sum rule $\frac{2}{\tilde m_2} = \frac{1}{\tilde m_1} + \frac{1}{\tilde m_3}$}

\begin{center}
\begin{figure}[h!]
\begin{tabular}{lcr}
\includegraphics[width=5cm]{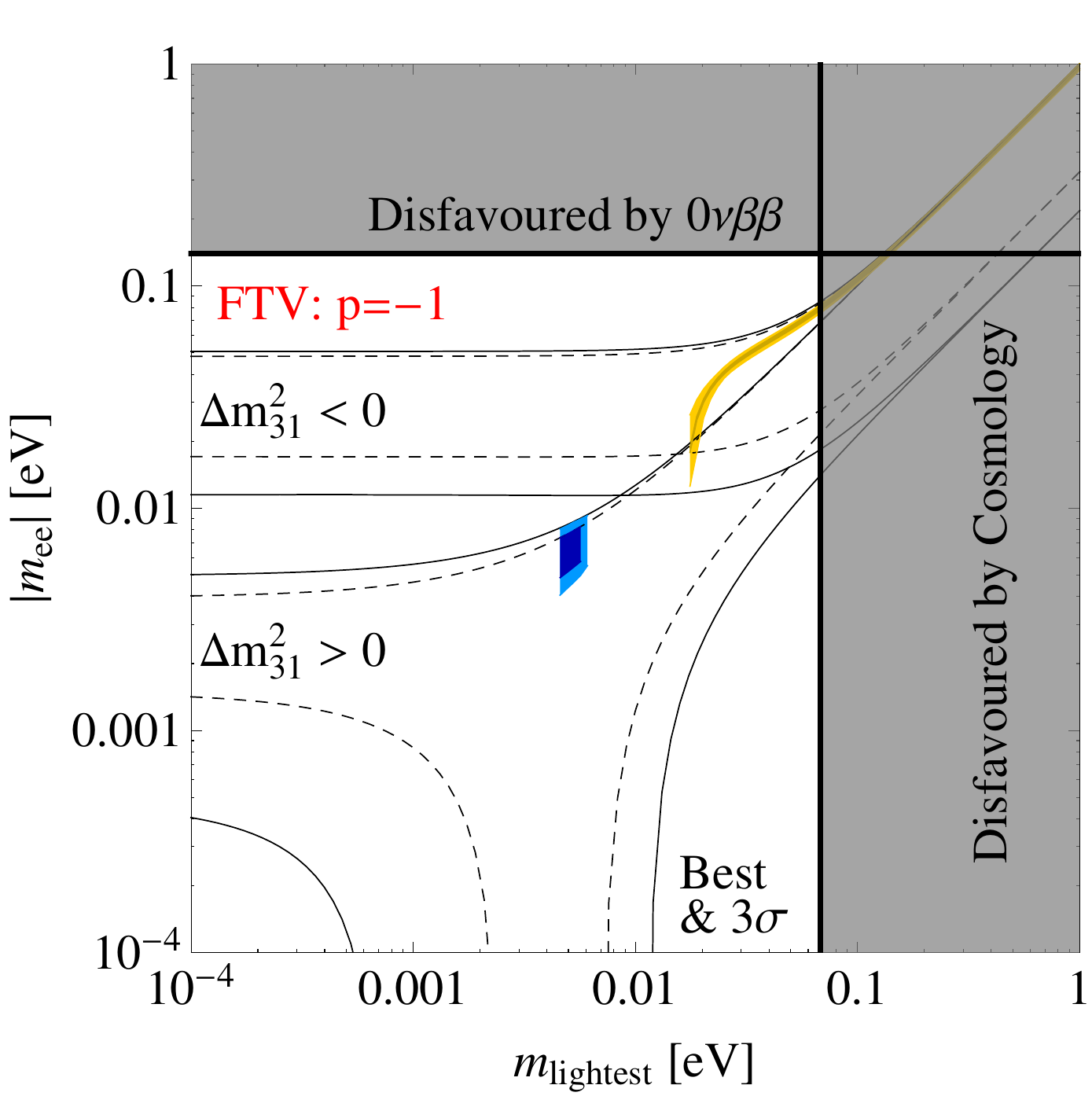} & \includegraphics[width=5cm]{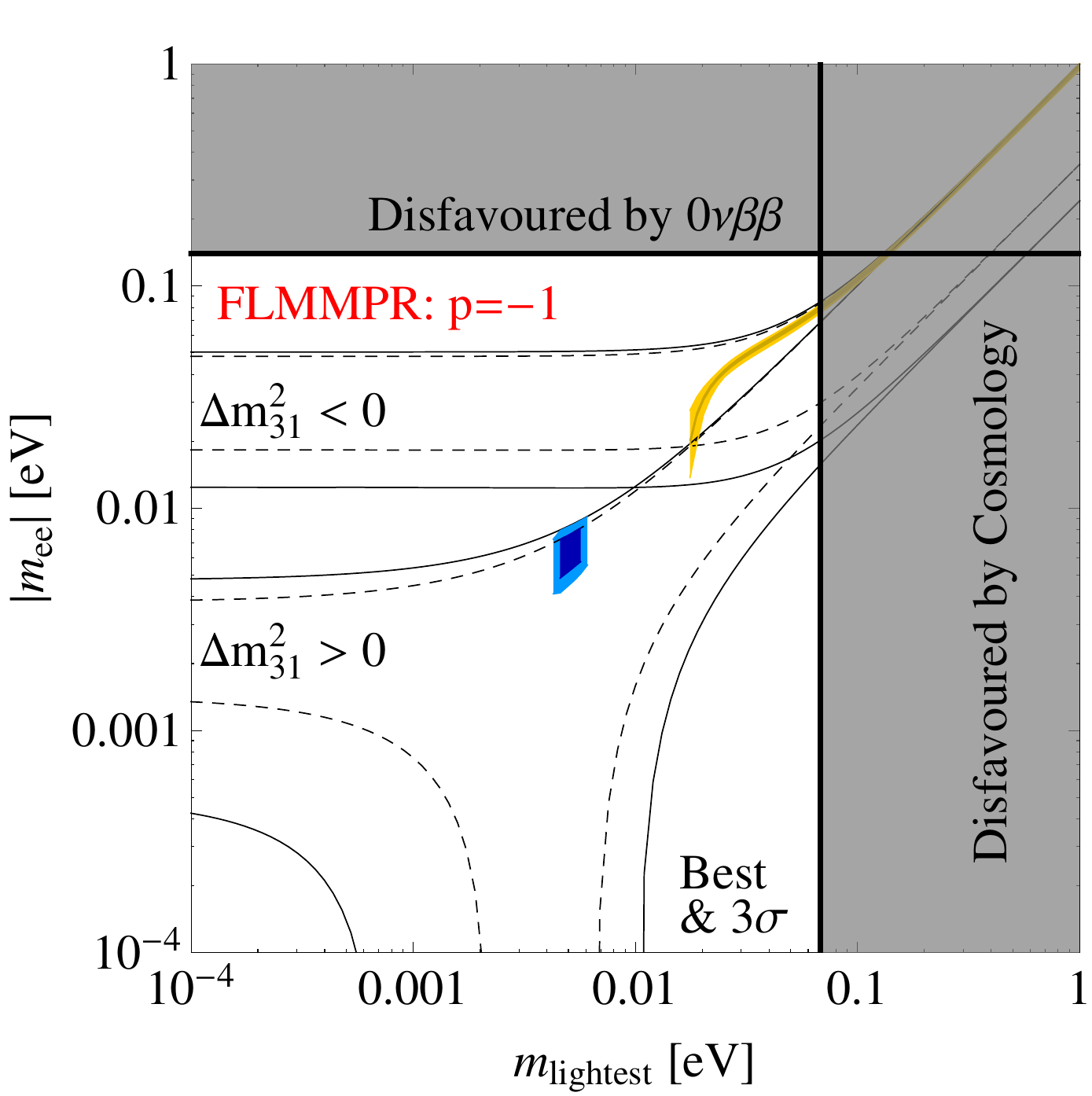} &
\includegraphics[width=5cm]{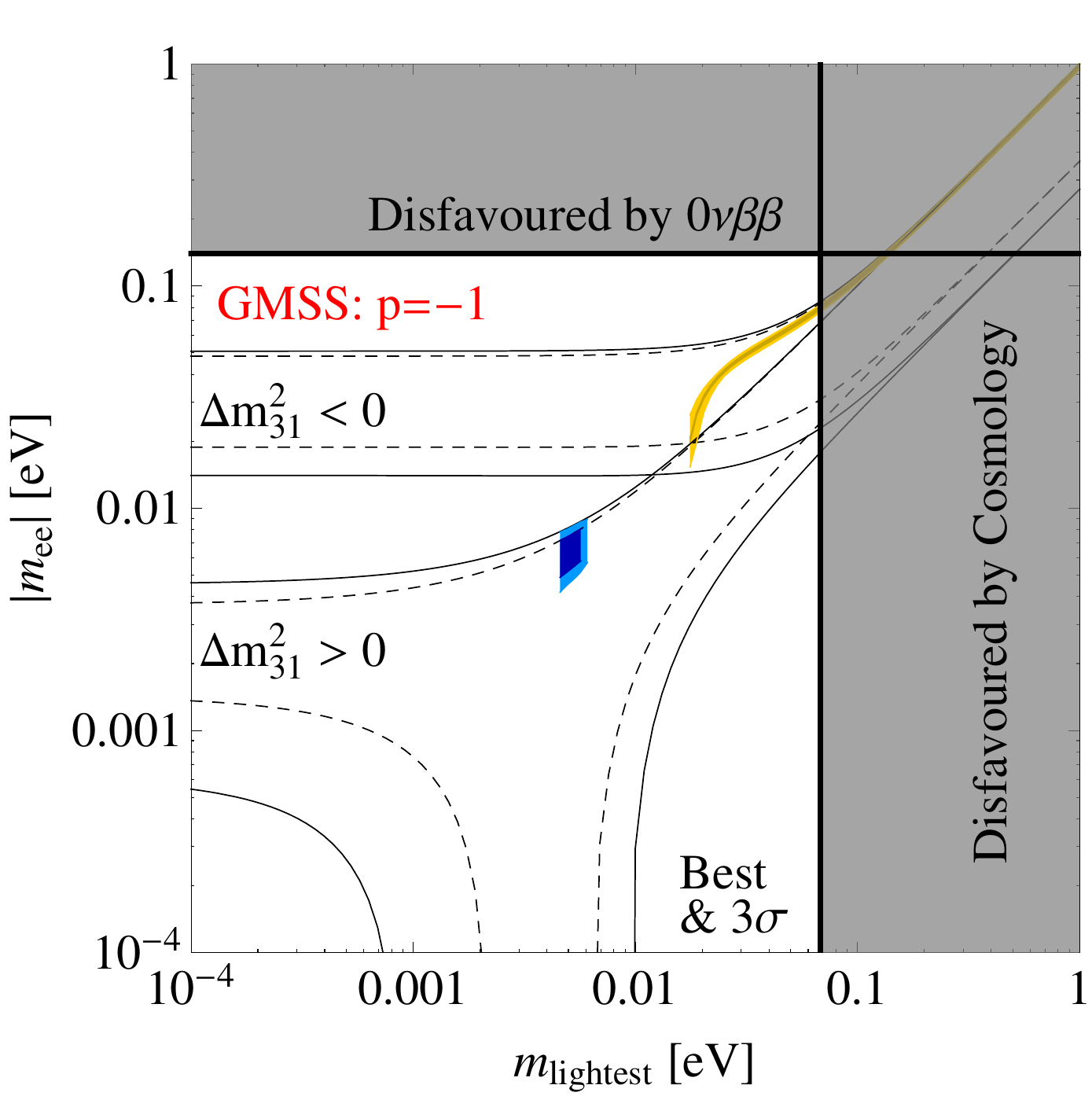}
\end{tabular}
\caption{\label{fig:mee12053442}Allowed regions for the sum rule $\frac{2}{\tilde m_2} = \frac{1}{\tilde m_1} + \frac{1}{\tilde m_3}$.}
\end{figure}
\end{center}

This sum rule has been found using either $A_4$~\cite{He:2006dk,Berger:2009tt} or $T'$~\cite{Lavoura:2012cv} symmetries. It reads
\begin{equation}
 \frac{2}{\tilde m_2} = \frac{1}{\tilde m_1} + \frac{1}{\tilde m_3},
 \label{eq:TpZ2_1}
\end{equation}
which implies, in terms of our parameters,
\begin{equation}
 p=-1,\ \ B_2 = 2,\ \  B_3 = 1,\ \ \Delta \chi_{21} = \pi,\ \ {\rm and}\ \ \Delta \chi_{31} = 0.
 \label{eq:TpZ2_2}
\end{equation}
The corresponding allowed regions for the effective mass are displayed in Fig.~\ref{fig:mee12053442}.
The general tendency of the plot is the same as in the similar sum rule discussed in Sec.~\ref{sec:concrete_Del96}. In fact, the analytical proof of the strange-looking behaviour given in that section does not rely on the values of $\Delta \chi_{21,31}$, and it hence carries over to the case presented here. Nevertheless, while the qualitative behaviour is the same, the sum rule presented in this section reveals a different functional dependence of the allowed region of $|m_{ee}|$ on $m_{\rm lightest}$, which is to be expected due to the different values of the phase differences $\Delta \chi_{21,31}$. This is an interesting point to mention, since the similarity between Eqs.~\eqref{eq:TpZ2_1} and~\eqref{eq:Del96_1} illustrates that very different flavour symmetries can lead to very similar sum rules.

\subsection{\label{sec:concrete_Del54}The sum rule $\tilde m_1 + \tilde m_2 = \tilde m_3$}

\begin{center}
\begin{figure}[h!]
\begin{tabular}{lcr}
\includegraphics[width=5cm]{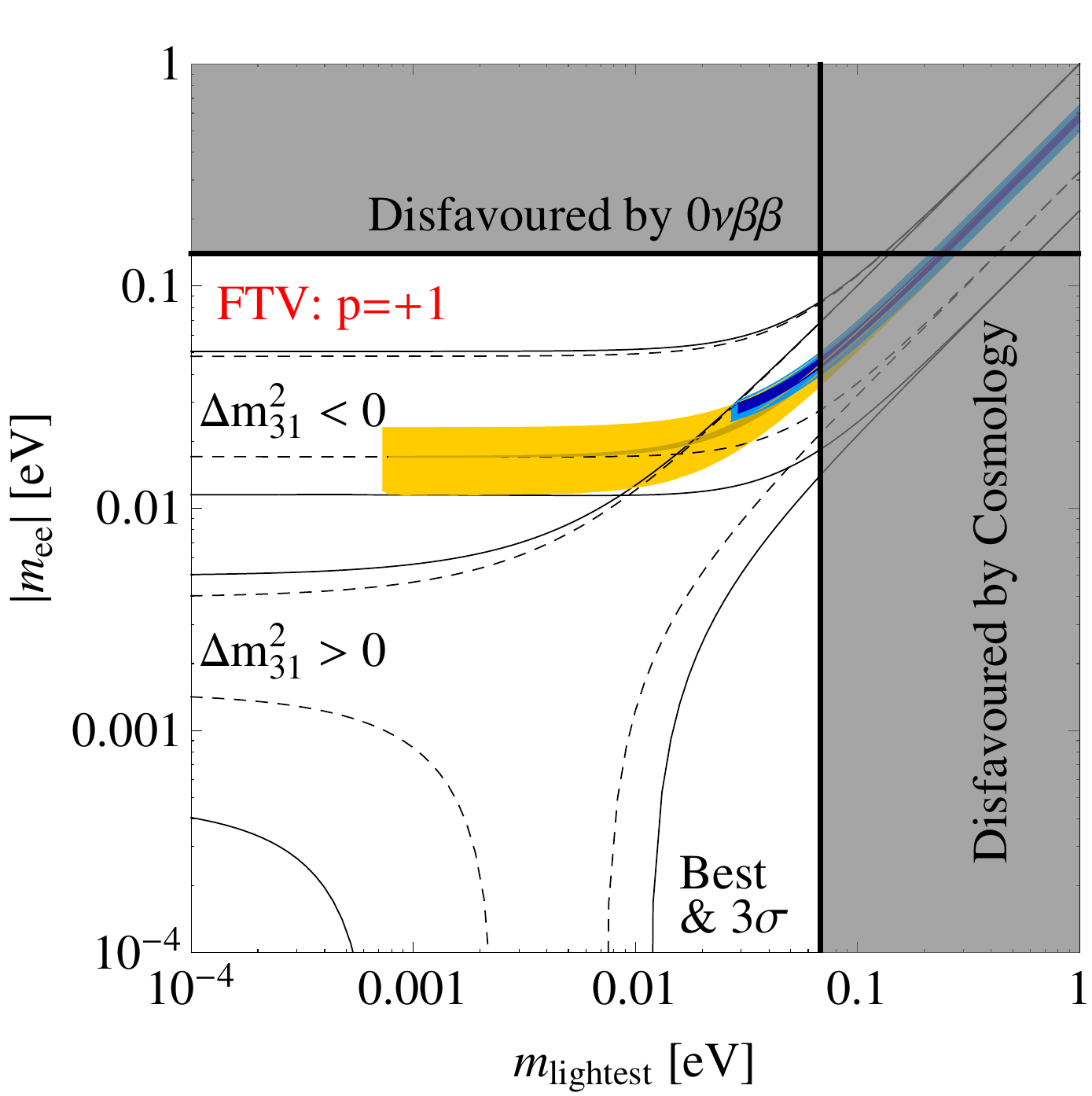} & \includegraphics[width=5cm]{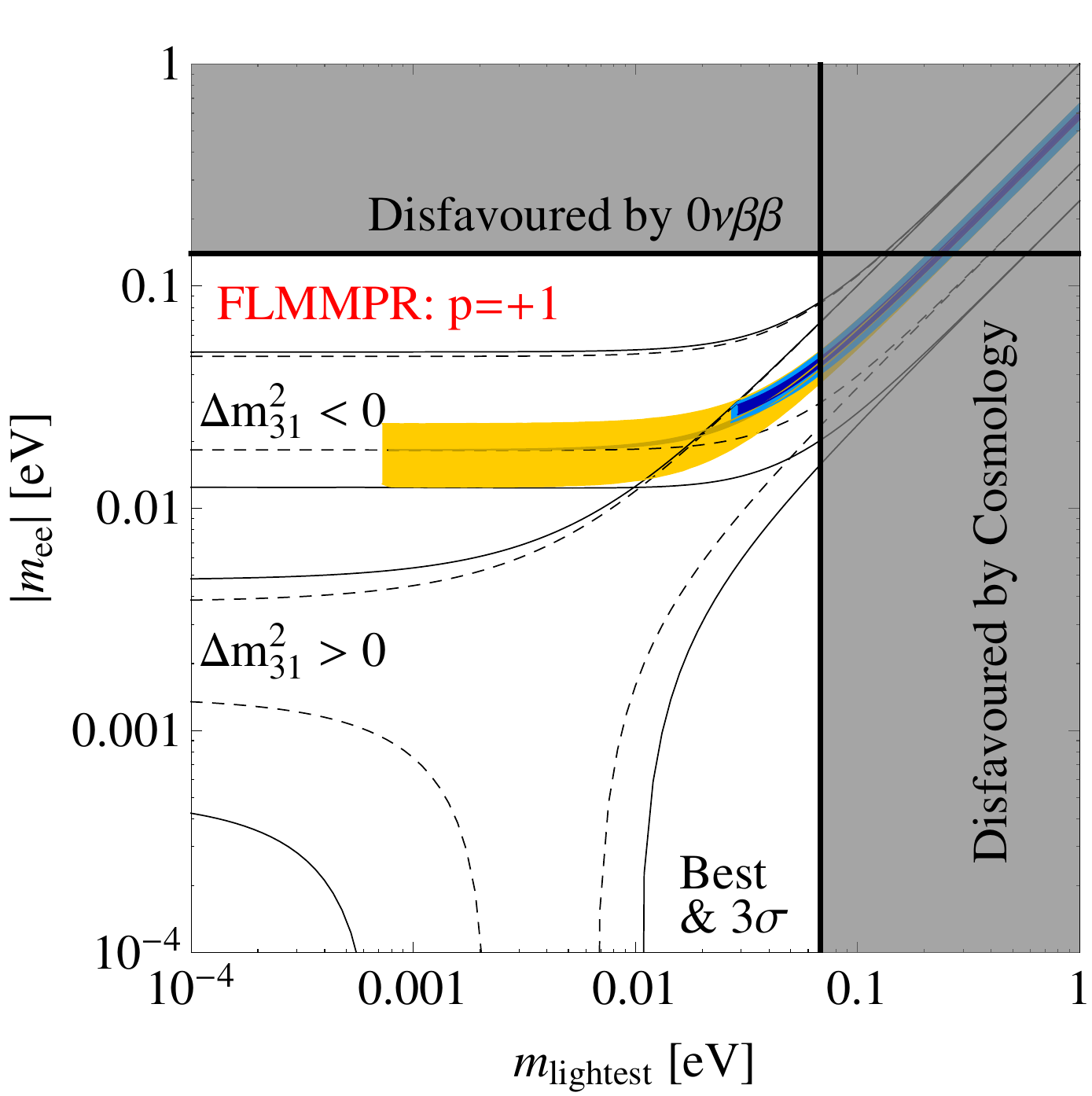} &
\includegraphics[width=5cm]{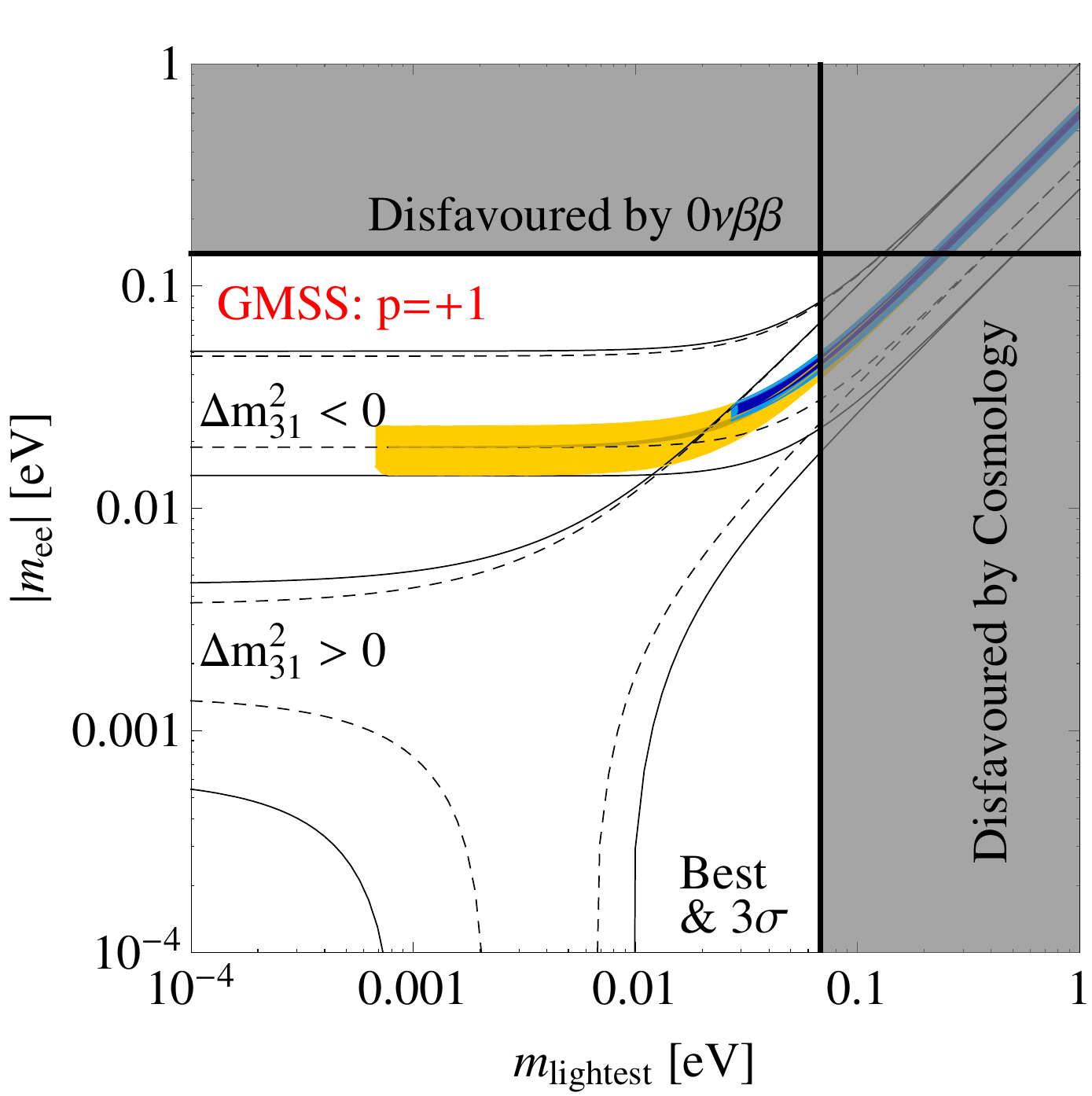}
\end{tabular}
\caption{\label{fig:mee12044733}Allowed regions for the sum rule $\tilde m_1 + \tilde m_2 = \tilde m_3$.}
\end{figure}
\end{center}

The next sum rule is probably the easiest one to imagine. Correspondingly, it has been found in many frameworks, based on various symmetries such as $A_4$~\cite{Barry:2010zk,Ma:2005sha,Ma:2006wm,Honda:2008rs,Brahmachari:2008fn}, $A_5$~\cite{Everett:2008et}, $S_4$~\cite{Bazzocchi:2009pv,Bazzocchi:2009da}, or $\Delta(54)$~\cite{Boucenna:2012qb}. Thereby, in particular Ref.~\cite{Boucenna:2012qb} is worth mentioning as a recent example model which cannot only lead to a sum rule but at the same time also accommodate for a non-zero value of the leptonic mixing angle $\theta_{13}$. The explicit rule is given by:
\begin{equation}
 \tilde m_1 + \tilde m_2 = \tilde m_3,
 \label{eq:Del54_1}
\end{equation}
or, in terms of our general parameters,
\begin{equation}
 p=1,\ \ B_2 = B_3 = 1,\ \ \Delta \chi_{21} = 0,\ \ {\rm and}\ \ \Delta \chi_{31} = \pi.
 \label{eq:Del54_2}
\end{equation}
The corresponding allowed regions for the effective mass are displayed in Fig.~\ref{fig:mee12044733}.
This should be compared to Fig.~1 in Ref.~\cite{Boucenna:2012qb} which seems to contradict our result at first sight, in particular when looking at the IO band. However, when looking more closely, it is visible that our IO band cuts off for some value $m_{\rm lightest} < 10^{-3}$~eV, which is actually consistent with Fig.~1 in Ref.~\cite{Boucenna:2012qb}, since that figure only goes down to $m_{\rm lightest} = 10^{-3}$~eV. Also the position of the IO allowed band at the bottom of the general allowed region seems to yield the same result.

Nevertheless there is a visible difference for NO. This difference must again come from the handling of the phases in the third term inside $|m_{ee}|$, which is proportional to $m_3$. While this term is small for the case of IO, due to $m_3$ being the smallest mass, it is non-negligible for NO. This explains the qualitative difference between our result and the one from Ref.~\cite{Boucenna:2012qb}. 

In more detail, we can first estimate the cutoff of the sum rule for NO: as done in Sec.~\ref{sec:concrete_hypothetical}, one can use Eq.~\eqref{eq:orderings} and the triangle inequality for the longest side $m_3$ of the triangle, which leads (again neglecting $\Delta m_\odot^2$) to a smallest mass of
\begin{equation}
 m_{\rm lightest} \simeq \sqrt{\frac{\Delta m_A^2}{3}}.
 \label{eq:Del54_3}
\end{equation}
This is roughly $0.029$~eV for the FTV- and GMSS-fits, while it is $0.028$~eV for the FLMMPR-fit, and it agrees in the case of NO with boths, our plot and also Fig.~1 in Ref.~\cite{Boucenna:2012qb}. However, what about the minimum value of $|m_{ee}|$? First, note that Eq.~\eqref{eq:orderings} implies that, in the limit of Eq.~\eqref{eq:Del54_3},
\begin{equation}
 m_1 \simeq m_2 \simeq \sqrt{\frac{\Delta m_A^2}{3}}\ \ \ {\rm and}\ \ \ m_3 \simeq 2 \sqrt{\frac{\Delta m_A^2}{3}}.
 \label{eq:Del54_4}
\end{equation}
Inserting this into Eq.~\eqref{eq:mee_2} and \emph{naively} minimising over the two phases $\alpha_{21}$ and $(\alpha_{31} - 2\delta)$ would, due to $m_1 c_{12}^2 c_{13}^2 > m_2 s_{12}^2 c_{13}^2 + m_3 s_{13}^2$, yield
\begin{equation}
 |m_{ee}|_{\rm min}^{\rm naive} \simeq |m_1 c_{12}^2 c_{13}^2 - m_2 s_{12}^2 c_{13}^2 - m_3 s_{13}^2| \simeq \sqrt{\frac{\Delta m_A^2}{3}} [\cos (2 \theta_{12}) c_{13}^2 - 2 s_{13}^2],
 \label{eq:Del54_5}
\end{equation}
which numerically turns out to be $0.009$~eV (FTV and FLMMPR) or $0.010$~eV (GMSS). This indeed looks very much like the result displayed in Fig.~1 of Ref.~\cite{Boucenna:2012qb}.

However, this value cannot be entirely correct, since the sum rule given in Eq.~\eqref{eq:Del54_1} contains \emph{complex} masses and hence it yields \emph{two} constraints. In particular, it also constrains the Majorana phases. Multiplying Eq.~\eqref{eq:Del54_1} by $e^{-i \phi_1}$ and taking the real and imaginary parts gives us the two resulting constraints,
\begin{equation}
 m_1 + m_2 \cos \alpha_{21} = m_3 \cos \alpha_{31}\ \ \ {\rm and}\ \ \ m_2 \sin \alpha_{21} = m_3 \sin \alpha_{31}.
 \label{eq:Del54_6}
\end{equation}
In the approximation of Eq.~\eqref{eq:Del54_4}, these two equations imply $\cos \alpha_{21, 31} \simeq 1$, such that
\begin{equation}
 \alpha_{21, 31} \simeq 0.
 \label{eq:Del54_7}
\end{equation}
Hence, it is \emph{not possible} to simply vary the phases $\alpha_{21}$ and $(\alpha_{31} - 2\delta)$ to yield the result from Eq.~\eqref{eq:Del54_5}. Instead, one would need to insert the condition Eq.~\eqref{eq:Del54_7} into Eq.~\eqref{eq:mee_2}, and one can then choose $\delta = \pi/2$ to yield the true minimum value,
\begin{equation}
 |m_{ee}|_{\rm min}^{\rm true} \simeq |m_1 c_{12}^2 c_{13}^2 + m_2 s_{12}^2 c_{13}^2 - m_3 s_{13}^2| \simeq \sqrt{\frac{\Delta m_A^2}{3}} [c_{13}^2 - 2 s_{13}^2].
 \label{eq:Del54_8}
\end{equation}
This quantity is numerically given by $0.027$~eV (FTV and GMSS) or $0.026$~eV (FLMMPR), in excellent agreement with our plots. Note also that Fig.~1(b) of Ref.~\cite{Bazzocchi:2009da} further backs up our result.

\subsection{\label{sec:concrete_A4TpSAME}The sum rule $\tilde m_1^{-1} - 2 \tilde m_2^{-1} - \tilde m_3^{-1} = 0$}

\begin{center}
\begin{figure}[h!]
\begin{tabular}{lcr}
\includegraphics[width=5cm]{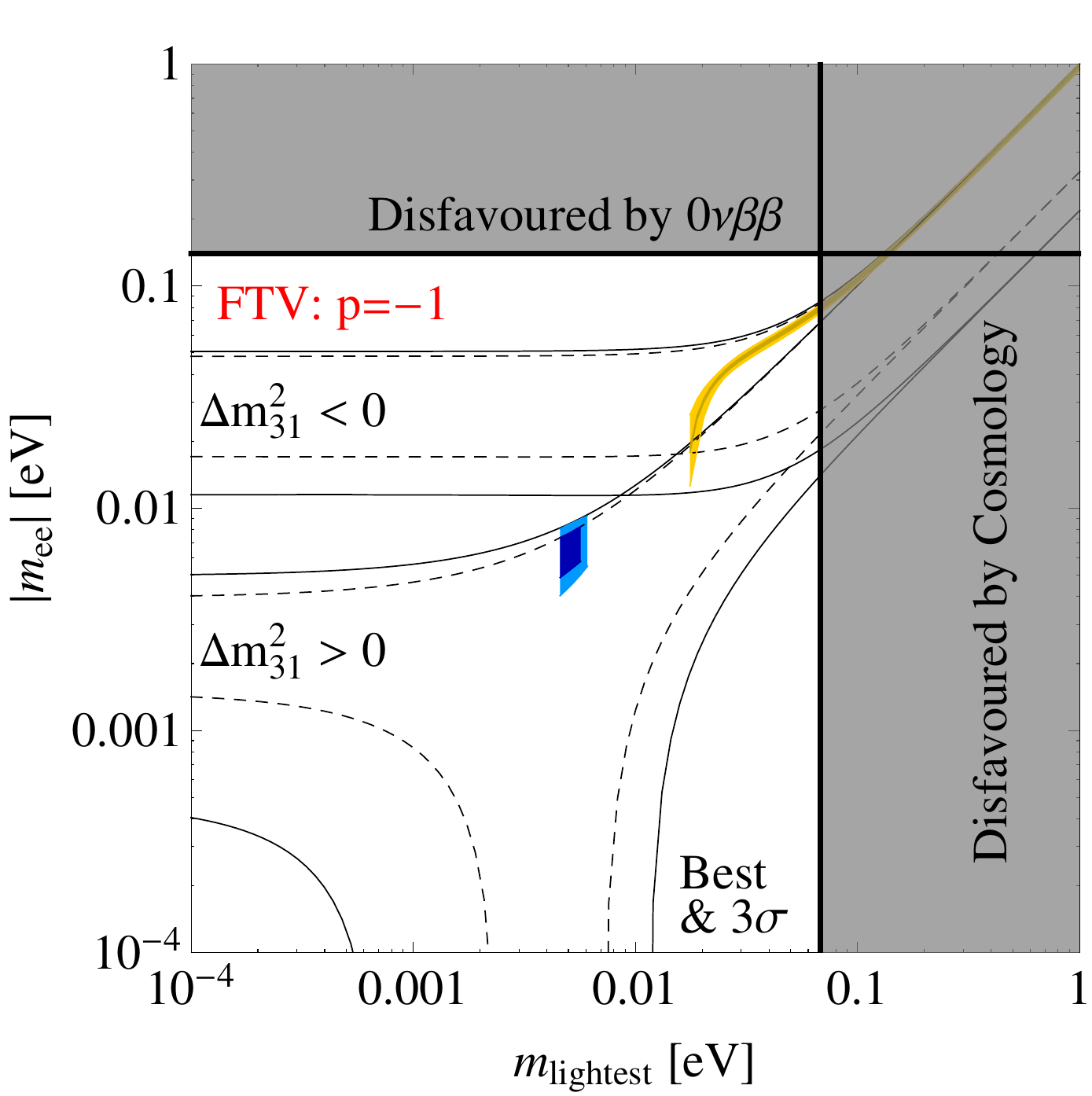} & \includegraphics[width=5cm]{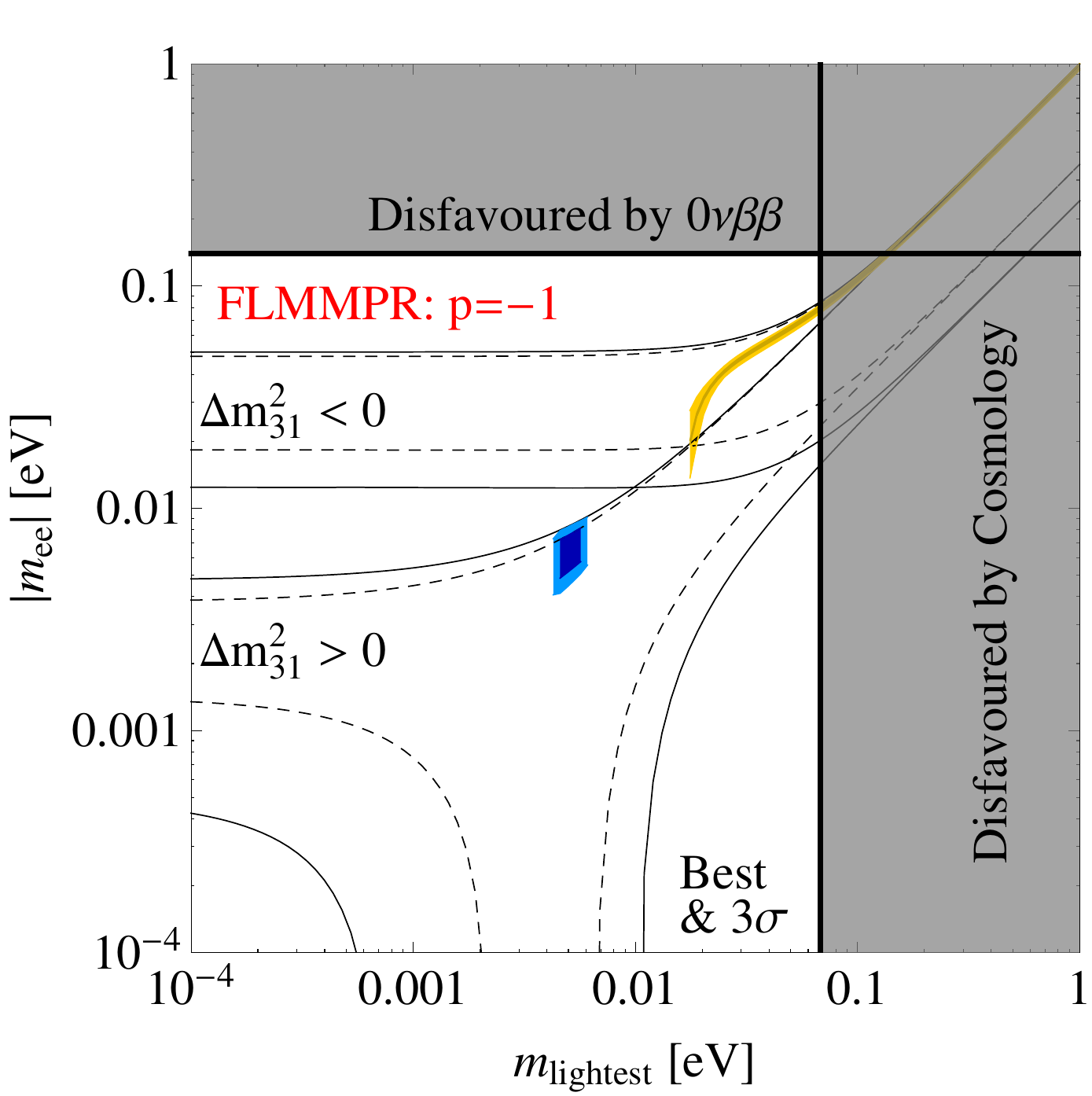} &
\includegraphics[width=5cm]{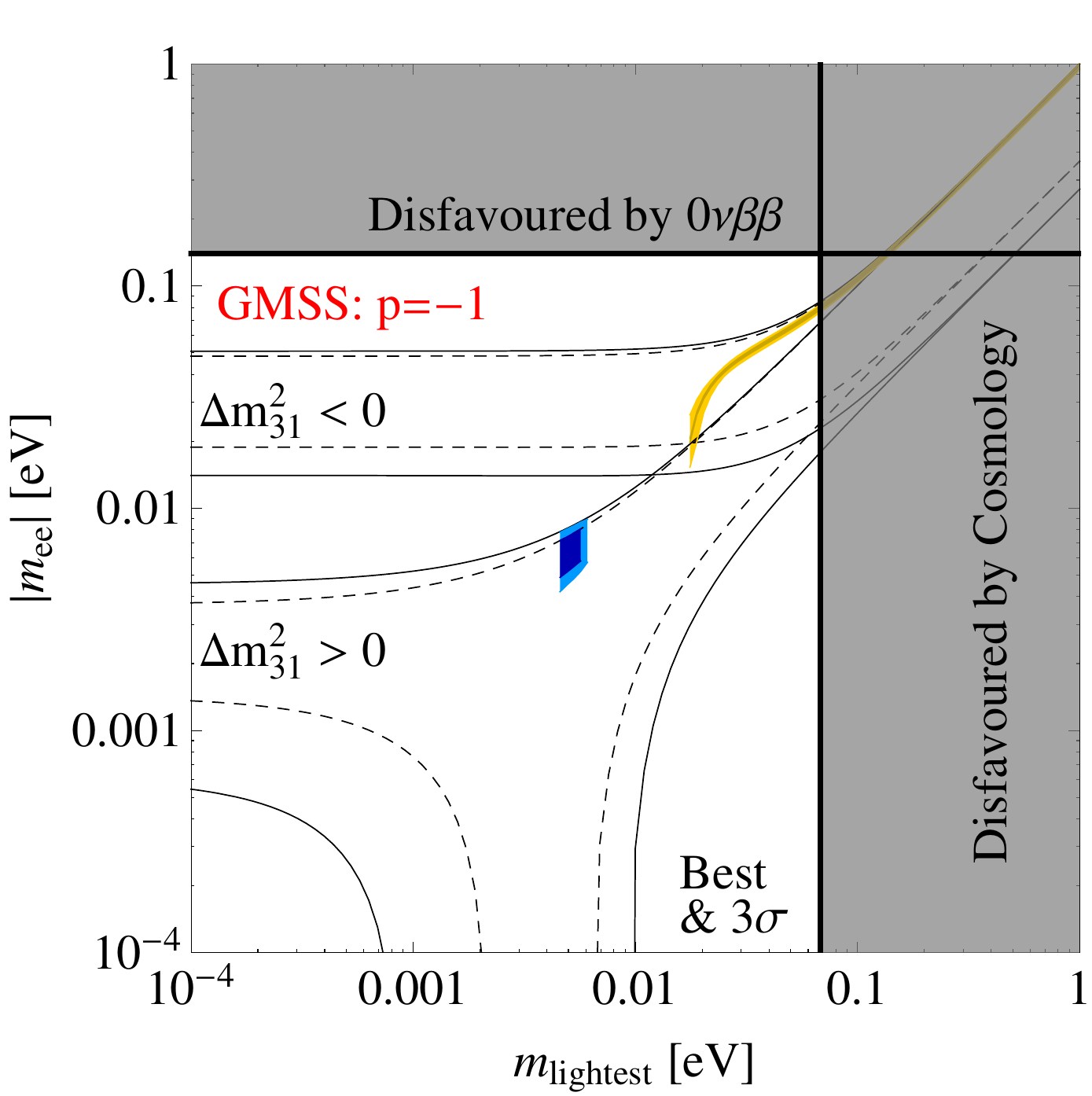}
\end{tabular}
\caption{\label{fig:mee0702034}Allowed regions for the sum rule $\tilde m_1^{-1} - 2 \tilde m_2^{-1} - \tilde m_3^{-1} = 0$.}
\end{figure}
\end{center}

This sum rule has been derived in many references, either based on an $A_4$~\cite{Barry:2010zk,Morisi:2007ft,Altarelli:2008bg,Adhikary:2008au,Csaki:2008qq,Altarelli:2009kr,Hagedorn:2009jy,Burrows:2009pi,Ding:2009gh,Mitra:2009jj,delAguila:2010vg,Burrows:2010wz,Altarelli:2005yx,Chen:2009um} or on a $T'$~\cite{Chen:2009gy} flavour symmetry. It is given by
\begin{equation}
 \frac{1}{\tilde m_1} - \frac{2}{\tilde m_2} - \frac{1}{\tilde m_3} = 0,
 \label{eq:A4TpSAME_1}
\end{equation}
leading to the parameter values
\begin{equation}
 p=-1,\ \ B_2 = 2,\ \  B_3 = 1,\ \ {\rm and}\ \ \Delta \chi_{21} = \Delta \chi_{31} = \pi.
 \label{eq:A4TpSAME_2}
\end{equation}
The corresponding allowed regions are displayed in Fig.~\ref{fig:mee0702034} (cf.\ Fig.~4(a) in Ref.~\cite{Bazzocchi:2009da}).
Just as in the sum rule discussed in Sec.~\ref{sec:concrete_Del96}, this sum rule predicts a very distinct region for NO, which is only about one order of magnitude below the near future sensitivity of $0\nu\beta\beta$ experiments on $|m_{ee}|$.

Note that, interestingly, this sum rule seems to lead to exactly the same allowed regions as the sum rule from Eq.~\eqref{eq:TpZ2_1} discussed in Sec.~\ref{sec:concrete_TpZ2}. Indeed, comparing the coefficients, the two sum rules only differ by one sign, cf.\ Eqs.~\eqref{eq:A4TpSAME_2} and~\eqref{eq:TpZ2_2}. This behaviour can be understood analytically by looking in detail at the constraints imposed by the respective sum rules.

Writing the sum rule from Eq.~\eqref{eq:A4TpSAME_1} on top, the real parts of the two sum rules imply the constraints,
\begin{equation}
 {\rm Re} \Rightarrow \left\{
 \begin{matrix}
 \frac{1}{m_1} - \frac{2}{m_2} \cos \alpha_{21} - \frac{1}{m_3} \cos \alpha_{31} = 0,\\
 \frac{1}{m_1} - \frac{2}{m_2} \cos \alpha_{21} + \frac{1}{m_3} \cos \alpha_{31} = 0,
 \end{matrix}
 \right.
 \label{eq:A4TpSAME_3}
\end{equation}
while the imaginary parts impose the constraints,
\begin{equation}
 {\rm Im} \Rightarrow \left\{
 \begin{matrix}
 \frac{2}{m_2} \sin \alpha_{21} + \frac{1}{m_3} \sin \alpha_{31} = 0,\\
 \frac{2}{m_2} \sin \alpha_{21} - \frac{1}{m_3} \sin \alpha_{31} = 0.
 \end{matrix}
 \right.
 \label{eq:A4TpSAME_4}
\end{equation}
While the upper and lower equations look differently in each case, it is easy to see that one can apply the identities $\sin \alpha_{31} = - \sin ( \alpha_{31} + \pi )$ and $\cos \alpha_{31} = - \cos ( \alpha_{31} + \pi )$, along with the redefinition $\alpha_{31} + \pi \to \alpha_{31}$ to the respective lower equations to prove that the constraints implied by the sum rules are perfectly identical. Note, however, that this latter redefinition is only possible as long as the phases are not known by some complementary source (e.g.\ by a hypothetical future experiment which would be able to measure the Majorana phases or by a model which gives clear and concrete predictions for the phases).

\subsection{\label{sec:concrete_FREQUENT}The sum rule $\tilde m_1 = 2 \tilde m_2 + \tilde m_3$}

\begin{center}
\begin{figure}[h!]
\begin{tabular}{lcr}
\includegraphics[width=5cm]{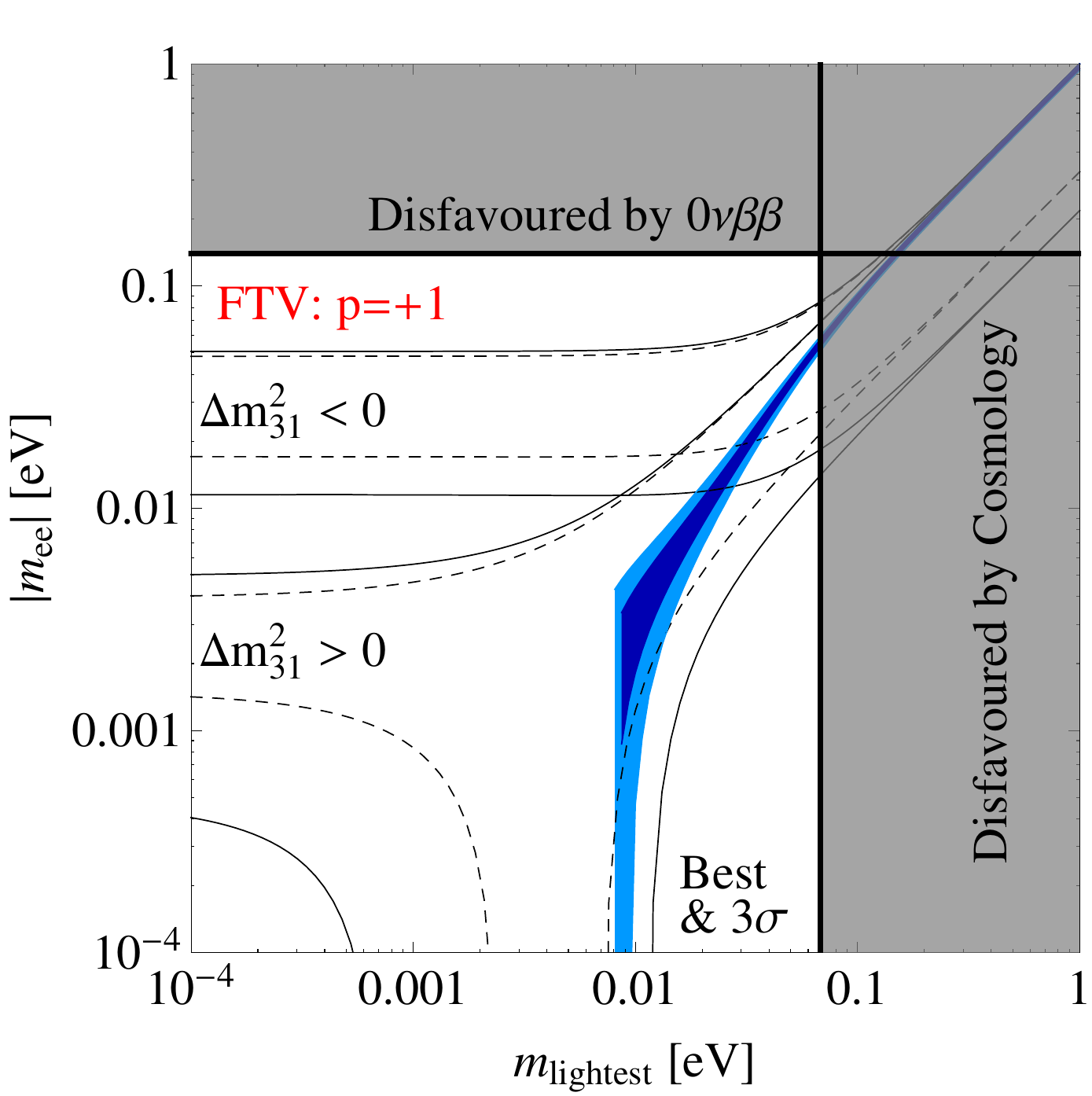} & \includegraphics[width=5cm]{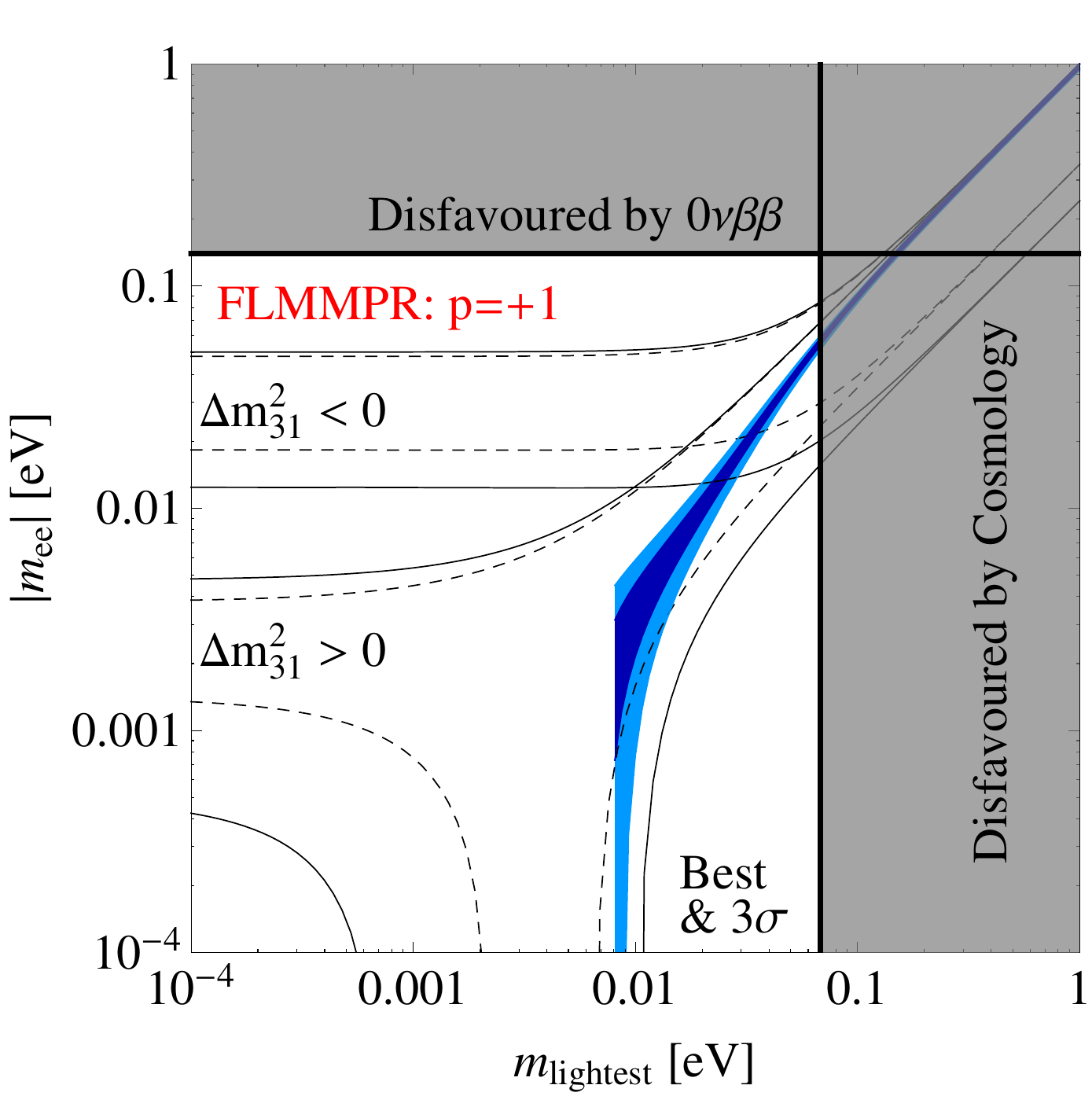} &
\includegraphics[width=5cm]{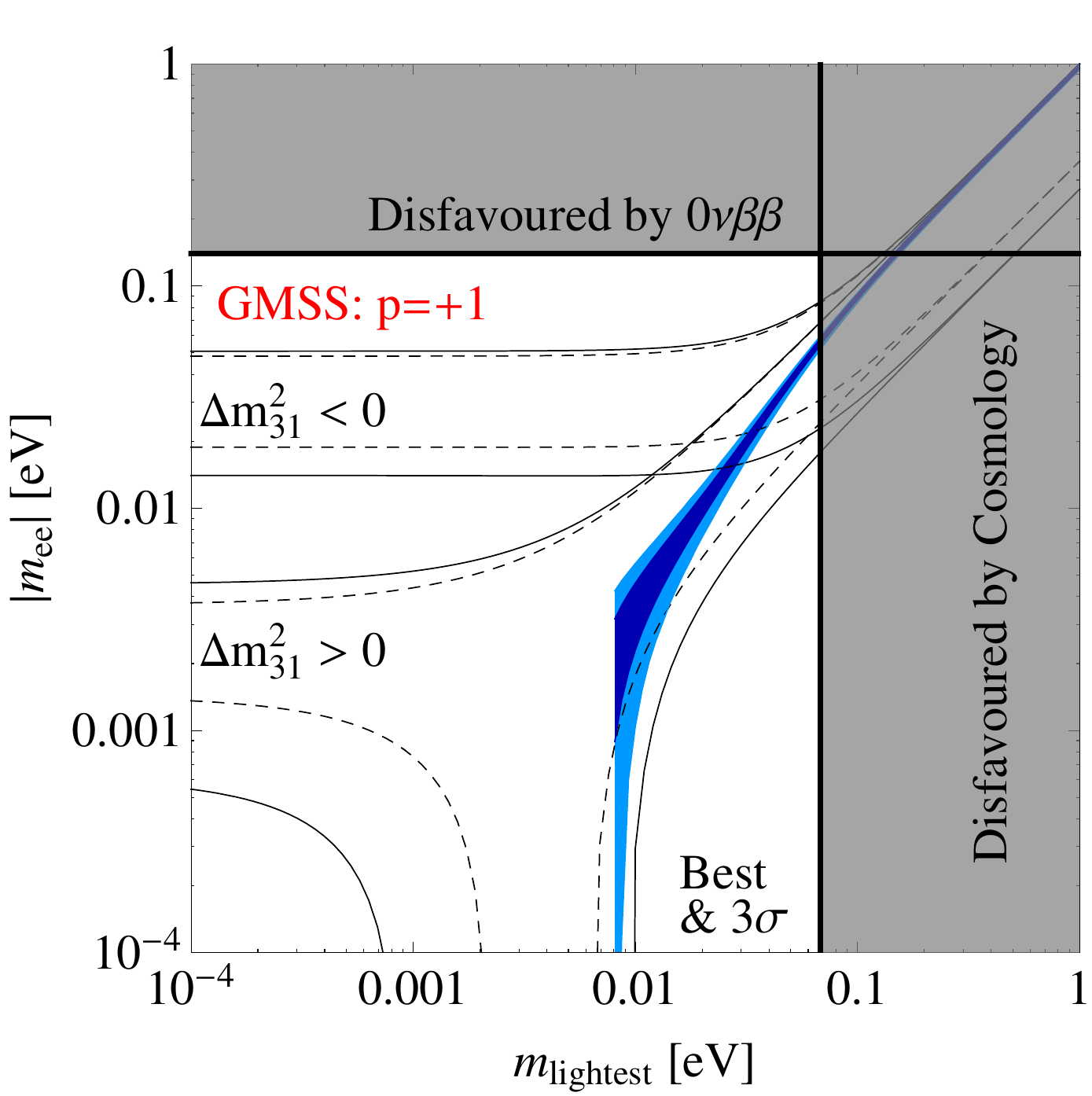}
\end{tabular}
\caption{\label{fig:mee11081795}Allowed regions for the sum rule $\tilde m_1 = 2 \tilde m_2 + \tilde m_3$.}
\end{figure}
\end{center}

This sum rule is one of the most frequent ones in the literature, and it has been found using $A_4$~\cite{Barry:2010zk,Altarelli:2005yp,Altarelli:2006kg,Ma:2006vq,Bazzocchi:2007na,Bazzocchi:2007au,Lin:2008aj,Ma:2009wi,Ciafaloni:2009qs,Fukuyama:2010mz,Ma:2005sha,Ma:2006wm,Honda:2008rs,Brahmachari:2008fn,Altarelli:2005yx,Chen:2009um}, $S_4$~\cite{Bazzocchi:2008ej}, $T'$~\cite{Chen:2007afa,Ding:2008rj,Chen:2009gf,Feruglio:2007uu,Merlo:2011hw,Chen:2009gy}, or $T_7$~\cite{Luhn:2012bc} symmetries. Explicitly, it reads
\begin{equation}
 \tilde m_1 = 2 \tilde m_2 + \tilde m_3,
 \label{eq:FREQUENT_1}
\end{equation}
leading to the parameter values
\begin{equation}
 p=1,\ \ B_2 = 2,\ \  B_3 = 1,\ \ {\rm and}\ \ \Delta \chi_{21} = \Delta \chi_{31} = \pi.
 \label{eq:FREQUENT_2}
\end{equation}
The allowed regions are displayed in Fig.~\ref{fig:mee11081795} (cf.\ Fig.~4(b) in Ref.~\cite{Bazzocchi:2009da}).
As one can see easily, the absolute value of the LHS of Eq.~\eqref{eq:FREQUENT_1} is always smaller than its RHS for IO, which is why this mass ordering is forbidden and not colored in the plots.

\subsection{\label{sec:concrete_S4Rabi}The sum rule $\tilde m_1 = \tilde m_3 - 2 \tilde m_2$}

\begin{center}
\begin{figure}[h!]
\begin{tabular}{lcr}
\includegraphics[width=5cm]{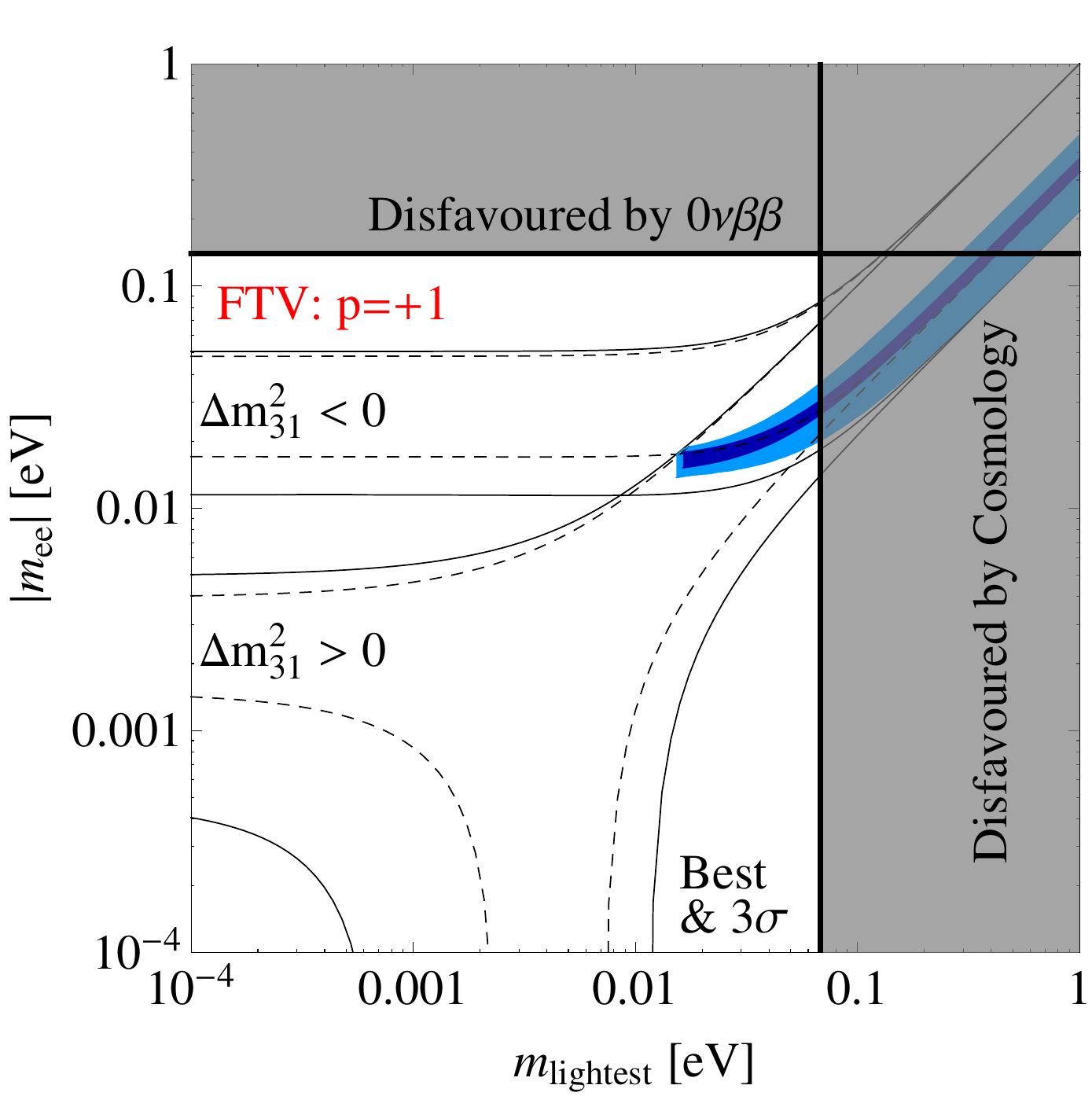} & \includegraphics[width=5cm]{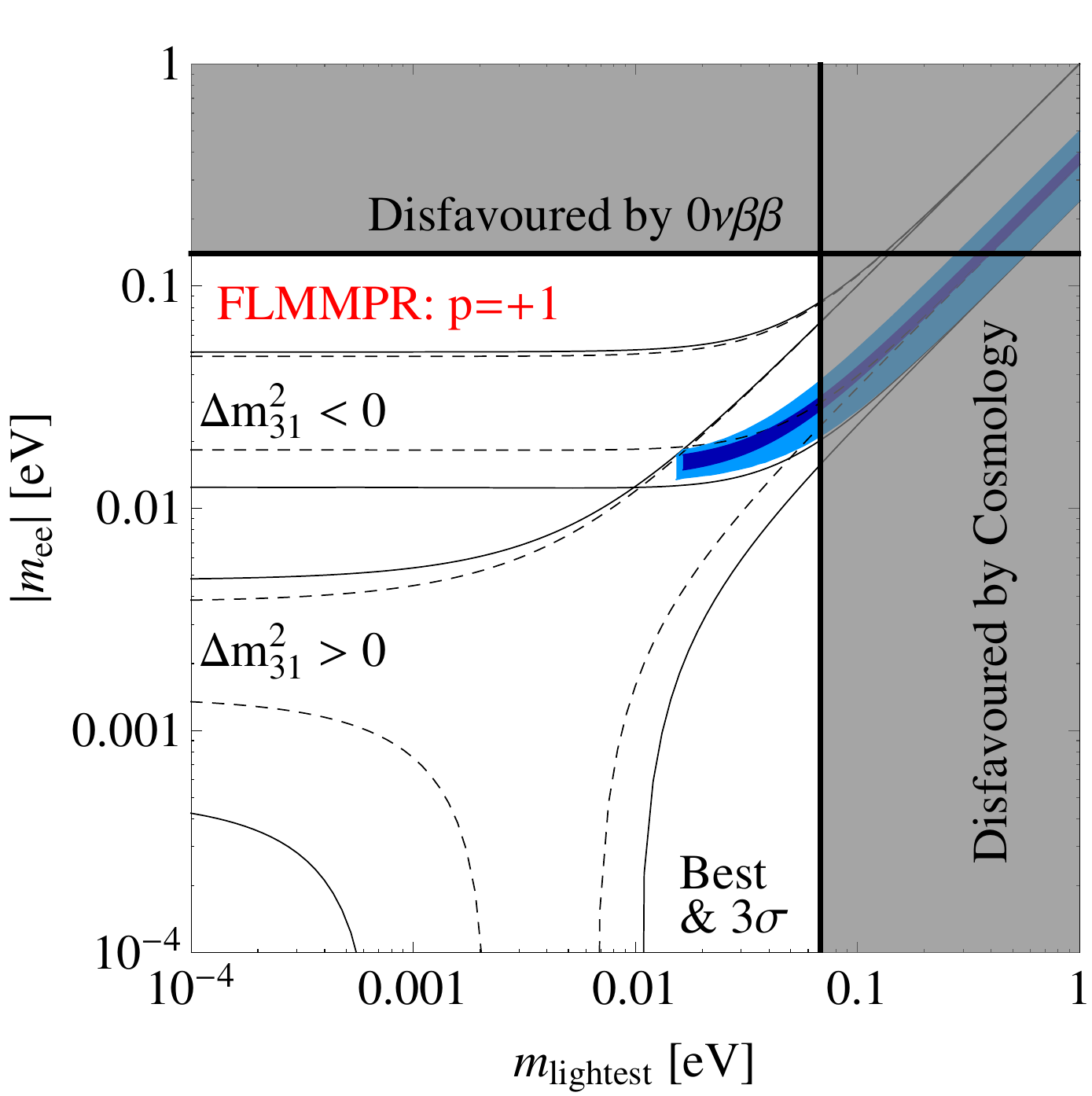} &
\includegraphics[width=5cm]{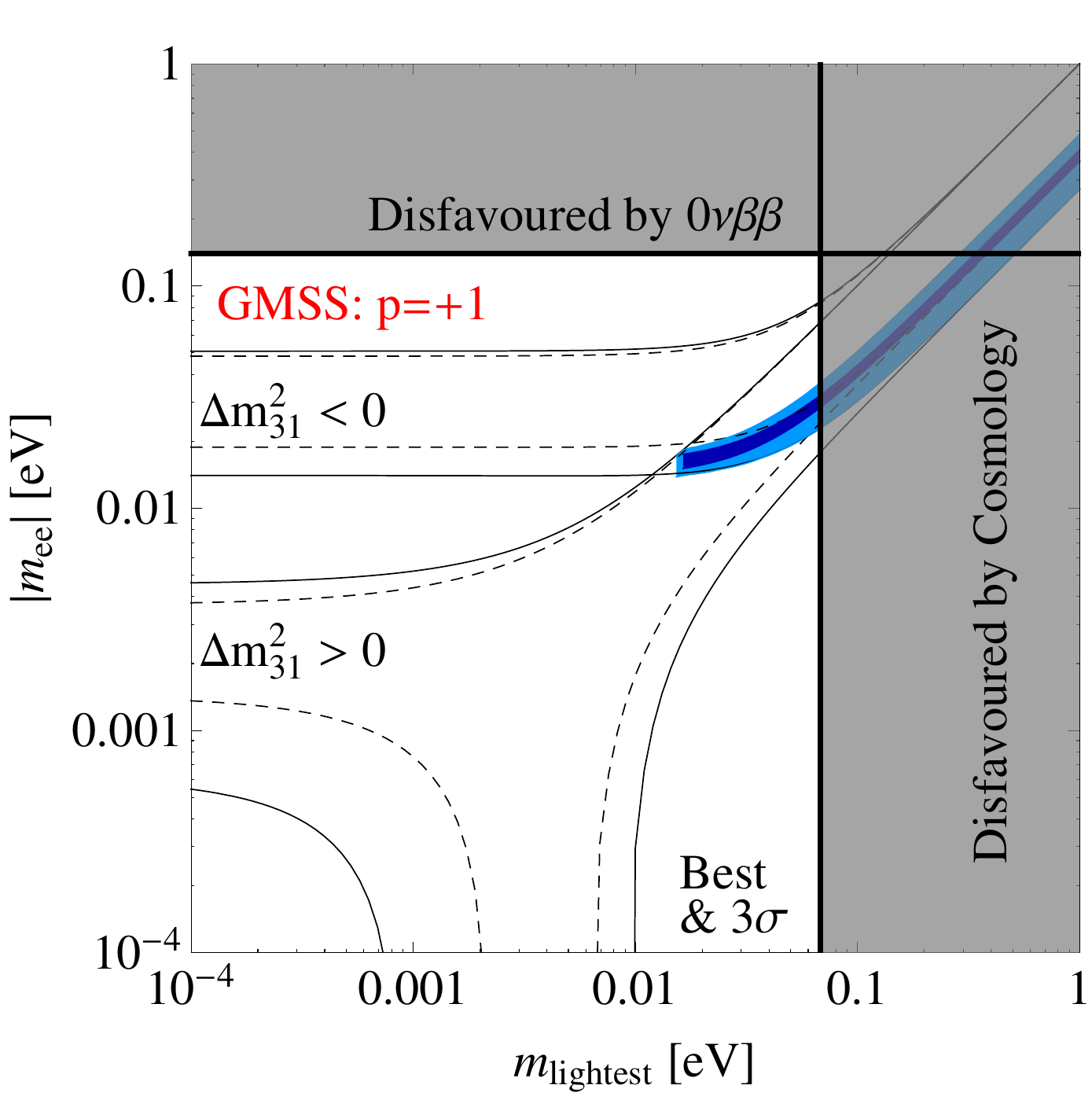}
\end{tabular}
\caption{\label{fig:mee12082875}Allowed regions for the sum rule $\tilde m_1 = \tilde m_3 - 2 \tilde m_2$.}
\end{figure}
\end{center}

This sum rule can be derived from the model in Ref.~\cite{Mohapatra:2012tb}, which is based on an $S_4$ flavour symmetry. Explicitly, the sum rule reads
\begin{equation}
 \tilde m_1 = \tilde m_3 - 2 \tilde m_2,
 \label{eq:S4Rabi_1}
\end{equation}
leading to the parameter values
\begin{equation}
 p=1,\ \ B_2 = 2,\ \  B_3 = 1,\ \ \Delta \chi_{21} = 0,\ \ {\rm and}\ \ \Delta \chi_{31} = \pi.
 \label{eq:S4Rabi_2}
\end{equation}
Similar to some of the previous cases, the absolute value of the LHS of Eq.~\eqref{eq:S4Rabi_1} is always larger than the absolute value of the RHS, which is confirmed by our numerical results displayed in Fig.~\ref{fig:mee12082875}.

\subsection{\label{sec:concrete_TpClaudia}The sum rule $\tilde m_1 + \tilde m_3 = 2 \tilde m_2$}

\begin{center}
\begin{figure}[h!]
\begin{tabular}{lcr}
\includegraphics[width=5cm]{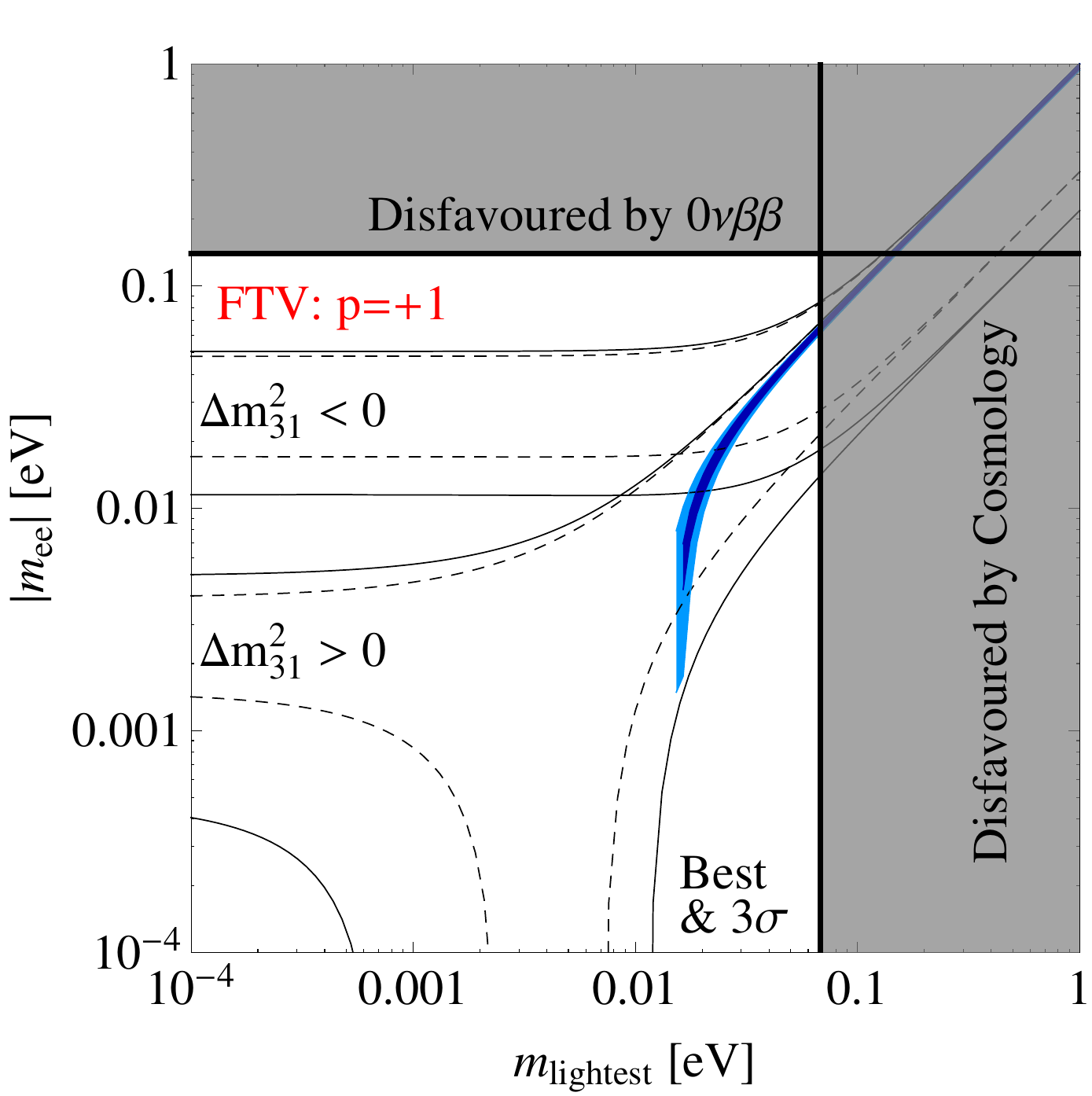} & \includegraphics[width=5cm]{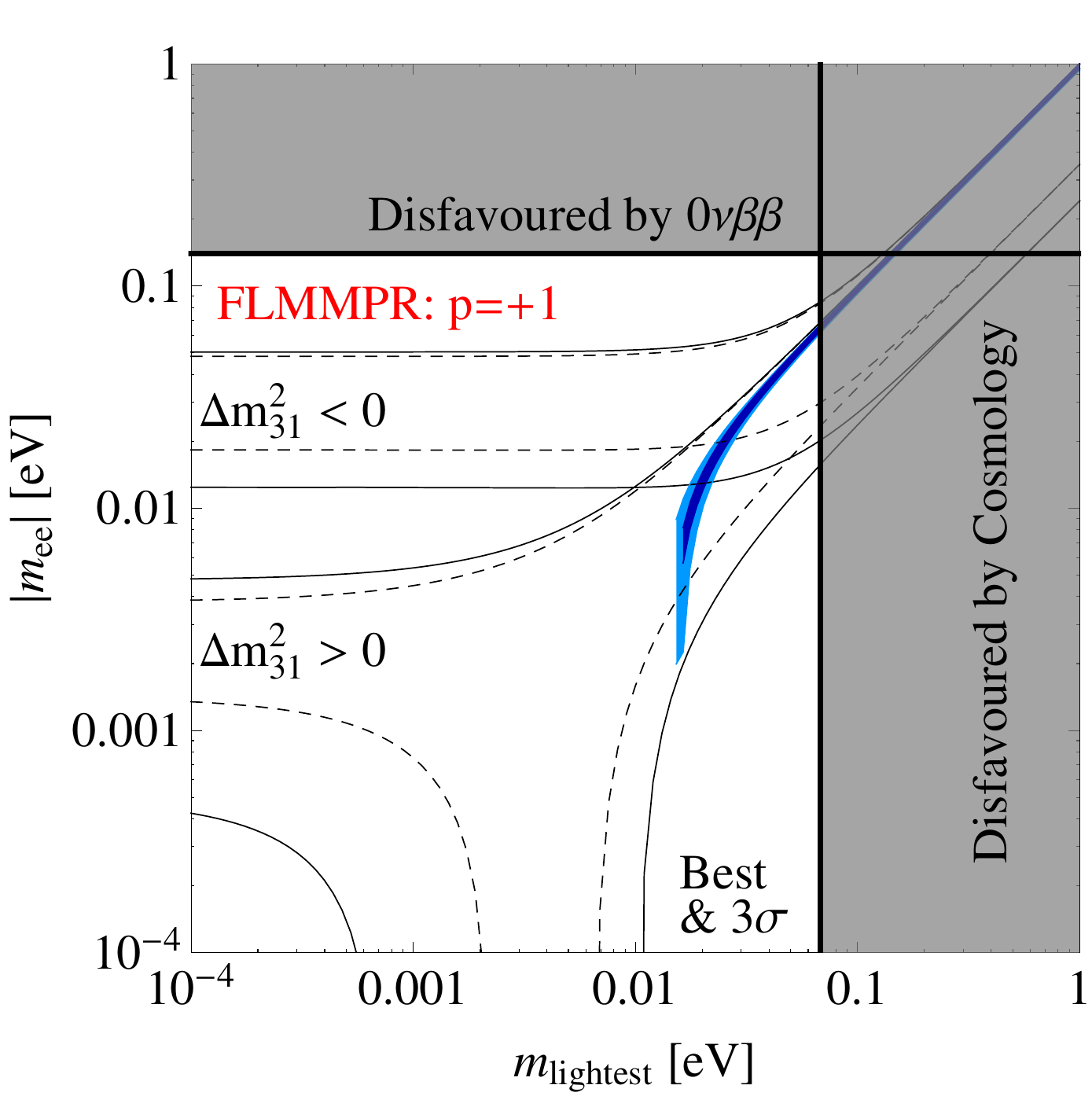} &
\includegraphics[width=5cm]{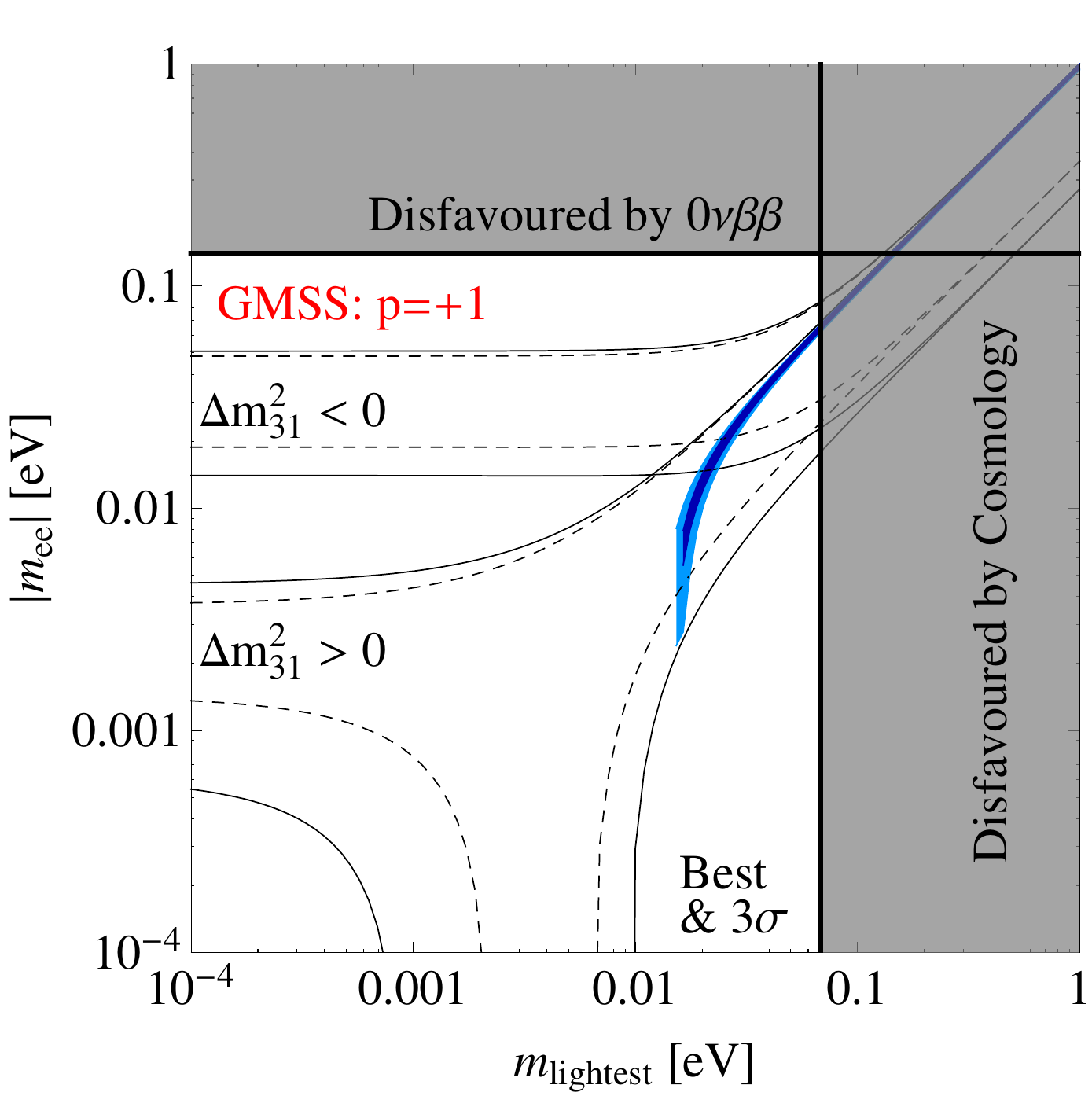}
\end{tabular}
\caption{\label{fig:mee13037178}Allowed regions for the sum rule $\tilde m_1 + \tilde m_3 = 2 \tilde m_2$.}
\end{figure}
\end{center}

Ref.~\cite{Feruglio:2013hia} presented a sum rule in a model based on an $S_4$ symmetry, given by
\begin{equation}
 \tilde m_1 + \tilde m_3 = 2 \tilde m_2,
 \label{eq:TpClaudia_1}
\end{equation}
and leading to the parameter values
\begin{equation}
 p=1,\ \ B_2 = 2,\ \  B_3 = 1,\ \ \Delta \chi_{21} = \pi,\ \ {\rm and}\ \ \Delta \chi_{31} = 0.
 \label{eq:TpClaudia_2}
\end{equation}
Note that this sum rule only differs from the one discussed in Sec.~\ref{sec:concrete_S4Rabi} by relative signs, cf.\ Eqs.~\eqref{eq:TpClaudia_2} and~\eqref{eq:S4Rabi_2}. According to Ref.~\cite{Barry:2010zk}, such relative signs should play no role, but our numerical results in Fig.~\ref{fig:mee13037178} seem to illustrate a counter example to that statement.
While for the same reasons as the sum rule in the previous section this sum rule forbids IO, the allowed regions for NO clearly look different. Trying to understand this behaviour analytically, we can again look at the real and imaginary parts of the sum rules. Writing the expressions corresponding to Eq.~\eqref{eq:TpClaudia_1} on top, we can derive the real
\begin{equation}
 {\rm Re} \Rightarrow \left\{
 \begin{matrix}
 m_1 - 2 m_2 \cos \alpha_{21} + m_3 \cos \alpha_{31} = 0,\\
 m_1 + 2 m_2 \cos \alpha_{21} - m_3 \cos \alpha_{31} = 0,
 \end{matrix}
 \right.
 \label{eq:TpClaudia_3}
\end{equation}
and imaginary parts,
\begin{equation}
 {\rm Im} \Rightarrow \left\{
 \begin{matrix}
 - 2 m_2 \sin \alpha_{21} + m_3 \sin \alpha_{31} = 0,\\
 + 2 m_2 \sin \alpha_{21} - m_3 \sin \alpha_{31} = 0.
 \end{matrix}
 \right.
 \label{eq:TpClaudia_4}
\end{equation}
Apparently, the constraints arising from the imaginary parts of the two sum rules are \emph{identical}, while the ones derived from the real parts are different. Accordingly, in this case, one \emph{cannot} find any phase redefinition (as done in Sec.~\ref{sec:concrete_A4TpSAME}) which could render the resulting conditions equal. Indeed, this example explicitly verifies that relative signs can in fact play a role, and they are \emph{not} always negligible. However, it is not so easy to generalize this statement for an arbitrary value of the power $p$, so that it is safest to investigate it case by case.

\subsection{\label{sec:concrete_Dirac}The sum rule $\tilde{m}_1+\tilde{m}_2=2\tilde{m}_3$}

\begin{center}
\begin{figure}[h!]
\begin{tabular}{lcr}
\includegraphics[width=5cm]{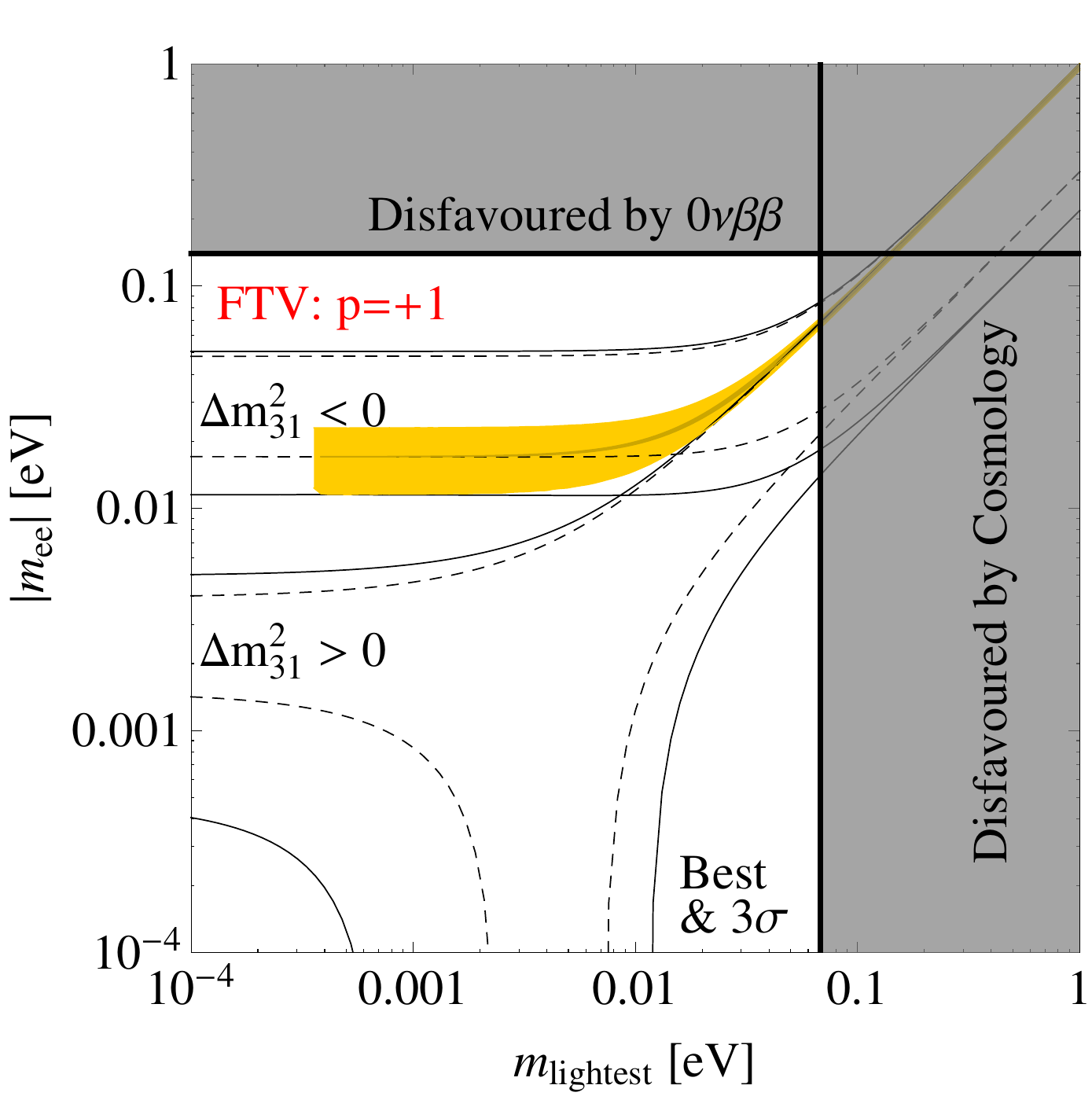} & \includegraphics[width=5cm]{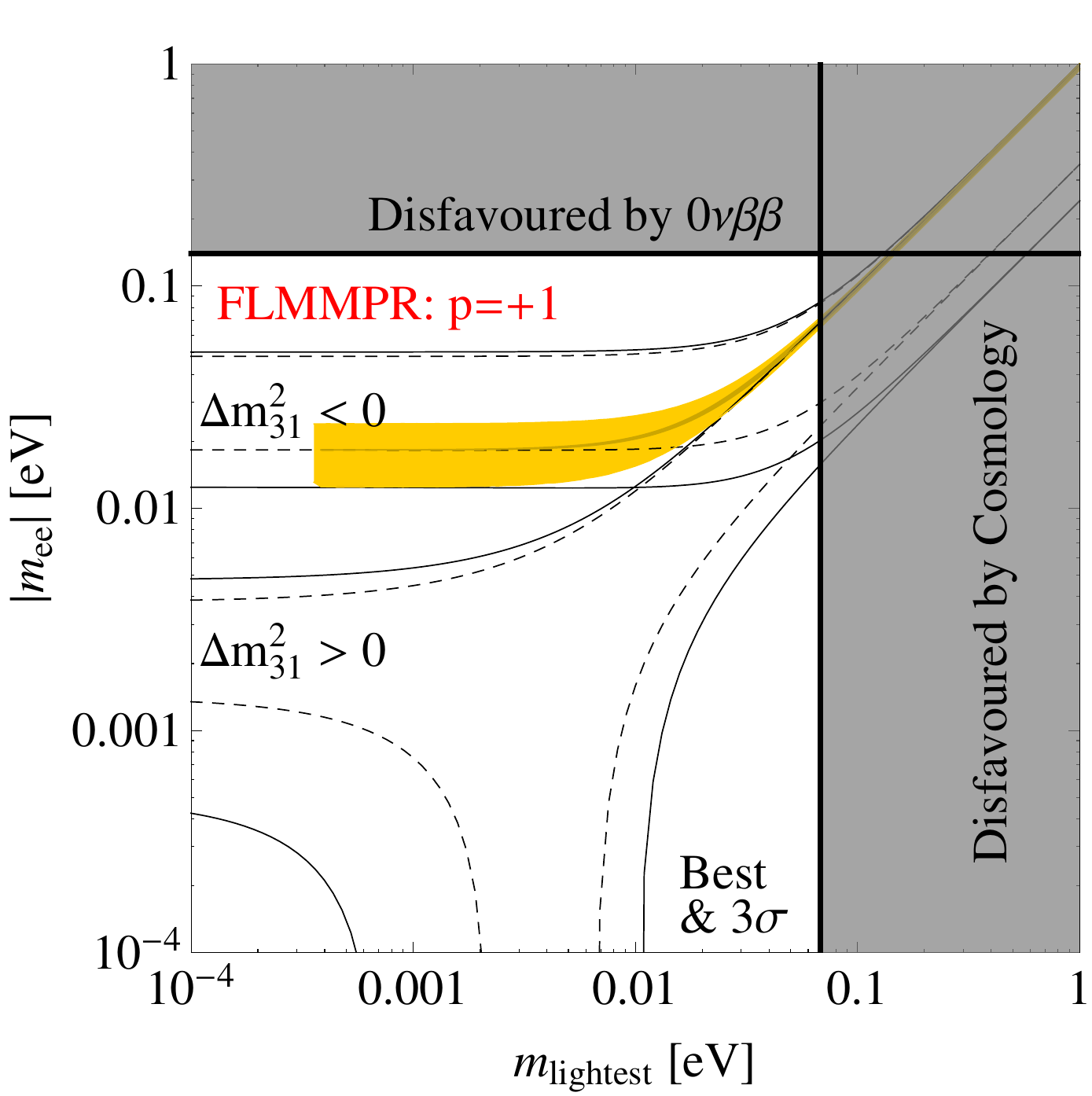} &
\includegraphics[width=5cm]{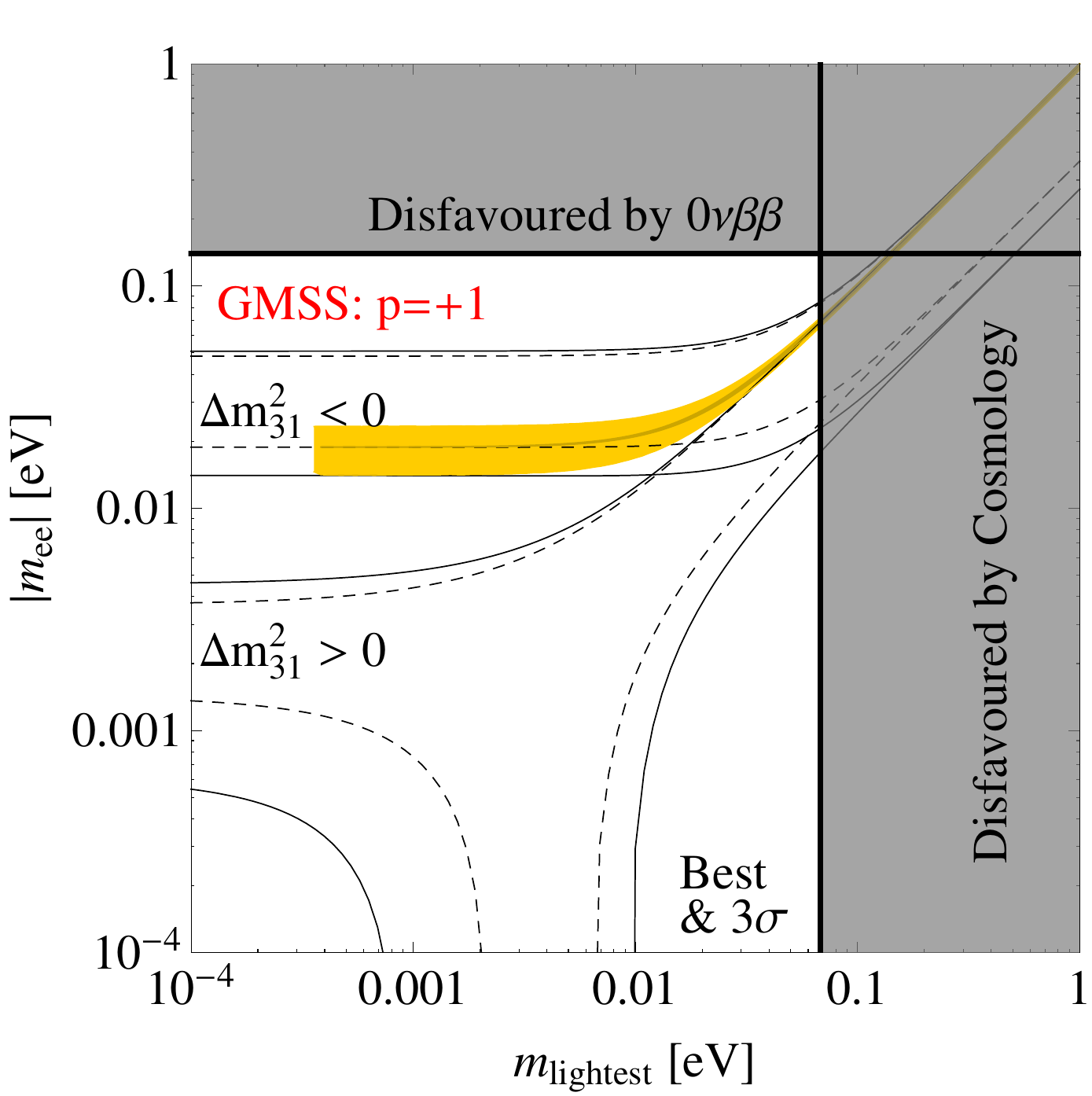}
\end{tabular}
\caption{\label{fig:mee13042645}Allowed regions for the sum rule $\tilde{m}_1+\tilde{m}_2=2\tilde{m}_3$.}
\end{figure}
\end{center}

This sum rule has been derived in the framework of an extra-dimensional $S_4$ model in Ref.~\cite{Ding:2013eca}. However, in that case it was actually a Dirac (real) mass sum rule, which in particular would not lead to $0\nu\beta\beta$. On the other hand, it has also been found in a seesaw type~II framework with a $L_e - L_\mu - L_\tau$ symmetry~\cite{Lindner:2010wr}, which can be seen as \emph{complex generalisation} of the Dirac sum rule. Explicitly, the rule is given by
\begin{equation}
 \tilde{m}_1+\tilde{m}_2=2\tilde{m}_3,
 \label{eq:Dirac_1}
\end{equation}
leading to the parameter values
\begin{equation}
 p=1,\ \ B_2 = 1,\ \  B_3 = 2,\ \ \Delta \chi_{21} = 0,\ \ {\rm and}\ \ \Delta \chi_{31} = \pi.
 \label{eq:Dirac_2}
\end{equation}
The resulting allowed regions are displayed in Fig.~\ref{fig:mee13042645}.
Remarkably, this is one of only two sum rules we have found in the literature which forbid NO. Indeed, in the case of $m_1 < m_2 < m_3$, the LHS of Eq.~\eqref{eq:Dirac_1} is always smaller than the RHS, while the rule can easily be fulfilled in the case of IO.

\subsection{\label{sec:concrete_scotogenic}The sum rule $\tilde{m}_1^{1/2}+\tilde{m}_3^{1/2}=2\tilde{m}_2^{1/2}$}

\begin{center}
\begin{figure}[h!]
\begin{tabular}{lcr}
\includegraphics[width=5cm]{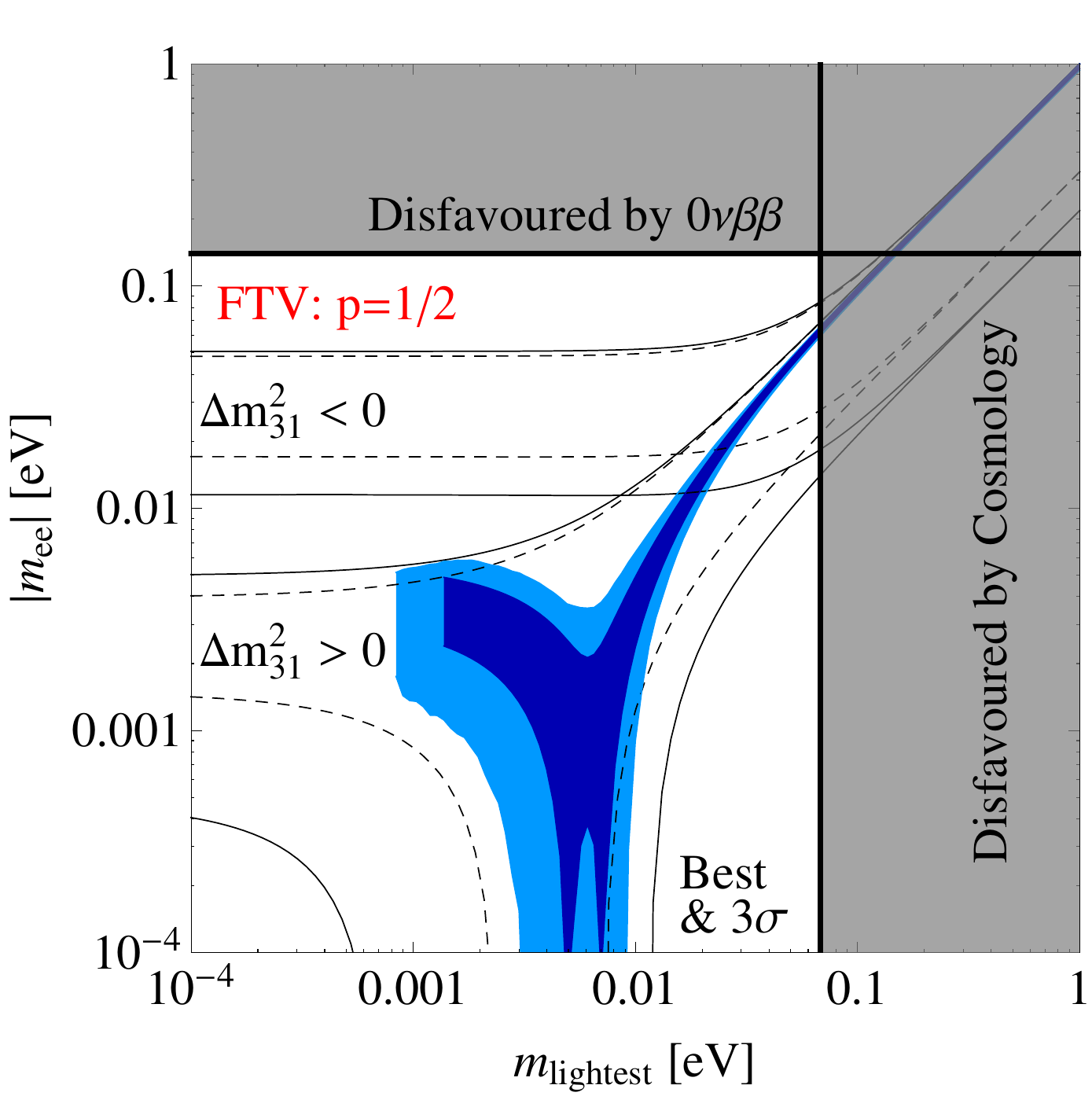} & \includegraphics[width=5cm]{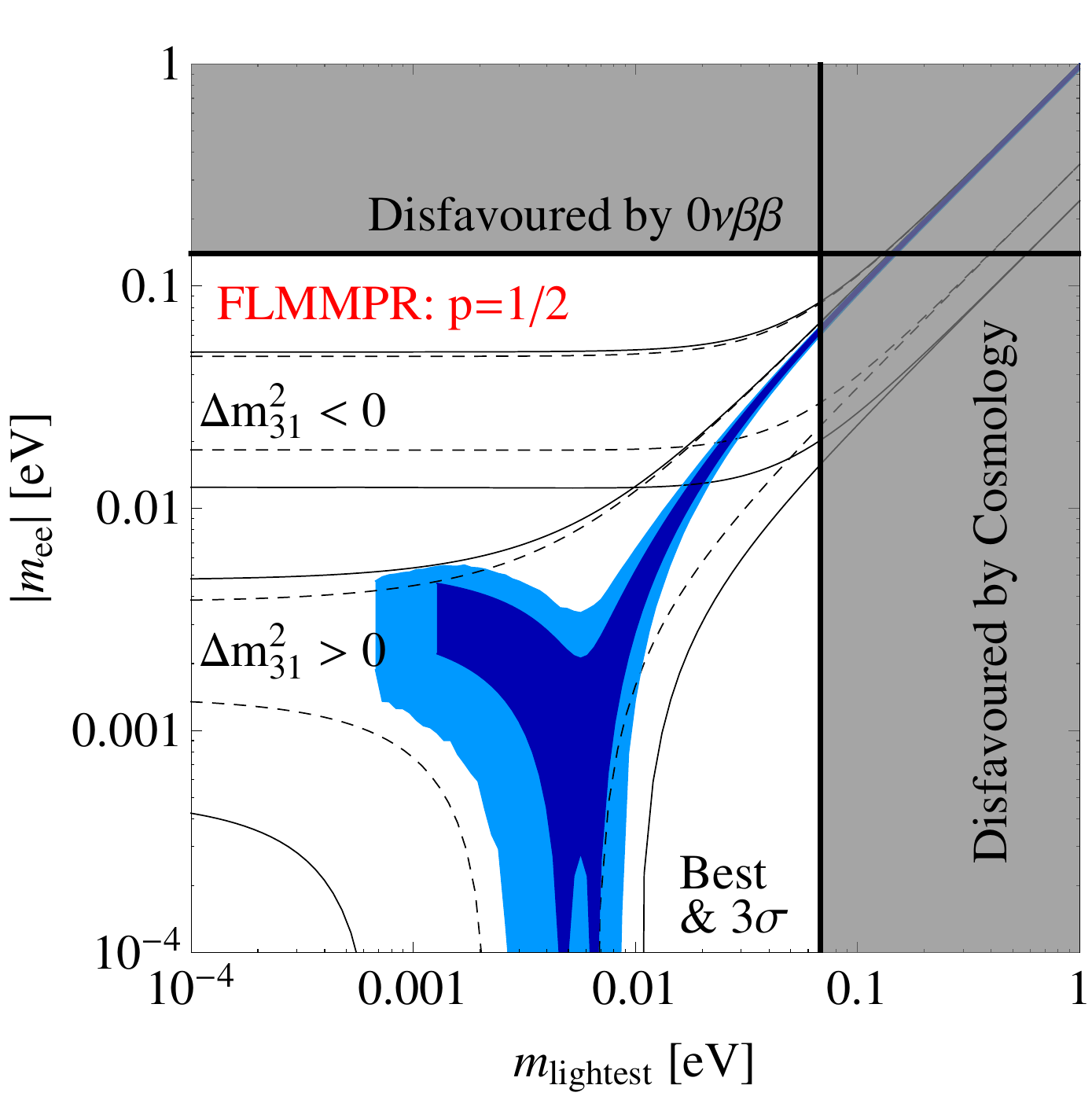} &
\includegraphics[width=5cm]{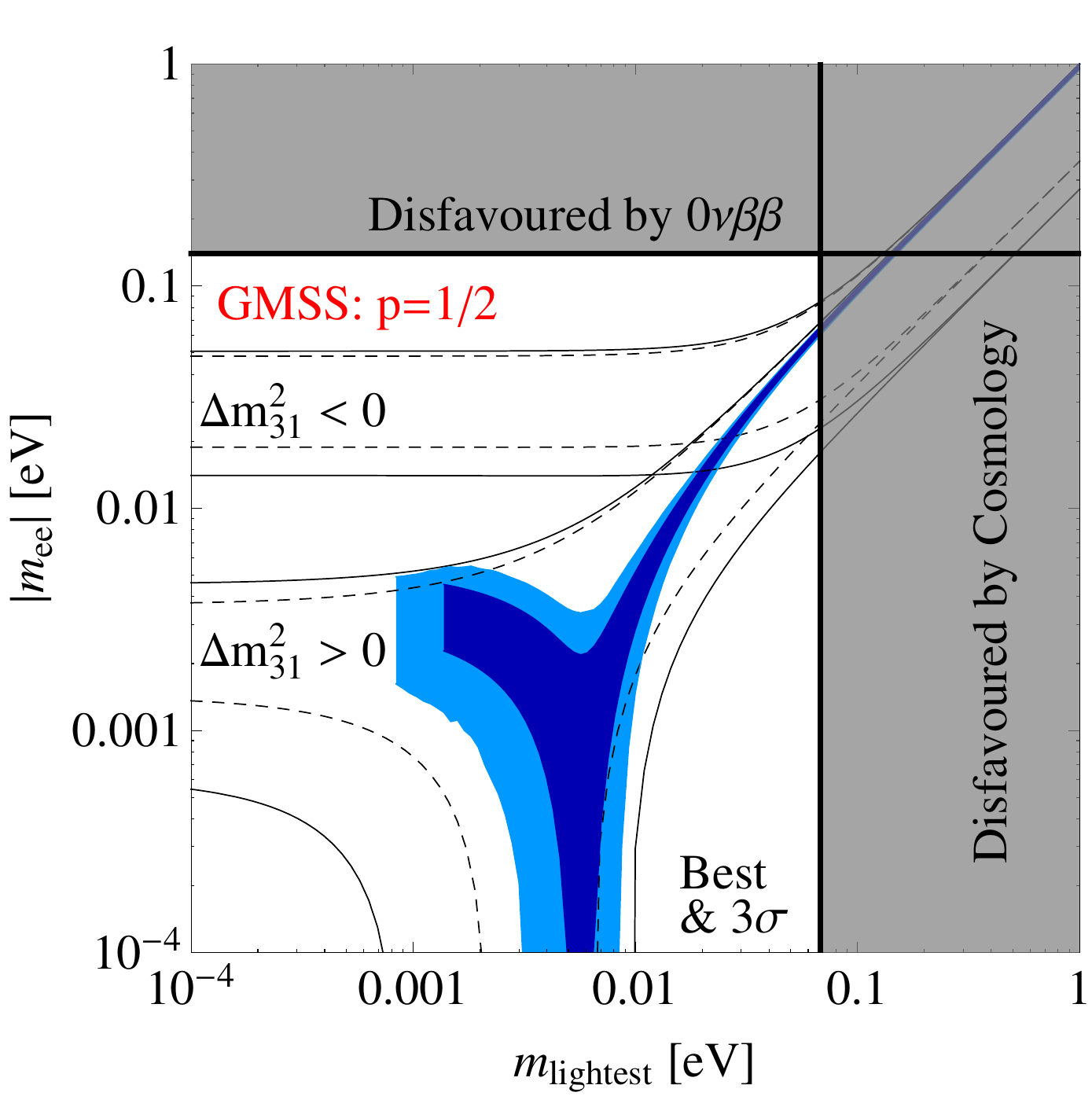}
\end{tabular}
\caption{\label{fig:mee09072147}Allowed regions for the sum rule $\tilde{m}_1^{1/2}+\tilde{m}_3^{1/2}=2\tilde{m}_2^{1/2}$.}
\end{figure}
\end{center}

A sum rule which is very similar to the one discussed in Sec.~\ref{sec:concrete_A4Z2} can be derived from the model presented in Ref.~\cite{Adulpravitchai:2009gi}, which is based on an $A_4$ symmetry. It is the only example we have found which generates the neutrino mass at 1-loop level and at the same time yields a mass sum rule. This rule is given by 
\begin{equation}
 \tilde{m}_1^{1/2}+\tilde{m}_3^{1/2}=2\tilde{m}_2^{1/2},
 \label{eq:scoto_1}
\end{equation}
leading to the parameter values
\begin{equation}
 p=1/2,\ \ B_2 = 2,\ \  B_3 = 1,\ \ \Delta \chi_{21} = \pi,\ \ {\rm and}\ \ \Delta \chi_{31} = 0.
 \label{eq:scoto_2}
\end{equation}
It should be easy by now to see that this sum rule forbids IO, and the corresponding allowed regions are displayed in Fig.~\ref{fig:mee09072147}.

\subsection{\label{sec:concrete_irrational}The sum rule $\tilde{m}_1- \frac{\sqrt{3}-1}{2} \tilde{m}_2 + \frac{\sqrt{3}+1}{2} \tilde{m}_3 = 0$}

\begin{center}
\begin{figure}[h!]
\begin{tabular}{lcr}
\includegraphics[width=5cm]{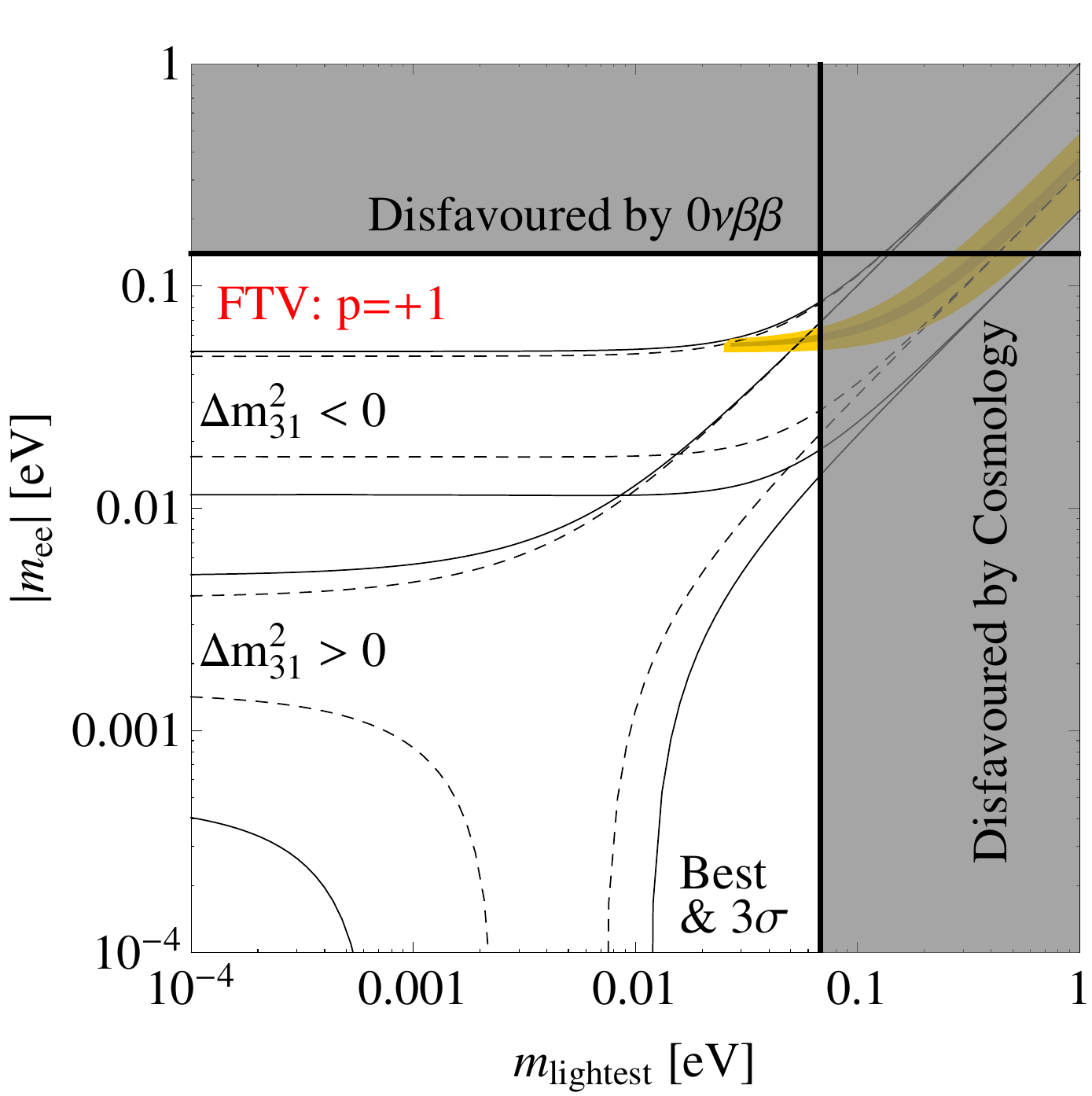} & \includegraphics[width=5cm]{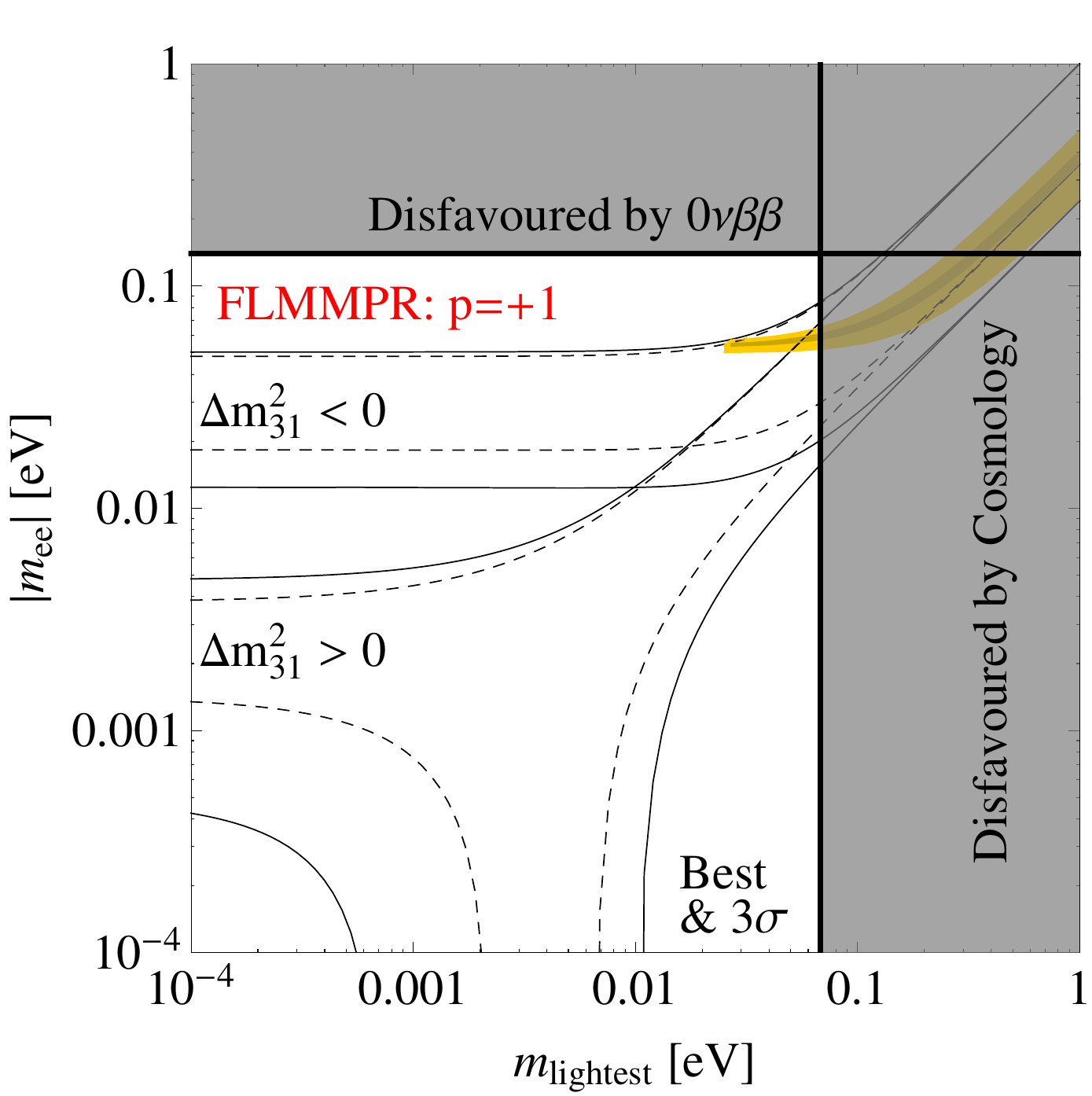} &
\includegraphics[width=5cm]{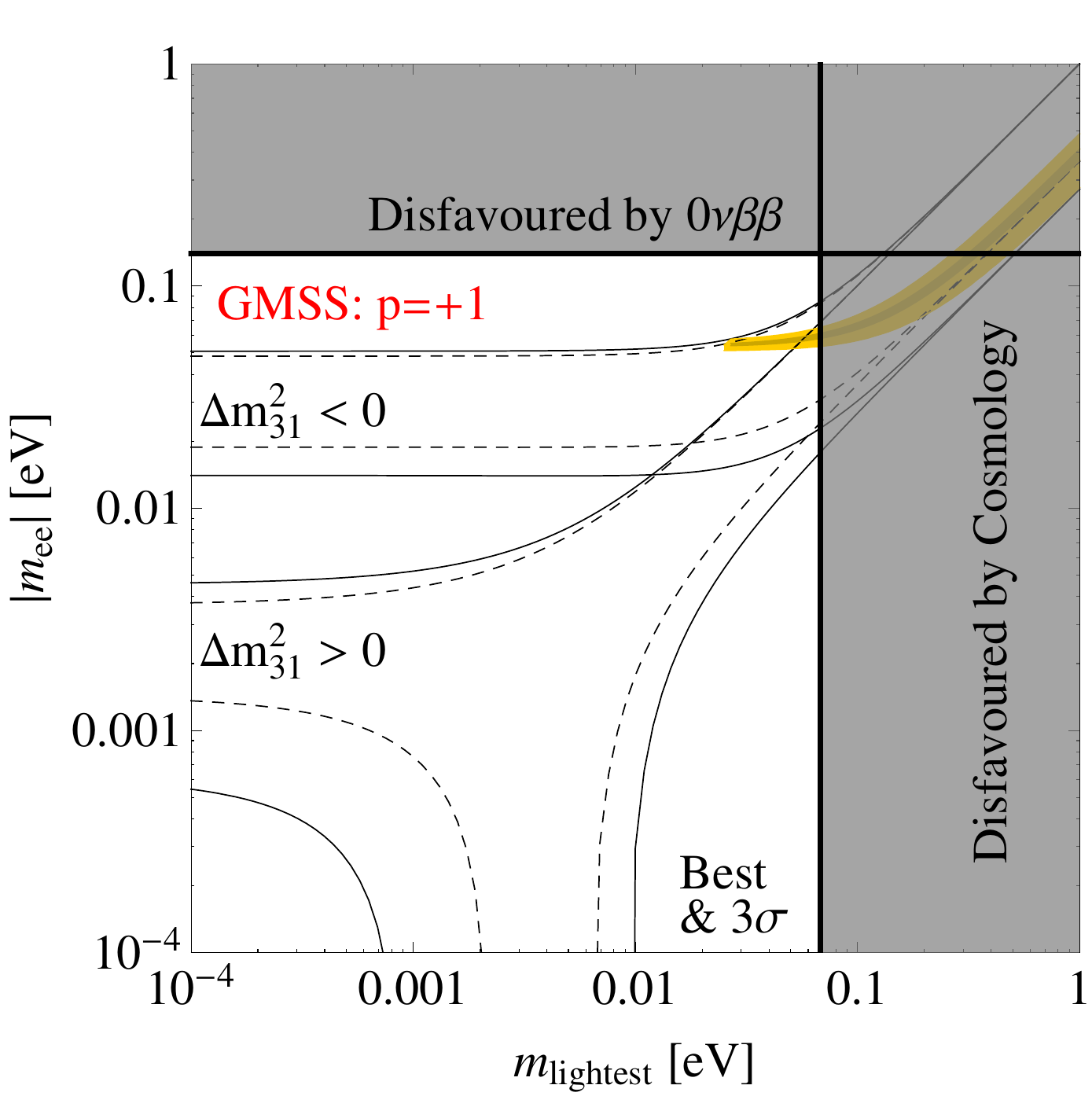}
\end{tabular}
\caption{\label{fig:mee11103640}Allowed regions for the sum rule $\tilde{m}_1- \frac{\sqrt{3}-1}{2} \tilde{m}_2 + \frac{\sqrt{3}+1}{2} \tilde{m}_3 = 0$.}
\end{figure}
\end{center}

Finally, the only sum rule with non-rational coefficients that we have found can be derived from the model presented in Ref.~\cite{Hashimoto:2011tn}, which is based on an $A_5'$ symmetry and which uses the Weinberg operator to generate the light neutrino mass.\footnote{Note, however, that this sum rule is \emph{not} unique, since the the model from Ref.~\cite{Hashimoto:2011tn} does not a priori specify which neutrino mass eigenstate corresponds to which generation.} This rule is given by 
\begin{equation}
 \tilde{m}_1- \frac{\sqrt{3}-1}{2} \tilde{m}_2 + \frac{\sqrt{3}+1}{2} \tilde{m}_3 = 0,
 \label{eq:irrational_1}
\end{equation}
leading to the parameter values
\begin{equation}
 p=+1,\ \ B_2 = \frac{\sqrt{3}-1}{2},\ \  B_3 = \frac{\sqrt{3}+1}{2},\ \ \Delta \chi_{21} = \pi,\ \ {\rm and}\ \ \Delta \chi_{31} = 0.
 \label{eq:irrational_2}
\end{equation}
This is another sum rule we found which forbids NO, as can be seen from the plots in Fig.~\ref{fig:mee11103640}.
This behaviour is again easy to understand analytically. Slightly rewriting Eq.~\eqref{eq:irrational_1},
\begin{equation}
 \tilde{m}_1 + \frac{\sqrt{3}+1}{2} \tilde{m}_3 = \frac{\sqrt{3}-1}{2} \tilde{m}_2,
 \label{eq:irrational_3}
\end{equation}
one can compare the absolute values of both sides. Since $\sqrt{3} > 1$ and hence $\frac{\sqrt{3}+1}{2} > 1$, we obtain for NO:
\begin{eqnarray}
 && |{\rm LHS}| \geq \frac{\sqrt{3}+1}{2} m_3 - m_1 > \frac{\sqrt{3}+1}{2} m_2 - m_1 = \frac{\sqrt{3}-1 + 2}{2} m_2 - m_1 \nonumber\\
 && = \frac{\sqrt{3}-1}{2} m_2 + \underbrace{(m_2 - m_1)}_{>0} > \frac{\sqrt{3}-1}{2} m_2 = |{\rm RHS}|.
 \label{eq:irrational_3}
\end{eqnarray}
Indeed, NO is not possible for this sum rule.

\subsection{\label{sec:concrete_summary}A summary table of all the sum rules we have found}

We have in Tab.~\ref{tab:models} compiled a summary of all the different models leading to sum rules. We have always indicated the respective sum rule, which is the decisive piece of information from an experimental point of view, and we have grouped the respective references according to the flavour symmetry they are based on. These groups of models can be probed simultaneously by the respective sum rule.\footnote{Note that we did not include the hypothetical sum rule from Sec.~\ref{sec:concrete_hypothetical}, as currently there is no corresponding model known.}

\begin{table}[ht]
\begin{tabular}{|c|c|c|c|}
\hline
Sum Rule & Group & Seesaw Type& Matrix\\\hline
$\tilde{m}_1+\tilde{m}_2=\tilde{m}_3$ & $A_4$\cite{Barry:2010zk,Ding:2010pc,Ma:2005sha,Ma:2006wm,Honda:2008rs,Brahmachari:2008fn}; $S_4$\cite{Bazzocchi:2009pv}; $A_5$\cite{Everett:2008et}$^*$ &Weinberg& $M_{\nu}$\\
$\tilde{m}_1+\tilde{m}_2=\tilde{m}_3$  &$\Delta(54)$\cite{Boucenna:2012qb}; $S_4$\cite{Bazzocchi:2009da} &Type II&$M_L$\\\hline
$\tilde{m}_1+2\tilde{m}_2 = \tilde{m}_3$ & $S_4$\cite{Mohapatra:2012tb}&Type II&$M_L$\\\hline
$2\tilde{m}_2+\tilde{m}_3=\tilde{m}_1$ & $A_4$\cite{Barry:2010zk,Altarelli:2005yp,Altarelli:2006kg,Ma:2006vq,Bazzocchi:2007na,Bazzocchi:2007au,Lin:2008aj,Ma:2009wi,Ciafaloni:2009qs,Ma:2005sha,Ma:2006wm,Honda:2008rs,Brahmachari:2008fn,Altarelli:2005yx,Chen:2009um} &Weinberg&$M_{\nu}$\\
  & $S_4$\cite{Bazzocchi:2008ej,Feruglio:2013hia}$^\dagger$; $T'$\cite{Chen:2007afa,Ding:2008rj,Chen:2009gf,Feruglio:2007uu,Merlo:2011hw,Chen:2009gy}; $T_7$\cite{Luhn:2012bc} & & \\
$2\tilde{m}_2+\tilde{m}_3=\tilde{m}_1$ & $A_4$\cite{Fukuyama:2010mz} & Type II&$M_L$\\\hline
$\tilde{m}_1+\tilde{m}_2=2\tilde{m}_3$ & $S_4$\cite{Ding:2013eca}$^\ddagger$& Dirac$^\ddagger$ &$M_{\nu}$\\
$\tilde{m}_1+\tilde{m}_2=2\tilde{m}_3$ & $L_e - L_\mu - L_\tau$\cite{Lindner:2010wr} &Type II&$M_L$\\\hline
$\tilde{m}_1 + \frac{\sqrt{3}+1}{2} \tilde{m}_3 = \frac{\sqrt{3}-1}{2} \tilde{m}_2$ & $A_5'$\cite{Hashimoto:2011tn} &Weinberg&$M_{\nu}$\\\hline
$\tilde{m}_1^{-1}+\tilde{m}_2^{-1}=\tilde{m}_3^{-1}$ & $A_4$\cite{Barry:2010zk}; $S_4$\cite{Bazzocchi:2009da,Ding:2010pc}; $A_5$\cite{Ding:2011cm,Cooper:2012bd}&Type I&$M_R$\\
$\tilde{m}_1^{-1}+\tilde{m}_2^{-1}=\tilde{m}_3^{-1}$&$S_4$\cite{Bazzocchi:2009da}&Type III&$M_{\Sigma}$\\\hline
$2\tilde{m}_2^{-1}+\tilde{m}_3^{-1}=\tilde{m}_1^{-1}$ &$A_4$\cite{Barry:2010zk,Morisi:2007ft,Altarelli:2008bg,Adhikary:2008au,Lin:2009bw,Csaki:2008qq,Altarelli:2009kr,Hagedorn:2009jy,Burrows:2009pi,Ding:2009gh,Mitra:2009jj,delAguila:2010vg,Burrows:2010wz,Altarelli:2005yx,Chen:2009um}; $T'$\cite{Chen:2009gy}&Type I&$M_R$\\
$\tilde{m}_1^{-1}+\tilde{m}_3^{-1}=2\tilde{m}_2^{-1}$ & $A_4$\cite{He:2006dk,Berger:2009tt,Kadosh:2010rm}; $T'$\cite{Lavoura:2012cv}&Type I&$M_R$\\\hline
$\tilde{m}_3^{-1} \pm 2i\tilde{m}_2^{-1}=\tilde{m}_1^{-1}$ & $\Delta(96)$\cite{King:2012in}& Type I&$M_R$\\\hline
$\tilde{m}_1^{1/2}- \tilde{m}_3^{1/2}=2\tilde{m}_2^{1/2}$ & $A_4$\cite{Hirsch:2008rp}&Type I &$M_{D}$\\
$\tilde{m}_1^{1/2}+\tilde{m}_3^{1/2}=2\tilde{m}_2^{1/2}$ & $A_4$\cite{Adulpravitchai:2009gi}&Scotogenic&$h_{\nu}$\\\hline
$\tilde{m}_1^{-1/2}+\tilde{m}_2^{-1/2}=2\tilde{m}_3^{-1/2}$ & $S_4$\cite{Dorame:2012zv}&Inverse&$M_{RS}$\\\hline
\end{tabular}
\caption{\label{tab:models}A sample of the various sum rules found in the literature and the groups generating them, where the sume rules which are grouped together give identical predictions. $^*$In this reference, the sum rule was only used as a consistency relation. $^\dagger$ Note that in Ref.~\cite{Feruglio:2013hia} the Majorana phases were predicted to have trivial values, which is why the prediction of that concrete model is stronger than our general prediction from the sum rule only. $^\ddagger$Even though this reference predicts a sum rule, it features Dirac neutrinos.}
\end{table}

We have furthermore indicated how the light neutrino mass is generated in the respective types of models and which mass matrix leads to the sum rule. The different mass mechanisms (and the decisive mass matrices) used in the literature to obtain neutrino mass sum rules are the Weinberg operator~\cite{Weinberg:1979sa} (light neutrino mass matrix $M_{\nu}$), the type~I seesaw mechanism~\cite{Minkowski:1977sc,Ramond:1979py,Yanagida:1979as,GellMann:1980vs,Glashow:1979nm,Mohapatra:1979ia} (Dirac neutrino mass matrix $M_{D}$ or right-handed Majorana mass matrix $M_R$), the type~II seesaw mechanism~\cite{Magg:1980ut,Lazarides:1980nt} (left-handed Majorana mass matrix $M_L$), the type~III seesaw mechanism~\cite{Foot:1988aq,Ma:2002pf} (fermion triplet Majorana mass matrix $M_{\Sigma}$), the scotogenic 1-loop diagram~\cite{Ma:2006km} (Dirac Yukawa coupling matrix $h_\nu$), and the inverse seesaw mechanism~\cite{Mohapatra:1986bd,GonzalezGarcia:1988rw} (the matrix $M_{RS}$ mixing the right-handed neutrinos with the additional singlet neutrinos).

\section{\label{sec:experiments}Experimental perspectives of and nuclear physics impact on $0\nu\beta\beta$}

We now want to discuss the experimental status and prospects of $0\nu\beta\beta$. The general problem is that a potential observation of $0\nu\beta\beta$ would only give us an experimental measurement of the \emph{decay rate} or the \emph{half-life}, while what we actually would like to know is the \emph{amplitude}, and in particular the value of the quantity $|m_{ee}|$. If indeed light neutrino exchange (as discussed above) dominates the decay rate, then the half-life $T_{1/2}^{0\nu}$ can be obtained by the following equation~\cite{Doi:1985dx},
\begin{equation}
 \frac{1}{T_{1/2}^{0\nu}} = G_{0\nu} |\mathcal{M}_{0\nu}|^2 \left( \frac{|m_{ee}|}{m_e} \right)^2,
 \label{eq:exp_1}
\end{equation}
where $G_{0\nu}$ is a phase space factor which can be easily computed for any isotope under consideration (we will later on make use of the values from Ref.~\cite{Suhonen:1998ck}, which were slightly updated by Ref.~\cite{Rodejohann:2011mu}) and $\mathcal{M}_{0\nu}$ is the so-called \emph{nuclear matrix element} (NME), which encodes all nuclear physics that goes into the process.

There are at least two practical problems associated with the correct interpretation of a positive signal of $0\nu\beta\beta$ and a resulting measurement of $T_{1/2}^{0\nu}$:
\begin{enumerate}

\item First of all, the computation of the NME is extremely involved~\cite{Suhonen:1998ck,Simkovic:2007vu,Faessler:2009zz}. Not only do different computational methods lead to somewhat different results, we also have at the moment no direct way to probe the NMEs for $0\nu\beta\beta$. To do this, we would in principle need a positive observation of the process itself~\cite{Bilenky:2012zz}, whose interpretation would again suffer from the lack of knowledge on the NMEs. One way to disentangle this degeneracy is to observed $0\nu\beta\beta$ in different nuclei~\cite{Fogli:2009py}. Alternatively, e.g.\ so-called charge exchange reactions could be used to get some information on the physics underlying the NMEs~\cite{Rodin:2009hy,Rodin:2010jw}.

\item Second, even if we had perfect knowledge of the nuclear physics, there is still the possibility to have further particle physics contributions on top of the standard exchange of light neutrinos (see, e.g., Ref.~\cite{Prezeau:2003xn} for a treatment of such contributions in effective field theory). In practice, such additional effects cannot be disentangled from the standard mechanism if $0\nu\beta\beta$ is observed for one isotope only (see, e.g., Refs.~\cite{Pas:1997fx,Pas:2000vn,Bergstrom:2011dt}). Again, a comparison of positive signals in more than one isotope might help in this respect~\cite{Fogli:2009py,Deppisch:2006hb,Gehman:2007qg,Deppisch:2012nb}, in particular if performed within one and the same experiment in order to avoid systematic differences between experiments blurring the differences induced by physics. An example future experiment capable of doing that job would be SuperNEMO~\cite{Arnold:2010tu}.

\end{enumerate}

In the study presented here, we will focus on the first point, i.e., we assume the standard mechanism to dominate over all other contributions. We will discuss how the different computations of NMEs can affect the derived value of $|m_{ee}|$ if a signal of $0\nu\beta\beta$ is seen. An excellent and very up to date collection of NME values for the two isotopes ${}^{76}$Ge and ${}^{136}$Xe has been presented in Ref.~\cite{Dev:2013vxa}, which focused on the resulting constraints on light neutrino masses and on the potential contribution of heavy right-handed neutrinos in a left-right symmetric setting. We will extend this collection with a particular focus on the existing and on the near future limits.

While a detailed discussion of all the possible computational methods to determine NMEs is beyond the scope of this paper, we will at least mention which methods do exist. A relatively complete discussion of the different methods is provided in Ref.~\cite{Vergados:2012xy}. However, it should be noted that a detailed understanding of the underlying nuclear physics is necessary to fully appreciate the principal difficulties in the NME computations. We leave it to the nuclear physics experts to decide which method they consider to be more credible and/or reliable, while we treat all of them on the same footing (as typically done by particle physicists). As long as no generally accepted method exists, this treatment is certainly a fair viewpoint to take.

The nine methods M1 -- M9 to compute the NME values which we have used to derive our mass limits are the same ones as listed in Ref.~\cite{Dev:2013vxa}, where for each method we have also listed the isotopes under consideration here (in case the NME for the particular isotopes was given in the respective reference):
\begin{itemize}

\item M1: Energy Density Functional Method~\cite{Rodriguez:2010mn} ($^{76}$Ge, $^{82}$Se, $^{100}$Mo, $^{130}$Te, $^{136}$Xe, $^{150}$Nd)

\item M2: Interacting Shell Model~\cite{Menendez:2008jp} ($^{76}$Ge, $^{82}$Se, $^{130}$Te, $^{136}$Xe)

\item M3: Microscopic Interacting Boson Model~\cite{Barea:2013bz} ($^{76}$Ge, $^{82}$Se, $^{100}$Mo, $^{130}$Te, $^{136}$Xe, $^{150}$Nd)

\item M4: Proton-Neutron Quasiparticle Random-Phase Approximation~\cite{Suhonen:2010zzc} ($^{76}$Ge, $^{82}$Se, $^{130}$Te, $^{136}$Xe)

\item M5: Self-Consistent Renormalized Quasiparticle Random Phase Approximation (realistic charge-dependent Bonn potential)~\cite{Meroni:2012qf} ($^{76}$Ge, $^{82}$Se, $^{100}$Mo, $^{130}$Te, $^{136}$Xe)

\item M6: Self-Consistent Renormalized Quasiparticle Random Phase Approximation (Argonne V18 potential)~\cite{Meroni:2012qf} ($^{76}$Ge, $^{82}$Se, $^{100}$Mo, $^{130}$Te, $^{136}$Xe)

\item M7: Quasiparticle Random Phase Approximation (realistic charge-dependent Bonn potential)~\cite{Simkovic:2013qiy} ($^{76}$Ge, $^{82}$Se, $^{100}$Mo, $^{130}$Te, $^{136}$Xe)

\item M8: Quasiparticle Random Phase Approximation (Argonne V18 potential)~\cite{Simkovic:2013qiy} ($^{76}$Ge, $^{82}$Se, $^{100}$Mo, $^{130}$Te, $^{136}$Xe)

\item M9: Deformed Self--Consistent Skyrme Quasiparticle Random Phase Approximation~\cite{Mustonen:2013zu} ($^{76}$Ge, $^{130}$Te, $^{136}$Xe, $^{150}$Nd)

\end{itemize}
We have extracted the NMEs from the references given for M1 -- M9. Note that, for definiteness, we have taken the standard value $g_A = 1.25$ for the axial vector coupling, which means that we had to rescale some of the NMEs we have used~\cite{Menendez:2008jp,Simkovic:2013qiy}. Also the phase space factors used~\cite{Suhonen:1998ck,Rodejohann:2011mu} are the ones for $g_A = 1.25$. However, note that not in all cases the NMEs had been calculated in the references for the all the isotopes we consider. Note further that, in case where different versions of a calculation have been available, we have for definiteness always chosen the \emph{most optimistic} result. For example, Ref.~\cite{Meroni:2012qf} reports results for both, an intermediate size model space and a large size single particle space, the latter of which cases tends to result into larger values of the NME. Hence, we have decided to use the large size results. This partially differs from the choices made in Ref.~\cite{Dev:2013vxa}, where in some cases the smaller and in others the larger value has been chosen. Nevertheless, none of the treatments is wrong, in the sense that at the moment we still have to live with certain nuclear physics uncertainties, and there simply exists no way to decide which value is closer to the true value.

The derived current limits and future sensitivities on $|m_{ee}|$ are displayed in Tab.~\ref{tab:mee-derived}. Note that we have made extensive use of the recent experimental review given in Ref.~\cite{Barabash:2011mf}, but we have also updated some of the values given there by more recently reported bounds~\cite{Gando:2012zm,Auger:2012ar}. Since the derived values strongly depend on the NME values used, we indicate in every case the minimum and maximum value of $|m_{ee}|$ together with the type of calculation which leads to this number. In some cases, one or two methods lead to very different results, which are far away from those obtained by any other type of calculation. These ``outliers'' are indicated as well, along with the respective method. Note that we do not attempt to judge in any way which method may be more or less reliable. Instead, we chose to simply report our results and leave it to the reader to decide which method they prefer. However, we nevertheless chose to split off the outliers, simply because including them into the range would blur the agreement between the other methods to calculate the NME.

\begin{table}[t]
 \hspace{-1.0cm}
 \begin{tabular}{|c||c|c|c|c|c|}\hline
 Isotope & Current Limits & $T_{1/2}^{0\nu\beta\beta}$~[y] & $|m_{ee}|$ [eV] & $|m_{ee}|$ outliers & Refs.\\ \hline \hline
 $^{76}$Ge & HD-Moscow  & $>1.9\cdot 10^{25}$  & 0.26(M5)--0.32(M1) & 0.65(M2) & \cite{KlapdorKleingrothaus:2000sn} \\
  & IGEX  & $>1.57\cdot 10^{25}$  & 0.28(M5)--0.36(M1) & 0.71(M2) & \cite{Aalseth:2002rf} \\
  & GERDA I  & $>2.1\cdot 10^{25}$  & 0.24(M5)--0.31(M1) & 0.61(M2) & \cite{Agostini:2013mzu} \\ \hline
 $^{82}$Se & NEMO-3 & $>3.2\cdot 10^{23}$  & 0.97(M5)--1.3(M1) & 2.5(M2) & \cite{Simard:2012gb} \\ \hline
 $^{100}$Mo & NEMO-3 & $>1.1\cdot 10^{24}$  & 0.41(M7)--0.49(M6) & 0.70(M3) & \cite{Simard:2012gb} \\ \hline
 $^{130}$Te & Cuoricino  & $>2.8\cdot 10^{24}$  & 0.29(M1)--0.40(M8) & 0.71(M2),1.1(M9) & \cite{Andreotti:2010vj} \\ \hline
 $^{136}$Xe & KamLAND-Zen  & $>1.9\cdot 10^{25}$  & 0.13(M1)--0.34(M9) & ----- & \cite{Gando:2012zm} \\
 & EXO-200  & $>1.6\cdot 10^{25}$  & 0.15(M1)--0.37(M9) & ----- & \cite{Auger:2012ar} \\ \hline \hline
 Isotope & Future Limits & $T_{1/2}^{0\nu\beta\beta}$~[y] & $|m_{ee}|$ [eV] & $|m_{ee}|$ outliers & Refs.\\ \hline \hline
 $^{76}$Ge & GERDA II  & $>2\cdot 10^{26}$  & 0.079(M5)--0.10(M1) & 0.20(M2) & \cite{Abt:2004yk,JanicskoCsathy:2009zz} \\
 & GERDA III  & $>6\cdot 10^{27}$  & 0.014(M5)--0.018(M1) & 0.036(M2) & \cite{Abt:2004yk,JanicskoCsathy:2009zz} \\
 & Majorana  & $>2\cdot 10^{26}$  & 0.079(M5)--0.10(M1) & 0.20(M2) & \cite{Gaitskell:2003zr,Guiseppe:2008aa} \\ \hline
 $^{82}$Se & SuperNEMO & $>2\cdot 10^{26}$  & 0.039(M5)--0.052(M1) & 0.10(M2) & \cite{Barabash:2002ps,Chauveau:2009zz} \\ \hline
 $^{130}$Te & CUORE  & $>6.5\cdot 10^{26}$  & 0.019(M1)--0.026(M8) & 0.047(M2),0.072(M9) & \cite{Arnaboldi:2002du,Bandac:2008zz} \\ \hline
 $^{136}$Xe & KamLAND-Zen  & $>1\cdot 10^{27}$  & 0.019(M1)--0.046(M9) & ----- & \cite{Barabash:2011mf} \\
 & EXO-1000  & $>8\cdot 10^{26}$  & 0.021(M1)--0.052(M9) & ----- & \cite{Danilov:2000pp,Gornea:2009zz} \\ \hline
 $^{150}$Nd & SNO+  & $>3\cdot 10^{25}$  & 0.068(M9)--0.12(M1) & ----- & \cite{Barabash:2011mf} \\ \hline
 \end{tabular}
 \caption{\label{tab:mee-derived} The derived ranges for $|m_{ee}|$ for several current experimental limits and future experimental sensitivities~\cite{Barabash:2011mf}. For each isotope, we report the name the respective experiments and the 90\% C.L.-limits for the half-life $T_{1/2}^{0\nu\beta\beta}$. For the effective mass we always indicate the minimal and maximal values (depending on the NME). In case some NMEs result into values for $|m_{ee}|$ which differ considerably from the ones obtained by the other methods, we list these results separately as ``outliers''.}
\end{table}

As also reported by the corresponding experimental collaborations, the current best upper limit on $|m_{ee}|$ is just above $0.1$~eV, cf.\ Tab.~\ref{tab:mee-derived}. Depending on the performance of the planned future experiments it might be possible to push this limit further down by about one order of magnitude, to come close to the $0.01$~eV mark which could potentially rule out inverted mass ordering in general (at least in case that our knowledge on the NMEs is increased). Note that, however, pushing the effective mass by one order of magnitude requires an increased experimental performance by two orders of magnitude, due to the quadratic dependence of the decay rate on $|m_{ee}|$, cf.\ Eq.~\eqref{eq:exp_1}. Note also that, depending on the values of the NME and of the phase space factors, a larger limit on the half-life of a certain isotope may nevertheless yield a weaker limit on $|m_{ee}|$ than a lower limit for another isotope. A good example is the current values for the Heidelberg-Moscow ($^{76}$Ge) and KamLAND-Zen ($^{136}$Xe) experiments: although both experiments report the same lower limit of $T_{1/2}^{0\nu\beta\beta} >1.9\cdot 10^{25}$ at 90\%C.L.\ on the $0\nu\beta\beta$ half-life, the resulting upper limits on $|m_{ee}|$ could be smaller for either of them, depending on the exact value of the NME. This example makes it particularly apparent that we would actually need to have a comparison of several measurements of $0\nu\beta\beta$ on different isotopes in order to fully disentangle the nuclear physics complications, and also to decide which isotopes to choose for future experiments.

\section{\label{sec:predictions}Predictions of the different sum rules}

\begin{table}
\footnotesize
\hspace{-1cm}
\begin{tabular}{ll}
{
\begin{tabular}{|l||c|c|} \hline
$\tilde m_1 + \tilde m_2 = \tilde m_3$ & NO, 3$\sigma$ & IO, 3$\sigma$ \\ \hline \hline
$m_{\rm lightest}$ & $\gtrsim 0.027$ & $\gtrsim 0.00071$ \\ \hline
$|m_{ee}|$ (FTV)          & $>0.025$ & $>0.012$ \\ \hline
$|m_{ee}|$ (FLMMPR) & $>0.024$ & $>0.013$ \\ \hline
$|m_{ee}|$ (GMSS)      & $>0.025$ & $>0.015$ \\ \hline
\end{tabular}
} &
{
\begin{tabular}{|l||c|c|} \hline
$\tilde m_1 + \tilde m_3 = 2\tilde m_2$ & NO, 3$\sigma$ & IO, 3$\sigma$ \\ \hline \hline
$m_{\rm lightest}$ & $\gtrsim 0.015$ & forbidden \\ \hline
$|m_{ee}|$ (FTV)    & $>0.0015$ & forbidden \\ \hline
$|m_{ee}|$ (FLMMPR) & $>0.0020$ & forbidden \\ \hline
$|m_{ee}|$ (GMSS)   & $>0.0024$ & forbidden \\ \hline
\end{tabular}
}\\
& \\
{
\begin{tabular}{|l||c|c|} \hline
$2 \tilde m_2 + \tilde m_3 = \tilde m_1$ & NO, 3$\sigma$ & IO, 3$\sigma$ \\ \hline \hline
$m_{\rm lightest}$ & $\gtrsim 0.0081$ & forbidden \\ \hline
$|m_{ee}|$ (FTV)    & usual & forbidden \\ \hline
$|m_{ee}|$ (FLMMPR) & usual & forbidden \\ \hline
$|m_{ee}|$ (GMSS)   & usual & forbidden \\ \hline
\end{tabular}
} &
{
\begin{tabular}{|l||c|c|} \hline
$\tilde m_1 + \tilde m_2 = 2 \tilde m_3$ & NO, 3$\sigma$ & IO, 3$\sigma$ \\ \hline \hline
$m_{\rm lightest}$ & forbidden & $\gtrsim 0.00036$ \\ \hline
$|m_{ee}|$ (FTV)    & forbidden & $>0.012$ \\ \hline
$|m_{ee}|$ (FLMMPR) & forbidden & $>0.013$ \\ \hline
$|m_{ee}|$ (GMSS)   & forbidden & $>0.015$ \\ \hline
\end{tabular}
}
\\
& \\
{
\begin{tabular}{|l||c|c|} \hline
$\tilde{m}_1 + \frac{\sqrt{3}+1}{2} \tilde{m}_3 = \frac{\sqrt{3}-1}{2} \tilde{m}_2$ & NO, 3$\sigma$ & IO, 3$\sigma$ \\ \hline \hline
$m_{\rm lightest}$ & forbidden & $\gtrsim 0.0025$ \\ \hline
$|m_{ee}|$ (FTV)    & forbidden & $> 0.051$ \\ \hline
$|m_{ee}|$ (FLMMPR) & forbidden & $> 0.050$ \\ \hline
$|m_{ee}|$ (GMSS)   & forbidden & $> 0.051$ \\ \hline
\end{tabular}
} &
{

}
\end{tabular}
\caption{\label{tab:p_p1}Predictions for $m_{\rm lightest}$ and $|m_{ee}|$ in eV for the known $p=+1$ sum rules.}
\vspace{-0.1cm}
\end{table}

\begin{table}
\footnotesize
\hspace{-1cm}
\begin{tabular}{ll}
{
\begin{tabular}{|l||c|c|} \hline
$\tilde m_1^{-1} + \tilde m_2^{-1} = \tilde m_3^{-1}$ & NO, 3$\sigma$ & IO, 3$\sigma$ \\ \hline \hline
$m_{\rm lightest}$ & $\gtrsim 0.011$ & $\gtrsim 0.029$ \\ \hline
$|m_{ee}|$ (FTV)    & $>0.00011$ & $>0.046$ \\ \hline
$|m_{ee}|$ (FLMMPR) & $>0.00045$ & $>0.046$ \\ \hline
$|m_{ee}|$ (GMSS)   & $>0.00085$ & $>0.047$ \\ \hline
\end{tabular}
} &
{
\begin{tabular}{|l||c|c|} \hline
$2 \tilde m_2^{-1} + \tilde m_3^{-1} = \tilde m_1^{-1}$ & NO, 3$\sigma$ & IO, 3$\sigma$ \\ \hline \hline
$m_{\rm lightest}$ & $\sim [0.0045,0.0061]$ & $\gtrsim 0.018$ \\ \hline
$|m_{ee}|$ (FTV)    & $[0.0040,0.0092]$ & $>0.013$ \\ \hline
$|m_{ee}|$ (FLMMPR) & $[0.0041,0.0091]$ & $>0.014$ \\ \hline
$|m_{ee}|$ (GMSS)   & $[0.0042,0.0089]$ & $>0.015$ \\ \hline
\end{tabular}
}\\
& \\
{
\begin{tabular}{|l||c|c|} \hline
$\tilde m_1^{-1} + \tilde m_3^{-1} = 2 \tilde m_2^{-1}$ & NO, 3$\sigma$ & IO, 3$\sigma$ \\ \hline \hline
$m_{\rm lightest}$ & $\sim [0.0045,0.0061]$ & $\gtrsim 0.018$ \\ \hline
$|m_{ee}|$ (FTV)    & $[0.0041,0.0093]$ & $>0.013$ \\ \hline
$|m_{ee}|$ (FLMMPR) & $[0.0041,0.0091]$ & $>0.014$ \\ \hline
$|m_{ee}|$ (GMSS)   & $[0.0042,0.0089]$ & $>0.015$ \\ \hline
\end{tabular}
} &
{
\begin{tabular}{|l||c|c|} \hline
$\tilde m_3^{-1} \pm 2 i \tilde m_2^{-1} = \tilde m_1^{-1}$ & NO, 3$\sigma$ & IO, 3$\sigma$ \\ \hline \hline
$m_{\rm lightest}$ & $\sim [0.0045,0.0061]$ & $\gtrsim 0.018$ \\ \hline
$|m_{ee}|$ (FTV)    & $[0.0023,0.0071]$ & $>0.012$ \\ \hline
$|m_{ee}|$ (FLMMPR) & $[0.0024,0.0070]$ & $>0.013$ \\ \hline
$|m_{ee}|$ (GMSS)   & $[0.0024,0.0068]$ & $>0.015$ \\ \hline
\end{tabular}
}
\end{tabular}
\caption{\label{tab:p_m1}Predictions for $m_{\rm lightest}$ and $|m_{ee}|$ in eV for the known $p=-1$ sum rules.}
\vspace{-0.1cm}
\end{table}

\begin{table}
\footnotesize
\hspace{-1cm}
\begin{tabular}{ll}
{
\begin{tabular}{|l||c|c|} \hline
$\tilde m_1^{1/2} - \tilde m_3^{1/2} = 2 \tilde m_2^{1/2}$ & NO, 3$\sigma$ & IO, 3$\sigma$ \\ \hline \hline
$m_{\rm lightest}$ & $\gtrsim 0.00071$ & forbidden \\ \hline
$|m_{ee}|$ (FTV)    & usual & forbidden \\ \hline
$|m_{ee}|$ (FLMMPR) & usual & forbidden \\ \hline
$|m_{ee}|$ (GMSS)   & usual & forbidden \\ \hline
\end{tabular}
} &
{
\begin{tabular}{|l||c|c|} \hline
$\tilde m_1^{1/2} + \tilde m_3^{1/2} = 2 \tilde m_2^{1/2}$ & NO, 3$\sigma$ & IO, 3$\sigma$ \\ \hline \hline
$m_{\rm lightest}$ & $\gtrsim 0.00071$ & forbidden \\ \hline
$|m_{ee}|$ (FTV)    & usual & forbidden \\ \hline
$|m_{ee}|$ (FLMMPR) & usual & forbidden \\ \hline
$|m_{ee}|$ (GMSS)   & usual & forbidden \\ \hline
\end{tabular}
}\\
& \\
{
\begin{tabular}{|l||c|c|} \hline
$\tilde m_1^{-1/2} + \tilde m_2^{-1/2} = 2 \tilde m_3^{-1/2}$ & NO, 3$\sigma$ & IO, 3$\sigma$ \\ \hline \hline
$m_{\rm lightest}$ & $\gtrsim 0.0026$ & forbidden \\ \hline
$|m_{ee}|$ (FTV)    & $>0.0026$ & forbidden \\ \hline
$|m_{ee}|$ (FLMMPR) & $>0.0026$ & forbidden \\ \hline
$|m_{ee}|$ (GMSS)   & $>0.0027$ & forbidden \\ \hline
\end{tabular}
} & \\
\end{tabular}
\caption{\label{tab:p_12}Predictions for $m_{\rm lightest}$ and $|m_{ee}|$ in eV for the known $p= \pm 1/2$ sum rules.}
\vspace{-0.1cm}
\end{table}

In the following tables, we will list the predictions for the range of $|m_{ee}|$ as well as an approximate\footnote{Our numerical procedure renders the estimate of $m_{\rm lightest}$ a little less accurate than that of $|m_{ee}|$.} range for the smallest mass $m_{\rm lightest}$ as obtained from the sum rules from Sec.~\ref{sec:concrete_summary}. Note that, whenever a certain sum rule does not lead to a restricted range compared to the standard case displayed in Fig.~\ref{fig:mee}, we will indicate this by ``usual'' (e.g., if the minimum of $|m_{ee}|$ is smaller than $0.0001$~eV). When a certain ordering is not at all possible for the sum rule under consideration, we instead use the term ``forbidden''. Note that we have decided to report only the 3$\sigma$ predictions, since the best-fit values show a considerable variation depending on which gobal fit is used. This is a reflection of the fact that the region where the effective mass tends to zero arises due to a delicate cancellations between the different contributions~\cite{Lindner:2005kr}, so that slightly altered best-fit values could lead to quite different results (as can be seen by comparing the best-fit regions obtained with the three different fits in, e.g., Secs.~\ref{sec:concrete_A4Z2} and~\ref{sec:concrete_scotogenic}). The 3$\sigma$ results, on the other hand, are much more stable and reliable predictions. We have obtained the predictions (which are always given in eV) displayed in Tabs.~\ref{tab:p_p1}, \ref{tab:p_m1}, and~\ref{tab:p_12}.

\begin{figure}[t]
\centering
\includegraphics[width=12cm]{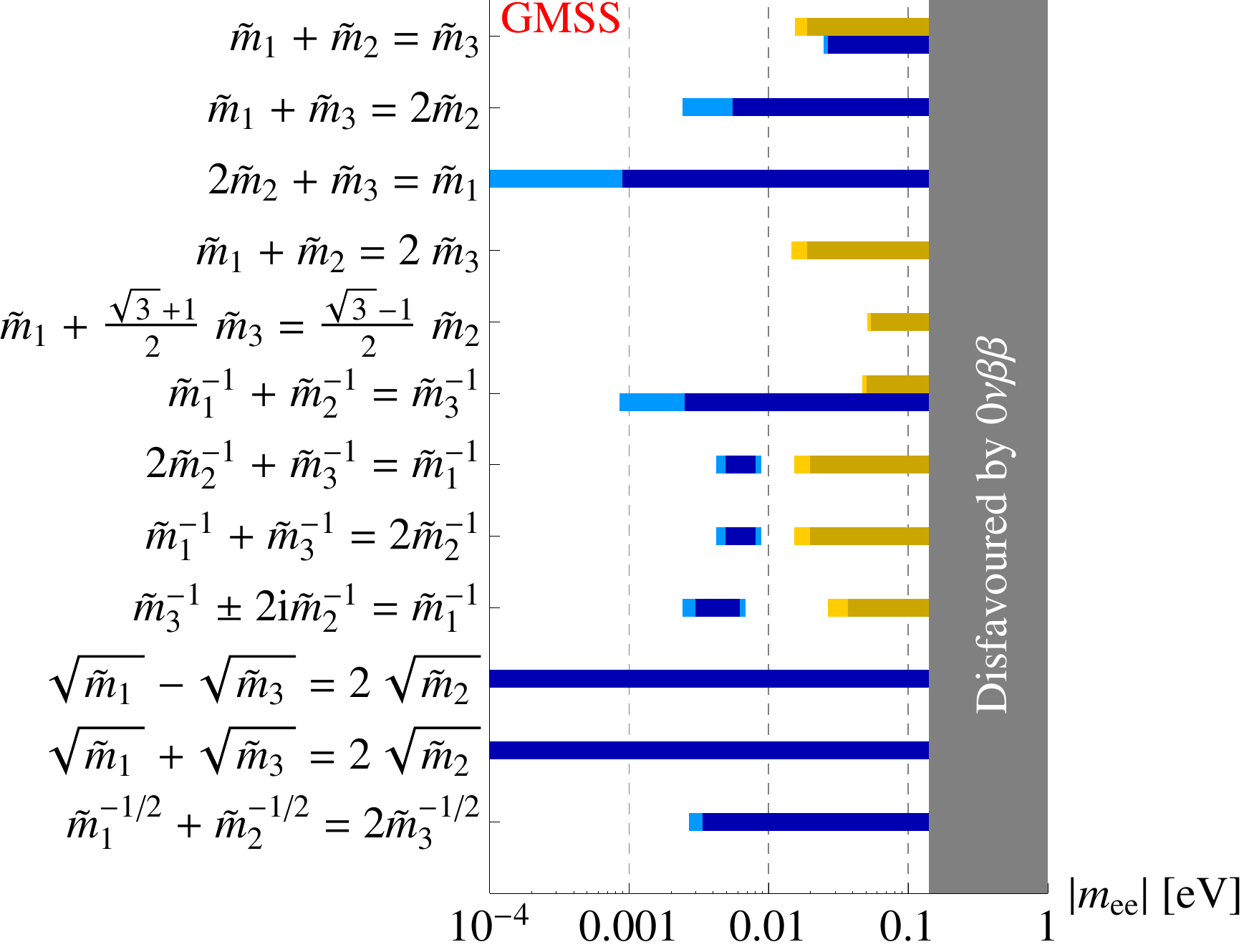}
\caption{\label{fig:predictions}Graphical representation of the predictions of the different sum rules.}
\end{figure}

Finally, we have graphically represented the predictions of all sum rule for the effective mass $|m_{ee}|$ in Fig.~\ref{fig:predictions}. Together with the information on the experimental sensitivities and on the ranges of the NME calculations given in Sec.~\ref{sec:experiments}, these predictions allow to determine whether a certain sum rule can be fully or partially probed by a certain experiment, even if nuclear physics uncertainties blur the picture. We have illustrated in Fig.~\ref{fig:power} that this is indeed possible in some cases. However, different scientific opinions exist about one or the other experiment, about one or the other global fit, or about one or the other method to calculate nuclear matrix elements. Thus we leave it to the reader to decide which rules they consider to be falsifiable with a given experiment. We have with this paper delivered the facts, and the resulting interpretation could be different, depending on the reader's scientific opinion. Nevertheless we have tried to open the door to invite model builders, phenomenologists, and experimentalists to think about the power of the sum rules, and to use our study to draw their own conclusions.

\section{\label{sec:conc}Conclusions}

Neutrino mass sum rules relate the three neutrino masses within generic classes of flavour models, leading to restrictions on the effective mass parameter derived from the observation of neutrinoless double beta decay, as a function of the lightest neutrino mass. After providing an illustration of how to obtain such sum rules in flavour models, we have presented a careful discussion of how to parametrise the effective mass and how to include constraints arising from neutrino mass sum rules necessary to derive predictions from sum rules. We have then performed a comprehensive study of the implications of such neutrino mass sum rules, which provides a link between model building, phenomenology, and experiments.  

We have discussed a large number of examples both numerically, using all three global fits available for the neutrino oscillation data, and analytically wherever possible. In some cases, our results disagreed with part of those in the literature for reasons that we have explained. We have also classified the different types of sum rules and derived some general properties. All the mass sum rules we are aware of have been investigated in varying detail, resulting in a complete classification of more than 50~known flavour models based on (mainly) discrete symmetries.

We have discussed the experimental prospects for many current and near-future experiments, with a particular focus on the uncertainties induced by the unknown nuclear physics involved. We find that, in many cases, the power of the neutrino mass sum rules is so strong as to allow certain classes of models to be fully tested by the next generation of neutrinoless double beta decay experiments. Finally, a list of numerical predictions of all sum rules discussed is given, which will enable the reader to decide about the prospects for a given experiment. Clearly the results in this paper can serve as both a guideline and a theoretical motivation for future experimental studies. 

In summary, neutrino mass sum rules provide a strong link between neutrino mass models and neutrinoless double beta decay experiments. Here we have provided a comprehensive phenomenological study, based on a broad survey of existing and possible future models, classified according to the power $p$ of the neutrino mass appearing in the sum rule. We have been very careful in extracting both the restrictions on the effective mass and in relating our results to experiments, including a discussion of the possible intrinsic uncertainties. We hope that this extensive study will be useful for both the theoretical and the experimental communities, and that it will contribute in some small way to advancing our understanding of neutrinos and of their fascinating properties.

\section*{\label{sec:ack}Acknowledgements}

We would like to thank L.~Dorame and S.~Morisi for their comments on the manuscript. SFK and AJS acknowledge support from the STFC Consolidated ST/J000396/1 grant. SFK acknowledges support from EU ITN UNILHC PITN-GA-2009-237920. AM acknowledges support by a Marie Curie Intra-European Fellowship within the 7th European Community Framework Programme FP7-PEOPLE-2011-IEF, contract PIEF-GA-2011-297557. All three authors acknowledge partial support from the European Union FP7 ITN-INVISIBLES (Marie Curie Actions, PITN-GA-2011-289442).

\bibliographystyle{./apsrev}
\bibliography{./Sum_Rules}

\begin{thebibliography}{100}
\expandafter\ifx\csname bibnamefont\endcsname\relax
  \def\bibnamefont#1{#1}\fi
\expandafter\ifx\csname bibfnamefont\endcsname\relax
  \def\bibfnamefont#1{#1}\fi
\expandafter\ifx\csname url\endcsname\relax
  \def\url#1{\texttt{#1}}\fi
\expandafter\ifx\csname urlprefix\endcsname\relax\def\urlprefix{URL }\fi
\providecommand{\bibinfo}[2]{#2}
\providecommand{\eprint}[2][]{\url{#2}}

\bibitem{Fukuda:1998mi}
\bibinfo{author}{\bibfnamefont{Y.}~\bibnamefont{Fukuda}} \emph{et~al.}
  (\bibinfo{collaboration}{Super-Kamiokande Collaboration}),
  \bibinfo{journal}{Phys. Rev. Lett.} \textbf{\bibinfo{volume}{81}},
  \bibinfo{pages}{1562} (\bibinfo{year}{1998}), \eprint{hep-ex/9807003}.

\bibitem{An:2012eh}
\bibinfo{author}{\bibfnamefont{F.~P.} \bibnamefont{An}} \emph{et~al.}
  (\bibinfo{collaboration}{DAYA-BAY Collaboration}), \bibinfo{journal}{Phys.
  Rev. Lett.} \textbf{\bibinfo{volume}{108}}, \bibinfo{pages}{171803}
  (\bibinfo{year}{2012}), \eprint{1203.1669}.

\bibitem{Ahn:2012nd}
\bibinfo{author}{\bibfnamefont{J.~K.} \bibnamefont{Ahn}} \emph{et~al.}
  (\bibinfo{collaboration}{RENO collaboration}), \bibinfo{journal}{Phys. Rev.
  Lett.} \textbf{\bibinfo{volume}{108}}, \bibinfo{pages}{191802}
  (\bibinfo{year}{2012}), \eprint{1204.0626}.

\bibitem{Beringer:1900zz}
\bibinfo{author}{\bibfnamefont{J.}~\bibnamefont{Beringer}} \emph{et~al.}
  (\bibinfo{collaboration}{Particle Data Group}), \bibinfo{journal}{Phys. Rev.}
  \textbf{\bibinfo{volume}{D86}}, \bibinfo{pages}{010001}
  (\bibinfo{year}{2012}).

\bibitem{King:2013eh}
\bibinfo{author}{\bibfnamefont{S.~F.} \bibnamefont{King}} \bibnamefont{and}
  \bibinfo{author}{\bibfnamefont{C.}~\bibnamefont{Luhn}},
  \bibinfo{journal}{Rept. Prog. Phys.} \textbf{\bibinfo{volume}{76}},
  \bibinfo{pages}{056201} (\bibinfo{year}{2013}), \eprint{1301.1340}.

\bibitem{Morisi:2012fg}
\bibinfo{author}{\bibfnamefont{S.}~\bibnamefont{Morisi}} \bibnamefont{and}
  \bibinfo{author}{\bibfnamefont{J.~W.~F.} \bibnamefont{Valle}},
  \bibinfo{journal}{Fortsch. Phys.} \textbf{\bibinfo{volume}{61}},
  \bibinfo{pages}{466} (\bibinfo{year}{2013}), \eprint{1206.6678}.

\bibitem{Grimus:2011fk}
\bibinfo{author}{\bibfnamefont{W.}~\bibnamefont{Grimus}} \bibnamefont{and}
  \bibinfo{author}{\bibfnamefont{P.~O.} \bibnamefont{Ludl}},
  \bibinfo{journal}{J. Phys.} \textbf{\bibinfo{volume}{A45}},
  \bibinfo{pages}{233001} (\bibinfo{year}{2012}), \eprint{1110.6376}.

\bibitem{Altarelli:2010gt}
\bibinfo{author}{\bibfnamefont{G.}~\bibnamefont{Altarelli}} \bibnamefont{and}
  \bibinfo{author}{\bibfnamefont{F.}~\bibnamefont{Feruglio}},
  \bibinfo{journal}{Rev. Mod. Phys.} \textbf{\bibinfo{volume}{82}},
  \bibinfo{pages}{2701} (\bibinfo{year}{2010}), \eprint{1002.0211}.

\bibitem{Haba:2000be}
\bibinfo{author}{\bibfnamefont{N.}~\bibnamefont{Haba}} \bibnamefont{and}
  \bibinfo{author}{\bibfnamefont{H.}~\bibnamefont{Murayama}},
  \bibinfo{journal}{Phys. Rev.} \textbf{\bibinfo{volume}{D63}},
  \bibinfo{pages}{053010} (\bibinfo{year}{2001}), \eprint{hep-ph/0009174}.

\bibitem{Adulpravitchai:2009re}
\bibinfo{author}{\bibfnamefont{A.}~\bibnamefont{Adulpravitchai}},
  \bibinfo{author}{\bibfnamefont{M.}~\bibnamefont{Lindner}},
  \bibinfo{author}{\bibfnamefont{A.}~\bibnamefont{Merle}}, \bibnamefont{and}
  \bibinfo{author}{\bibfnamefont{R.~N.} \bibnamefont{Mohapatra}},
  \bibinfo{journal}{Phys. Lett.} \textbf{\bibinfo{volume}{B680}},
  \bibinfo{pages}{476} (\bibinfo{year}{2009}), \eprint{0908.0470}.

\bibitem{King:2005bj}
\bibinfo{author}{\bibfnamefont{S.~F.} \bibnamefont{King}},
  \bibinfo{journal}{JHEP} \textbf{\bibinfo{volume}{0508}}, \bibinfo{pages}{105}
  (\bibinfo{year}{2005}), \eprint{hep-ph/0506297}.

\bibitem{Masina:2005hf}
\bibinfo{author}{\bibfnamefont{I.}~\bibnamefont{Masina}},
  \bibinfo{journal}{Phys. Lett.} \textbf{\bibinfo{volume}{B633}},
  \bibinfo{pages}{134} (\bibinfo{year}{2006}), \eprint{hep-ph/0508031}.

\bibitem{Antusch:2005kw}
\bibinfo{author}{\bibfnamefont{S.}~\bibnamefont{Antusch}} \bibnamefont{and}
  \bibinfo{author}{\bibfnamefont{S.~F.} \bibnamefont{King}},
  \bibinfo{journal}{Phys. Lett.} \textbf{\bibinfo{volume}{B631}},
  \bibinfo{pages}{42} (\bibinfo{year}{2005}), \eprint{hep-ph/0508044}.

\bibitem{Antusch:2007rk}
\bibinfo{author}{\bibfnamefont{S.}~\bibnamefont{Antusch}},
  \bibinfo{author}{\bibfnamefont{P.}~\bibnamefont{Huber}},
  \bibinfo{author}{\bibfnamefont{S.~F.} \bibnamefont{King}}, \bibnamefont{and}
  \bibinfo{author}{\bibfnamefont{T.}~\bibnamefont{Schwetz}},
  \bibinfo{journal}{JHEP} \textbf{\bibinfo{volume}{0704}}, \bibinfo{pages}{060}
  (\bibinfo{year}{2007}), \eprint{hep-ph/0702286}.

\bibitem{Altarelli:2009kr}
\bibinfo{author}{\bibfnamefont{G.}~\bibnamefont{Altarelli}} \bibnamefont{and}
  \bibinfo{author}{\bibfnamefont{D.}~\bibnamefont{Meloni}},
  \bibinfo{journal}{J. Phys.} \textbf{\bibinfo{volume}{G36}},
  \bibinfo{pages}{085005} (\bibinfo{year}{2009}), \eprint{0905.0620}.

\bibitem{Chen:2009um}
\bibinfo{author}{\bibfnamefont{M.-C.} \bibnamefont{Chen}} \bibnamefont{and}
  \bibinfo{author}{\bibfnamefont{S.~F.} \bibnamefont{King}},
  \bibinfo{journal}{JHEP} \textbf{\bibinfo{volume}{0906}}, \bibinfo{pages}{072}
  (\bibinfo{year}{2009}), \eprint{0903.0125}.

\bibitem{BarryRodejohann-Classification}
\bibinfo{author}{\bibfnamefont{J.}~\bibnamefont{Barry}} \bibnamefont{and}
  \bibinfo{author}{\bibfnamefont{W.}~\bibnamefont{Rodejohann}},
  \bibinfo{journal}{Phys. Rev.} \textbf{\bibinfo{volume}{D81}},
  \bibinfo{pages}{093002} (\bibinfo{year}{2010}), \eprint{1003.2385}.

\bibitem{Altarelli:2008bg}
\bibinfo{author}{\bibfnamefont{G.}~\bibnamefont{Altarelli}},
  \bibinfo{author}{\bibfnamefont{F.}~\bibnamefont{Feruglio}}, \bibnamefont{and}
  \bibinfo{author}{\bibfnamefont{C.}~\bibnamefont{Hagedorn}},
  \bibinfo{journal}{JHEP} \textbf{\bibinfo{volume}{0803}}, \bibinfo{pages}{052}
  (\bibinfo{year}{2008}), \eprint{0802.0090}.

\bibitem{Hirsch:2008rp}
\bibinfo{author}{\bibfnamefont{M.}~\bibnamefont{Hirsch}},
  \bibinfo{author}{\bibfnamefont{S.}~\bibnamefont{Morisi}}, \bibnamefont{and}
  \bibinfo{author}{\bibfnamefont{J.~W.~F.} \bibnamefont{Valle}},
  \bibinfo{journal}{Phys. Rev.} \textbf{\bibinfo{volume}{D78}},
  \bibinfo{pages}{093007} (\bibinfo{year}{2008}), \eprint{0804.1521}.

\bibitem{Bazzocchi:2009da}
\bibinfo{author}{\bibfnamefont{F.}~\bibnamefont{Bazzocchi}},
  \bibinfo{author}{\bibfnamefont{L.}~\bibnamefont{Merlo}}, \bibnamefont{and}
  \bibinfo{author}{\bibfnamefont{S.}~\bibnamefont{Morisi}},
  \bibinfo{journal}{Phys. Rev.} \textbf{\bibinfo{volume}{D80}},
  \bibinfo{pages}{053003} (\bibinfo{year}{2009}), \eprint{0902.2849}.

\bibitem{Barry:2010zk}
\bibinfo{author}{\bibfnamefont{J.}~\bibnamefont{Barry}} \bibnamefont{and}
  \bibinfo{author}{\bibfnamefont{W.}~\bibnamefont{Rodejohann}},
  \bibinfo{journal}{Nucl. Phys.} \textbf{\bibinfo{volume}{B842}},
  \bibinfo{pages}{33} (\bibinfo{year}{2011}), \eprint{1007.5217}.

\bibitem{Dorame:2011eb}
\bibinfo{author}{\bibfnamefont{L.}~\bibnamefont{Dorame}},
  \bibinfo{author}{\bibfnamefont{D.}~\bibnamefont{Meloni}},
  \bibinfo{author}{\bibfnamefont{S.}~\bibnamefont{Morisi}},
  \bibinfo{author}{\bibfnamefont{E.}~\bibnamefont{Peinado}}, \bibnamefont{and}
  \bibinfo{author}{\bibfnamefont{J.~W.~F.} \bibnamefont{Valle}},
  \bibinfo{journal}{Nucl. Phys.} \textbf{\bibinfo{volume}{B861}},
  \bibinfo{pages}{259} (\bibinfo{year}{2012}), \eprint{1111.5614}.

\bibitem{Agostini:2013mzu}
\bibinfo{author}{\bibfnamefont{M.}~\bibnamefont{Agostini}} \emph{et~al.}
  (\bibinfo{collaboration}{GERDA Collaboration})  (\bibinfo{year}{2013}),
  \eprint{1307.4720}.

\bibitem{Abt:2004yk}
\bibinfo{author}{\bibfnamefont{I.}~\bibnamefont{Abt}},
  \bibinfo{author}{\bibfnamefont{M.~F.} \bibnamefont{Altmann}},
  \bibinfo{author}{\bibfnamefont{A.}~\bibnamefont{Bakalyarov}},
  \bibinfo{author}{\bibfnamefont{I.}~\bibnamefont{Barabanov}},
  \bibinfo{author}{\bibfnamefont{C.}~\bibnamefont{Bauer}}, \emph{et~al.}
  (\bibinfo{year}{2004}), \eprint{hep-ex/0404039}.

\bibitem{JanicskoCsathy:2009zz}
\bibinfo{author}{\bibfnamefont{J.}~\bibnamefont{Janicsko-Csathy}}
  (\bibinfo{collaboration}{GERDA Collaboration}), \bibinfo{journal}{Nucl. Phys.
  Proc. Suppl.} \textbf{\bibinfo{volume}{188}}, \bibinfo{pages}{68}
  (\bibinfo{year}{2009}).

\bibitem{Tortola:2012te}
\bibinfo{author}{\bibfnamefont{D.~V.} \bibnamefont{Forero}},
  \bibinfo{author}{\bibfnamefont{M.}~\bibnamefont{Tortola}}, \bibnamefont{and}
  \bibinfo{author}{\bibfnamefont{J.~W.~F.} \bibnamefont{Valle}},
  \bibinfo{journal}{Phys. Rev.} \textbf{\bibinfo{volume}{D86}},
  \bibinfo{pages}{073012} (\bibinfo{year}{2012}), \eprint{1205.4018}.

\bibitem{Fogli:2012ua}
\bibinfo{author}{\bibfnamefont{G.~L.} \bibnamefont{Fogli}},
  \bibinfo{author}{\bibfnamefont{E.}~\bibnamefont{Lisi}},
  \bibinfo{author}{\bibfnamefont{A.}~\bibnamefont{Marrone}},
  \bibinfo{author}{\bibfnamefont{D.}~\bibnamefont{Montanino}},
  \bibinfo{author}{\bibfnamefont{A.}~\bibnamefont{Palazzo}}, \emph{et~al.},
  \bibinfo{journal}{Phys. Rev.} \textbf{\bibinfo{volume}{D86}},
  \bibinfo{pages}{013012} (\bibinfo{year}{2012}), \eprint{1205.5254}.

\bibitem{GonzalezGarcia:2012sz}
\bibinfo{author}{\bibfnamefont{M.~C.} \bibnamefont{Gonzalez-Garcia}},
  \bibinfo{author}{\bibfnamefont{M.}~\bibnamefont{Maltoni}},
  \bibinfo{author}{\bibfnamefont{J.}~\bibnamefont{Salvado}}, \bibnamefont{and}
  \bibinfo{author}{\bibfnamefont{T.}~\bibnamefont{Schwetz}},
  \bibinfo{journal}{JHEP} \textbf{\bibinfo{volume}{1212}}, \bibinfo{pages}{123}
  (\bibinfo{year}{2012}), \eprint{1209.3023}.

\bibitem{Lindner:2005kr}
\bibinfo{author}{\bibfnamefont{M.}~\bibnamefont{Lindner}},
  \bibinfo{author}{\bibfnamefont{A.}~\bibnamefont{Merle}}, \bibnamefont{and}
  \bibinfo{author}{\bibfnamefont{W.}~\bibnamefont{Rodejohann}},
  \bibinfo{journal}{Phys. Rev.} \textbf{\bibinfo{volume}{D73}},
  \bibinfo{pages}{053005} (\bibinfo{year}{2006}), \eprint{hep-ph/0512143}.

\bibitem{Weinberg:1979sa}
\bibinfo{author}{\bibfnamefont{S.}~\bibnamefont{Weinberg}},
  \bibinfo{journal}{Phys. Rev. Lett.} \textbf{\bibinfo{volume}{43}},
  \bibinfo{pages}{1566} (\bibinfo{year}{1979}).

\bibitem{Harrison:2002er}
\bibinfo{author}{\bibfnamefont{P.~F.} \bibnamefont{Harrison}},
  \bibinfo{author}{\bibfnamefont{D.~H.} \bibnamefont{Perkins}},
  \bibnamefont{and} \bibinfo{author}{\bibfnamefont{W.~G.} \bibnamefont{Scott}},
  \bibinfo{journal}{Phys. Lett.} \textbf{\bibinfo{volume}{B530}},
  \bibinfo{pages}{167} (\bibinfo{year}{2002}), \eprint{hep-ph/0202074}.

\bibitem{Minkowski:1977sc}
\bibinfo{author}{\bibfnamefont{P.}~\bibnamefont{Minkowski}},
  \bibinfo{journal}{Phys. Lett.} \textbf{\bibinfo{volume}{B67}},
  \bibinfo{pages}{421} (\bibinfo{year}{1977}).

\bibitem{Ramond:1979py}
\bibinfo{author}{\bibfnamefont{P.}~\bibnamefont{Ramond}} pp.
  \bibinfo{pages}{265--280} (\bibinfo{year}{1979}), \eprint{hep-ph/9809459}.

\bibitem{Yanagida:1979as}
\bibinfo{author}{\bibfnamefont{T.}~\bibnamefont{Yanagida}},
  \bibinfo{journal}{Conf. Proc.} \textbf{\bibinfo{volume}{C7902131}},
  \bibinfo{pages}{95} (\bibinfo{year}{1979}).

\bibitem{GellMann:1980vs}
\bibinfo{author}{\bibfnamefont{M.}~\bibnamefont{Gell-Mann}},
  \bibinfo{author}{\bibfnamefont{P.}~\bibnamefont{Ramond}}, \bibnamefont{and}
  \bibinfo{author}{\bibfnamefont{R.}~\bibnamefont{Slansky}},
  \bibinfo{journal}{Conf. Proc.} \textbf{\bibinfo{volume}{C790927}},
  \bibinfo{pages}{315} (\bibinfo{year}{1979}), \bibinfo{note}{to be published
  in Supergravity, P. van Nieuwenhuizen \& D. Z. Freedman (eds.), North Holland
  Publ. Co., 1979}.

\bibitem{Glashow:1979nm}
\bibinfo{author}{\bibfnamefont{S.}~\bibnamefont{Glashow}},
  \bibinfo{journal}{NATO Adv. Study Inst. Ser. B Phys.}
  \textbf{\bibinfo{volume}{59}}, \bibinfo{pages}{687} (\bibinfo{year}{1980}),
  \bibinfo{note}{preliminary version given at Colloquium in Honor of A.
  Visconti, Marseille-Luminy Univ., Jul 1979}.

\bibitem{Mohapatra:1979ia}
\bibinfo{author}{\bibfnamefont{R.~N.} \bibnamefont{Mohapatra}}
  \bibnamefont{and}
  \bibinfo{author}{\bibfnamefont{G.}~\bibnamefont{Senjanovic}},
  \bibinfo{journal}{Phys. Rev. Lett.} \textbf{\bibinfo{volume}{44}},
  \bibinfo{pages}{912} (\bibinfo{year}{1980}).

\bibitem{King:2012in}
\bibinfo{author}{\bibfnamefont{S.~F.} \bibnamefont{King}},
  \bibinfo{author}{\bibfnamefont{C.}~\bibnamefont{Luhn}}, \bibnamefont{and}
  \bibinfo{author}{\bibfnamefont{A.~J.} \bibnamefont{Stuart}},
  \bibinfo{journal}{Nucl. Phys.} \textbf{\bibinfo{volume}{B867}},
  \bibinfo{pages}{203} (\bibinfo{year}{2013}), \eprint{1207.5741}.

\bibitem{Cooper:2012bd}
\bibinfo{author}{\bibfnamefont{I.~K.} \bibnamefont{Cooper}},
  \bibinfo{author}{\bibfnamefont{S.~F.} \bibnamefont{King}}, \bibnamefont{and}
  \bibinfo{author}{\bibfnamefont{A.~J.} \bibnamefont{Stuart}}
  (\bibinfo{year}{2012}), \eprint{1212.1066}.

\bibitem{Schechter:1981bd}
\bibinfo{author}{\bibfnamefont{J.}~\bibnamefont{Schechter}} \bibnamefont{and}
  \bibinfo{author}{\bibfnamefont{J.~W.~F.} \bibnamefont{Valle}},
  \bibinfo{journal}{Phys. Rev.} \textbf{\bibinfo{volume}{D25}},
  \bibinfo{pages}{2951} (\bibinfo{year}{1982}).

\bibitem{Duerr:2011zd}
\bibinfo{author}{\bibfnamefont{M.}~\bibnamefont{Duerr}},
  \bibinfo{author}{\bibfnamefont{M.}~\bibnamefont{Lindner}}, \bibnamefont{and}
  \bibinfo{author}{\bibfnamefont{A.}~\bibnamefont{Merle}},
  \bibinfo{journal}{JHEP} \textbf{\bibinfo{volume}{1106}}, \bibinfo{pages}{091}
  (\bibinfo{year}{2011}), \eprint{1105.0901}.

\bibitem{Pal:2010ih}
\bibinfo{author}{\bibfnamefont{P.~B.} \bibnamefont{Pal}}
  (\bibinfo{year}{2010}), \eprint{1006.1718}.

\bibitem{Rodejohann:2011mu}
\bibinfo{author}{\bibfnamefont{W.}~\bibnamefont{Rodejohann}},
  \bibinfo{journal}{Int. J. Mod. Phys.} \textbf{\bibinfo{volume}{E20}},
  \bibinfo{pages}{1833} (\bibinfo{year}{2011}), \eprint{1106.1334}.

\bibitem{King:2002nf}
\bibinfo{author}{\bibfnamefont{S.~F.} \bibnamefont{King}},
  \bibinfo{journal}{JHEP} \textbf{\bibinfo{volume}{0209}}, \bibinfo{pages}{011}
  (\bibinfo{year}{2002}), \eprint{hep-ph/0204360}.

\bibitem{Rodejohann:2011vc}
\bibinfo{author}{\bibfnamefont{W.}~\bibnamefont{Rodejohann}} \bibnamefont{and}
  \bibinfo{author}{\bibfnamefont{J.}~\bibnamefont{Valle}},
  \bibinfo{journal}{Phys. Rev.} \textbf{\bibinfo{volume}{D84}},
  \bibinfo{pages}{073011} (\bibinfo{year}{2011}), \eprint{1108.3484}.

\bibitem{Auger:2012ar}
\bibinfo{author}{\bibfnamefont{M.}~\bibnamefont{Auger}} \emph{et~al.}
  (\bibinfo{collaboration}{EXO Collaboration}), \bibinfo{journal}{Phys. Rev.
  Lett.} \textbf{\bibinfo{volume}{109}}, \bibinfo{pages}{032505}
  (\bibinfo{year}{2012}), \eprint{1205.5608}.

\bibitem{Ade:2013lta}
\bibinfo{author}{\bibfnamefont{P.~A.~R.} \bibnamefont{Ade}} \emph{et~al.}
  (\bibinfo{collaboration}{Planck Collaboration})  (\bibinfo{year}{2013}),
  \eprint{1303.5076}.

\bibitem{Faessler:2009zz}
\bibinfo{author}{\bibfnamefont{A.}~\bibnamefont{Faessler}},
  \bibinfo{journal}{Nucl. Phys. Proc. Suppl.} \textbf{\bibinfo{volume}{188}},
  \bibinfo{pages}{20} (\bibinfo{year}{2009}).

\bibitem{Vergados:2012xy}
\bibinfo{author}{\bibfnamefont{J.~D.} \bibnamefont{Vergados}},
  \bibinfo{author}{\bibfnamefont{H.}~\bibnamefont{Ejiri}}, \bibnamefont{and}
  \bibinfo{author}{\bibfnamefont{F.}~\bibnamefont{Simkovic}},
  \bibinfo{journal}{Rept. Prog. Phys.} \textbf{\bibinfo{volume}{75}},
  \bibinfo{pages}{106301} (\bibinfo{year}{2012}), \eprint{1205.0649}.

\bibitem{Suhonen:2012ii}
\bibinfo{author}{\bibfnamefont{J.}~\bibnamefont{Suhonen}} \bibnamefont{and}
  \bibinfo{author}{\bibfnamefont{O.}~\bibnamefont{Civitarese}},
  \bibinfo{journal}{J. Phys.} \textbf{\bibinfo{volume}{G39}},
  \bibinfo{pages}{124005} (\bibinfo{year}{2012}).

\bibitem{Rodin:2012gp}
\bibinfo{author}{\bibfnamefont{V.}~\bibnamefont{Rodin}}, \bibinfo{journal}{J.
  Phys. Conf. Ser.} \textbf{\bibinfo{volume}{375}}, \bibinfo{pages}{042025}
  (\bibinfo{year}{2012}).

\bibitem{Maneschg:2008sf}
\bibinfo{author}{\bibfnamefont{W.}~\bibnamefont{Maneschg}},
  \bibinfo{author}{\bibfnamefont{A.}~\bibnamefont{Merle}}, \bibnamefont{and}
  \bibinfo{author}{\bibfnamefont{W.}~\bibnamefont{Rodejohann}},
  \bibinfo{journal}{Europhys. Lett.} \textbf{\bibinfo{volume}{85}},
  \bibinfo{pages}{51002} (\bibinfo{year}{2009}), \eprint{0812.0479}.

\bibitem{Dorame:2012zv}
\bibinfo{author}{\bibfnamefont{L.}~\bibnamefont{Dorame}},
  \bibinfo{author}{\bibfnamefont{S.}~\bibnamefont{Morisi}},
  \bibinfo{author}{\bibfnamefont{E.}~\bibnamefont{Peinado}},
  \bibinfo{author}{\bibfnamefont{J.~W.~F.} \bibnamefont{Valle}},
  \bibnamefont{and} \bibinfo{author}{\bibfnamefont{A.~D.} \bibnamefont{Rojas}},
  \bibinfo{journal}{Phys. Rev.} \textbf{\bibinfo{volume}{D86}},
  \bibinfo{pages}{056001} (\bibinfo{year}{2012}), \eprint{1203.0155}.

\bibitem{Ding:2010pc}
\bibinfo{author}{\bibfnamefont{G.-J.} \bibnamefont{Ding}},
  \bibinfo{journal}{Nucl. Phys.} \textbf{\bibinfo{volume}{B846}},
  \bibinfo{pages}{394} (\bibinfo{year}{2011}), \eprint{1006.4800}.

\bibitem{Ding:2011cm}
\bibinfo{author}{\bibfnamefont{G.-J.} \bibnamefont{Ding}},
  \bibinfo{author}{\bibfnamefont{L.~L.} \bibnamefont{Everett}},
  \bibnamefont{and} \bibinfo{author}{\bibfnamefont{A.~J.}
  \bibnamefont{Stuart}}, \bibinfo{journal}{Nucl. Phys.}
  \textbf{\bibinfo{volume}{B857}}, \bibinfo{pages}{219} (\bibinfo{year}{2012}),
  \eprint{1110.1688}.

\bibitem{Mohapatra:1986bd}
\bibinfo{author}{\bibfnamefont{R.~N.} \bibnamefont{Mohapatra}}
  \bibnamefont{and} \bibinfo{author}{\bibfnamefont{J.~W.~F.}
  \bibnamefont{Valle}}, \bibinfo{journal}{Phys. Rev.}
  \textbf{\bibinfo{volume}{D34}}, \bibinfo{pages}{1642} (\bibinfo{year}{1986}).

\bibitem{GonzalezGarcia:1988rw}
\bibinfo{author}{\bibfnamefont{M.~C.} \bibnamefont{Gonzalez-Garcia}}
  \bibnamefont{and} \bibinfo{author}{\bibfnamefont{J.~W.~F.}
  \bibnamefont{Valle}}, \bibinfo{journal}{Phys. Lett.}
  \textbf{\bibinfo{volume}{B216}}, \bibinfo{pages}{360} (\bibinfo{year}{1989}).

\bibitem{He:2006dk}
\bibinfo{author}{\bibfnamefont{X.-G.} \bibnamefont{He}},
  \bibinfo{author}{\bibfnamefont{Y.-Y.} \bibnamefont{Keum}}, \bibnamefont{and}
  \bibinfo{author}{\bibfnamefont{R.~R.} \bibnamefont{Volkas}},
  \bibinfo{journal}{JHEP} \textbf{\bibinfo{volume}{0604}}, \bibinfo{pages}{039}
  (\bibinfo{year}{2006}), \eprint{hep-ph/0601001}.

\bibitem{Berger:2009tt}
\bibinfo{author}{\bibfnamefont{J.}~\bibnamefont{Berger}} \bibnamefont{and}
  \bibinfo{author}{\bibfnamefont{Y.}~\bibnamefont{Grossman}},
  \bibinfo{journal}{JHEP} \textbf{\bibinfo{volume}{1002}}, \bibinfo{pages}{071}
  (\bibinfo{year}{2010}), \eprint{0910.4392}.

\bibitem{Lavoura:2012cv}
\bibinfo{author}{\bibfnamefont{L.}~\bibnamefont{Lavoura}},
  \bibinfo{author}{\bibfnamefont{S.}~\bibnamefont{Morisi}}, \bibnamefont{and}
  \bibinfo{author}{\bibfnamefont{J.~W.~F.} \bibnamefont{Valle}},
  \bibinfo{journal}{JHEP} \textbf{\bibinfo{volume}{1302}}, \bibinfo{pages}{118}
  (\bibinfo{year}{2013}), \eprint{1205.3442}.

\bibitem{Ma:2005sha}
\bibinfo{author}{\bibfnamefont{E.}~\bibnamefont{Ma}}, \bibinfo{journal}{Phys.
  Rev.} \textbf{\bibinfo{volume}{D72}}, \bibinfo{pages}{037301}
  (\bibinfo{year}{2005}), \eprint{hep-ph/0505209}.

\bibitem{Ma:2006wm}
\bibinfo{author}{\bibfnamefont{E.}~\bibnamefont{Ma}}, \bibinfo{journal}{Mod.
  Phys. Lett.} \textbf{\bibinfo{volume}{A21}}, \bibinfo{pages}{2931}
  (\bibinfo{year}{2006}), \eprint{hep-ph/0607190}.

\bibitem{Honda:2008rs}
\bibinfo{author}{\bibfnamefont{M.}~\bibnamefont{Honda}} \bibnamefont{and}
  \bibinfo{author}{\bibfnamefont{M.}~\bibnamefont{Tanimoto}},
  \bibinfo{journal}{Prog. Theor. Phys.} \textbf{\bibinfo{volume}{119}},
  \bibinfo{pages}{583} (\bibinfo{year}{2008}), \eprint{0801.0181}.

\bibitem{Brahmachari:2008fn}
\bibinfo{author}{\bibfnamefont{B.}~\bibnamefont{Brahmachari}},
  \bibinfo{author}{\bibfnamefont{S.}~\bibnamefont{Choubey}}, \bibnamefont{and}
  \bibinfo{author}{\bibfnamefont{M.}~\bibnamefont{Mitra}},
  \bibinfo{journal}{Phys. Rev.} \textbf{\bibinfo{volume}{D77}},
  \bibinfo{pages}{073008} (\bibinfo{year}{2008}), \eprint{0801.3554}.

\bibitem{Everett:2008et}
\bibinfo{author}{\bibfnamefont{L.~L.} \bibnamefont{Everett}} \bibnamefont{and}
  \bibinfo{author}{\bibfnamefont{A.~J.} \bibnamefont{Stuart}},
  \bibinfo{journal}{Phys. Rev.} \textbf{\bibinfo{volume}{D79}},
  \bibinfo{pages}{085005} (\bibinfo{year}{2009}), \eprint{0812.1057}.

\bibitem{Bazzocchi:2009pv}
\bibinfo{author}{\bibfnamefont{F.}~\bibnamefont{Bazzocchi}},
  \bibinfo{author}{\bibfnamefont{L.}~\bibnamefont{Merlo}}, \bibnamefont{and}
  \bibinfo{author}{\bibfnamefont{S.}~\bibnamefont{Morisi}},
  \bibinfo{journal}{Nucl. Phys.} \textbf{\bibinfo{volume}{B816}},
  \bibinfo{pages}{204} (\bibinfo{year}{2009}), \eprint{0901.2086}.

\bibitem{Boucenna:2012qb}
\bibinfo{author}{\bibfnamefont{M.~S.} \bibnamefont{Boucenna}},
  \bibinfo{author}{\bibfnamefont{S.}~\bibnamefont{Morisi}},
  \bibinfo{author}{\bibfnamefont{E.}~\bibnamefont{Peinado}},
  \bibinfo{author}{\bibfnamefont{Y.}~\bibnamefont{Shimizu}}, \bibnamefont{and}
  \bibinfo{author}{\bibfnamefont{J.~W.~F.} \bibnamefont{Valle}},
  \bibinfo{journal}{Phys. Rev.} \textbf{\bibinfo{volume}{D86}},
  \bibinfo{pages}{073008} (\bibinfo{year}{2012}), \eprint{1204.4733}.

\bibitem{Morisi:2007ft}
\bibinfo{author}{\bibfnamefont{S.}~\bibnamefont{Morisi}},
  \bibinfo{author}{\bibfnamefont{M.}~\bibnamefont{Picariello}},
  \bibnamefont{and}
  \bibinfo{author}{\bibfnamefont{E.}~\bibnamefont{Torrente-Lujan}},
  \bibinfo{journal}{Phys. Rev.} \textbf{\bibinfo{volume}{D75}},
  \bibinfo{pages}{075015} (\bibinfo{year}{2007}), \eprint{hep-ph/0702034}.

\bibitem{Adhikary:2008au}
\bibinfo{author}{\bibfnamefont{B.}~\bibnamefont{Adhikary}} \bibnamefont{and}
  \bibinfo{author}{\bibfnamefont{A.}~\bibnamefont{Ghosal}},
  \bibinfo{journal}{Phys. Rev.} \textbf{\bibinfo{volume}{D78}},
  \bibinfo{pages}{073007} (\bibinfo{year}{2008}), \eprint{0803.3582}.

\bibitem{Csaki:2008qq}
\bibinfo{author}{\bibfnamefont{C.}~\bibnamefont{Csaki}},
  \bibinfo{author}{\bibfnamefont{C.}~\bibnamefont{Delaunay}},
  \bibinfo{author}{\bibfnamefont{C.}~\bibnamefont{Grojean}}, \bibnamefont{and}
  \bibinfo{author}{\bibfnamefont{Y.}~\bibnamefont{Grossman}},
  \bibinfo{journal}{JHEP} \textbf{\bibinfo{volume}{0810}}, \bibinfo{pages}{055}
  (\bibinfo{year}{2008}), \eprint{0806.0356}.

\bibitem{Hagedorn:2009jy}
\bibinfo{author}{\bibfnamefont{C.}~\bibnamefont{Hagedorn}},
  \bibinfo{author}{\bibfnamefont{E.}~\bibnamefont{Molinaro}}, \bibnamefont{and}
  \bibinfo{author}{\bibfnamefont{S.}~\bibnamefont{Petcov}},
  \bibinfo{journal}{JHEP} \textbf{\bibinfo{volume}{0909}}, \bibinfo{pages}{115}
  (\bibinfo{year}{2009}), \eprint{0908.0240}.

\bibitem{Burrows:2009pi}
\bibinfo{author}{\bibfnamefont{T.~J.} \bibnamefont{Burrows}} \bibnamefont{and}
  \bibinfo{author}{\bibfnamefont{S.~F.} \bibnamefont{King}},
  \bibinfo{journal}{Nucl. Phys.} \textbf{\bibinfo{volume}{B835}},
  \bibinfo{pages}{174} (\bibinfo{year}{2010}), \eprint{0909.1433}.

\bibitem{Ding:2009gh}
\bibinfo{author}{\bibfnamefont{G.-J.} \bibnamefont{Ding}} \bibnamefont{and}
  \bibinfo{author}{\bibfnamefont{J.-F.} \bibnamefont{Liu}},
  \bibinfo{journal}{JHEP} \textbf{\bibinfo{volume}{1005}}, \bibinfo{pages}{029}
  (\bibinfo{year}{2010}), \eprint{0911.4799}.

\bibitem{Mitra:2009jj}
\bibinfo{author}{\bibfnamefont{M.}~\bibnamefont{Mitra}},
  \bibinfo{journal}{JHEP} \textbf{\bibinfo{volume}{1011}}, \bibinfo{pages}{026}
  (\bibinfo{year}{2010}), \eprint{0912.5291}.

\bibitem{delAguila:2010vg}
\bibinfo{author}{\bibfnamefont{F.}~\bibnamefont{del Aguila}},
  \bibinfo{author}{\bibfnamefont{A.}~\bibnamefont{Carmona}}, \bibnamefont{and}
  \bibinfo{author}{\bibfnamefont{J.}~\bibnamefont{Santiago}},
  \bibinfo{journal}{JHEP} \textbf{\bibinfo{volume}{1008}}, \bibinfo{pages}{127}
  (\bibinfo{year}{2010}), \eprint{1001.5151}.

\bibitem{Burrows:2010wz}
\bibinfo{author}{\bibfnamefont{T.~J.} \bibnamefont{Burrows}} \bibnamefont{and}
  \bibinfo{author}{\bibfnamefont{S.~F.} \bibnamefont{King}},
  \bibinfo{journal}{Nucl. Phys.} \textbf{\bibinfo{volume}{B842}},
  \bibinfo{pages}{107} (\bibinfo{year}{2011}), \eprint{1007.2310}.

\bibitem{Altarelli:2005yx}
\bibinfo{author}{\bibfnamefont{G.}~\bibnamefont{Altarelli}} \bibnamefont{and}
  \bibinfo{author}{\bibfnamefont{F.}~\bibnamefont{Feruglio}},
  \bibinfo{journal}{Nucl. Phys.} \textbf{\bibinfo{volume}{B741}},
  \bibinfo{pages}{215} (\bibinfo{year}{2006}), \eprint{hep-ph/0512103}.

\bibitem{Chen:2009gy}
\bibinfo{author}{\bibfnamefont{M.-C.} \bibnamefont{Chen}},
  \bibinfo{author}{\bibfnamefont{K.}~\bibnamefont{Mahanthappa}},
  \bibnamefont{and} \bibinfo{author}{\bibfnamefont{F.}~\bibnamefont{Yu}},
  \bibinfo{journal}{Phys. Rev.} \textbf{\bibinfo{volume}{D81}},
  \bibinfo{pages}{036004} (\bibinfo{year}{2010}), \eprint{0907.3963}.

\bibitem{Altarelli:2005yp}
\bibinfo{author}{\bibfnamefont{G.}~\bibnamefont{Altarelli}} \bibnamefont{and}
  \bibinfo{author}{\bibfnamefont{F.}~\bibnamefont{Feruglio}},
  \bibinfo{journal}{Nucl. Phys.} \textbf{\bibinfo{volume}{B720}},
  \bibinfo{pages}{64} (\bibinfo{year}{2005}), \eprint{hep-ph/0504165}.

\bibitem{Altarelli:2006kg}
\bibinfo{author}{\bibfnamefont{G.}~\bibnamefont{Altarelli}},
  \bibinfo{author}{\bibfnamefont{F.}~\bibnamefont{Feruglio}}, \bibnamefont{and}
  \bibinfo{author}{\bibfnamefont{Y.}~\bibnamefont{Lin}},
  \bibinfo{journal}{Nucl. Phys.} \textbf{\bibinfo{volume}{B775}},
  \bibinfo{pages}{31} (\bibinfo{year}{2007}), \eprint{hep-ph/0610165}.

\bibitem{Ma:2006vq}
\bibinfo{author}{\bibfnamefont{E.}~\bibnamefont{Ma}}, \bibinfo{journal}{Mod.
  Phys. Lett.} \textbf{\bibinfo{volume}{A22}}, \bibinfo{pages}{101}
  (\bibinfo{year}{2007}), \eprint{hep-ph/0610342}.

\bibitem{Bazzocchi:2007na}
\bibinfo{author}{\bibfnamefont{F.}~\bibnamefont{Bazzocchi}},
  \bibinfo{author}{\bibfnamefont{S.}~\bibnamefont{Kaneko}}, \bibnamefont{and}
  \bibinfo{author}{\bibfnamefont{S.}~\bibnamefont{Morisi}},
  \bibinfo{journal}{JHEP} \textbf{\bibinfo{volume}{0803}}, \bibinfo{pages}{063}
  (\bibinfo{year}{2008}), \eprint{0707.3032}.

\bibitem{Bazzocchi:2007au}
\bibinfo{author}{\bibfnamefont{F.}~\bibnamefont{Bazzocchi}},
  \bibinfo{author}{\bibfnamefont{S.}~\bibnamefont{Morisi}}, \bibnamefont{and}
  \bibinfo{author}{\bibfnamefont{M.}~\bibnamefont{Picariello}},
  \bibinfo{journal}{Phys. Lett.} \textbf{\bibinfo{volume}{B659}},
  \bibinfo{pages}{628} (\bibinfo{year}{2008}), \eprint{0710.2928}.

\bibitem{Lin:2008aj}
\bibinfo{author}{\bibfnamefont{Y.}~\bibnamefont{Lin}}, \bibinfo{journal}{Nucl.
  Phys.} \textbf{\bibinfo{volume}{B813}}, \bibinfo{pages}{91}
  (\bibinfo{year}{2009}), \eprint{0804.2867}.

\bibitem{Ma:2009wi}
\bibinfo{author}{\bibfnamefont{E.}~\bibnamefont{Ma}}, \bibinfo{journal}{Mod.
  Phys. Lett.} \textbf{\bibinfo{volume}{A25}}, \bibinfo{pages}{2215}
  (\bibinfo{year}{2010}), \eprint{0908.3165}.

\bibitem{Ciafaloni:2009qs}
\bibinfo{author}{\bibfnamefont{P.}~\bibnamefont{Ciafaloni}},
  \bibinfo{author}{\bibfnamefont{M.}~\bibnamefont{Picariello}},
  \bibinfo{author}{\bibfnamefont{A.}~\bibnamefont{Urbano}}, \bibnamefont{and}
  \bibinfo{author}{\bibfnamefont{E.}~\bibnamefont{Torrente-Lujan}},
  \bibinfo{journal}{Phys. Rev.} \textbf{\bibinfo{volume}{D81}},
  \bibinfo{pages}{016004} (\bibinfo{year}{2010}), \eprint{0909.2553}.

\bibitem{Fukuyama:2010mz}
\bibinfo{author}{\bibfnamefont{T.}~\bibnamefont{Fukuyama}},
  \bibinfo{author}{\bibfnamefont{H.}~\bibnamefont{Sugiyama}}, \bibnamefont{and}
  \bibinfo{author}{\bibfnamefont{K.}~\bibnamefont{Tsumura}},
  \bibinfo{journal}{Phys. Rev.} \textbf{\bibinfo{volume}{D82}},
  \bibinfo{pages}{036004} (\bibinfo{year}{2010}), \eprint{1005.5338}.

\bibitem{Bazzocchi:2008ej}
\bibinfo{author}{\bibfnamefont{F.}~\bibnamefont{Bazzocchi}} \bibnamefont{and}
  \bibinfo{author}{\bibfnamefont{S.}~\bibnamefont{Morisi}},
  \bibinfo{journal}{Phys. Rev.} \textbf{\bibinfo{volume}{D80}},
  \bibinfo{pages}{096005} (\bibinfo{year}{2009}), \eprint{0811.0345}.

\bibitem{Chen:2007afa}
\bibinfo{author}{\bibfnamefont{M.-C.} \bibnamefont{Chen}} \bibnamefont{and}
  \bibinfo{author}{\bibfnamefont{K.}~\bibnamefont{Mahanthappa}},
  \bibinfo{journal}{Phys. Lett.} \textbf{\bibinfo{volume}{B652}},
  \bibinfo{pages}{34} (\bibinfo{year}{2007}), \eprint{0705.0714}.

\bibitem{Ding:2008rj}
\bibinfo{author}{\bibfnamefont{G.-J.} \bibnamefont{Ding}},
  \bibinfo{journal}{Phys. Rev.} \textbf{\bibinfo{volume}{D78}},
  \bibinfo{pages}{036011} (\bibinfo{year}{2008}), \eprint{0803.2278}.

\bibitem{Chen:2009gf}
\bibinfo{author}{\bibfnamefont{M.-C.} \bibnamefont{Chen}} \bibnamefont{and}
  \bibinfo{author}{\bibfnamefont{K.~T.} \bibnamefont{Mahanthappa}},
  \bibinfo{journal}{Phys. Lett.} \textbf{\bibinfo{volume}{B681}},
  \bibinfo{pages}{444} (\bibinfo{year}{2009}), \eprint{0904.1721}.

\bibitem{Feruglio:2007uu}
\bibinfo{author}{\bibfnamefont{F.}~\bibnamefont{Feruglio}},
  \bibinfo{author}{\bibfnamefont{C.}~\bibnamefont{Hagedorn}},
  \bibinfo{author}{\bibfnamefont{Y.}~\bibnamefont{Lin}}, \bibnamefont{and}
  \bibinfo{author}{\bibfnamefont{L.}~\bibnamefont{Merlo}},
  \bibinfo{journal}{Nucl. Phys.} \textbf{\bibinfo{volume}{B775}},
  \bibinfo{pages}{120} (\bibinfo{year}{2007}), \eprint{hep-ph/0702194}.

\bibitem{Merlo:2011hw}
\bibinfo{author}{\bibfnamefont{L.}~\bibnamefont{Merlo}},
  \bibinfo{author}{\bibfnamefont{S.}~\bibnamefont{Rigolin}}, \bibnamefont{and}
  \bibinfo{author}{\bibfnamefont{B.}~\bibnamefont{Zaldivar}},
  \bibinfo{journal}{JHEP} \textbf{\bibinfo{volume}{1111}}, \bibinfo{pages}{047}
  (\bibinfo{year}{2011}), \eprint{1108.1795}.

\bibitem{Luhn:2012bc}
\bibinfo{author}{\bibfnamefont{C.}~\bibnamefont{Luhn}},
  \bibinfo{author}{\bibfnamefont{K.~M.} \bibnamefont{Parattu}},
  \bibnamefont{and}
  \bibinfo{author}{\bibfnamefont{A.}~\bibnamefont{Wingerter}},
  \bibinfo{journal}{JHEP} \textbf{\bibinfo{volume}{1212}}, \bibinfo{pages}{096}
  (\bibinfo{year}{2012}), \eprint{1210.1197}.

\bibitem{Mohapatra:2012tb}
\bibinfo{author}{\bibfnamefont{R.~N.} \bibnamefont{Mohapatra}}
  \bibnamefont{and} \bibinfo{author}{\bibfnamefont{C.~C.} \bibnamefont{Nishi}},
  \bibinfo{journal}{Phys. Rev.} \textbf{\bibinfo{volume}{D86}},
  \bibinfo{pages}{073007} (\bibinfo{year}{2012}), \eprint{1208.2875}.

\bibitem{Feruglio:2013hia}
\bibinfo{author}{\bibfnamefont{F.}~\bibnamefont{Feruglio}},
  \bibinfo{author}{\bibfnamefont{C.}~\bibnamefont{Hagedorn}}, \bibnamefont{and}
  \bibinfo{author}{\bibfnamefont{R.}~\bibnamefont{Ziegler}}
  (\bibinfo{year}{2013}), \eprint{1303.7178}.

\bibitem{Ding:2013eca}
\bibinfo{author}{\bibfnamefont{G.-J.} \bibnamefont{Ding}} \bibnamefont{and}
  \bibinfo{author}{\bibfnamefont{Y.-L.} \bibnamefont{Zhou}}
  (\bibinfo{year}{2013}), \eprint{1304.2645}.

\bibitem{Lindner:2010wr}
\bibinfo{author}{\bibfnamefont{M.}~\bibnamefont{Lindner}},
  \bibinfo{author}{\bibfnamefont{A.}~\bibnamefont{Merle}}, \bibnamefont{and}
  \bibinfo{author}{\bibfnamefont{V.}~\bibnamefont{Niro}},
  \bibinfo{journal}{JCAP} \textbf{\bibinfo{volume}{1101}}, \bibinfo{pages}{034}
  (\bibinfo{year}{2011}), \eprint{1011.4950}.

\bibitem{Adulpravitchai:2009gi}
\bibinfo{author}{\bibfnamefont{A.}~\bibnamefont{Adulpravitchai}},
  \bibinfo{author}{\bibfnamefont{M.}~\bibnamefont{Lindner}}, \bibnamefont{and}
  \bibinfo{author}{\bibfnamefont{A.}~\bibnamefont{Merle}},
  \bibinfo{journal}{Phys. Rev.} \textbf{\bibinfo{volume}{D80}},
  \bibinfo{pages}{055031} (\bibinfo{year}{2009}), \eprint{0907.2147}.

\bibitem{Hashimoto:2011tn}
\bibinfo{author}{\bibfnamefont{K.}~\bibnamefont{Hashimoto}} \bibnamefont{and}
  \bibinfo{author}{\bibfnamefont{H.}~\bibnamefont{Okada}}
  (\bibinfo{year}{2011}), \eprint{1110.3640}.

\bibitem{Lin:2009bw}
\bibinfo{author}{\bibfnamefont{Y.}~\bibnamefont{Lin}}, \bibinfo{journal}{Nucl.
  Phys.} \textbf{\bibinfo{volume}{B824}}, \bibinfo{pages}{95}
  (\bibinfo{year}{2010}), \eprint{0905.3534}.

\bibitem{Kadosh:2010rm}
\bibinfo{author}{\bibfnamefont{A.}~\bibnamefont{Kadosh}} \bibnamefont{and}
  \bibinfo{author}{\bibfnamefont{E.}~\bibnamefont{Pallante}},
  \bibinfo{journal}{JHEP} \textbf{\bibinfo{volume}{1008}}, \bibinfo{pages}{115}
  (\bibinfo{year}{2010}), \eprint{1004.0321}.

\bibitem{Magg:1980ut}
\bibinfo{author}{\bibfnamefont{M.}~\bibnamefont{Magg}} \bibnamefont{and}
  \bibinfo{author}{\bibfnamefont{C.}~\bibnamefont{Wetterich}},
  \bibinfo{journal}{Phys. Lett.} \textbf{\bibinfo{volume}{B94}},
  \bibinfo{pages}{61} (\bibinfo{year}{1980}).

\bibitem{Lazarides:1980nt}
\bibinfo{author}{\bibfnamefont{G.}~\bibnamefont{Lazarides}},
  \bibinfo{author}{\bibfnamefont{Q.}~\bibnamefont{Shafi}}, \bibnamefont{and}
  \bibinfo{author}{\bibfnamefont{C.}~\bibnamefont{Wetterich}},
  \bibinfo{journal}{Nucl. Phys.} \textbf{\bibinfo{volume}{B181}},
  \bibinfo{pages}{287} (\bibinfo{year}{1981}).

\bibitem{Foot:1988aq}
\bibinfo{author}{\bibfnamefont{R.}~\bibnamefont{Foot}},
  \bibinfo{author}{\bibfnamefont{H.}~\bibnamefont{Lew}},
  \bibinfo{author}{\bibfnamefont{X.~G.} \bibnamefont{He}}, \bibnamefont{and}
  \bibinfo{author}{\bibfnamefont{G.~C.} \bibnamefont{Joshi}},
  \bibinfo{journal}{Z. Phys.} \textbf{\bibinfo{volume}{C44}},
  \bibinfo{pages}{441} (\bibinfo{year}{1989}).

\bibitem{Ma:2002pf}
\bibinfo{author}{\bibfnamefont{E.}~\bibnamefont{Ma}} \bibnamefont{and}
  \bibinfo{author}{\bibfnamefont{D.~P.} \bibnamefont{Roy}},
  \bibinfo{journal}{Nucl. Phys.} \textbf{\bibinfo{volume}{B644}},
  \bibinfo{pages}{290} (\bibinfo{year}{2002}), \eprint{hep-ph/0206150}.

\bibitem{Ma:2006km}
\bibinfo{author}{\bibfnamefont{E.}~\bibnamefont{Ma}}, \bibinfo{journal}{Phys.
  Rev.} \textbf{\bibinfo{volume}{D73}}, \bibinfo{pages}{077301}
  (\bibinfo{year}{2006}), \eprint{hep-ph/0601225}.

\bibitem{Doi:1985dx}
\bibinfo{author}{\bibfnamefont{M.}~\bibnamefont{Doi}},
  \bibinfo{author}{\bibfnamefont{T.}~\bibnamefont{Kotani}}, \bibnamefont{and}
  \bibinfo{author}{\bibfnamefont{E.}~\bibnamefont{Takasugi}},
  \bibinfo{journal}{Prog. Theor. Phys. Suppl.} \textbf{\bibinfo{volume}{83}},
  \bibinfo{pages}{1} (\bibinfo{year}{1985}).

\bibitem{Suhonen:1998ck}
\bibinfo{author}{\bibfnamefont{J.}~\bibnamefont{Suhonen}} \bibnamefont{and}
  \bibinfo{author}{\bibfnamefont{O.}~\bibnamefont{Civitarese}},
  \bibinfo{journal}{Phys. Rept.} \textbf{\bibinfo{volume}{300}},
  \bibinfo{pages}{123} (\bibinfo{year}{1998}).

\bibitem{Simkovic:2007vu}
\bibinfo{author}{\bibfnamefont{F.}~\bibnamefont{Simkovic}},
  \bibinfo{author}{\bibfnamefont{A.}~\bibnamefont{Faessler}},
  \bibinfo{author}{\bibfnamefont{V.}~\bibnamefont{Rodin}},
  \bibinfo{author}{\bibfnamefont{P.}~\bibnamefont{Vogel}}, \bibnamefont{and}
  \bibinfo{author}{\bibfnamefont{J.}~\bibnamefont{Engel}},
  \bibinfo{journal}{Phys. Rev.} \textbf{\bibinfo{volume}{C77}},
  \bibinfo{pages}{045503} (\bibinfo{year}{2008}), \eprint{0710.2055}.

\bibitem{Bilenky:2012zz}
\bibinfo{author}{\bibfnamefont{S.~M.} \bibnamefont{Bilenky}} \bibnamefont{and}
  \bibinfo{author}{\bibfnamefont{F.}~\bibnamefont{Simkovic}},
  \bibinfo{journal}{Phys. Part. Nucl. Lett.} \textbf{\bibinfo{volume}{9}},
  \bibinfo{pages}{220} (\bibinfo{year}{2012}).

\bibitem{Fogli:2009py}
\bibinfo{author}{\bibfnamefont{G.~L.} \bibnamefont{Fogli}},
  \bibinfo{author}{\bibfnamefont{E.}~\bibnamefont{Lisi}}, \bibnamefont{and}
  \bibinfo{author}{\bibfnamefont{A.~M.} \bibnamefont{Rotunno}},
  \bibinfo{journal}{Phys. Rev.} \textbf{\bibinfo{volume}{D80}},
  \bibinfo{pages}{015024} (\bibinfo{year}{2009}), \eprint{0905.1832}.

\bibitem{Rodin:2009hy}
\bibinfo{author}{\bibfnamefont{V.}~\bibnamefont{Rodin}} \bibnamefont{and}
  \bibinfo{author}{\bibfnamefont{A.}~\bibnamefont{Faessler}},
  \bibinfo{journal}{Phys. Rev.} \textbf{\bibinfo{volume}{C80}},
  \bibinfo{pages}{041302} (\bibinfo{year}{2009}), \eprint{0906.1759}.

\bibitem{Rodin:2010jw}
\bibinfo{author}{\bibfnamefont{V.}~\bibnamefont{Rodin}} \bibnamefont{and}
  \bibinfo{author}{\bibfnamefont{A.}~\bibnamefont{Faessler}},
  \bibinfo{journal}{Prog. Part. Nucl. Phys.} \textbf{\bibinfo{volume}{66}},
  \bibinfo{pages}{441} (\bibinfo{year}{2011}), \eprint{1012.5176}.

\bibitem{Prezeau:2003xn}
\bibinfo{author}{\bibfnamefont{G.}~\bibnamefont{Prezeau}},
  \bibinfo{author}{\bibfnamefont{M.}~\bibnamefont{Ramsey-Musolf}},
  \bibnamefont{and} \bibinfo{author}{\bibfnamefont{P.}~\bibnamefont{Vogel}},
  \bibinfo{journal}{Phys. Rev.} \textbf{\bibinfo{volume}{D68}},
  \bibinfo{pages}{034016} (\bibinfo{year}{2003}), \eprint{hep-ph/0303205}.

\bibitem{Pas:1997fx}
\bibinfo{author}{\bibfnamefont{H.}~\bibnamefont{Pas}},
  \bibinfo{author}{\bibfnamefont{M.}~\bibnamefont{Hirsch}},
  \bibinfo{author}{\bibfnamefont{S.~G.} \bibnamefont{Kovalenko}},
  \bibnamefont{and} \bibinfo{author}{\bibfnamefont{H.~V.}
  \bibnamefont{Klapdor-Kleingrothaus}}  (\bibinfo{year}{1997}),
  \eprint{hep-ph/9804374}.

\bibitem{Pas:2000vn}
\bibinfo{author}{\bibfnamefont{H.}~\bibnamefont{Pas}},
  \bibinfo{author}{\bibfnamefont{M.}~\bibnamefont{Hirsch}},
  \bibinfo{author}{\bibfnamefont{H.~V.} \bibnamefont{Klapdor-Kleingrothaus}},
  \bibnamefont{and} \bibinfo{author}{\bibfnamefont{S.~G.}
  \bibnamefont{Kovalenko}}, \bibinfo{journal}{Phys. Lett.}
  \textbf{\bibinfo{volume}{B498}}, \bibinfo{pages}{35} (\bibinfo{year}{2001}),
  \eprint{hep-ph/0008182}.

\bibitem{Bergstrom:2011dt}
\bibinfo{author}{\bibfnamefont{J.}~\bibnamefont{Bergstrom}},
  \bibinfo{author}{\bibfnamefont{A.}~\bibnamefont{Merle}}, \bibnamefont{and}
  \bibinfo{author}{\bibfnamefont{T.}~\bibnamefont{Ohlsson}},
  \bibinfo{journal}{JHEP} \textbf{\bibinfo{volume}{1105}}, \bibinfo{pages}{122}
  (\bibinfo{year}{2011}), \eprint{1103.3015}.

\bibitem{Deppisch:2006hb}
\bibinfo{author}{\bibfnamefont{F.}~\bibnamefont{Deppisch}} \bibnamefont{and}
  \bibinfo{author}{\bibfnamefont{H.}~\bibnamefont{Pas}},
  \bibinfo{journal}{Phys. Rev. Lett.} \textbf{\bibinfo{volume}{98}},
  \bibinfo{pages}{232501} (\bibinfo{year}{2007}), \eprint{hep-ph/0612165}.

\bibitem{Gehman:2007qg}
\bibinfo{author}{\bibfnamefont{V.~M.} \bibnamefont{Gehman}} \bibnamefont{and}
  \bibinfo{author}{\bibfnamefont{S.~R.} \bibnamefont{Elliott}},
  \bibinfo{journal}{J. Phys.} \textbf{\bibinfo{volume}{G34}},
  \bibinfo{pages}{667} (\bibinfo{year}{2007}), \eprint{hep-ph/0701099}.

\bibitem{Deppisch:2012nb}
\bibinfo{author}{\bibfnamefont{F.~F.} \bibnamefont{Deppisch}},
  \bibinfo{author}{\bibfnamefont{M.}~\bibnamefont{Hirsch}}, \bibnamefont{and}
  \bibinfo{author}{\bibfnamefont{H.}~\bibnamefont{Pas}}, \bibinfo{journal}{J.
  Phys.} \textbf{\bibinfo{volume}{G39}}, \bibinfo{pages}{124007}
  (\bibinfo{year}{2012}), \eprint{1208.0727}.

\bibitem{Arnold:2010tu}
\bibinfo{author}{\bibfnamefont{R.}~\bibnamefont{Arnold}} \emph{et~al.}
  (\bibinfo{collaboration}{SuperNEMO Collaboration}), \bibinfo{journal}{Eur.
  Phys. J.} \textbf{\bibinfo{volume}{C70}}, \bibinfo{pages}{927}
  (\bibinfo{year}{2010}), \eprint{1005.1241}.

\bibitem{Dev:2013vxa}
\bibinfo{author}{\bibfnamefont{P.~S.} \bibnamefont{Bhupal~Dev}},
  \bibinfo{author}{\bibfnamefont{S.}~\bibnamefont{Goswami}},
  \bibinfo{author}{\bibfnamefont{M.}~\bibnamefont{Mitra}}, \bibnamefont{and}
  \bibinfo{author}{\bibfnamefont{W.}~\bibnamefont{Rodejohann}}
  (\bibinfo{year}{2013}), \eprint{1305.0056}.

\bibitem{Rodriguez:2010mn}
\bibinfo{author}{\bibfnamefont{T.~R.} \bibnamefont{Rodriguez}}
  \bibnamefont{and}
  \bibinfo{author}{\bibfnamefont{G.}~\bibnamefont{Martinez-Pinedo}},
  \bibinfo{journal}{Phys. Rev. Lett.} \textbf{\bibinfo{volume}{105}},
  \bibinfo{pages}{252503} (\bibinfo{year}{2010}), \eprint{1008.5260}.

\bibitem{Menendez:2008jp}
\bibinfo{author}{\bibfnamefont{J.}~\bibnamefont{Menendez}},
  \bibinfo{author}{\bibfnamefont{A.}~\bibnamefont{Poves}},
  \bibinfo{author}{\bibfnamefont{E.}~\bibnamefont{Caurier}}, \bibnamefont{and}
  \bibinfo{author}{\bibfnamefont{F.}~\bibnamefont{Nowacki}},
  \bibinfo{journal}{Nucl. Phys.} \textbf{\bibinfo{volume}{A818}},
  \bibinfo{pages}{139} (\bibinfo{year}{2009}), \eprint{0801.3760}.

\bibitem{Barea:2013bz}
\bibinfo{author}{\bibfnamefont{J.}~\bibnamefont{Barea}},
  \bibinfo{author}{\bibfnamefont{J.}~\bibnamefont{Kotila}}, \bibnamefont{and}
  \bibinfo{author}{\bibfnamefont{F.}~\bibnamefont{Iachello}},
  \bibinfo{journal}{Phys. Rev.} \textbf{\bibinfo{volume}{C87}},
  \bibinfo{pages}{014315} (\bibinfo{year}{2013}), \eprint{1301.4203}.

\bibitem{Suhonen:2010zzc}
\bibinfo{author}{\bibfnamefont{J.}~\bibnamefont{Suhonen}} \bibnamefont{and}
  \bibinfo{author}{\bibfnamefont{O.}~\bibnamefont{Civitarese}},
  \bibinfo{journal}{Nucl. Phys.} \textbf{\bibinfo{volume}{A847}},
  \bibinfo{pages}{207} (\bibinfo{year}{2010}).

\bibitem{Meroni:2012qf}
\bibinfo{author}{\bibfnamefont{A.}~\bibnamefont{Meroni}},
  \bibinfo{author}{\bibfnamefont{S.~T.} \bibnamefont{Petcov}},
  \bibnamefont{and} \bibinfo{author}{\bibfnamefont{F.}~\bibnamefont{Simkovic}},
  \bibinfo{journal}{JHEP} \textbf{\bibinfo{volume}{1302}}, \bibinfo{pages}{025}
  (\bibinfo{year}{2013}), \eprint{1212.1331}.

\bibitem{Simkovic:2013qiy}
\bibinfo{author}{\bibfnamefont{F.}~\bibnamefont{Simkovic}},
  \bibinfo{author}{\bibfnamefont{V.}~\bibnamefont{Rodin}},
  \bibinfo{author}{\bibfnamefont{A.}~\bibnamefont{Faessler}}, \bibnamefont{and}
  \bibinfo{author}{\bibfnamefont{P.}~\bibnamefont{Vogel}},
  \bibinfo{journal}{Phys. Rev.} \textbf{\bibinfo{volume}{C87}},
  \bibinfo{pages}{045501} (\bibinfo{year}{2013}), \eprint{1302.1509}.

\bibitem{Mustonen:2013zu}
\bibinfo{author}{\bibfnamefont{M.~T.} \bibnamefont{Mustonen}} \bibnamefont{and}
  \bibinfo{author}{\bibfnamefont{J.}~\bibnamefont{Engel}}
  (\bibinfo{year}{2013}), \eprint{1301.6997}.

\bibitem{Barabash:2011mf}
\bibinfo{author}{\bibfnamefont{A.~S.} \bibnamefont{Barabash}},
  \bibinfo{journal}{Phys. Atom. Nucl.} \textbf{\bibinfo{volume}{74}},
  \bibinfo{pages}{603} (\bibinfo{year}{2011}), \eprint{1104.2714}.

\bibitem{Gando:2012zm}
\bibinfo{author}{\bibfnamefont{A.}~\bibnamefont{Gando}} \emph{et~al.}
  (\bibinfo{collaboration}{KamLAND-Zen Collaboration}), \bibinfo{journal}{Phys.
  Rev. Lett.} \textbf{\bibinfo{volume}{110}}, \bibinfo{pages}{062502}
  (\bibinfo{year}{2013}), \eprint{1211.3863}.

\bibitem{KlapdorKleingrothaus:2000sn}
\bibinfo{author}{\bibfnamefont{H.~V.} \bibnamefont{Klapdor-Kleingrothaus}},
  \bibinfo{author}{\bibfnamefont{A.}~\bibnamefont{Dietz}},
  \bibinfo{author}{\bibfnamefont{L.}~\bibnamefont{Baudis}},
  \bibinfo{author}{\bibfnamefont{G.}~\bibnamefont{Heusser}},
  \bibinfo{author}{\bibfnamefont{I.~V.} \bibnamefont{Krivosheina}},
  \emph{et~al.}, \bibinfo{journal}{Eur. Phys. J.}
  \textbf{\bibinfo{volume}{A12}}, \bibinfo{pages}{147} (\bibinfo{year}{2001}),
  \eprint{hep-ph/0103062}.

\bibitem{Aalseth:2002rf}
\bibinfo{author}{\bibfnamefont{C.~E.} \bibnamefont{Aalseth}} \emph{et~al.}
  (\bibinfo{collaboration}{IGEX Collaboration}), \bibinfo{journal}{Phys. Rev.}
  \textbf{\bibinfo{volume}{D65}}, \bibinfo{pages}{092007}
  (\bibinfo{year}{2002}), \eprint{hep-ex/0202026}.

\bibitem{Simard:2012gb}
\bibinfo{author}{\bibfnamefont{L.}~\bibnamefont{Simard}}
  (\bibinfo{collaboration}{NEMO-3 Collaboration}), \bibinfo{journal}{J. Phys.
  Conf. Ser.} \textbf{\bibinfo{volume}{375}}, \bibinfo{pages}{042011}
  (\bibinfo{year}{2012}).

\bibitem{Andreotti:2010vj}
\bibinfo{author}{\bibfnamefont{E.}~\bibnamefont{Andreotti}},
  \bibinfo{author}{\bibfnamefont{C.}~\bibnamefont{Arnaboldi}},
  \bibinfo{author}{\bibfnamefont{F.~T.} \bibnamefont{Avignone}},
  \bibinfo{author}{\bibfnamefont{M.}~\bibnamefont{Balata}},
  \bibinfo{author}{\bibfnamefont{I.}~\bibnamefont{Bandac}}, \emph{et~al.},
  \bibinfo{journal}{Astropart. Phys.} \textbf{\bibinfo{volume}{34}},
  \bibinfo{pages}{822} (\bibinfo{year}{2011}), \eprint{1012.3266}.

\bibitem{Gaitskell:2003zr}
\bibinfo{author}{\bibfnamefont{R.}~\bibnamefont{Gaitskell}} \emph{et~al.}
  (\bibinfo{collaboration}{Majorana Collaboration})  (\bibinfo{year}{2003}),
  \eprint{nucl-ex/0311013}.

\bibitem{Guiseppe:2008aa}
\bibinfo{author}{\bibfnamefont{V.~E.} \bibnamefont{Guiseppe}} \emph{et~al.}
  (\bibinfo{collaboration}{Majorana Collaboration}), \bibinfo{journal}{IEEE
  Nucl. Sci. Symp. Conf. Rec.} \textbf{\bibinfo{volume}{2008}},
  \bibinfo{pages}{1793} (\bibinfo{year}{2008}), \eprint{0811.2446}.

\bibitem{Barabash:2002ps}
\bibinfo{author}{\bibfnamefont{A.~S.} \bibnamefont{Barabash}}
  (\bibinfo{collaboration}{NEMO Collaboration}), \bibinfo{journal}{Czech. J.
  Phys.} \textbf{\bibinfo{volume}{52}}, \bibinfo{pages}{575}
  (\bibinfo{year}{2002}).

\bibitem{Chauveau:2009zz}
\bibinfo{author}{\bibfnamefont{E.}~\bibnamefont{Chauveau}}
  (\bibinfo{collaboration}{SuperNEMO Collaboration}), \bibinfo{journal}{AIP
  Conf. Proc.} \textbf{\bibinfo{volume}{1180}}, \bibinfo{pages}{26}
  (\bibinfo{year}{2009}).

\bibitem{Arnaboldi:2002du}
\bibinfo{author}{\bibfnamefont{C.}~\bibnamefont{Arnaboldi}} \emph{et~al.}
  (\bibinfo{collaboration}{CUORE Collaboration}), \bibinfo{journal}{Nucl.
  Instrum. Meth.} \textbf{\bibinfo{volume}{A518}}, \bibinfo{pages}{775}
  (\bibinfo{year}{2004}), \eprint{hep-ex/0212053}.

\bibitem{Bandac:2008zz}
\bibinfo{author}{\bibfnamefont{I.~C.} \bibnamefont{Bandac}}
  (\bibinfo{collaboration}{CUORE Collaboration}), \bibinfo{journal}{J. Phys.
  Conf. Ser.} \textbf{\bibinfo{volume}{110}}, \bibinfo{pages}{082001}
  (\bibinfo{year}{2008}).

\bibitem{Danilov:2000pp}
\bibinfo{author}{\bibfnamefont{M.}~\bibnamefont{Danilov}},
  \bibinfo{author}{\bibfnamefont{R.}~\bibnamefont{DeVoe}},
  \bibinfo{author}{\bibfnamefont{A.}~\bibnamefont{Dolgolenko}},
  \bibinfo{author}{\bibfnamefont{G.}~\bibnamefont{Giannini}},
  \bibinfo{author}{\bibfnamefont{G.}~\bibnamefont{Gratta}}, \emph{et~al.},
  \bibinfo{journal}{Phys. Lett.} \textbf{\bibinfo{volume}{B480}},
  \bibinfo{pages}{12} (\bibinfo{year}{2000}), \eprint{hep-ex/0002003}.

\bibitem{Gornea:2009zz}
\bibinfo{author}{\bibfnamefont{R.}~\bibnamefont{Gornea}}
  (\bibinfo{collaboration}{EXO Collaboration}), \bibinfo{journal}{J. Phys.
  Conf. Ser.} \textbf{\bibinfo{volume}{179}}, \bibinfo{pages}{012004}
  (\bibinfo{year}{2009}).

\end{thebibliography}

\end{document}